\documentstyle[12pt]{article}

\makeatletter
\def\numberbysection{\@addtoreset{equation}{section}
	\def\theequation{\thesection.\arabic{equation}}}
\makeatother

\newcommand{\algg}{{\rm \bf g}}
\newcommand{\algt}{{\rm \bf t}}
\newcommand{\algb}{{\rm \bf b}}
\newcommand{\algh}{{\rm \bf h}}
\newcommand{\algm}{{\rm \bf m}}
\newcommand{\ZZ}{{\rm \bf Z}}
\newcommand{\CC}{{\rm \bf C}}
\newcommand{\RR}{{\rm \bf R}}
\newcommand{\alggs}{{\rm \bf g}^*}

\newcommand{\cd}{{\cal D}}

\newcommand{\mysection}{\setcounter{equation}{0}\section}
\renewcommand{\theequation}{\thesection.\arabic{equation}}

\newcommand{\bge}{\begin{equation}}
\newcommand{\ee}{\end{equation}}
\newcommand{\sss}{\scriptscriptstyle}
\newcommand{\pdv}{\partial}
\newcommand{\wb}{W_b}
\newcommand{\wbc}{W_{b,c}}
\newcommand{\bc}{(b, ic)}
\newcommand{\hb}{{\cal H}_b}
\newcommand{\di}{{\rm Diff}S^1 / S^1}
\newcommand{\dis}{{\rm Diff}S^1 / SL^{(n)}(2,\RR)}
\newcommand{\dif}{{\rm Diff}S^1}
\newcommand{\vv}{{\rm V}}
\newcommand{\dio}{{\rm Diff}_0 S^1}
\newcommand{\vvec}{\widehat{{\rm Vect}S^1}}
\newcommand{\vc}{{\rm Vect}S^1}

\newcommand{\lb}{{\cal L}_b}
\newcommand{\lbc}{{\cal L}_{b,c}}
\newcommand{\vir}{\widehat{{\rm Diff}S^1}}
\newcommand{\kl}{K\"{a}hler }
\newcommand{\aads}{{\rm Ad}^*}
\newcommand{\ad}{{\rm ad}}
\newcommand{\ads}{{\rm ad}^*}
\newcommand{\aadsg}{{\rm Ad}^*_g}
\newcommand{\aad}{{\rm Ad}}
\newcommand{\aadg}{{\rm Ad}_g}
\newcommand{\tu}{\tilde{u}}
\newcommand{\tv}{\tilde{v}}
\newcommand{\hu}{\hat{u}}
\newcommand{\hv}{\hat{v}}
\newcommand{\hj}{\hat{J}}

\newcommand{\hl}{\hat{L}}
\newcommand{\pu}{\Phi_u}

\newcommand{\xu}{\xi_u}
\newcommand{\xn}{\xi_a}
\newcommand{\bb}{{\bf b}}
\newcommand{\bbo}{{\bf b}}
\newcommand{\bo}{b}
\newcommand{\eps}{\epsilon}
\newcommand{\azz}{a(z, \bar{z})}
\newcommand{\bzz}{\beta(z, \bar{z})}

\newcommand{\mzz}{\mu_n(z, \bar{z})}
\newcommand{\rzz}{\rho(z, \bar{z})}
\newcommand{\gzz}{\gamma(z, \bar{z})}
\newcommand{\co}{{\cal O}}
\newcommand{\nul}{|\:\rangle}

\newcommand{\mixten}[3]{{#1}_{#2}^{\phantom{{#2}} #3}}

\newcommand{\jj}[3]{#1_{#2 (#3)}}
\newcommand{\ip}[2]{\langle#1,#2\rangle}
\newcommand{\derivative}[1]{\frac{\partial}{\partial #1}}
\newcommand{\derivetwo}[1]{\frac{\partial^2}{\partial #1^2}}

\newtheorem{prop}{Proposition}
\newtheorem{corollary}{Corollary}

\newcommand{\nuls}{\langle \: |}

\def\committee{Committee in charge:\par
    \list{}{\itemindent 0.25in \let\makelabel\relax}}

\begin{document}

\pagestyle{empty}

%
%
\begin{titlepage}
%
%
 \begin{flushright}
{LBL-34507}
 \end{flushright}

\begin{center}
\begin{large}			
Coadjoint Orbits \\
and Conformal Field Theory \\[\stretch{0.75}]
\end{large}
by\\[\stretch{0.75}]		
Washington Taylor IV
\end{center}
\begin{center}
\vspace{\stretch{1.0}}		
B. A.  (Stanford University) 1986 \\
M. S.  (University of California at Berkeley) 1990 \\
\vspace{\stretch{1.0}}		
%
%
\renewcommand{\baselinestretch}{2.0}\large\normalsize
A thesis submitted in partial satisfaction of the\\
requirements for the degree Doctor of Philosophy\\
in Physics\\
in the \\
{\large GRADUATE DIVISION \\}
of the \\
{\large UNIVERSITY OF CALIFORNIA \\ BERKELEY \par}
\end{center}
\vspace{\stretch{1.0}}		
\begin{committee}
\item Professor Orlando Alvarez, Chair
\item Professor Bruno Zumino
\item Professor Vaughan Jones
\end{committee}
\vspace{\stretch{0.25}}		
\begin{center}
{\large 1993 \par}		
\end{center}
\end{titlepage}
\cleardoublepage

\pagenumbering{arabic}
\pagestyle{plain}
\begin{center}
\begin{large}
Coadjoint Orbits and Conformal Field Theory
\end{large}
\end{center}

\begin{center}
by \\
Washington Taylor IV \\
Doctor of Philosophy in Physics \\
University of California at Berkeley \\
Professor Orlando Alvarez, Chair
\end{center}

\begin{bf}
\begin{center}
Abstract
\end{center}
\end{bf}


This thesis describes a new approach to conformal field theory.  This
approach combines the method of coadjoint orbits with the algebraic
structures of resolutions and chiral vertex operators to give a
construction of the correlation functions of conformal field theories
in terms of geometrically defined objects.  The coadjoint orbit method
is used to construct representations of the Virasoro and affine
algebras.  Explicit formulae are given for these representations in
terms of a local gauge choice on the line bundle associated with
geometric quantization of a given coadjoint orbit; these formulae
define a new set of explicit bosonic realizations of the Virasoro
algebra and affine algebras.  Unlike the Feigin-Fuchs and Wakimoto
realizations, which correspond to ``infinitely twisted'' Verma
modules, the coadjoint orbit realizations take the form of dual Verma
modules.  Thus, the realizations described here can be used to
construct irreducible Virasoro and affine algebra representations with
one-sided resolutions.  This makes it possible to avoid the technical
difficulties associated with the two-sided resolutions which arise
{}from Feigin-Fuchs and Wakimoto realizations.  Formulae are given for
screening and intertwining operators on the coadjoint orbit
representations.  Chiral vertex operators between Virasoro modules are
constructed, and related directly to Virasoro algebra generators in
certain cases.  From the point of view taken in this thesis, vertex
operators have a geometric interpretation as differential operators
taking sections of one line bundle to sections of another.  It is
shown that vertex operators acting on the coadjoint orbit
representations can be characterized by an infinite set of equations.
However, this approach is not yet practical for the computation of
interesting correlation functions in conformal field theory due to the
lack of a closed-form expression for the vertex operators.  A
suggestion is made that by connecting this description with recent
work deriving field theory actions from coadjoint orbits, a deeper
understanding of the geometry of conformal field theory might be
achieved.

\cleardoublepage

\pagenumbering{roman}
\pagestyle{plain}
\setcounter{page}{3}

\Large

\noindent {\bf Acknowledgments}
\vspace*{.2in}

\normalsize
I would like to thank my advisor, Orlando Alvarez, who initiated my
interest in the questions addressed in this thesis, who was always
happy to answer my questions, some difficult and many silly, who
helped me to separate the good ideas from the dead ends, and who gave
me the freedom and encouragement to pursue whatever problems and ideas
I found interesting.

Thanks also to Korkut Bardak\c{c}i, Dan Freed, Vaughan Jones, Nolan
Wallach, and Bruno Zumino for helpful conversations about aspects of
the subject matter of this thesis.

This work was supported in part by the Director, Office of Energy
Research, Office of High Energy and Nuclear Physics, Division of High
Energy Physics of the U.S. Department of Energy under Contract
DE-AC03-76SF00098, in part by the National Science Foundation under
grant PHY90-21139, and in part by a Department of Education
fellowship.  I thank the University of California, Berkeley and
Lawrence Berkeley Laboratories for their assistance.

\vspace{.4in}

Thanks to Aram Antaramian, Karyn Apfeldorf, Anton Kast, and Eve
Ostriker for many interesting discussions about physics and life in
general.

Thanks also to the many professors, postdocs, and other graduate
students who helped make life here interesting and enjoyable.  In
particular, thanks to Eugene Commins, Marty Halpern, J. D. Jackson,
Rich Lebed, Robert Littlejohn, Harry Morrison, Dan Rokhsar, Alexander
Sevrin, and Paul Watts.

I would also like to thank Bruce Boghosian and David Gross for
involving me in other interesting projects and helping to keep me
intellectually stimulated.

Throughout my time in Berkeley, there were many people who helped with
countless logistical, bureaucratic, and academic matters.  In
particular, I would like to express thanks to Sandy Ewing, Mary
Gorman, Ken Miller, Betty Moura, Luanne Neumann, Donna Sakima, and
Anne Takizawa, who were never too busy to help or chat.

A special expression of gratitude is due to the people at Dragon
Systems, who developed the speech recognition system with which this
thesis was composed.  Without their software, this thesis could never
have been written, and I would now be entering a new career in the
field of grape-crushing.  Thanks also to the people at the Free
Software Foundation, for developing an arbitrarily powerful virtual
environment (emacs) which I could easily modify to my own eclectic
standards.

Thanks to my parents Vivian Taylor and Washington Taylor III, and
early teachers Kevin Sullivan, Jim Wisdom, and Kathy Rehfield, who
helped me to learn to pursue the things which interested me, and to
have faith in my own point of view.

Finally, no thanks are enough for Kara Swanson, now my wife, who
provided me with a constant point of reference, support, and
distraction through the experience of graduate school.
\cleardoublepage

\tableofcontents

\listoffigures

\listoftables
\cleardoublepage

\pagenumbering{arabic}
\pagestyle{headings}

\mysection{Introduction}

\subsection{Results}

This thesis is primarily a study of certain aspects of the geometric
and algebraic structure of coadjoint orbit representations of
infinite-dimensional Lie groups.  The goal of this work is to use
coadjoint orbit representations to construct conformal field theories,
in a fashion analogous to the free-field constructions of conformal
field theories.

The new results which are presented in this thesis are as follows:
First, an explicit set of formulae are derived giving an algebraic
realization of coadjoint orbit representations in terms of
differential operators acting on a polynomial Fock space.  These
representations are equivalent to dual Verma module representations.
Next, intertwiners are explicitly constructed which allow the
construction of resolutions for irreducible representations using
these Fock space realizations.  Finally, vertex operators between
these irreducible representations are explicitly constructed as chain
maps between the resolutions; these vertex operators allow the
construction of rational conformal field theories according to an
algebraic prescription.

{}From the point of view presented in this thesis, the space of states
associated with each primary field of a conformal field theory is
described by a Hilbert space of holomorphic sections of a certain line
bundle over a complex homogeneous space.  The vertex operators are
simply differential operators taking sections of one line bundle to
sections of another line bundle.  For conformal field theories with a
simple Virasoro symmetry, there is a BRST complex of line bundles
connected by differential operators for each primary state, and
physical states are those states in the BRST cohomology of this
complex.  In the case of conformal field theories with affine algebra
symmetries such as the WZW model, the restriction to holomorphic
sections of the line bundles automatically restricts to the space of
physical states, so no BRST complex is necessary.

The infinite-dimensional groups which are studied in this thesis are
the centrally extended loop groups $\widehat{L}G$ where $G$ is a
compact simple finite-dimensional Lie group, and the Virasoro group
$\vir$.  In the case of the loop groups, the coadjoint orbit
construction is effectively equivalent to the Borel-Weil theory
describing irreducible representations, and is already fairly well
understood \cite{PS}.  For the Virasoro group, on the other hand,
there is not yet a complete understanding of the structure of
coadjoint orbit representations; the results presented here are a
modest step towards such an understanding.  For all these
infinite-dimensional groups, the explicit realization in terms of
differential operators given here is new.

The construction of rational conformal field theories using
resolutions and vertex operators on dual Verma module representations
is algebraically very similar to the related construction using free
field theories and Feigin-Fuchs or Wakimoto modules \cite{Feld,bmp}.
The approach developed here has several advantages over the free field
construction.  Because of the geometric nature of the
coadjoint orbit construction, the dual Verma module resolutions are
more naturally geometrically motivated.  Additionally, the
resolutions in terms of coadjoint orbit representations are one-sided,
and therefore avoid some of the complications associated with the
two-sided resolutions arising from Feigin-Fuchs and Wakimoto modules.
Finally, the known relationship between coadjoint orbits and actions
for conformal field theories may indicate that the vertex operators
described here naturally arise in some unifying geometric approach to
conformal field theory.  On the other hand, some of the advantages of
the free field realizations are absent in the coadjoint orbit
construction.  The most significant of these features is the field
theory interpretation of the free field construction.  As yet, we have
no analogous field theoretic interpretation for the coadjoint orbit
representations, which presents a major obstacle to the physical
interpretation of this construction.  Also, the free field
construction is based on well-studied techniques from string theory,
which lead to powerful computational methods for the conformal field
theory correlation functions.  The construction presented here in
terms of coadjoint orbit representations is not yet sufficiently well
developed to allow the computation of correlation functions of any
significant complexity.

\subsection{Background}

The method of coadjoint orbits was originated by Kirillov and Kostant
twenty years ago \cite{Kir}.  This approach has proven to be a
valuable tool in investigating geometrical aspects of the
representation theory of Lie groups.  The Kirillov-Kostant approach is
essentially a generalization of the Borel-Weil theorem, which
constructs irreducible unitary representations of a finite-dimensional
compact semi-simple Lie group $G$ as spaces of holomorphic sections of
complex line bundles over the homogeneous space $G/T$, where $T$ is a
maximal subtorus of $G$.  In the coadjoint orbit approach, one begins
with a group $G$, with Lie algebra ${\algg}$.  The group $G$ has a
natural coadjoint action on the dual space $\algg^*$.  Choosing an
element $b$ in $\algg^*$, one considers the coadjoint orbit $W_b$ of $b$
in $\algg^*$.  For any $b$, the space $W_b$ has a natural symplectic
form $\omega$.  For those $b$ with the property that a complex line
bundle ${\cal L}_b$ can be constructed over $W_b$ with curvature form
$i \omega$, one attempts to relate an appropriate space of sections of
${\cal L}_b$ to an irreducible unitary representation of $G$ by using
the technique of geometric quantization on the space $W_b$.  For
finite-dimensional compact semi-simple $G$, the representations
produced by this construction are equivalent to those given by the
Borel-Weil theory.  The coadjoint orbit approach is particularly
useful in the case of non-compact groups, where the Borel-Weil theory
does not apply.  It is possible to apply the Borel-Weil approach to
certain infinite-dimensional groups such as the centrally extended
loop groups $\widehat{L}G$ \cite{PS}.  For other infinite-dimensional
groups, such as the (orientation-preserving) diffeomorphism group of
the circle ${\rm Diff} S^1$, and its central extension $\vir$, the
Virasoro group, there are difficulties with applying even the more
general coadjoint orbit theory.  Many of the Virasoro coadjoint orbits
do not admit a \kl structure, so that it is difficult to geometrically
quantize these spaces.  Also, it is known that the Virasoro group has
rather peculiar mathematical properties, such as the fact that the
exponential map on the Lie algebra is neither onto nor 1-1 in the
vicinity of the identity.  Due to these difficulties, a full
understanding of the coadjoint orbit representations for this group
has not yet been attained, although there are some partial results in
this direction \cite{k1,lp,segal,Witt1}.  Achieving a full
understanding of the geometry of the coadjoint orbit representations
of the Virasoro group could be a valuable step in the general study of
Virasoro representations and conformal field theory.

A more direct relationship between irreducible representations of
infinite-dimensional Lie groups and conformal field theory (CFT)
is embodied in the free field approach to CFT's.   The free field
approach was fundamental in early developments in string theory
\cite{gsw}.  It was eventually shown that
the operators associated with free fields acting on a bosonic Fock
space could be used to construct all irreducible unitary
representations of the Virasoro algebra.  The ``Feigen-Fuchs'' free
field representations were described in the work of Dotsenko and
Fateev \cite{DF} using a Coulomb gas-like free field theory with
background charge.  In this work, these representations were used to
calculate correlation functions in Virasoro minimal models.  The
structure of the Fock space in the Feigen-Fuchs representations was
originally described in \cite{FF}.  It was subsequently shown by
Felder \cite{Feld} that the irreducible representations in the $c \leq
1$ discrete series could be described in terms of the Feigin-Fuchs
representations using a BRST-type screening operator which had
previously been introduced by Thorn \cite{Thorn}.  A similar
construction for conformal field theories with affine algebra
symmetries has also been carried out \cite{bf,gmmos,w}.  The resulting
free field representations of the affine algebras are known as
``Wakimoto'' modules.  Recently, the free field approach to conformal
field theory has been described in the algebraic language of modules
and resolutions, and put into a systematic algebraic formalism.  From
this point of view, the algebraic structure of the Feigin-Fuchs
representations is similar to that of a twisted Verma module.  For a
review of this approach, see \cite{bmp}.

\subsection{Summary}

The structure of this thesis is as follows.  In Chapter 2, we review
the coadjoint orbit description of irreducible representations of Lie
groups, and construct an explicit algebraic realization of the Lie
algebra of an arbitrary Lie group by taking local coordinates on a
coadjoint orbit and performing a gauge fixing on the complex line
bundle over that coadjoint orbit.  This construction gives a set of
representations of the Lie algebra which are expressed in terms of
first-order differential operators acting on a polynomial Fock space.
These representations are studied in detail for finite-dimensional Lie
groups, loop groups, and the Virasoro group, and examples of explicit
realizations are given for each of these types of groups.  By
considering global aspects of these representations, we show that in
the cases of finite-dimensional groups and loop groups this
construction gives rise to irreducible unitary representations, and
that for such groups all irreducible unitary representations can be
constructed using this method.  In the case of the Virasoro group,
these representations are reducible representations, with irreducible
subspaces corresponding to the irreducible unitary representations of
the Virasoro group.  In Chapter 3, we consider in more detail the
module structure of these representations.  We show that all the
coadjoint orbit representations we have constructed are locally
equivalent to dual Verma modules, and describe resolutions for
irreducible representations in terms of explicitly defined formulae
for intertwiners between coadjoint orbit representations.  We go on in
Chapter 4 to describe vertex operators which allow these irreducible
representations to be combined into rational conformal field theories.
In Chapter 5, we review our results, and discuss open questions and
relationships between this research and other recent work.  In
particular, we describe briefly a recently active area of research in
which coadjoint orbits are used to construct conformal field theory
actions.  We speculate that there may be an underlying connection
between the structure of these actions and the conformal field theory
formalism described in this thesis.

\mysection{Coadjoint Orbit Representations}

In this chapter, the geometric and algebraic structures of coadjoint
orbit representations are presented and analyzed.  In Section 2.1, we
describe the class of groups to which the analysis in this chapter is
applicable and establish notation for Lie groups and algebras which
will be used throughout this thesis.  For each of the groups in this
section, a description is given of a local coordinate system near the
identity on a homogeneous quotient space of the group which will be
later identified with a coadjoint orbit space.  In Section 2.2, we
review the geometric construction of coadjoint orbit representations
as sections of line bundles ${\cal L}_b $ over the coadjoint orbit
spaces associated with elements $b$ in the dual of the Lie algebra.
In Section 2.3, we prove a pair of general propositions in which
sufficient conditions are given for a set of functions to describe a
local gauge-fixed connection on the line bundle ${\cal L}_b $.  These
propositions are used in Section 2.4 to construct formulae for the
coadjoint orbit representations of the group algebra in terms of
differential operators on the space of locally holomorphic functions
at the point $b$ in the orbit space.  In Section 2.5, we give examples
of these realizations for the groups $SU(2), SU(3),
\widehat{L} SU(2),$ and the Virasoro group.  In Section 2.6, we
discuss global questions about these representations; in particular we
consider the question of which sections of the line bundle ${\cal L}_b
$ are globally holomorphic, and we discuss the existence of Hermitian
structures on the line bundles ${\cal L}_b$.  Both of these questions
are addressed from the point of view of the local formulae derived in
the earlier sections.

\subsection{Lie groups and algebras}
\label{sec:groups}

There are three main types of Lie groups which, along with their
associated Lie algebras, will be of interest in this thesis.  The
compact simple Lie groups are the most familiar and best understood
examples in the theory of continuous groups and their representations.
We will use these groups as a class of elementary examples and as a
reference point with which to compare our results for
infinite-dimensional groups.  The centrally-extended loop groups,
whose associated Lie algebras are affine algebras, are the second
class of groups which we will study.  Although not as well understood
as the finite-dimensional simple Lie groups, these groups share many
of the properties of groups in that class.  The affine algebras are
important in physics, where they appear as current algebras in
two-dimensional field theories and provide a primary tool for the
understanding of the algebraic structure of a large class of conformal
field theories.  Finally, we will study the representations of the
Virasoro group, which is of central importance in the physics of
conformal field theories.  This group is even less understood from a
geometric perspective than the loop groups -- hopefully some of the
work in this thesis will help to provide a basis for a deeper
understanding of the geometric structure of the representations of
this group.

In this section, we define these three types of Lie groups and develop
the notation which will be used throughout the thesis.  We also derive
or state some important theorems about these groups which will be used
in later developments.  In particular, we give a description in terms
of local coordinates of a natural quotient space of each Lie group by
a maximal abelian subgroup.  In the case of the Virasoro group, a
rigorous proof of the validity of this coordinate system is not given;
however, arguments are given for the plausibility of this coordinate
system in a well-defined mathematical context.  General references for
the background material in this section are
\cite{Adams,Humphreys,Kac,Knapp,Milnor,PS}.

\subsubsection{Compact simple Lie groups}
\label{sec:compactgroups}

A finite-dimensional Lie group $G$ is a manifold of finite dimension
which has a group structure
\begin{equation}
\cdot:G\times G \rightarrow G.
\end{equation}
In this subsection, we assume that all Lie groups are
finite-dimensional.  Associated with every Lie group is an algebraic
structure on the tangent space $\algg$ to $G$ at the identity.  This
algebra, called the Lie algebra of $G$, is defined by the product
\begin{equation}
[\xi, \eta] = \lim_{s,t \rightarrow 0} \frac{1}{st}
\{{\rm e}^{s \xi} {\rm e}^{t \eta} {\rm e}^{- s \xi} {\rm e}^{- t \eta} -1\},
\end{equation}
where the exponential map ${\rm e}: \algg \rightarrow G$ is defined by
taking ${\rm e}^{t \xi}$ to be a one-parameter subgroup of $G$ with
${\rm d} {\rm e}^{t \xi}/{\rm d}t = \lim_{t \rightarrow 0}{\rm e}^{t
\xi}/t = \xi$.  When $G$ is
finite-dimensional, it can be shown that this exponential map exists
and is locally 1-1.  The Lie algebra product is antisymmetric and
satisfies the Jacobi identity
\begin{equation}
[\xi,[\eta, \zeta]] +[\eta,[\zeta, \xi]] +[\zeta,[\xi, \eta]] = 0.
\end{equation}

A Lie group is compact if its underlying manifold is compact.  A Lie
group is complex if its manifold is a complex manifold and the group
composition law is holomorphic.  A Lie algebra is complex if the
vector space is complex and the Lie algebra product is
complex-bilinear.  The Lie algebra associated with a complex Lie group
is a complex Lie algebra.  A complexification $G_{\CC}$ of a Lie group
$G$ is a complex Lie group whose Lie algebra is the complexification
$\algg_{\CC}= \algg \otimes_{\RR} \CC$ of the Lie algebra $\algg$ of
$G$.  When $G$ is a compact group, such a complexification always
exists \cite{PS}.

A simple Lie group is a nonabelian Lie group with no continuous normal
subgroups.  It follows that a simple Lie algebra is a nonabelian Lie
algebra with no nontrivial ideals.  The compact simple Lie groups have
been classified, and consist of the groups $SU(N)$, $SO(N)$, $Sp (N)$,
the exceptional groups $G_2, F_4, E_6, E_7, E_8$, and groups which are
related to any of these groups by taking a quotient through a finite
normal subgroup \cite{Humphreys}.
For the remainder of this subsection, we assume that all Lie groups are
compact and simple.  Most of the results for simple groups can be
easily generalized to the class of semisimple Lie groups, which are
groups locally isomorphic to a product of simple groups (with no
abelian factors); we will not bother, however, to explicitly discuss
these generalizations, in order to keep the presentation relatively simple.

A representation of a Lie group $G$ (Lie algebra $\algg$) is a linear
action of $G$ ($\algg$) on a vector space $V$.  A representation is
irreducible if $V$ does not contain a nontrivial subspace $W$ which is
mapped into itself by $G$ ($\algg$).  It is a theorem that all
irreducible representations of compact Lie groups are
finite-dimensional \cite{PS}.

The group $G$ has a natural representation on $\algg$ known as the
{\em adjoint action}; for each $g \in G$
\begin{equation}
	{\rm Ad}_g: \algg \rightarrow \algg,
\end{equation}
where
\begin{equation}
	{\rm Ad}_g:u \mapsto  \frac{{\rm d}}{{\rm d} t}
		\{g {\rm e}^{tu} g^{-1}-1\}.
\end{equation}
There is a related {\em coadjoint action} of $G$ on the dual space to
$\algg$, $\algg^*$.  The coadjoint action is denoted by ${\rm Ad}^*$,
and given by
\bge \ip{{\rm Ad}_g^* b}{ u } = \ip{b}{ {\rm Ad}_{g^{-1}}
u }, \;\;\;{\rm for}\; b \in \algg^*, u \in \algg.
\ee
The derivative of the adjoint action gives an action of
$\algg$ on $\algg$, denoted by $\ad$, where $\ad_u v = [u,v]$, for all $u,
v \in \algg$.  Similarly, the infinitesimal coadjoint action of $\algg$ on
$\alggs$ is denoted $\ads$, and is given by
\bge
\ip{\ads_v b}{ u}  = \ip{b}{ [u,v] } , \;\;\;{\rm
for}\; b \in \algg^*, u,v \in \algg.
\label{eq:coadjointaction}
\ee

In order to study the representations of a Lie group $G$ and its
algebra $\algg$, it is useful to describe a {\em root space
decomposition} of the complexified Lie algebra $\algg_\CC$.  One first
picks a maximal subtorus $T$ of $G$.  (A maximal subtorus of $G$ is a
continuous abelian subgroup of $G$ which is not contained in a
continuous abelian subgroup of higher dimension.)  The
complexification $\algt_\CC$ of the Lie algebra of $T$ is a {\em
Cartan subalgebra} of $\algg_\CC$.  The dimension of $T$ is defined to
be the {\em rank} $r$ of $G$.  All maximal subtori of $G$ are
conjugate \cite{Adams}, so the rank is well-defined.  The adjoint
action of $\algg$ on $\algg$ extends naturally to an action of $\algg$
on $\algg_\CC$.  Since the Lie algebra $\algt$ is abelian, the
restriction of the adjoint action on $\algg_\CC$ to $\algt$ is
simultaneously diagonalizable, and we can write
\begin{equation}
\algg_\CC = \algt_\CC\oplus \bigoplus_{\alpha \in \Phi} \algg_\alpha,
\label{eq:decompositionf}
\end{equation}
where $\Phi  \subset \algt_\CC^*$ is the set of nonzero {\em roots} of $G$
(which we will also refer to as roots of $\algg$), and
\begin{equation}
\algg_\alpha =\{u \in \algg_\CC |
[h,u] = \ip{\alpha}{h} u \;\forall h \in \algt\}.
\end{equation}
An analysis of this root space decomposition reveals that the
root spaces $\algg_\alpha$ are one-dimensional and that $\alpha
\in \Phi\Rightarrow - \alpha \in \Phi$.   It is possible to choose
elements $e_\alpha \in \algg_\alpha$
and $h_\alpha \in \algt_\CC$ such that when $f_\alpha = e_{- \alpha}$,
we have
\begin{eqnarray}
\left[e_\alpha, f_\alpha\right] & = & h_\alpha   \nonumber \\
\left[h_\alpha, e_\alpha\right] & = & 2e_\alpha \label{eq:defineh}\\
\left[h_\alpha, f_\alpha\right] & = & -2 f_\alpha.\nonumber
\end{eqnarray}
In this basis, when $\alpha \neq - \beta$ and $[e_\alpha, e_\beta]\neq
0$, it follows that $[e_\alpha, e_\beta]\subset \algg_{\alpha +
\beta}$.
It is furthermore possible \cite{Humphreys} to choose a {\em
base} $\Delta$ for $\Phi$, where $\Delta
\subset \Phi$ is defined to be a base when $\Delta =\{ \alpha_1,
\ldots, \alpha_r\}$ gives a basis for
the linear space $\algt_\CC^*$ spanned by $\Phi$, and each root $\beta \in
\Phi$ has a (unique) decomposition
\begin{equation}
\beta = \sum_{\alpha \in \Delta} k_\alpha \alpha,
\label{eq:simpledecomposition}
\end{equation}
with $k_\alpha$ integral and with all coefficients $k_\alpha$ being
either nonnegative or nonpositive.  The roots in $\Delta$ are called
{\em simple} roots.  If all the coefficients $k_\alpha$ in the
decomposition (\ref{eq:simpledecomposition}) of a root $\beta$ are
nonnegative, $\beta$ is defined to be a {\em positive} root, and we
write $\beta \succ 0$.  Similarly, if all coefficients are
nonpositive, we define $\beta$ to be a {\em negative} root.  Because
the root spaces are one-dimensional, we can directly associate
generators $e_\alpha$ with roots; thus, we will often refer to
generators as being simple or positive when the associated root has
such a property.  We denote by $\Phi_+=\{\alpha \in \Phi | \alpha
\succ 0\}$ the set of all positive roots.  In physics terminology,
positive and negative roots are often referred to as ``annihilation''
and ``creation'' operators on a representation space, respectively.
The {\em root lattice} $\Lambda$ is defined to be the sublattice of
$\algt^*_{\CC}$ spanned by the simple roots $\Delta$.

We denote the generators associated with simple roots by $e_j =
e_{\alpha_j}$ for $1 \leq j \leq r$; similarly, we define $f_j = e_{-
\alpha_j}$, $h_j = h_{\alpha_j}$.
The generators $\{e_j,h_j,f_j:1
\leq j \leq r\}$ give rise to a complete basis for the Lie algebra
$\algg_\CC$, under the relations
\begin{eqnarray}
\left[e_i,f_i\right] & = &  h_i \nonumber\\
\left[e_j,f_k\right] & = &  0 \; \; \; {\rm when} \; \; \; j
\neq k \nonumber\\
\left[h_i,e_j\right] & = & A_{ij} e_j \label{eq:cartanrelations}\\
\left[h_i, f_j\right] & = & -A_{ij} f_j \nonumber\\
({\rm ad}\; e_i)^{1 - A_{ij}} e_j & = & 0 \; \; \; {\rm when} \; \; \;
i \neq j\nonumber\\
({\rm ad}\;  f_i)^{1 - A_{ij}} f_j & = & 0 \; \; \; {\rm when} \; \; \;
i \neq j.\nonumber
\end{eqnarray}
This basis for the Lie algebra is called a {\em Chevalley basis} for the
algebra, and the matrix $A$ is the {\em Cartan matrix} for the group
$G$.
Note that the algebra elements $e_j$ associated with simple
roots generate the algebra of positive roots in $\algg_\CC$;
similarly, the negatives of the simple roots, $f_j$, generate the
algebra of negative roots.  Both of these subalgebras are closed
subalgebras of $\algg_\CC$.
Note also that the algebra elements $h_j$ form a basis $\Theta$ for
the Cartan subalgebra $\algt_\CC$.

We now turn our attention to irreducible representations of simple
groups.  To begin with, it is a well-known result that all irreducible
representations of simple groups admit unitary structures.  That is,
on any complex representation space $V$ of a finite-dimensional
compact simple group $G$, it is possible to construct a
positive-definite inner product which is invariant under the action of
$G$.  The existence of such a unitary structure can be seen by taking
an arbitrary positive-definite inner product on $V$, and averaging
over $G$ with respect to an invariant measure (such a measure, called
a Haar measure, exists on all compact finite-dimensional groups).  In
most physical systems where group representations play an important
role, the understanding of a unitary structure on the representation
space is essential.  It is customary for physicists, when dealing with
unitarity group representations, to take a basis for the Lie algebra
of the form $iJ_a$ where $J_a$ are Hermitian operators.  In terms of
this basis, the Lie algebra is written in terms of the {\em structure
constants} $\mixten{f}{ab}{c}$ as
\begin{equation}
[J_a,J_b] = i\mixten{f}{ab}{c} J_c.
\label{eq:Liealgebra}
\end{equation}
We will freely switch between mathematical and physical notations for
Lie groups and representations in this thesis.

In any (complex) vector space $V$ carrying a representation of $G$, it
is possible to choose a basis with respect to which the Cartan algebra
$\algt_\CC$ is diagonal.  The eigenvalues of $\algt_\CC$ on a basis
element are then described by a {\it weight} $w \in \algt_\CC^*$.
The weights $w$ of a representation lie on the  {\em weight lattice}
$\Lambda_w
\subset \algt_\CC^*$, which contains the root lattice $\Lambda$ as a
sublattice.  Clearly, acting on a basis element of weight $w$ with a
generator $e_\alpha$ gives a vector of weight $w + \alpha$ in $V$.  A
representation of $G$ on a vector space $V$ is defined to be {\em
highest weight} if there exists a vector $v \in V$ which is
annihilated by all positive roots; {\it i.e.,} $e_\alpha v= 0$ for all
$\alpha \in \Phi_+$.  For $v$ to be a highest weight vector, it clearly
suffices for all simple roots $e_j$ to annihilate $v$.  It follows
{}from the fact that all irreducible representations of a compact Lie
group are finite-dimensional, that all irreducible representations of
such a group are highest weight representations.

Given a choice $\Delta$ of base for $G$, the associated Borel
subalgebra $\algb^+ \subset \algg_\CC$ is defined to be the subalgebra
generated by the Cartan subalgebra and the positive roots,
\begin{equation}
\algb^+ = \algt_\CC \oplus \bigoplus_{\alpha \in \Phi_+} \algg_\alpha.
\end{equation}
Similarly, the Borel
subalgebra $\algb^- \subset \algg_\CC$ is defined to be the subalgebra
generated by the Cartan subalgebra and the negative roots,
\begin{equation}
\algb^- = \algt_\CC \oplus \bigoplus_{\alpha \prec 0} \algg_\alpha.
\end{equation}
The Borel subalgebra which is customarily used in representation
theory is the positive subalgebra $\algb^+$.  For the purposes of this
thesis, however, we will find it more constructive to use the negative
subalgebra $\algb^-$.  It is a theorem that these two Borel subalgebras
are conjugate \cite{Humphreys}; therefore the analysis using
$\algb^-$ is equivalent to that using $\algb^+$.  (In fact, generally a Borel
subalgebra is defined as a maximal solvable subalgebra; the theorem
states that all Borel subalgebras are conjugate.)  The Borel subgroup
$B^-$ of $G_\CC$ is the subgroup whose Lie algebra is spanned by
$\algb^-$.  The space $G_\CC/B^-$ is a compact K\"{a}hler homogeneous
space for $G_\CC$; as long as the group $G$ is compact, we have
$G_\CC/B^- = G/T$ \cite{PS}.  This space can be used to construct a
general irreducible representation of $G$ according to the Borel-Weil
theory \cite{Bott}, which we now review briefly, following
\cite{PS}.

It is a general fact that every irreducible representation of $G$ is
uniquely determined by the associated representation of $T$ on the
highest weight state
\cite{Adams}.  The representations of $T$ are the one-dimensional
representations on a space of fixed weight, and
are simply the products of
representations of the circle group $S^1 =\{{\rm e}^{i \theta}:0 \leq
\theta \leq 2 \pi\}$,
\begin{equation}
{\rm e}^{i \theta} \mapsto {\rm e}^{n i \theta},\; \; \; n \in \ZZ.
\end{equation}
Given a representation $\lambda:T \rightarrow S^1$ of $T$, it is
possible to construct a line bundle ${\cal L}_\lambda$ over the homogeneous
space $G/T$.  This is done by considering the product bundle $G \times
\CC$, and modding out by the equivalence relation $\sim$ defined by
\begin{equation}
(g,z) \sim (gt, \lambda (t^{-1})z) \forall t \in T.
\label{eq:equivalence}
\end{equation}
The line bundle is thus defined by
\begin{equation}
{\cal L}_\lambda = G\times \CC/ \sim.
\end{equation}
An alternative description of this line bundle is given by extending
the representation $\lambda$ holomorphically to the Borel subgroup
$B^-$.  This gives a holomorphic representation of $B^-$ taking values
in $\CC \setminus
\{0\}$.  The line bundle ${\cal L}_\lambda$ can then be defined by
\begin{equation}
{\cal L}_\lambda = G_\CC \times \CC/ \sim,
\label{eq:Borel-Weil}
\end{equation}
where the equivalence relation (\ref{eq:equivalence}) is extended to
all elements $t \in B^-$.  From this point of view, it follows
immediately that ${\cal L}_\lambda$ is a homogeneous complex line
bundle and that the group $G_\CC$ has a natural action on the space
${\cal H}_\lambda$ of holomorphic sections of ${\cal L}_\lambda$.  The
Borel-Weil theorem states that the resulting representation of $G$ on
${\cal H}_\lambda$ is an irreducible representation which reduces to
the representation $\lambda$ of $T$ on a one-dimensional subspace.  All
irreducible representations of $G$ can be obtained in this fashion;
however, an irreducible representation of $G$ may arise from several
distinct representations of $T$ which are equivalent under a Weyl
group symmetry.  In particular, however, choosing
$\lambda$ to be the representation of $T$  associated with the highest
weight subspace of a particular representation of $G$ will reproduce
that representation of $G$ in the action on ${\cal H}_{\lambda}$.

Most of the analysis in this thesis will be done locally, in a
coordinate patch around the identity of $G$.  In order to perform
local calculations, it will be useful to have a general system of
coordinates in a neighborhood of the identity both on $G$ and on
$G/T$.  It is clearly desirable to choose these coordinates to be
holomorphic on $G/T$.  Such a coordinate system is a familiar tool in
physics; see e.g., \cite{Zum,Alv1}.  We can define a natural set of
coordinates $\{z_\alpha, a_\alpha, \beta_\alpha | \alpha \in
\Phi_+\}$ on $G_\CC$ by writing an arbitrary element near the identity
in the form
\begin{equation}
g = \exp \left[ \sum_{\alpha \in \Phi_+} z_\alpha e_{\alpha} \right]
\exp \left[ \sum_{\alpha \in \Phi_+} a_\alpha e_{-\alpha} \right]
\exp \left[ \sum_{\alpha \in \Delta} \beta_\alpha h_\alpha  \right],
\label{eq:generalelement}
\end{equation}
Where $h_\alpha$ are the basis for the Cartan subalgebra $\algt_\CC$
defined by (\ref{eq:defineh}).  Clearly, $z_\alpha$ are a
set of coordinates on $G_\CC/B^-$, since modding out by the right
action of $B^-$ simply corresponds to dropping the coordinates
$a_\alpha$, $\beta_\alpha$.  It follows that $z_\alpha$ can also be
taken as coordinates on $G/T$.  This result can be made
more explicit by observing that (\ref{eq:generalelement}) can be
constrained to be an element of $G$, which gives $a_\alpha (z,
\bar{z})$ and $b_\alpha (z,
\bar{z})= {\rm Re}\; \beta_\alpha$ as functions of the coordinates
$z_\alpha, \bar{z}_\alpha$.  These functions are single-valued in a
neighborhood of the identity, and can be calculated in a perturbative
expansion about $z_\alpha = 0$ by applying the
Baker-Campbell-Hausdorff (BCH) formula
\cite{Kir}
\bge
{\rm e}^X {\rm e}^Y = {\rm e}^{X + Y + \frac{1}{2}[X,Y] + \cdots},
\label{eq:BCH}\ee
which expresses
the product of two exponentiated elements of a Lie algebra in terms of
a single exponentiated element of the algebra as a formal power
series.  (The ellipses in this formula denote third- and higher-order
commutators between $X$ and $Y$.)  It can be seen that the coordinates
$z_\alpha$ define a $G$-invariant complex structure on $G/T$ by
multiplying (\ref{eq:generalelement}) on the left by an arbitrary
element $g' \in G$; expressing the result again in the form
(\ref{eq:generalelement}), one finds that the coordinates $z_\alpha(g'
g)$ of the resulting point in $G/T$ depend only on $g'$ and the
holomorphic coordinates $z_\alpha (g)$ and not on the antiholomorphic
coordinates $\bar{z}_\alpha (g)$ \cite{Zum}.  Although this complex
structure is defined locally, it can be extended to a global complex
structure on $G/T$ by translating under the left action of $G$.

We now conclude this subsection with a pair of examples of
finite-dimensional simple Lie groups, $SU(2)$ and $SU(3)$, which we
will use throughout this thesis as canonical examples with which to
compare results from infinite-dimensional groups.  We will follow the
customary practice of using lower-case characters to denote Lie
algebras, so that the algebras of $SU(2)$ and $SU(3)$ are written
${\rm su}(2)$, ${\rm su}(3)$.

First we consider the group $SU(2)$ of unitary $2 \times 2$ matrices.
As usual (for physicists), we take the generators of the algebra $\algg
= {\rm su}(2)$ to be $\{iJ_k: k=1,2,3\}$, where $[J_j, J_k] = i \eps_{jkl}
J_l$.  The structure constants of ${\rm su}(2)$ are thus $\mixten{f}{ab}{c}=
\epsilon_{abc}$.  ${\rm su}(2)$ is a three-dimensional real vector space.
Taking coordinates $x_1, x_2, x_3$ on $\algg$, an arbitrary element $u
\in \algg$ can be written as $u = i \Sigma x_k J_k$.  An arbitrary
element $g$ of $G$ can be written as $g = {\rm e}^{u}$, where $u \in
\algg$.  We will take the maximal subtorus $T$ to be $\{{\rm e}^{2 i
\theta J_3}:0
\leq \theta \leq 2  \pi\}$.  A root space decomposition of $\algg_\CC$
is given by defining the generators
\bge
J_\pm = J_1 \pm i J_2.\ee These generators satisfy $[J_3, J_\pm] = \pm
J_\pm$, and $[J_+, J_-] = 2J_3$.  The basis elements $e_\alpha,
h_\alpha$ can be described in terms of these generators by $e_+= J_+,
e_- = f_+ = J_-$, and $h_+ =2J_3$.  We can now define $\Delta
=\{J_+\}$ to be a base, so that $J_+$ is the unique positive root.
The resulting Chevalley basis is $e_1 = e_+, f_1 = e_-, h_1 = h_+$, and
the Cartan matrix is the $1 \times 1$ matrix $(2)$.

\begin{figure}
\centering
\footnotesize
\begin{picture}(200,40)(- 100,-20)
\put(- 80,0){\vector(1,0){160}}
\put(0,- 10){\line(0,1){ 20}}
\multiput(- 60,- 5)( 30,0){5}{\line(0,1){ 10}}
\multiput(- 45,0)(30,0){4}{\circle*{5}}
\put(- 45,- 8){\makebox(0,0){ $- 3/2$}}
\put(- 15,- 8){\makebox(0,0){ $- 1/2$}}
\put( 15,- 8){\makebox(0,0){ $1/2$}}
\put( 45,- 8){\makebox(0,0){ $ 3/2$}}
\put(3,10){\makebox(0,0){ $0$}}
\end{picture}
\caption[$SU(2)$ representation]{\footnotesize
 Spin 3/2 representation of $SU(2)$}
\label{f:su2}
\end{figure}
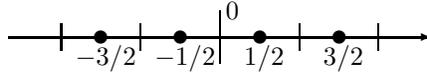

The finite-dimensional irreducible representations of ${\rm su}(2)$ are
representations with highest weight $j \in \ZZ/2$, corresponding to
the eigenvalue of $J_3$ on the highest weight vector.
The set of
weights for the ``spin'' $j$ representation is given by $\{j,j -1,
\ldots,1 - j, - j\}$; to each weight there corresponds a single vector
in the representation space $V_j$.  To each finite-dimensional
irreducible representation of ${\rm su}(2)$ there corresponds a
representation of $SU(2)$.  The weight space of a typical $SU(2)$
representation (spin 3/2) is shown in Figure~\ref{f:su2}.

With each irreducible representation $j$ of $SU(2)$, there is an
associated representation of the maximal subtorus  $T$.
The homogeneous space $SU(2)/T$ is
$S^2$ with the usual complex structure.  The bundle structure of this
quotient space is exactly that of the well-known Hopf fibration $S^1
\rightarrow S^3 \rightarrow S^2$
\cite{Spanier}.  Each representation $j$ of $T$ defines a complex line
bundle ${\cal L}_j$ over $S^2$ by the Borel-Weil theory
(\ref{eq:Borel-Weil}).  The bundle associated with the representation
$j$ has a first Chern class with integral $2j$, and $2j +1$ linearly
independent holomorphic sections which transform under the
representation $j$ of $SU(2)$.  The geometric structure of this
representation is equivalent to the coadjoint orbit construction for
$SU(2)$ and will be discussed in more detail in Section 2.5.

We next consider the group $SU(3)$ of unitary $3 \times 3$
matrices.  The generators of ${\rm su}(3)$ are customarily taken to be
$\{J_a:1 \leq a \leq 8\}$, which in the fundamental representation are
related to the Hermitian Gell-Mann matrices $\{\lambda_a\}$ by $J_a =
\lambda_a/2$.  The Gell-Mann matrices are
\begin{equation}
\begin{array}{rclrcl}
\lambda_1 & = &
\left(\begin{array}{ccc}
0 & 1 & 0\\1 & 0 & 0\\0 & 0 & 0
\end{array}\right)
& \lambda_2 & = &
\left(\begin{array}{ccc}
0 &  - i & 0\\ i & 0 & 0\\0 & 0 & 0
\end{array}\right)\\
& & & & &\\
\lambda_3 & = &
\left(\begin{array}{ccc}
1 & 0 & 0\\0 & -1 & 0\\0 & 0 & 0
\end{array}\right)
& \lambda_4 & = &
\left(\begin{array}{ccc}
0 &  0 & 1\\  0 & 0 & 0\\1 & 0 & 0
\end{array}\right)\\
& & & & &\\
\lambda_5 & = &
\left(\begin{array}{ccc}
0 & 0 &  - i\\0 & 0 & 0\\i & 0 & 0
\end{array}\right)
& \lambda_6 & = &
\left(\begin{array}{ccc}
0 &  0 & 0\\ 0 & 0 & 1\\0 &  1 & 0
\end{array}\right)\\
& & & & &\\
\lambda_7 & = &
\left(\begin{array}{ccc}
0 & 0 & 0\\0 & 0 & - i\\0 & i& 0
\end{array}\right)
& \lambda_8 & = &
\frac{1}{\sqrt{3}} \left(\begin{array}{ccc}
1 &   0 & 0\\  0 & 1 & 0\\0 & 0 &  -2
\end{array}\right)
\end{array}
\label{eq:GellMann}
\end{equation}

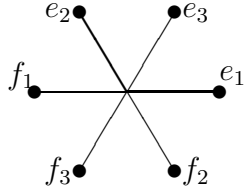
\begin{figure}
\centering
\begin{picture}(200,100)(- 100,- 50)
\thicklines
\put(0,0){\line(1,0){35}}
\put(0,0){\line(-3,5){18}}
\thinlines
\put(0,0){\line(3,5){18}}
\put(0,0){\line(3,-5){18}}
\put(0,0){\line(-3,-5){18}}
\put(0,0){\line(-1,0){35}}
\put(-35,0){\circle*{5}}
\put(35,0){\circle*{5}}
\put(18,30){\circle*{5}}
\put(-18,30){\circle*{5}}
\put(18,-30){\circle*{5}}
\put(-18,-30){\circle*{5}}
\put(40,6){\makebox(0,0){$e_1$}}
\put(-40,6){\makebox(0,0){$f_1$}}
\put(26,30){\makebox(0,0){$e_3$}}
\put(-26,30){\makebox(0,0){$e_2$}}
\put(26,-30){\makebox(0,0){$f_2$}}
\put(-26,-30){\makebox(0,0){$f_3$}}
\end{picture}
\caption[Roots of $SU(3)$]{\footnotesize
 Roots of $SU(3)$; simple roots are bold.}
\label{f:roots3}
\end{figure}

The generators $J_3,J_8$ form a maximal abelian subalgebra of ${\rm su}(3)$
and can be taken to be the generators of a maximal subtorus
\begin{equation}
T =\{{\rm e}^{2 i \theta J_3}{\rm e}^{2 \sqrt{3} i \psi J_8}:0 \leq \theta,
\psi <2 \pi\}.
\end{equation}
A root space decomposition of $\algg_\CC$ is given by the generators
\begin{equation}
\begin{array}{rclrclrcl}
e_t & = & J_1 + iJ_2 & e_u & = & J_6 + iJ_7 & e_v & = & J_4 + iJ_5\\
f_t & = & J_1 - iJ_2 & f_u & = & J_6 - iJ_7 & f_v & = & J_4 - iJ_5 \\
h_t & = & 2J_3 &  h_u & = & (\sqrt{3}J_8 - J_3)
 &  h_v & = & (\sqrt{3}J_8 + J_3),
\end{array}
\end{equation}
where the roots $t,u,v$ are given by $t = (\ip{t}{J_3 },
\ip{t}{J_8 })=(1,0)$, $u = (- 1/2, \sqrt{3}/2)$, $v = (
1/2, \sqrt{3}/2)$.  In terms of this basis, we can choose a base
$\Delta$ for ${\rm su}(3)$ to be $\Delta =\{e_t,e_u\}$.  The resulting
Chevalley basis is given by
\begin{equation}
\begin{array}{rclrcl}
e_1 & = & e_t\hspace*{1in} & e_2 & = & e_u  \\
f_1 & = & f_t & f_2 & = & f_u  \\
h_1 & = & h_t & h_2 & = & h_u  ,
\end{array}
\end{equation}
and the Cartan matrix is
\begin{equation}
A = \left(\begin{array}{cc} 2 & -1 \\ -1 & 2 \end{array} \right).
\end{equation}
In terms of this Chevalley basis, the remaining generators are $e_3
=[e_1,e_2]= e_v$ and $f_3 =[f_2,f_1]= f_v$.  The generators are
graphed according to their roots in Figure~\ref{f:roots3}.

\begin{figure}
\centering
\begin{picture}(200,190)(- 100,- 100)
\multiput(- 72,60)( 36,0){5}{\circle*{5}}
\multiput(- 72,0)( 36,0){5}{\circle*{5}}
\multiput(- 54,- 30)( 36,0){4}{\circle*{5}}
\multiput(- 36,- 60)( 36,0){3}{\circle*{5}}
\multiput(- 18,- 90)( 36,0){2}{\circle*{5}}
\multiput(- 90,30)( 36,0){6}{\circle*{5}}
\put(80, 68){\makebox(0,0){ (4,1)}}
\put(- 80,0){\vector(1,0){160}}
\put(90,5){\makebox(0,0){ $h_1$}}
\put(0,-95){\vector(0,1){170}}
\put(5,80){\makebox(0,0){ $\frac{1}{\sqrt{3}}(2 h_2 +  h_1) $}}

\end{picture}
\caption[$SU(3)$ representation]{\footnotesize
 Representation (4,1) of $SU(3)$}
\label{f:representation3}
\end{figure}
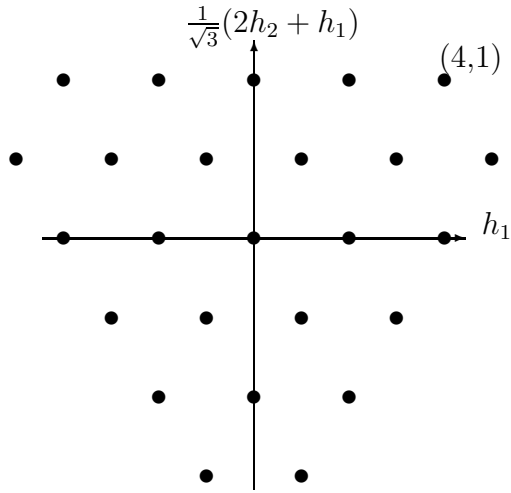

Irreducible representations of $SU(3)$ are labeled by the
eigenvectors
$(p,q)$ of the highest weight vector with respect to the Cartan
algebra elements $h_1,h_2$.
The weights of a typical highest weight representation
$(4,1)$ of $SU(3)$ are shown in Figure~\ref{f:representation3}.  Note
that there are in general multiple linearly independent vectors in the
representation space with a given weight.  As in the case of $SU(2)$,
all the irreducible representations of $SU(3)$ can be realized via the
Borel-Weil theory in terms of the left action of $SU(3)$ on spaces of
holomorphic sections of line bundles over the homogeneous space
$SU(3)/T$.

\subsubsection{Loop groups}
\label{sec:loopgroups}

In the previous subsection we described a class of finite-dimensional
Lie groups.  It is possible to develop a similar theory for
infinite-dimensional Lie groups \cite{Milnor}.  In general, an
infinite-dimensional Lie group is a group whose manifold is an
infinite-dimensional manifold modeled on a complete, locally convex
topological vector space (a manifold is modeled on a vector space $V$
when the coordinate charts give diffeomorphisms between neighborhoods
in the manifold and open sets in $V$).  When the group is modeled on a
Banach space, there is a fairly well-developed theory, and several
theorems are true which do not hold in a more general context.  We
will primarily be concerned with local algebraic properties of certain
infinite-dimensional groups, so we will not develop the general theory
of infinite-dimensional groups in any detail, but will simply quote
results as necessary.

One of the main properties which we will need in the analysis of
infinite-dimensional groups is the existence of a real analytic
structure on the group.  A sufficient condition for a group to have a
real analytic structure is that the exponential map from the Lie
algebra onto the Lie group is locally 1-1 and onto.  The existence of
a real analytic structure on a Lie group is in turn sufficient for the
group to have the BCH property, which means that the
Baker-Campbell-Hausdorff formula (\ref{eq:BCH}) is valid in a
neighborhood of the identity \cite{Milnor}.  Groups with this property
are completely described in a neighborhood of the identity by the Lie
algebra structure.  Since we will be concentrating on analyzing the
representations of the algebra, this property is necessary to validate
our analysis in a more global context.

It is possible to generalize the concepts of a complex structure and
differential forms on a manifold to the infinite-dimensional case,
provided that certain conditions are satisfied.  These generalizations
are possible for all the manifolds we consider here \cite{PS} .

The first class of infinite-dimensional groups which we will consider
are the centrally extended loop groups $\widehat{L}G$.  The loop group
$LG$ is defined for an arbitrary finite-dimensional Lie group $G$ by
the set of smooth maps from $S^1$ to $G$
\begin{equation}
LG=\{ f: S^1 \rightarrow G|f \in C^\infty (S^1, G)\}.
\end{equation}
The composition law in a loop group is defined pointwise by
\begin{equation}
(f \cdot g) (t) = f (t) \cdot g (t) \;\forall t \in S^1.
\end{equation}
For simplicity, in this thesis we will restrict attention to loop
groups $LG$ where $G$ is simple; as in the finite-dimensional case, an
extension to general semisimple Lie groups is straightforward.

The Lie algebra $L \algg$ is simply the algebra of smooth maps from
$S^1$ to the Lie algebra $\algg$.  Again, the Lie algebra product of
$L \algg$ is defined pointwise on $S^1$.  It is not hard to show
\cite{Milnor}  that the loop groups $LG$ admit real analytic
structures, and thus have the BCH property (the argument is more
generally valid for arbitrary map groups; {\it i.e.}, groups of maps from an
arbitrary compact manifold $M$ to $G$).  The existence of a real
analytic structure in this case follows from the fact that the
exponential map
\begin{equation}
\exp:L \algg \rightarrow LG
\end{equation}
is a local homeomorphism.  If the Lie group $G$ has a complexification
$G_\CC$, there is a natural complexification $LG_\CC$ of the loop
group $LG$, which gives a natural complexification $L \algg_\CC$ of
the Lie algebra.  The Lie algebra $L \algg_\CC$ has a natural basis
$\jj{J}{a}{n}$ given by the set of Fourier modes relative to a basis $J_a$ of
$\algg_\CC$ (actually, technically this is a basis for the related
algebra of real-analytic maps from $S^1$ to $\algg$; in this thesis we
will ignore this distinction, for details see \cite{PS}).
The function from $S^1 \rightarrow \algg$ described by an element of
this Fourier basis is
\begin{equation}
\jj{J}{a}{n} (\theta) = J_a {\rm e}^{i n \theta}.
\end{equation}
If the
structure constants of $\algg$ are given by $\mixten{f}{ab}{c}$, then the Lie
algebra of the loop group is given by
\begin{equation}
[\jj{J}{a}{n},\jj{J}{b}{m}] = i \mixten{f}{ab}{c} \jj{J}{c}{n + m}.
\end{equation}

The loop groups $LG$ appear in physics as the symmetry groups of
certain conformal field theories (CFT's); see for example
\cite{Gins}.  For example, the conformal field theory with $n$
free real fermionic fields has a natural $LSO(n)$ symmetry, and the
CFT with $n$ free complex fermionic fields has $L SU(N)$ symmetry.
These symmetries apply to the classical field theories.  When the
quantum field theories associated with these classical theories are
constructed, the classical symmetry group acquires a central extension
due to a quantum anomaly.  For other conformal field theories, such as
the WZW model, the classical symmetry group has already a central
extension.  Thus, we now discuss the modification to a loop group
caused by a central extension.

A central extension $\widehat{G}$ of a group $G$ is topologically a
circle bundle over $G$.  The algebraic structure of a central
extension is described by the exact sequence
\begin{equation}
S^1 \rightarrow  \widehat{G} \rightarrow G.
\end{equation}
In a neighborhood of the identity, we can choose a
coordinate system in which the circle bundle of the central extension
is locally trivial.  In such a coordinate system, an element of
$\widehat{G}$ can be written as $(g,z)$, $g \in G,z \in S^1$; the
group product is given by
\begin{equation}
(g,z) \cdot (g', z') = (gg', c (g,g')zz'),
\end{equation}
where $c (g,g')\in S^1$ is a group cocycle satisfying
\begin{equation}
c (g,g') c (gg', g'') = c (g,g' g'') c (g', g'').
\label{eq:groupcocycle}
\end{equation}

The Lie algebra of a centrally extended group $\widehat{G}$ has a
similar description as $\widehat{\algg}= \algg \oplus \RR$.  An
element of $\widehat{\algg}$ is given by a pair $(u,t)$ where $u \in
\algg$ and $t \in \RR$.  The Lie algebra product is given by
\begin{equation}
[(u,t), (v,s)] = ([u,v], \omega (u,v)),
\end{equation}
where $\omega$ is a skew-symmetric Lie algebra cocycle satisfying
\begin{equation}
\omega ([u,v], w) +\omega ([v,w], u)+\omega ([w,u], v)= 0.
\label{eq:algebracocycle}
\end{equation}
The cocycle conditions (\ref{eq:groupcocycle}) and
(\ref{eq:algebracocycle}) follow directly from the condition of
associativity on the group $\widehat{G}$ and the Jacobi identity on
the algebra $\widehat{\algg}$, respectively.

It is a straightforward algebraic exercise to demonstrate that in
terms of the Fourier basis $\jj{J}{a}{n}$, the most general central extension
of $L \algg$ has a complexification with a Lie algebra described by
\begin{equation}
[\jj{J}{a}{n},\jj{J}{b}{m}] = i \mixten{f}{ab}{c} \jj{J}{c}{n + m} + n
\delta_{n, - m}
\gamma ( J_a,J_b ) C,
\label{eq:currentalgebra}
\end{equation}
where $C$ is the (Hermitian) central generator and $\gamma ( J_a,J_b
)$ is a symmetric invariant bilinear in $\algg_\CC$\cite{PS}.
In particular, since $G$ is simple, there is a unique symmetric
invariant bilinear form in $\algg_\CC$ (up to scalar multiplication).
This is the {\it Killing form}
\begin{equation}
\eta_{ab} = \mixten{f}{ad}{c} \mixten{f}{bc}{d}.
\end{equation}
It follows that $\gamma ( J_a,J_b )$ is proportional to $\eta_{ab}$.
By choosing a basis in which the structure constants
$\mixten{f}{ab}{c}$ are completely antisymmetric, the Killing form can
be put in diagonal form; often physicists write the algebra
(\ref{eq:currentalgebra}) in terms of the diagonal form $\delta_{ab}$.

In order to extend the algebra (\ref{eq:currentalgebra}) to a global
central extension of $LG$, it is necessary to make a choice of
$\gamma$ such that the cocycle $\omega$ can be extended to a central
extension of the group.  A necessary and sufficient condition for this
to be possible is that $\gamma ( h_\alpha, h_\alpha )$ be an even
integer for every root $\alpha$ of $G$.  It is fairly easy to see that
this condition is at least necessary.  Consider the subalgebra of
(\ref{eq:currentalgebra}) generated by
\begin{eqnarray}
I_\pm & = & \jj{e}{{\pm \alpha}}{0},\label{eq:subalgebra1} \\
I_3 & = &  \jj{h}{\alpha}{0}/2 \nonumber
\end{eqnarray}
This subalgebra is isomorphic to the (complexified) Lie algebra
${\rm su}(2)$, which follows from $[I_3,I_\pm]= \pm I_\pm$ and $[I_+, I_-]
=2 I_3$.  The central terms vanish, since they are proportional to the
mode number which in this case is 0.  We now consider the subalgebra
generated by
\begin{eqnarray}
\tilde{I}_\pm & = & \jj{e}{{\mp \alpha}}{{\pm 1}}, \label{eq:subalgebra2} \\
\tilde{I}_3 & = &  \frac{ x  C}{4} - \jj{h}{\alpha}{0}/2,\nonumber
\end{eqnarray}
where $x = \gamma ( h_\alpha, h_\alpha )$.  This subalgebra is also an
${\rm su}(2)$ subalgebra, as can be seen from computing
\begin{eqnarray}
\left[\tilde{I}_3,\tilde{I}_\pm\right] & = & \pm  \tilde{I}_\pm\\
\left[\tilde{I}_+, \tilde{I}_-\right] & = & 2 \tilde{I}_3 -  \frac{x C}{2}  +
 \gamma ( f_\alpha, e_\alpha ) C, \nonumber
\end{eqnarray}
and observing that the invariance of $\gamma$ under $G$
implies that
\begin{eqnarray}
\gamma ( f_\alpha, e_\alpha )  & = &
\frac{1}{2} \gamma ( [f_\alpha, h_\alpha], e_\alpha )\\
& = &\frac{1}{2} \gamma ( h_\alpha,[e_\alpha, f_\alpha ])  =   \frac{x}{2}.
\nonumber
\end{eqnarray}
{}From the subalgebra (\ref{eq:subalgebra1}), we  know that
that $\exp (4 \pi i  I_3)= 1$ in the group $LG$.
Similarly, for (\ref{eq:currentalgebra}) to extend to a global group
structure, we must have $\exp (4 \pi i \tilde{I}_3)= 1$.  From the
definition of $C$ as the generator of the central extension satisfying
$\exp (2 \pi i C) = 1$, it follows that $x \in 2 \ZZ$.

It can be shown that the above condition is also sufficient to
guarantee that the algebra (\ref{eq:currentalgebra}) can be globally
extended to a central extension $\widehat{L}G$ of the group $LG$
\cite{PS}.  The condition on the algebra can be rewritten as a global
topological condition on the differential form associated with the
cocycle $\omega$; we will discuss the geometry of this condition
further in Section \ref{sec:coadjointexamples2}.  It is easy to see
that there must be a minimum multiple of the Killing form which has
the desired property.  We will denote this minimum multiple by
$g_{ab}= \gamma_m ( J_a,J_b )$.  In terms of $g_{ab}$, the centrally
extended algebra (\ref{eq:currentalgebra}) is written
\begin{equation}
[\jj{J}{a}{n},\jj{J}{b}{m}] = i \mixten{f}{ab}{c} \jj{J}{c}{n + m} + k n
\delta_{n, - m}  g_{ab} C,
\label{eq:currentalgebra2}
\end{equation}
where $k$ is integral.

At this point, it is necessary to discuss a difference between
physical and mathematical notation for current algebras.  In general,
physicists take the convention that $C = 1$, and define the algebra
(\ref{eq:currentalgebra2}) to be the current algebra, or (untwisted)
affine algebra, on $G$ at level $k$.  Mathematicians, on the other
hand, restrict attention to the case $k = 1$, and retain the
description of $C$ as a (commuting) algebra element.  The group for $k
= 1$ is referred to as the universal central extension of $LG$, and is
written as $\widehat{L}G$.  This extension is called universal because
all other extensions can be realized as quotients of $\widehat{L}G$ by
a finite cyclic group.  We will primarily use the mathematical
description here, because it allows for a simpler and more consistent
description of the coadjoint orbit representations of centrally
extended groups.  We will, however, denote the eigenvalue of the
operator $C$ by $k$, so that the notation is still ostensibly
equivalent to that used by physicists.  Note that the centrally
extended loop groups have locally 1-1 exponential maps from the Lie
algebra, which follows directly from the fact that the groups without
central extensions have this property.  The exponential map is also
locally surjective, so it follows that the groups $\widehat{L}G$ have
the BCH property.  (Note that in general, the exponential map is not
globally surjective for loop groups.)

We will now describe the adjoint and coadjoint action of the centrally
extended loop group $\widehat{L}G$ on the algebra $\widehat{L}\algg$.
Denoting an arbitrary element of $\widehat{L}\algg$ by
\begin{equation}
(f,ia) = f (t) + iaC, \; \; \; f: S^1 \rightarrow \algg, a \in \RR,
\end{equation}
we can describe the adjoint action of an element $g: S^1
\rightarrow G$ of $L G$ on $(f,ia)$ by
\begin{equation}
\aad_g (f,ia) = \left(\aad_g (f),ia -
\frac{i}{2 \pi} \int_{0}^{2 \pi} {\rm d} \theta
\gamma_m (g^{-1}(\theta) g'(\theta), f(\theta))\right).
\label{eq:loopadjoint}
\end{equation}
(The adjoint action of the central element $S^1$ is trivial.)
To prove that (\ref{eq:loopadjoint}) is the correct formula for the
adjoint action, it suffices to demonstrate that  taking $g$ close to
the identity reproduces (\ref{eq:currentalgebra2}) and that the action
is indeed a group action in the sense that $\aad_h \aad_g = \aad_{hg}$.
The first of these conditions is easily verified; when
\begin{equation}
g = 1 + \epsilon \jj{J}{a}{n},
\end{equation}
and $f = \jj{J}{b}{m}$, (\ref{eq:loopadjoint}) reduces to
\begin{eqnarray}
\frac{{\rm d}}{{\rm d} \epsilon}  |_{\epsilon = 0}
\aad_g (f,ia) & = & \left(i\mixten{f}{ab}{c}\jj{J}{c}{n + m},
-\frac{i}{2 \pi} \int_{0}^{2 \pi} {\rm d} \theta
\gamma_m (in \jj{J}{a}{n}, \jj{J}{b}{m})\right) \\
 & = &  \left( i\mixten{f}{ab}{c}\jj{J}{c}{n + m},
n \delta_{n, - m} g_{ab} \right). \nonumber
\end{eqnarray}
Although this calculation is performed in the complexified algebra,
clearly the restriction to real linear combinations satisfies the same
equation.  Recall that we are using the mathematician's convention of
$k = 1$.  To demonstrate that the action is a group action, we write
\begin{equation}
\aad_h \aad_g (f,ia) = \left(\aad_h \aad_g (f),ia -  x \right),
\end{equation}
where
\begin{eqnarray}
x & = &
\frac{i}{2 \pi} \int_{0}^{2 \pi} {\rm d} \theta \;
\left[\gamma_m (g^{-1}(\theta) g'(\theta), f(\theta))
+\gamma_m (h^{-1} (\theta) h' (\theta), \aad_{g (\theta)}f (\theta))\right]
\nonumber\\
& = &
\frac{i}{2 \pi} \int_{0}^{2 \pi} {\rm d} \theta  \;
\gamma_m (g^{-1}(\theta) g'(\theta) +
g^{-1} (\theta)h^{-1} (\theta) h' (\theta) g (\theta), f(\theta))
\\
& = & \frac{i}{2 \pi}  \int_{0}^{2 \pi}
{\rm d} \theta \; \gamma_m ((hg)^{-1} (\theta) (hg)' (\theta), f
(\theta)), \nonumber
\end{eqnarray}
so
\begin{equation}
\aad_h \aad_g (f,ia) = \left(\aad_{hg} (f),ia -
\frac{i}{2 \pi} \int_{0}^{2 \pi} {\rm d} \theta
\gamma_m ((hg)^{-1}(\theta) (hg)'(\theta), f(\theta))\right).
\end{equation}

The dual space to $\widehat{L}\algg$ is $L (\algg^*)\oplus \RR$.  We
can write elements of this space as pairs $(b, - it)$ where $b: S^1
\rightarrow \algg^*$ and $t \in \RR$.  The dual pairing between the
extended loop algebra and dual space is given by
\begin{equation}
\ip{ (b, - it)}{ (f,ia)} = \frac{1}{2 \pi}  \int_0^{2 \pi} {\rm d}
\theta \; \ip{ b (\theta)}{ f (\theta)} + at.
\label{eq:loopinnerproduct}
\end{equation}
It is straightforward to verify that in order for this dual pairing to
be preserved under the  group action, the coadjoint action of an
arbitrary element $g \in LG$ on $(b, - it)$ must be given by
\begin{equation}
\aads_g (b, - it) = (\aad_g^* b +  t (g' g^{-1})^*, - it),
\label{eq:loopcoadjointaction}
\end{equation}
where we have defined a dual map $*$ from $\algg$ to $\algg^*$ such
that
\begin{equation}
\ip{ g^*}{f} = \gamma_m (g,f), \; \; \; \forall f,g \in \algg.
\end{equation}
This map is well-defined since the Killing metric is nondegenerate for
simple $G$.  From (\ref{eq:loopcoadjointaction}) we can calculate the
coadjoint action of the algebra; we find that
\begin{equation}
\ip{\ads_u (b, - it)}{(f,ia)} =
\frac{1}{2 \pi}  \int_0^{2\pi} {\rm d} \theta \;
 \left[ \ip{ b (\theta)}{[f (\theta), u (\theta)]}
+ t \gamma_m( u' (\theta), f (\theta)) \right],
\label{eq:loopalgebracoadjoint}
\end{equation}
in agreement with (\ref{eq:currentalgebra2}).

Just as for finite-dimensional groups, we will find it useful to
define simple roots for the centrally extended algebra $\widehat{L}
\algg$.  If $G$ is a simple finite-dimensional group of rank $r$, and
we denote the highest root of $\algg$ by $\psi$, then we define the
generators associated with the
$r + 1$ simple roots of $\widehat{L} \algg$ to be
\begin{equation}
e_j =
\left\{\begin{array}{ll}
\jj{e}{j}{0}, &1 \leq j \leq r \\
\jj{e}{- \psi}{1}, &  j = r + 1
\end{array}\right.\label{eq:affinesimple}
\end{equation}
The space of roots for $\widehat{L} \algg$ is naturally described
by $\ZZ \times \Phi$, where the integral parameter is the mode number
and $\Phi$ is the root space of $\algg$.
(Often, an auxiliary operator generating rotations about $S^1$ is
introduced, for which the mode number is the eigenvalue; this
formalism has the advantage of keeping the interpretation of a root as
an element of the dual space to the Cartan subalgebra, where the dual
of $C$ is 0 and the new Cartan subalgebra is $\RR \otimes \RR C
\otimes \algt_{\CC (0)}$.  We will forgo the extra complications in
notation which would be involved in this formalism, and we will simply
take the mode number as an extra parameter in the root space.)  In
this space, we define a positive root $(n, \alpha)\succ 0$ to be any
root which can be written as a sum of the simple roots with
nonnegative coefficients.  For a general representation, positive
weights can be defined similarly.  We have then a natural
decomposition of the algebra $\widehat{L}
\algg_\CC$,
\begin{equation}
\widehat{L} \algg_\CC = \algt_{\CC (0)} \oplus \CC C \oplus\bigoplus_{(n,
\alpha)\succ  0}  \algg_{\alpha(n)} \oplus\bigoplus_{(n,
\alpha)\prec  0} \algg_{\alpha(n)}.
\end{equation}
Note that the central element has weight $(0,0)$ in the adjoint
representation.  Note also, that the adjoint representation is not
highest weight.  Just as in (\ref{eq:affinesimple}), we can define the
rest of a Chevalley basis for $\widehat{L} \algg_\CC$ by
\begin{equation}
f_j =
\left\{\begin{array}{ll}
\jj{f}{j}{0}, &1 \leq j \leq r \\
\jj{e}{{\psi}}{{-1}}, &  j = r + 1
\end{array}\right.
\end{equation}
and
\begin{equation}
h_j =
\left\{\begin{array}{ll}
\jj{h}{j}{0}, &1 \leq j \leq r \\
\frac{1}{2}\gamma_m ( h_\psi, h_\psi ) C -\jj{h}{{\psi}}{0}, &  j = r + 1
\end{array}\right.
\end{equation}
Using this basis, the Cartan relations (\ref{eq:cartanrelations}) are
satisfied, with a singular Cartan matrix $A$.  Using the Cartan matrix
approach, a general class of affine algebras can be defined which
contains in particular the centrally extended current algebras
described here \cite{Kac}.

The main result we will need about simple roots is that just as for
finite-dimensional groups, the simple roots of $\widehat{L}
\algg$ generate the subalgebra $S$ of positive roots.  This result
can easily be seen, as follows.  Certainly, if $\alpha \in \Phi_+$ is a
positive root of $\algg$ then $\jj{e}{\alpha}{0}$ is in $S$, since
$e_\alpha$ is in the subalgebra of positive roots of $\algg$, which is
generated by the simple roots of $\algg$.  The root
$\jj{e}{{-\psi}}{{1}}$ is a simple root of $\widehat{L}
\algg$, and is thus in $S$ automatically.  Since the roots of $\algg$
form an irreducible representation of $\algg$ (the adjoint) with
lowest weight $-\psi$, we can reach any root $\jj{e}{\beta}{{1}}$ by
commuting positive roots $\jj{e}{\alpha}{0}$ with
$\jj{e}{{-\psi}}{{1}}$.  Since $\psi \neq 0$, there is an element $h =
-h_\psi/2$ of the Cartan subalgebra of $\algg_\CC$ which satisfies
$[h, e_{-\psi}]= e_{-\psi}$.  The first Fourier mode of this operator,
$\jj{h}{}{1}$, is an element of $S$ from the above argument, so
$[\jj{h}{}{1}, \jj{e}{{-\psi}}{{1}}] =
\jj{e}{{-\psi}}{{2}}\in
S$.  Continuing in this fashion, we see that all generators associated
with positive roots are in $S$.  Similarly, the negatives of the
simple roots generate the algebra of negative roots, again as in the
finite-dimensional case.

We now review briefly the representation theory for centrally extended
loop groups and their algebras.  In the case of finite-dimensional
groups, we had the result that all irreducible representations are
highest weight representations.  This is not the case for the groups
$\widehat{L}G$.  We shall restrict attention, however, to a class of
irreducible representations which are of physical interest; all the
representations we shall consider are highest weight.  To begin with,
we will restrict attention to representations on vector spaces
$V$ which can be written as direct sums of weight spaces
\begin{equation}
V =\bigoplus_{(n, \psi)\in \ZZ \times \algt_\CC^*} V_{ \psi (n)}
\label{eq:decompositionr}
\end{equation}
Not all representations are of this form, since some representation
spaces do not admit an action of the rotation group $S^1$ which
commutes with the constant generators $\jj{J}{\alpha}{0}$.  We will
further restrict to the class of representations for which the
decomposition (\ref{eq:decompositionr}) contains only weights $(n,
\psi)$ with $n < m$ for some integer $m$.  These representations are
referred to as ``positive energy'' representations, because the mode
number $m - n$ can be related to an energy operator of a physical
system.  From now on, unless otherwise noted
we will assume that all representations of
$\widehat{L}G$ and its algebra are positive energy representations
which admit the decomposition (\ref{eq:decompositionr}).  As an
example of a representation which does not fit in this category,
consider the adjoint representation of $\widehat{L}G$ on $\widehat{L}
\algg$, which has both positive and negative mode numbers of arbitrary
magnitude.
It can be shown that all irreducible representations in
the category of interest admit unitary structures \cite{PS}.  We shall
discuss these unitary structures further in Section \ref{sec:global}.

With the above restriction, the representation theory of
the groups $\widehat{L}G$ can be described in a very similar
fashion to that of the finite-dimensional simple groups.  In fact
\cite{PS}, it can be shown that all irreducible representations  of
the algebra $\widehat{L} \algg$ of the
desired type are highest weight representations.  Note that we have
defined highest weight representations with respect to the choice of
simple roots given in (\ref{eq:affinesimple}).  In some of the
literature, these representations are referred to as ``antidominant''
representations, due to a different convention for positive and
negative roots.  The fact that all irreducible representations are
highest weight can be seen fairly easily by showing that the subspace
of highest mode number $n$ must admit an irreducible representation of
the subalgebra $\algg_{(0)}$ of $L
\algg$ generated by the 0 modes $\jj{J}{a}{0}$; since this
representation is irreducible, it must be finite-dimensional, and
therefore must contain a highest weight vector which is annihilated by
all simple roots of $\widehat{L}G$.

Thus, we can describe every representation of $\widehat{L} \algg$ by
an integer $k$ (the eigenvalue of $C$, also called the level of the
representation) and a highest weight vector $v$, with weight $(n,
\lambda)$.  Since the mode number $n$ of the highest weight vector can
be shifted by an arbitrary integer, we will take $n = 0$ for all
representations.  Finally, by examining $SU(2)$ subgroups of
$\widehat{L}G$, it is possible to show that $k$ and $\lambda$ must
satisfy an additional relation for a highest weight representation of
$\widehat{L}\algg$ to extend to a representation of $\widehat{L}G$.
{}From (\ref{eq:subalgebra1}) and (\ref{eq:subalgebra2}), we know that
the eigenvalues of the operators $2 I_3$ and $2 \tilde{I}_3$ defined
in those equations must be nonnegative integers when acting on the
highest weight state $v$.  It follows that for all roots $\alpha$, we
have
\begin{equation}
0 \leq \ip{\lambda}{h_\alpha} \leq \frac{1}{2} k \gamma_m ( h_\alpha,
h_\alpha ).
\label{eq:constraint}
\end{equation}
(Recall that the weight $\lambda$ is naturally in the dual space
$\algt_\CC^*$.)
It can also be shown \cite{PS}  that for any $k, \lambda$ satisfying the
constraint (\ref{eq:constraint}) an irreducible representation of
$\widehat{L}G$ exists.  Note that the condition (\ref{eq:constraint})
indicates that for a given level $k$, there are a finite number of
distinct irreducible representations of $\widehat{L}G$.

The Borel-Weil theory which is used to describe irreducible
representations of finite-dimensional compact groups can be
generalized to cover the groups $\widehat{L}G$; however, the apparatus
used to implement the construction in the infinite-dimensional case is
considerably more sophisticated \cite{PS}.  This construction is again
closely related to the coadjoint orbit 	technique which we will
describe later in this thesis.  As in the finite-dimensional case, one
can construct a homogeneous space
\begin{equation}
\widehat{L}G/(T_{(0)} \times S^1) =
\widehat{L}G_\CC/B^-,
\end{equation}
where $T_{(0)}$ is the zero mode restriction of a maximal subtorus of
$G$, $S^1$ is the central extension, and $B^-$ is a Borel subgroup
formed just as in the finite-dimensional case from the group whose Lie
algebra contains the negative roots of $\widehat{L} \algg$ and the
zero modes of a Cartan subalgebra (and also $C$).  Because
$\widehat{L}G$ has an analytic structure, we can put a coordinate
system on this quotient space just as in (\ref{eq:generalelement})
(with an infinite number of coordinates $\{z_{\alpha (n)}: (n,
\alpha)\succ 0\}$).  We will use such a
coordinate system when describing coadjoint orbit representations of
loop groups in Section \ref{sec:coadjointexamples2}.

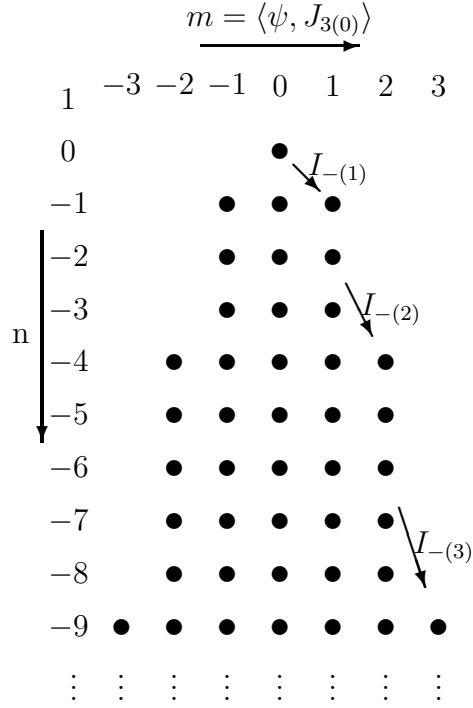
\begin{figure}
\centering
\begin{picture}(200,250)(- 100,- 210)
\thicklines
\put(- 30,40){\vector(1,0){ 60}}
\put(0,50){\makebox(0,0){$m =\ip{\psi}{\jj{J}{3}{0}}$}}
\put(- 90,- 30){\vector(0,-1){ 80}}
\put(- 100,- 70){\makebox(0,0){ n}}
\put(60,25){\makebox(0,0){$3$}}
\put(40,25){\makebox(0,0){$2$}}
\put(20,25){\makebox(0,0){$1$}}
\put(0,25){\makebox(0,0){$0$}}
\put(-20,25){\makebox(0,0){$-1$}}
\put(-40,25){\makebox(0,0){$-2$}}
\put(-60,25){\makebox(0,0){$-3$}}
\put(- 80,0){\makebox(0,0){$0$}}
\put(- 80,-20){\makebox(0,0){$-1$}}
\put(- 80,20){\makebox(0,0){$1$}}
\put(- 80,-40){\makebox(0,0){$-2$}}
\put(- 80,-60){\makebox(0,0){$-3$}}
\put(- 80,-80){\makebox(0,0){$-4$}}
\put(- 80,-100){\makebox(0,0){$-5$}}
\put(- 80,-120){\makebox(0,0){$-6$}}
\put(- 80,-140){\makebox(0,0){$-7$}}
\put(- 80,-160){\makebox(0,0){$-8$}}
\put(- 80,-180){\makebox(0,0){$-9$}}
\put(- 80,-200){\makebox(0,0){ $\vdots$}}
\put(0,0){\circle*{6}}
\multiput(- 20,- 20)( 20,0){3}{\circle*{6}}
\multiput(- 20,- 40)( 20,0){3}{\circle*{6}}
\multiput(- 20,- 60)( 20,0){3}{\circle*{6}}
\multiput(-40,- 80)( 20,0){5}{\circle*{6}}
\multiput(-40,- 100)( 20,0){5}{\circle*{6}}
\multiput(-40,- 120)( 20,0){5}{\circle*{6}}
\multiput(-40,- 140)( 20,0){5}{\circle*{6}}
\multiput(-40,- 160)( 20,0){5}{\circle*{6}}
\multiput(-60,- 180)( 20,0){7}{\circle*{6}}
\multiput(-60,- 200)( 20,0){7}{\makebox(0,0){$\vdots$}}
\put(5,- 5){\vector(1,-1){ 10}}
\put(25,- 50){\vector(1,-2){ 10}}
\put(45,- 135){\vector(1,-3){ 10}}
\put(22,- 7){\makebox(0,0){$\jj{I}{-}{1}$}}
\put(42,- 60){\makebox(0,0){$\jj{I}{-}{2}$}}
\put(62,- 150){\makebox(0,0){$\jj{I}{-}{3}$}}
\end{picture}
\caption[Weights of (0,1) representation of
$\widehat{L}SU(2)$]{\footnotesize Weights of representation $(j,k)= (0,1)$ of
$\widehat{L}SU(2)$}
\label{f:affinerepresentation}
\end{figure}

Finally, we will give a simple example of a representations of a loop
group.  Consider the group $\widehat{L} SU(2)$.  For $SU(2)$, weights
are given by a single integer $2j= \ip{\lambda}{h_+}=
\ip{\lambda}{2J_3}$.
The inner product $\gamma_m$ satisfies
$\gamma_m ( h_+, h_+ ) =2$
{}From (\ref{eq:constraint}) it follows that at level
$k$, the allowed highest weights $\lambda$ satisfy $j \leq \frac{k}{2}
$.  Thus, for $k = 1$, the only allowed representations of $\widehat{L}
SU(2)$ have $j = 0$ and $j = 1/2$.  Consider the representation with
$j = 0$.
For each $n$, we have an ${\rm su}(2)$ subalgebra of the current algebra
generated by
\begin{eqnarray}
\jj{I}{\pm}{n} & = & \jj{e}{{\mp}}{{\pm n}}, \label{eq:subalgebran}\\
\jj{I}{3}{n} & = &  \frac{n C}{2}  - \jj{h}{+}{0}/2. \nonumber
\end{eqnarray}
The eigenvalue of $\jj{I}{3}{n}$ on $v$ is $n/2$, so the operator
$\jj{I}{-}{n}^{n+ 1}$ must annihilate $v$.  Based on this observation,
we graph the weights of the representation $(j,k)= (0,1)$ in
Figure~\ref{f:affinerepresentation}\footnote{ Note that the weights of
this representation (called the fundamental representation) are
ostensibly given in \cite{PS} (Figure 3, Chapter 9; pp.  180).  The
weights in that figure are different from those shown here.  By
examining the character formula from \cite{PS} (Chapter 14; pp.  282),
it is clear that the weights given here are correct and those in the
reference are incorrect.  Apparently, the error in \cite{PS} is that
the inner product should be defined (in their notation) by $|| \mu
||^2 = \mu^2/2$, not by $||\mu||^2 = \mu^2$.}.  In general, highest
weight representations of $\widehat{L}G$ have this general structure.
The subspace with mode number 0 carries an irreducible representation
of the group $G$, and the subspaces with fixed mode number contain a
finite number of irreducible representations of $G$, with highest
weight vectors whose weights increase approximately as $\sqrt{n}$.  We
will discuss the detailed structure of these representations further
later in this thesis.

\subsubsection{The Virasoro group}
\label{sec:virasorogroup}

We will now consider our final example, the Virasoro group.  The
Virasoro group is the universal central extension of the group $\dif$
of smooth diffeomorphisms of the circle $S^1$.  Unlike the centrally
extended loop groups, the Virasoro group does not admit a locally 1-1
or onto exponential map from its Lie algebra.  We shall, however,
avoid this difficulty by analyzing a more well-behaved group which
has an equivalent algebraic structure.

The group $\dif$ is defined to be the group of all smooth
orientation-preserving diffeomorphisms $f$ of the circle $S^1$,
\begin{equation}
\dif =\{f: S^1 \stackrel{1:1}{\rightarrow} S^1| \; \; f \in C^\infty
(S^1, S^1)\}.
\end{equation}
The group multiplication is defined by the composition of maps, so $(g
\cdot f)(\theta)=f (g (\theta))$.  The Lie algebra of this group is
the algebra
$\vc$ of smooth vector fields on $S^1$.  Writing a vector field $f$ as
a function $f: S^1 \rightarrow \RR $, the Lie algebra product is given
by
\begin{equation}
[f (\theta),g (\theta)] = f (\theta) g' (\theta) - g (\theta) f' (\theta).
\end{equation}
Just as for central extensions of loop groups, it is possible to find
a single central extension $\vir$ of $\dif$ which is universal in the
sense that all other central extensions can be realized by taking a
quotient by a finite cyclic group.  This universal central extension
is the Virasoro group.  We denote its algebra by $\vvec$.  Elements of
$\vvec$ are of the form $(f, -i a)$, with $f(\theta) \pdv/ \pdv
\theta$ a vector field on $S^1$ and $a
\in \RR$.  The commutation relation between
elements of $\vvec$ is given by
\bge
[(f, -ia_1), (g, -ia_2)] = \left( f g' - gf', \frac{i}{48 \pi} \int_0^{2
\pi} (f(\theta) g'''(\theta) - g(\theta) f'''(\theta)) {\rm d} \theta
\right).\ee
Defining the (complex) vector fields $l_n = i {\rm e}^{i n \theta} \pdv /
\pdv \theta$ in $\vc$, we can define the usual Virasoro generators
by
\begin{eqnarray}
L_n & = & (l_n, 0); \;\;{\rm for} \; n \neq 0, \nonumber \\
L_0 & = & (l_0, \frac{1}{24}), \label{eq:vgen} \\
C & = & (0, 1). \nonumber \end{eqnarray}
The commutation relations then take the standard form
\begin{eqnarray}
[L_m, L_n] & = & (m - n) L_{m + n} + \frac{C}{12}(m^3 - m) \delta_{m
, - n},  \\
\lbrack C, L_n \rbrack & = & 0. \nonumber \end{eqnarray}
The Virasoro algebra is defined to be the complex Lie algebra spanned
by the generators (\ref{eq:vgen}).
As in the case of current algebras, it is traditional for physicists
to treat the operator $C$ as a c-number $c$,  whereas
mathematicians leave  $C$ in operator form and take $c$ to be its
eigenvalue in a specific representation.  For the purpose of
the coadjoint orbit description of representations, we will adhere to
the latter convention.

We will now briefly describe the adjoint and coadjoint action of
$\vvec$ on itself and its dual space.
The (smooth) dual space to $\vvec$
consists of pairs $(b, it)$, with $b(\theta) d \theta^2$ a quadratic
differential on $S^1$, and $t \in
\RR$.  The dual pairing between $(b, it)$ and an element $(f, -ia) \in
\vvec$ is given by
\bge
\langle (b, it), (f, -ia) \rangle = \int_0^{2 \pi} b(\theta) f(\theta)
d \theta + at.\ee
For this pairing to be invariant under the action of the algebra
$\vvec$, $(b, it)$ must transform under the coadjoint action by
\bge
\aads_{(f, -ia)} (b, it) = (2 b f' + b' f - \frac{t}{24 \pi} f''',
0).\label{eq:virasorocoadjoint}
\ee
Using these equations, it is possible to derive formulae  describing
the coadjoint action $\aads$ of the group $\vir$ on the dual space
$\vvec^*$; however, we will not need these formulae since the orbits
of the coadjoint action with which we will be concerned are
topologically trivial (see Section \ref{sec:coadjointexamples3}).

Unlike the centrally extended loop groups, the Virasoro group cannot
be described by the formalism of simple roots and the Cartan matrix.
In many ways, the Virasoro group is structurally similar to
finite-dimensional non-compact groups, while the loop groups are more
similar to compact groups.  Nonetheless, it is possible to make a
definition of positive and negative weights and roots for the Virasoro
algebra.  The Virasoro algebra has a natural decomposition
\begin{equation}
\vir =  (\CC C \oplus \CC L_0) \oplus V_+ \oplus V_-
\end{equation}
where
\begin{equation}
V_+ = \bigoplus_{n > 0} \CC L_n
\label{eq:Virasoropositive}
\end{equation}
and
\begin{equation}
V_- = \bigoplus_{n > 0} \CC L_{- n}
\end{equation}
are defined to be the spaces of positive and negative roots of $\vir$.
(Technically, these are the spaces of generators associated with the
positive and negative roots; we will simplify notation by referring to
these generators as themselves being the roots.  This will not lead to
confusion, since the roots and generators of this algebra are in a 1-1
correspondence,  just as is the case for compact groups.)
The operators $C$ and $L_0$ form a maximal commuting subalgebra of the
Virasoro algebra.  On any irreducible representation space $V$ of $\vir$,
the eigenvalue of $C$ is a constant  $c$, and the
representation space can be decomposed into subspaces with distinct
eigenvalues of $L_0$
\begin{equation}
V = \bigoplus_{n \in \ZZ} V_n
\label{eq:virasorodecomposition}
\end{equation}
where $L_0 v = n v$ for all $v \in V_n$.  We take $n$ to be the weight
of the space $V_n$; according to the definition
(\ref{eq:Virasoropositive}) of positive roots, a weight $n$ is
``positive'' when $n < 0$ (this unfortunate terminology arises from
the merging of well-established conventions in mathematics and physics
for highest weight representations and Virasoro generators).

At this point, one might be tempted to define $L_1$ to be the single
simple root of $\vir$.  Unfortunately, however, the generators
$\{L_0,L_{\pm 1}\}$ form a closed subalgebra of the Virasoro algebra,
isomorphic to the algebra of $SL (2, \CC)$.  Thus, it is impossible to
define the Virasoro algebra from this subalgebra with any choice of
Cartan matrix.  It is impossible to choose more than one simple root
for the Virasoro algebra, since then roots could not be uniquely
written as a linear combination of simple roots.  For our purposes,
however, the only result about simple roots which will be used in the
following development is the fact that the simple roots generate the
subalgebra of positive roots.  To this end, we can select the pair of
roots $L_1,L_2$ of the Virasoro algebra, and denote these roots to be
``quasi-simple'' roots.  These roots do indeed generate the algebra of
positive roots, which can be seen inductively by observing that
$[L_1,L_n]$ is nonzero and proportional to $L_{n + 1}$ for $n \geq 2$.
In the general analysis in the subsequent sections of this chapter,
whenever we are discussing a general Lie group and its simple roots,
we include under consideration the Virasoro group and its quasi-simple
roots $L_1,L_2$.

In discussing representations of the Virasoro group, we will again
restrict attention to a category of representations of particular
physical interest.  This category of representations consists of
highest weight representations which admit a unitary structure.  Since
some highest weight representations do not admit unitary structures,
this means that we will restrict attention to representations with a
certain set of values for the highest weight.

A highest weight representation of the Virasoro algebra $\vvec$ is
simply a representation which admits a decomposition
(\ref{eq:virasorodecomposition}) such that for some weight $h$ the
space $V_h$ is one-dimensional, and all spaces $V_{m}$ with $m< h$ are
zero-dimensional (recall the convention of positivity for the Virasoro
roots).  For all values of $h$ and $c$ ($c$ being the eigenvalue of
$C$), there exists a unique irreducible highest weight representation
of the Virasoro algebra.  These representations do not all admit
unitary structures, and cannot all be integrated to representations of
the Virasoro group $\vir$.  A theorem by Goodman and Wallach, however,
states that all unitary representations of the Virasoro algebra can be
integrated to a continuous unitary representation of the Virasoro
group \cite{GW}.  This theorem is based on a conjecture due to Kac
\cite{Kac2}.  Since we are actually only interested in unitary
representations, it will suffice to categorize representations of
$\vvec$ which admit unitary structures.

The set of values $(h,c)$ for which the Virasoro algebra admits an
irreducible unitary highest weight representation is a well-known
result to physicists.  We will simply quote the result here, although
in Section \ref{sec:global} we will briefly discuss the
formalism used for its proof.  The result is that unitary
representations break up into two categories.  In the first category
are all pairs $(h,c)$ with $c \geq 1$ and $h\geq 0$.  The second
category consists of the so-called ``discrete series'' of
representations, which have
\bge
c = 1 - \frac{6}{m(m+1)}, \label{eq:min1} \ee
and
\bge
h = \frac{[(m+1)p - mq]^2 - 1}{4 m (m+1)}, \label{eq:min2} \ee
for integers $m,p$ and $q$ satisfying $m > 2$ and $1 \leq q \leq p
\leq m-1$.

We conclude this subsection with a description of the geometry of the
Virasoro group and a set of coordinates on a quotient space analogous
to the homogeneous spaces $G/T$ for compact finite-dimensional groups.

As mentioned above, the Virasoro group has a poorly behaved exponential
map, which keeps us from defining coordinates in a neighborhood of the
origin as we did in the cases of compact groups and loop groups using
(\ref{eq:generalelement}).  The failure of the exponential map to be
1-1 and onto follows from two important features of this group, which
we will now describe, following Milnor \cite{Milnor}.  To begin with,
we note that because the extension of the diffeomorphism group $\dif$
is central, the failure of the exponential map must also occur for the
unextended group.  Thus, we will focus attention here  on
the unextended group.

The first aspect of the full diffeomorphism group which gives problems
for the exponential map is the inclusion of diffeomorphisms which have
no fixed-points.  From the existence of such diffeomorphisms in the
group, we can prove that the exponential map is neither locally 1-1 or
onto.  To see that the map is not locally 1-1, consider the
diffeomorphism $\phi: S^1 \rightarrow S^1$ given by $ \phi (\theta)=
\theta +2 \pi/n$ where $n$ is an integer.  For any vector field $f
\in\vc$ which is nowhere zero ($f (\theta)\neq 0 \; \forall \theta:0
\leq \theta < 2 \pi$) and which is periodic with frequency $n$ ($f
(\theta + 2 \pi/n)= f (\theta)$), there must exist a real number $t \in
\RR$ such that
\begin{equation}
\phi (\theta) = {\rm e}^{t f} (\theta).
\end{equation}
This follows because ${\rm e}^{tf}(\theta)$ satisfies
\begin{equation}
\frac{\partial}{ \partial  s}  {\rm e}^{sf} (\theta) =
f ({\rm e}^{sf}(\theta)),
\end{equation}
so
\begin{equation}
{\rm e}^{tf} (\theta) = \theta + \int_{0}^{t} {\rm d}
s f ({\rm e}^{sf}(\theta)),
\end{equation}
and when $f$ is periodic with frequency $n$ it suffices for
\begin{equation}
\int_{0}^{t} {\rm d} s f ({\rm e}^{sf}(\theta)) = \frac{2 \pi}{n}
\label{eq:condition1}
\end{equation}
to hold for any value of $\theta$ to guarantee that the condition
(\ref{eq:condition1}) holds for all values of $\theta$.
It follows that the periodic diffeomorphism $\phi$ lies on the
one-parameter families corresponding to the trajectories under the
exponential map of an infinite number of elements of the Lie algebra;
thus, the exponential map is $\infty$-1 at each point $\phi$.  Since
by choosing $n$ large, we can make the diffeomorphism $\phi$
arbitrarily small, the exponential map is certainly not locally 1-1.

In a similar fashion, we can use the existence of diffeomorphisms
without fixed-points to show that the exponential map is not
surjective.
Consider the diffeomorphism
\begin{equation}
\phi (\theta) = \theta + \frac{2 \pi}{n}  + \epsilon \sin^2 (n \theta),
\end{equation}
where $\epsilon \ll 1/n$.
The diffeomorphism $\phi^n$ clearly has fixed-points at $\theta =
\pi k/n$ for integral $k$, however no other points are fixed under
$\phi^n$.  Because
$\phi$ has no fixed points, if $\phi = {\rm e}^{f}$ for some vector
field $f$ it follows that $f$ can have no zeros.  But by the same
argument as above, if $f$ has no zeros and ${\rm e}^{nf}(0)= 0$ then
${\rm e}^{nf}(\theta)=
\theta$ for all $\theta$.  It follows that $\phi$ is not in the image
of the exponential map.  Since $\phi$ can be taken arbitrarily small
by taking $n$ arbitrarily large and $\epsilon$ arbitrarily small, it
follows that the exponential map on $\vc$ is not locally surjective
onto $\dif$.  (Actually, here we must be careful about our definition
of local; technically, $\dif$ as a group is modeled on a Frechet space
-- see below.  In fact, however, it is still true that any
neighborhood of the identity in $\dif$ contains some diffeomorphism of
the form of $\phi$.)

We see then, that the existence of diffeomorphisms without
fixed-points in the diffeomorphism group makes the exponential map
from the algebra of vector fields rather poorly behaved.  In fact,
still more problems arise from the inclusion of non-analytic vector
fields; this is the second problem referred to above.  As an example,
take $f$ to be a vector field which vanishes with all derivatives at
$\theta = 0$, but which is nonzero in a region $\pi/2 < \theta < 3
\pi/2$.  The constant vector field $I = - i l_0 = 1$ can be exponentiated to
form the rotation diffeomorphism $R_\psi$ satisfying $R_\psi (\theta)=
\theta + \psi$.  If the Lie group $\dif$ had a well-behaved real
analytic structure, the adjoint action of the group on the algebra
which gives
\begin{equation}
(R_\psi f R_{- \psi}) (\theta) = f (\theta + \psi)
\label{eq:exponential}
\end{equation}
could be described by an exponentiation of the adjoint action of the
algebra on itself
\begin{equation}
[I,f] = {\rm ad}_I f = f'.
\end{equation}
However, since $f$ and all its derivatives vanish at
$\theta = 0$ it follows that
\begin{equation}
{\rm e}^{\psi \; {\rm ad}_I} f (0) = 0 \; \; \forall \psi,
\end{equation}
which contradicts (\ref{eq:exponential}).  Thus, the analytic
structure of $\dif$ is further disrupted by the existence of
non-analytic diffeomorphisms.

In order to accomplish the coadjoint orbit construction of
representations of the Virasoro group, we will need to have
coordinates on a homogeneous space of the form $\di$.  Although by the
arguments above, such a coordinate system cannot directly be
constructed using a formula like (\ref{eq:generalelement}), it is
possible to proceed formally as though such a formula were applicable.
This approach has been successfully used in the past, and leads to
correct results for the curvature and other properties of the space
$\di$ \cite{Zum} (the curvature result was first obtained by
Bowick and Rajeev using a generalization of a method of Freed
\cite{BR,fr}).
We will make some attempt to justify the use of this type of
coordinate system here, by considering a closely related group which
is more nicely behaved.  Finally, however, the justification for the
formulas we will derive is that they successfully give algebraic
representations of the form expected from the coadjoint orbit
construction.  A more rigorous demonstration of the validity of the
methods used here is left as a project for further research.

We now sketch an argument for the validity of formula
(\ref{eq:generalelement}) on a group with an algebra isomorphic to the
Virasoro algebra.  To begin with, the homogeneous space $\di$ is the
space of the infinite-dimensional manifold of the group $\dio$ of
diffeomorphisms of $S^1$ which have a fixed point at $\theta = 0$.  By
restricting to this group, the problems associated with
diffeomorphisms without fixed-points are removed.  The only obvious
difficulty which remains is the existence of non-analytic
diffeomorphisms.  For our purposes, and as far as most physicists are
concerned, we can restrict to the subgroup of $\dio$ of real-analytic
diffeomorphisms, which we denote $D_0$.  This group has none of the
problems described above for its exponential map; however, it is still
not clear whether this group admits a real-analytic structure.  We
will now show that a closely related group with an isomorphic algebra
can be given a real-analytic structure; we will denote this group by
$\vv$, and we will show that $D_0 \subset\vv$.
Although this result does not prove conclusively that $D_0$ admits a
real-analytic structure, it indicates the plausibility of this
assertion.  For those purists unwilling to accept this statement
without a more rigorous proof, the discussion in the rest of this
chapter should be taken to apply to the formal group $\vv$, with the
homogeneous space $\di$ being the formal quotient of $\vv$ through the
action of $L_0$.

We define the group $\vv$ to be the set of all formal power series in
$x$ with leading term  proportional to $x$ and leading coefficient positive,
\begin{equation}
\vv = \{f =f_1 x + f_2 x^2 +  \cdots \in x \RR[[x]]:f_1 > 0\}.
\end{equation}
We can define a topology on the space of $\vv$ by treating $\vv$ as a
subspace of the vector space $ A =x\RR[[x]]$.  We take a local base around
$0 \in A$ for the topology of $A$ to be given by the
countable family of open sets
\begin{equation}
B (n,m) =\{f \in\vv:f_i < \frac{1}{n}\forall i \leq m \},
\end{equation}
where $n$ and $m$ are integers.  The topology on $\vv$ is then that
induced by the embedding in $A$.
Since the topology thus defined on
$\vv$ has a countable, convex local basis, it follows that $\vv$ has a
metric invariant under linear translations on the vector space
\cite{Rudin}.  It can be shown that $\vv$ is complete in this metric,
so $\vv$ is a Frechet space.  It is also possible to show that $A$ has
no bounded neighborhood of 0 with the resulting metric, so $\vv$
cannot be a Banach space.  The technical details of what type of space
$\vv$ is modeled on will not concern us here; however, it is
significant that $\vv$ is modeled on a Frechet space, since the
essential problem here is to show that a group modeled on a Frechet
space can have a real-analytic structure.

Having defined a topology on $\vv$, we can now define a group structure
given by composition.  Given elements $f = f_1 x + f_2 x^2 + \cdots$
and $g = g_1 x + g_2 x^2 + \cdots$, we define $g\cdot f \in\vv$ by
\begin{equation}
(g \cdot f) (x) = f (g (x)) = c_1 x + c_2 x^2+ \cdots,
\end{equation}
where
\begin{equation}
c_k = \sum f_s g_{l_1} g_{l_2} \cdots g_{l_s},
\end{equation}
with the sum taken over all $s, l_1, \ldots l_s > 0 $ with $k =
l_1 +\cdots + l_s$.

The inverse can be computed; if $g\cdot f = 1 $, then
\begin{eqnarray}
f_1 & = & 1/g_1  \\
f_k & = &- g_1^{-k} \left[\sum_{s < k} f_s g_{l_1} \cdots g_{l_s}\right]\; \;
\; k > 2, \nonumber
\end{eqnarray}
with the sum taken over the same range as before, however with the
additional restriction that $s < k$.

{}From these explicit expressions, it is easy to verify that
multiplication and inverses are continuous in the chosen topology, so
$\vv$ is a Lie group modeled on a Frechet space.
Furthermore, we can define a smooth ``square root'' operation in $\vv$.
If $f = g^2$, then we have
\begin{eqnarray}
g_1 & = &(f_1)^{1/2} \\
g_k &= &( f_k - \sum_{1 < s < k} g_s g_{l_1} \cdots g_{l_s}) / (g_1 +
g_1^k)\; \; \; k > 2, \nonumber
\end{eqnarray}
where the sum is again as before, except now we restrict $1 < s < k$.

Now, we can prove the further claim that the exponential map from the
Lie algebra of $\vv$ to $\vv$ is 1-1 and onto (globally).  The Lie
algebra of $\vv$ is exactly the space $A$, with the Lie
bracket of two vector fields  $u = u_1 x + u_2 x^2 + \cdots$ and $v =
v_1 x + v_2 x^2 + \cdots$ given by
\begin{equation}
[u,v] = w = w_1 x + w_2 x^2 + \cdots,
\end{equation}
with
\begin{eqnarray}
w_1 & = & 0\\
w_k & = &  \sum_{j = 1}^{k} j (u_{k - j+ 1} v_j - u_j v_{k - j + 1}).
\nonumber
\end{eqnarray}
The Lie bracket
operation on the algebra is clearly smooth in the same topology we had
before.

The proof that $\exp:A \rightarrow\vv$ is 1-1 and onto essentially
follows from iterating the square root map above.  If we define
$f^{(n)}$ by $f^{(0)} = f$, $f^{(n-1)} = f^{(n)} \cdot f^{(n)}$, then
by iterating the square root map we can show that  the limits
\begin{eqnarray}
g_1 & = &  \lim_{n \rightarrow \infty}\left(2^n (f_1^{(n)} - 1)\right)\\
g_k & = &  \lim_{n \rightarrow \infty}\left(2^n f_k^{(n)}\right) \; \;
\; k > 2\nonumber
\end{eqnarray}
exist.  We  then define $g = g_1 x + g_2 x^2 +\cdots $, and have ${\rm
e}^g = f$.
Since each function $f$ has a unique ``logarithm'' $g$, the
exponential map is 1-1 and surjective.

We have thus shown that the Lie group $\vv$, which is
modeled on a Frechet space, has a 1-1 and onto exponential map.  The
Lie group $D_0$ can be identified with the subgroup of $\vv$
consisting of power series which are a) convergent with all
derivatives for all $x$, b) periodic in $x$ in the sense that $f (x +
2 \pi)= f (x)+ 2
\pi$, and c) monotonically increasing in $x$ ($f' (x)> 0 \; \forall
x$).  The topology on $D_0$ induced by this embedding is different
from the usual topology, however the group operation is smooth under
both definitions.
In order to prove that the exponential map on $D_0$ is 1-1 and onto,
it would suffice to demonstrate that the square root operation defined
above for $\vv$ is closed on $D_0$.  In fact, we assert that it
suffices to show that if $f \in D_0 \subset\vv$ and $f = g \cdot g$ in
$\vv$ then condition (a) holds for $g$.  That is, if $g$ is convergent
for all $x$ conditions (b) and (c) on $g$ follow from the same
conditions on $f$.  A brief outline of this argument will now be given
to conclude this subsection.

Assume that $f$ and $g$ are as above, with $f = g^2$ in $\vv$ and $f$
satisfying conditions (a), (b), (c), and with $g$ satisfying condition
(a).  It follows from the fact that $g$ satisfies the differential
equation $g'(g(x)) = f'(x) / g'(x)$, that if $g$ has a bounded
continuous derivative, then $g$ is monotone increasing.  It is then
fairly easy to see that the following conditions must hold on $g$:
\begin{eqnarray}
f(x) > g(x) > x, & \;{\rm when}\;& f(x) > x  \nonumber\\
f(x) = g(x) = x, & \;{\rm when}\;& f(x) = x\\
f(x) < g(x) < x, & \;{\rm when}\;& f(x) < x. \nonumber
\end{eqnarray}
To see that these conditions must hold, assume that $x_0,x_1$ are
consecutive fixed-points of $g$ (and therefore also of $f$).  If $g
(x)> x$ for $x_0 < x < x_1$ then $g (g (x)) = f (x)> g (x)> x$ since
$g (x)< x_1$, which follows from the monotonicity of $g$.  The other
conditions can be proven in a similar fashion.  From these
inequalities, we see that $G(x) = g(x+2 \pi) - 2 \pi$ is also a
solution to $G(G(x)) = f(x)$.  If the solution $g$ is unique, which it
is in $\vv$, then $g = G$, and $g$ is periodic.  We have thus shown
everything except that the square root of an element of $D_0$ must be
convergent.  The exponential map on $D_0$ is thus at least 1-1;
however, the surjectivity of this map has not been shown rigorously.
To be certain of working with a group with the BCH property, one must
consider $\vv$; in the remainder of this thesis, however, we will
ignore the details of this question and simply calculate with the
Virasoro group and algebra as though it were well-behaved.

\subsection{Coadjoint orbit construction}

This section contains a description of the geometry of coadjoint
orbit representations.  The construction of coadjoint orbit
representations is originally due to Kirillov and Kostant \cite{Kir}.
The discussion here is similar to the presentations in
\cite{Kir,Witt1}; however, the notation is slightly different; in
particular, some difficulties with signs are dealt with here in an
internally consistent fashion.

As described in Section 1.1, given a Lie group $G$ with algebra
$\algg$, $G$ acts on the dual space $\algg^*$ by the coadjoint action
(\ref{eq:coadjointaction}).  For any $b \in \algg^*$, one can consider
its orbit $W_b$ in $\alggs$ under the coadjoint action of $G$.  It turns
out that $W_b$ admits a natural symplectic structure, which may be
defined as follows: There is a natural association between elements of
$\algg$ and tangent vectors to $\wb$ at $b$.  Given an element $u \in
\algg$, we define $\tu (b) \in T_b \wb$ to be the tangent vector to
$\wb$ at $b$ associated with $\ads_u b$.  (Note that $\tu(b) = 0$ when
$u$ is in the stabilizer of $b$; {\it i.e.}, when $\ads_u b = 0$.)  We can
define a 2-form $\omega$ on $\wb$ by
\bge\omega(\tu(b),\tilde{v}(b)) = \ip{b}{ [u,v] }.
\label{eq:omega} \ee
It can be verified that this 2-form is well-defined, closed,
$G$-invariant, and nondegenerate, and thus defines a $G$-invariant
symplectic structure on $W_b$.  $\omega$ also gives a Poisson bracket
structure to the space of functions on $\wb$.  In component notation,
the Poisson bracket of two functions $f$ and $g$ is given by
\bge
\lbrace f,g \rbrace = \omega^{ij} (\pdv_i f)(\pdv_j g),\ee
where $\omega^{ij}$ are the components of $\omega^{-1}$.  Every
function $f$ on $\wb$ generates a Hamiltonian vector field $v_f$ on
$\wb$, defined by
\bge v_f^i = \omega^{ij} (\pdv_j f).\ee
For any $u \in \algg$, there is a function $\pu$ on $W_b$ which generates
the Hamiltonian vector field $\tu$.  This function is given by
\bge
\pu(b) = - \ip{b}{ u } .\ee
To see that $\pu$ generates the vector field $\tu$, we use the fact
that for any $v \in \algg$,
\bge
\tilde{v}^j\pdv_j \pu(b) = - \ip{b}{ [u,v] } = \omega_{jk}
\tilde{v}^j \tu^k.\ee
Since the vector fields $\tv$ span the tangent space to $\wb$ at $b$,
we have
\bge
\pdv_j \pu(b) = \omega_{jk} \tu^k (b),\ee
so
\bge  \omega^{ij}\pdv_j \pu(b) = \tu^i.\ee
The functions $\pu$ also satisfy the equation
\bge
\lbrace \pu, \Phi_v \rbrace = \Phi_{[u,v]},
\label{eq:poiss} \ee
since
\begin{eqnarray}
\lbrace \pu, \Phi_v \rbrace & = & \omega^{ij} (\pdv_i \pu) (\pdv_j
\Phi_v)
= \omega^{ij} (\omega_{ik} \tu^k)( \omega_{jl} \tilde{v}^l) \nonumber \\
& = & \omega_{ik} \tu^k \tilde{v}^i
 =  \ip{b}{ [v,u] }
 =  \Phi_{[u,v]}.
\end{eqnarray}
In order to construct representations of $G$ using the coadjoint orbit
$\wb$, it is now necessary to quantize the manifold $\wb$ according to
the technique of geometric quantization \cite{Wood,Sni}.  The first
step in this procedure is to construct a complex line bundle $\lb$
over $\wb$ with curvature form $i \omega$.  This is known as
``prequantization''. For this step to be possible, it is necessary
that $\frac{\omega}{2\pi}$ be an integral cohomology class ({\it i.e.}, that
the integral of $\omega$ over any closed 2-surface in $\wb$ be an
integral multiple of $2\pi$.)  If such a line bundle $\lb$ exists,
then there is a natural homomorphism $\phi$ from the Lie algebra $\algg$
to the space of first-order differential operators on sections of
$\lb$, given by
\bge
\phi : u \mapsto \hu = -\nabla_{\tu} +i \pu,\label{eq:algebraaction}\ee
where $\nabla_{\tu}$ is the covariant derivative in $\lb$ in the
direction $\tu$.  Explicitly, written in component notation in a local
coordinate chart,
\bge
\hu = -\tu^i(b)(\pdv_i + A_i(b)) +i \pu(b),
\label{eq:oprep}
\ee
where $A_i$ is a connection on $\lb$ satisfying $\pdv_i A_j - \pdv_j
A_i = i \omega_{ij}$.  To verify that $\phi$ is a homomorphism, we
must check that
\bge
\lbrack \hu, \hv \rbrack = \widehat{\lbrack u,v \rbrack}.
\label{eq:comm}
\ee
We define $\xi_u$ to be the differential operator corresponding to the
vector field $-\tu$; {\it i.e.}, $\xi_u = -\tu^i \pdv_i$, and we define
$A_u = \tu^i A_i$.  With these definitions,
\bge
\hu = \xu - A_u +i \pu.\label{eq:oprep2} \ee
One can easily calculate
\bge
\lbrack \xu, \xi_v \rbrack = \xi_{[u,v]},\ee
and
\bge  \xu \Phi_v(b) = \Phi_{[u,v]}(b).\ee
One also finds that
\begin{eqnarray}
\xu A_v - \xi_v A_u & = & \widehat{[u,v]}^i A_i - \tu^i \tilde{v}^j
(\pdv_i A_j - \pdv_j A_i)  \label{eq:xacomm}\\
& = & A_{[u,v]} + i \Phi_{[u,v]}. \nonumber \end{eqnarray}
Note that since the vectors $\tu$ span the tangent space to $\wb$ at
each point, Equation \ref{eq:xacomm}, along with the conditions that
$A_u$ is linear in $u$ and that $A_u(b) = 0$ when $\ads_u b = 0$,
could have been taken as the definition of a connection $A_u$
associated with the derivative operators $\xu$.  It is now trivial to
compute the commutator
\begin{eqnarray}
[\hu, \hv] & = & [\xu - A_u +i \pu, \xi_v - A_v +i \Phi_v] \nonumber \\
& = & \xi_{[u,v]} - A_{[u,v]} +i \Phi_{[u,v]}  \\
& = & \widehat{[u,v]}. \nonumber
\end{eqnarray}
Thus $\phi$ is a
homomorphism, so we have determined that $\phi$ gives a representation
of $\algg$ on the space of smooth sections of $\lb$.  Unfortunately,
this representation is in general much too large to be irreducible;
this is where the second stage of geometric quantization enters,
which involves choosing a ``polarization''.  We will only be concerned
here with a specific type of polarization, the \kl polarization.  In
general, choosing a polarization restricts the space of
allowed smooth sections of $\lb$ to a subspace containing only those
sections which satisfy some local first-order differential equations.
A \kl polarization of $\wb$ exists when $\wb$ admits a $G$-invariant
\kl structure with $\omega$ as the associated $(1,1)$-form.  This
condition is equivalent to the condition that $\wb$ admits a
$G$-invariant complex structure with respect to which $\omega$ is a
$(1,1)$-form; {\it i.e.}, the only nonvanishing terms in $\omega$ have
one holomorphic and one antiholomorphic index.
In general, if $\wb$ does not admit a \kl polarization, and is not
equivalent to a cotangent bundle, there is no standard way to find a
polarization, and carrying out the geometric quantization program
becomes extremely difficult.  In case $\wb$ does admit a \kl
polarization, we can restrict the space of allowed sections of $\lb$
to the space $\hb$ of holomorphic sections.  When $\lb$ has a
Hermitian metric, then we can further restrict $\hb$ to be the Hilbert
space of square-integrable holomorphic sections of $\lb$.  According
to the general principles of Kirillov and Kostant, the action of $G$
on $\hb$ should give an irreducible unitary representation of $G$ for
every $b$ such that $\hb$ can be constructed.  This principle holds
fairly well for compact semi-simple finite-dimensional groups, and
even for loop groups; however, it does not seem to hold in complete
generality.  Some of the representations of $\vir$ constructed this
fashion are nonunitary, and some are reducible, as we demonstrate
below.  In the case of finite-dimensional compact simple Lie groups,
this construction is equivalent to the Borel-Weil construction
outlined in Section 1.1, and the K\"{a}hler structure compatible with
$\omega$ is equivalent to the complex structure defined in
(\ref{eq:generalelement}) for the homogeneous space $G/T$.

\subsection{Gauge fixing}

In this section, we prove several assertions
which will simplify the process of explicitly constructing the
coadjoint orbit
representations in local coordinates.  If one attempts to use Equation
\ref{eq:oprep2} to construct explicit formulae for the operators $\hu$
as differential operators on $\hb$, one encounters several obstacles.
First, it is necessary to calculate the functions $\pu$ in local
coordinates.  Second, one must find an explicit formula for
a connection $A_u$ which satisfies (\ref{eq:xacomm}).  Finding these
expressions in terms of a local set of holomorphic coordinates is in
general a somewhat nontrivial problem.  Note, however, that the
operator $\hu$ can be written as
\bge
\hu = \xu + f_u,\ee
where $\xu$ is the first-order differential operator defined above,
and $f_u$ is a function of the local coordinates satisfying
\bge
\xu f_v - \xi_v f_u = f_{[u,v]}.\label{eq:xfcomm} \ee
We will find it easiest to construct explicit expressions for the
operators $\hu$ by finding directly a set of functions $f_u$ which
satisfy (\ref{eq:xfcomm}), and which correspond to the representation in
question.  We find these functions $f_u$ by making a simplifying
assumption which amounts to choosing a simple gauge for the connection
$A_u$.  To ensure that the set of $f_u$'s we construct in this fashion
are equivalent to those we would get from (\ref{eq:xacomm}) by a
specific choice of gauge, we will need the following two propositions.

\begin{prop}
Given a coadjoint orbit $\wb$ of a group $G$, with $\lb$ a complex line bundle
over $\wb$ with curvature $i \omega$, and with $\xu$ and $\pu$ defined
as above, on a coordinate chart corresponding to a
local trivialization of $\lb$, if a set of functions $f_u$ on $\wb$
are linear in $u \in \algg$, and satisfy the conditions
\vspace{.1in} \newline
\makebox[.3in][r]{{\rm (}{\sl i}\/{\rm )}} \hspace{.08in} \parbox[t]{5.4in}
{$\xu f_v - \xi_v f_u = f_{[u,v]}$,}
\vspace{.1in} \newline
\makebox[.3in][r]{{\rm (}{\sl ii}\/{\rm )}} \hspace{.08in} \parbox[t]{5.4in}
{$f_u(b) = i \pu(b)$ when $\ads_u b = 0$,}
\vspace{.1in} \newline
then the operators $\hu = \xu + f_u$ are equal to the operators $\hu$
from Equation \ref{eq:oprep2} for some choice of connection $A_u$ on
$\lb$ satisfying {\rm (}\ref{eq:xacomm}{\rm )}.
\label{p:p1}
\end{prop}
\noindent {\it Proof.} \,\,
To prove this proposition, it will suffice to show that the functions
$A'_u(b) = - f_u(b) + i \Phi_u(b)$ satisfy (\ref{eq:xacomm}), are linear
in $u$, and are zero when $\ads_u b = 0$.  The last two conditions
follow immediately from the definition of $f_u$ and assumption ({\sl
ii}).  To see that $A'_u$ satisfies (\ref{eq:xacomm}) is a simple
calculation:
\begin{eqnarray}
\xu A'_v - \xi_v A'_u & = & \xu (- f_v + i\Phi_v) - \xi_v (- f_u + i \pu)
\nonumber \\
& = & - f_{[u,v]} + 2 i \Phi_{[u,v]}  \\
& = & A'_{[u,v]} + i \Phi_{[u,v]}.  \nonumber
\end{eqnarray}
Thus, $A'_u$ is a valid connection on $\lb$, and the proposition is proven.
$\Box$

\begin{prop}
With the same premises as Proposition \ref{p:p1}, when $G$ is path
connected the condition {\rm (ii)} can be replaced by the weaker condition
\vspace{.1in} \newline
\makebox[.3in][r]{{\rm (}{\sl ii$'$}\/{\rm )}} \hspace{.08in}
\parbox[t]{5.4in}
{For some point $b_0 \in \wb$, $f_u(b_0) = i \pu(b_0)$ for all $u$
such that $\ads_u b_0 = 0$,}
\vspace{.1in} \newline
and the result of proposition \ref{p:p1} still holds.
\label{p:p2}
\end{prop}
\noindent {\it Proof.} \,\,
We need to prove that when
$G$ is path connected, condition ({\sl ii$'$}) implies condition ({\sl
ii}).  Assume $\ads_u b = 0$ for some $u \in \algg, b \in \wb$.  Since
$b_0 \in \wb$, for some $g \in G$ we have $b = \aadsg b_0$.  If $u$
stabilizes $b$, then $u_0 = \aad_{g^{-1}} u$ must stabilize $b_0$.
But then we have
\bge
\ip{b}{ u} = \ip{\aadsg b_0 }{\aadg u_0} = \ip{b_0 }{u_0},\ee
so $\pu(b) = \Phi_{u_0}(b_0)$.  It remains to be shown that $f_u(b) =
f_{u_0}(b_0)$.  Since $G$ is path connected, we have a path $g(t)$ in
$G$ with $g(0) = 1$ and $g(1) = g$.  We
claim that
\bge
\frac{d}{dt} f_{u(t)}(b(t)) = 0,\ee
where $u(t) = \aad_{g(t)} u_0$, and $b(t) = \aads_{g(t)} b_0$.
Defining
\bge v(t) = \frac{d g(t)}{d t} g^{-1}(t) \in \algg,\ee
we have
\bge
\frac{d}{dt} b(t) = \ads_v b(t),\ee
and
\bge \frac{d}{dt} u(t) = \ad_v u(t).\ee
It follows that
\begin{eqnarray}
\frac{d}{dt} f_{u(t)}(b(t)) & = & -\xi_v f_{u(t)} (b(t)) +
f_{[v,u(t)]}(b(t)) \nonumber \\
& = & -\xi_v f_{u(t)} (b(t)) + \xi_{u(t)} f_{v} (b(t)) +
f_{[v,u(t)]}(b(t))  \\
& = & 0, \nonumber
\end{eqnarray}
where we have used the fact that $\tu(t)(b(t)) = 0$.  Thus, we have
shown that
\bge f_u(b) = f_{u_0}(b_0) = i \Phi_{u_0}(b_0) = i \pu(b).\ee
Since $u$ and $b$ were an arbitrary solution of $\ads_u b = 0$, we
have proven that condition ({\sl ii$'$}) implies condition ({\sl
ii}), and thus the proposition is proven.$\Box$

\subsection{Local formulae}
\label{sec:localformulae}

In this section, we derive a set of general formulae for the local
realization of a coadjoint orbit representation in a neighborhood of a
point $b$ in the coadjoint orbit $W_b$.  This is done by first
computing an exact local formula for the vector fields $\xi_u$ and
then performing a gauge-fixing to derive a general expression for a
set of functions $f_u$ satisfying the conditions of Proposition 2.
For finite-dimensional compact groups, the resulting formulae give a
representation of the Lie algebra $\algg$ in terms of first-order
differential operators acting on a ring $R$ of polynomials in a finite
number of complex variables.  When the group $G$ is an
infinite-dimensional group, such as a centrally extended loop group or
the Virasoro group, the realizations are again in terms of first-order
differential operators acting on a ring $R$ of polynomials in a set of
complex variables; however, the set of complex variables becomes
infinite, and the differential operators are described by infinite
series, with only a finite number of terms acting nontrivially on any
fixed polynomial in $R$.  One important feature of these explicit
realizations of Lie algebras in terms of differential operators is
that the differential operators associated with raising operators
$J_\alpha, \alpha \in \Phi_+$ in the Lie algebra are independent of
which representation of the algebra is being realized.  The
independence of these operators from the highest weight of the
relevant representation will be used in the following chapters to
simplify formulae for conformal field theory observables calculated
using these representations.

The first step in finding a local realization of the coadjoint orbit
representations is to choose a set of coordinates on the orbit space
$W_b$ in the vicinity of $b$.  In Section \ref{sec:groups} we
described such a set of coordinates on quotient spaces of the form
$G/T$ where $T$ is a maximal abelian subgroup of $G$, for all types of
groups under consideration in this thesis.  A given coadjoint orbit
space $W_b$ is homeomorphic to a quotient space of $G$ of the form
$G/S$ where $S$ is the stabilizing subgroup of $b$ in
$G$.  (The stabilizer of a point $p$ in a space $V$ which carries an
action of the group $G$ is the subgroup of $G$ which leaves $p$
invariant.)  All the coadjoint orbit spaces which we will consider
here arise from elements $b \in \algg^*$ whose stabilizer is a maximal
abelian subgroup $T$; thus, these coadjoint orbits are homeomorphic to
quotient spaces of the form $G/T$.  We will describe these coadjoint
orbits more explicitly for specific Lie groups in Section
\ref{sec:coadjointexamples}.  In general, however, we can use the
coordinates $\{z_\alpha: \alpha \in \Phi_+\}$ associated with the
positive roots of a group $G$, which are defined through
(\ref{eq:generalelement}), as complex coordinates on the coadjoint
orbit spaces of interest.

Once we have a set of coordinates on $W_b$, we can use explicit
formulae for the coadjoint action to calculate the value of the
symplectic form $\omega$ associated with the coadjoint orbit $W_b$ in
terms of the local coordinates $z_\alpha$.  As discussed above,
the form $\frac{\omega}{2\pi}$ must be an integral cohomology class in
order for the coadjoint orbit to admit a holomorphic line bundle $\lb$
with curvature $i \omega$.  For all the groups with which we are
concerned in this thesis, there is a set of coadjoint orbits which
have a symplectic form $\omega$ satisfying this condition.  These
coadjoint orbits have the additional property that with respect to the
complex structure described by the coordinates
(\ref{eq:generalelement}), the symplectic form $\omega$ is a $(1,1)$
form.  Thus, on all these coadjoint orbits we can take a K\"{a}hler
polarization of the space of sections of $\lb$ by restricting to the
space $\hb$ of holomorphic sections.  Locally, the holomorphic
sections of $\lb$ are described around the point $b$ by polynomials in
the variables $z_\alpha$.  The action (\ref{eq:algebraaction}) of
the Lie algebra on the space of holomorphic sections reduces to a
representation of $\algg$ in terms of first-order differential
operators $\hat{u}$ on the space of these polynomials.  The remainder
of this section is devoted to the derivation of a general formula for
these differential operators.

For the rest of this section, we assume that a specific group $G$ has
been chosen, and that a particular coadjoint orbit $W_b$ satisfying
the necessary conditions for quantization has also been selected.  We
will assume that the group $G$ has a set of simple roots which
generate the subalgebra associated with positive roots; as described
in \ref{sec:virasorogroup}, when $G$ is the Virasoro group, the set of
simple roots actually refers to the set $\{L_1,L_2\}$ of quasi-simple
roots, which also generates the subalgebra of positive roots.  We will
use a combination of physical and mathematical notation for the
generators of the algebra $\algg_\CC$; we write all generators in the form
$J_a$, where $a$ can either be a root $a = \alpha \in \Phi$, or an
element $a = h\in \Theta$ of the basis $\Theta =\{h_\alpha |
\alpha \in \Delta\}$ of the Cartan subalgebra.  When $J_a$ is in the
Cartan subalgebra, we write $a = h\approx 0$, since $a$ corresponds to a
weight of 0.  We use the physics notation $\mixten{f}{ab}{c}$
for algebra structure constants, so that
\begin{equation}
[J_a,J_b] = i \mixten{f}{ab}{c} J_c.
\end{equation}
In particular, in this notation we have
\begin{equation}
i\mixten{f}{h \alpha}{\alpha} = \ip{\alpha}{h} \; \; \; {\rm for}
\;\alpha \in \Phi, h \in \Theta.
\end{equation}
All the equations in this section could be rewritten in terms of a
Chevalley basis and Cartan matrix; however, this does not seem to
simplify the form of the results.

\subsubsection{Vector fields}

The first step in constructing the operators
\begin{equation}
\hat{J}_a = \xi_a + f_a
\end{equation}
which implement the action (\ref{eq:algebraaction}) of $\algg$ on
${\cal H}_b$ is to calculate the vector fields
\begin{equation}
\xi_a = - \tilde{u}_a^{\alpha} \partial/\partial z_\alpha
\label{eq:holomorphicfields}
\end{equation}
associated with the coadjoint action $\ads_{J_a}$ of $J_a$ on $W_b$.
Actually, these are vector fields in the complexification of the
tangent space to $\wb$, since the generators $J_a$ are in $\algg_\CC$
and not necessarily in $\algg$.  The vector fields in the actual
tangent space are constructed by taking the complex linear
combinations of $J_a$ in $\algg$; since the left action of $G$ on
$W_b$ leaves the complex structure invariant, we are only interested
in the holomorphic parts of the vector fields $ \xi_a$.  We are
constructing operators which will act on polynomials in the
holomorphic variables $z_\alpha$, so by taking the holomorphic vector
fields (\ref{eq:holomorphicfields}) we actually will construct a
representation of $\algg_\CC$ on the space of polynomials in ${\cal
H}_b$.

The holomorphic vector fields $\xi_a$ can be calculated by
multiplying an arbitrary element of the form (\ref{eq:generalelement})
on the left by the group element $\exp (\epsilon J_a)$.
The components $\tilde{u}_a^{\alpha}$   appear as the order
$\epsilon$ shifts to the holomorphic coordinates $z_\alpha$,  and  can
be derived by writing the product $\exp (\epsilon J_a)\exp (\sum
z_\alpha J_\alpha)$ in the form
\begin{equation}
{\rm e}^{\epsilon J_a} \exp (\sum_{\alpha \in \Phi_+}z_\alpha J_\alpha) =
\exp \left[\sum_{\alpha \in \Phi_+}(z_\alpha +\epsilon\tilde{u}_a^{\alpha})
J_\alpha \right] f (\{J_a | a\preceq 0\}) + {\cal O} (\epsilon^2)
\label{eq:generalcalculation}
\end{equation}
to first order in $\epsilon$, where $f$ is some function of the
generators $J_a$ corresponding to negative roots and the Cartan
subalgebra.  To explicitly compute these vector fields, we use the
infinitesimal forms of the BCH theorem,
\begin{eqnarray}
{\rm e}^{\eps X} {\rm e}^{Y} & = & \exp(Y + \eps \sum_{k \geq 0}
\frac{B_k}{k!} (\ad_Y)^k X) + \co(\eps^2), \label{eq:bch1} \\
{\rm e}^{Y + \eps Z + \eps X} & = & \exp\left(Y + \eps Z - \eps
\sum_{k \geq 1}
\frac{B_k}{k!} (-\ad_Y)^k X\right) \;\;{\rm e}^{\eps X} + \co(\eps^2),
\label{eq:bch2} \end{eqnarray}
and
\bge
{\rm e}^{\eps X} {\rm e}^Y = {\rm e}^{Y + \eps [X,Y]} {\rm e}^{\eps X}
+ \co(\eps^2), \label{eq:bch3}
\ee
where $B_k$ is the $k$th Bernoulli number (Table \ref{t:bernoulli}).
The first of these equations can be derived by writing
\begin{equation}
g(\epsilon) ={\rm e}^{\epsilon X} {\rm e}^{Y} =
{\rm e}^{Y + \epsilon Z}+ {\cal
O}(\epsilon^2)
\end{equation}
and defining
\begin{eqnarray}
g&=& g (0)= {\rm e}^{Y}, \\
\delta g &=& \frac{\partial}{\partial \epsilon}  |_{\epsilon = 0}
g (\epsilon)= X {\rm e}^{Y}. \nonumber
\end{eqnarray}
{}From these definitions it follows that
\begin{equation}
g^{-1} \delta g = {\rm e}^{-\ad_Y} X.
\label{eq:BCHderivation1}
\end{equation}
But writing
\begin{equation}
g_t (\epsilon) = {\rm e}^{t (Y + \epsilon Z)},
\end{equation}
we have
\begin{equation}
\frac{\partial}{\partial t}  (\delta g_t) = Z g_t + Y \delta g_t.
\end{equation}
It follows that
\begin{equation}
\frac{\partial}{\partial t}  (g_t^{-1}\delta g_t) = g_t^{-1}Z g_t,
\end{equation}
and thus that
\begin{equation}
g^{-1} \delta g = \int_{0}^{1} {\rm d} t \; {\rm e}^{- t\;\ad_Y} Z = \left(
\frac{1 - {\rm e}^{- \; \ad_Y}}{\ad_Y}  \right) Z.
\end{equation}
{}From this expression and (\ref{eq:BCHderivation1}), it follows  by
formally manipulating the power series in  $\ad_Y$ that
\begin{equation}
Z =\left(
\frac{\ad_Y}{{\rm e}^{\;\ad_Y}- 1}  \right)  X = \sum_{k = 0}^{
\infty}  \frac{B_k}{k!}   (\ad_Y)^k X,
\end{equation}
which gives (\ref{eq:bch1}).  Equations (\ref{eq:bch2}) and
(\ref{eq:bch3}) follow
immediately from the same type of argument.

We now give the general expression for the vector field components
which arise from performing the calculation
(\ref{eq:generalcalculation}) using (\ref{eq:bch1}), (\ref{eq:bch2}) and
(\ref{eq:bch3}).
\begin{prop}
The components $\tilde{u}_a^{\alpha}$  of the vector fields $\xn$ are
given by
\begin{equation}
\tilde{u}_a^{\alpha} = -\sum_{\scriptscriptstyle k \geq 0, A_a (k, \alpha)}
\beta_{k,\lambda} C_a(a_1, \ldots, a_k) z_{a_1} \ldots z_{a_k},
\end{equation}
where $\lambda$ is the minimum integer
such that $a + a_1 + a_2 + \ldots + a_{\lambda} \succ 0$ {\rm (}$\lambda = 0$
when $a \succ 0${\rm )},
\begin{equation}
A_a (k, \alpha) = \{(a_1, a_2, \ldots, a_k) : a_1,\ldots,a_k \succ 0 ,a +
a_1 + a_2 + \ldots + a_k  = \alpha\},
\end{equation}
\begin{equation}
C_a(a_1, \ldots, a_k) =  i^{k}
\sum_{b_1, \ldots b_{k - 1}}^{} \mixten{f}{a_1a}{b_1} \mixten{f}{a_2
b_1}{b_2} \ldots \mixten{f}{a_k b_{k - 1}}{\alpha},
\end{equation}
and
\bge \beta_{k, \lambda} = (-1)^{k+1} \sum_{l=0}^{k-\lambda}
\frac{B_l}{l! \: (k-l)!}.\ee
{\rm (}$B_l$ is the $l$th Bernoulli number, as in {\rm
(}\ref{eq:bch1}{\rm )} and {\rm (}\ref{eq:bern}{\rm )}.  The Bernoulli
numbers and values of $\beta_{k, \lambda}$ are tabulated for small
values of the subscripts $l,k, \lambda$ in tables \ref{t:beta},
\ref{t:bernoulli} at the end of this section.
{\rm )}
\label{p:xcalc}
\end{prop}
\noindent {\it Proof.} \,\,
We begin by noting the identities
\bge
(\ad_{(\sum z_\alpha J_\alpha)})^k J_a = \sum_{\sss a_1, \ldots, a_k \succ
0} C_a(a_1, \ldots, a_k) z_{a_1} \ldots z_{a_k} J_{a + a_1 + \ldots +
a_k},\label{eq:adk} \ee
and
\bge C_a(a_1, \ldots, a_s) C_{a+ a_1 + \ldots + a_s}(a_{s+1}, \ldots, a_k)
= C_a(a_1, \ldots, a_k). \label{eq:ccomb} \ee
Applying Equation \ref{eq:bch1} to $\exp(\eps J_a) \exp(\sum
z_\alpha J_\alpha)$, we have
\begin{eqnarray}
\lefteqn{\exp(\eps J_a) \exp(\sum_{\alpha \in \Phi_+} z_\alpha
J_\alpha) \sim}  \\
& & \exp \left(\sum_{\alpha \in \Phi_+} z_\alpha J_\alpha
+ \eps \sum_{\sss k \geq 0, a_1, \ldots, a_k \succ 0} \frac{B_k}{k!}
C_a(a_1, \ldots, a_k) z_{a_1} \ldots z_{a_k} J_{a + a_1 + \ldots +
a_k}\right),\nonumber
\end{eqnarray}
where by $ x \sim  y$ it is meant that $ x = yf + \co(\eps^2)$, with $f$
some function of the $J_a$'s with $a \preceq 0$.
Dividing the terms in the exponential into generators $J_a$ with $a \succ
0$ and $a \preceq 0$, this can be rewritten as
\begin{eqnarray}
\lefteqn{\exp(\eps J_a) \exp(\sum_{\alpha \in \Phi_+} z_\alpha
J_\alpha) \sim}
\label{eq:exp0}
\\ & &  \exp \left(\sum_{\alpha \in \Phi_+} z_\alpha J_\alpha
- \eps \sum_{\sss k \geq 0, A_a^+(k)} \beta_{k,\lambda}^{(0)}
C_a(a_1, \ldots, a_k) z_{a_1} \ldots z_{a_k} J_{a + a_1 + \ldots +
a_k} \right.\nonumber \\ & & \left.\;\;\;\;\;\;\; + \eps \!\!\!
\sum_{\sss l_1 \geq 0, A_a^-(l_1)}
\! \frac{B_{l_1}}{l_1 !} C_a(a_1, \ldots, a_{l_1}) z_{a_1} \ldots
z_{a_{l_1}} J_{a + a_1 + \ldots + a_{l_1}}\right),  \nonumber
\end{eqnarray}
 where
$\beta_{k, \lambda}^{(t)}$ is defined by
\bge \beta_{k, \lambda}^{(t)} = - \frac{B_k}{k!} - \!\!\! \sum_{\sss t
\geq s > 0, 0 \leq l_1
< l_2 < \ldots < l_s < \lambda} \!\!\! (-1)^{s + k - l_1}
\frac{B_{l_1}}{l_1 !} \frac{B_{l_2 - l_1}}{(l_2 - l_1) !} \ldots
\frac{B_{l_s - l_{s-1}}}{(l_s - l_{s-1}) !} \frac{B_{k - l_s}}{(k -
l_s) !}, \ee
and the sets $A_a^{\pm}(k)$ are defined by
\begin{equation}
A_a^+ (k) = \{(a_1, a_2, \ldots, a_k) : a_1,\ldots,a_k \succ 0 ,a +
a_1 + a_2 + \ldots + a_k \succ 0\},
\end{equation}
and
\begin{equation}
A_a^- (k) = \{(a_1, a_2, \ldots, a_k) : a_1,\ldots,a_k \succ 0 ,a +
a_1 + a_2 + \ldots + a_k \preceq 0\}.
\end{equation}
Applying (\ref{eq:bch2}) and (\ref{eq:ccomb}) to Equation
\ref{eq:exp0} $t$ times, we get
\begin{eqnarray}
\lefteqn{\exp(\eps J_a) \exp(\sum_{\alpha \in \Phi_+} z_\alpha J_\alpha)
\sim} \nonumber \\ & & \!\!\!\!\!\!\exp \left(\sum_{\alpha \in \Phi_+}
z_\alpha J_\alpha
- \eps \sum_{\sss k \geq 0, A_a^+(k)} \beta_{k,\lambda}^{(t)}
C_a(a_1, \ldots, a_k) z_{a_1} \ldots z_{a_k} J_{a + a_1 + \ldots +
a_k} \right. \\ & & \left.  + \eps \!\!\!\!\!\!\!\!\!\!
\sum_{\sss 0 \leq l_1 < \ldots < l_t < l, A_a^-(l)} \!\!\!\!\!\!\!\!\!
(-1)^{t+l-l_1}
\frac{B_{l_1}}{l_1 !}
\frac{B_{l_2 - l_1}}{(l_2 - l_1) !} \ldots
\frac{B_{l - l_{t}}}{(l - l_{t}) !}
C_a(a_1, \ldots, a_{l}) z_{a_1} \ldots z_{a_{l}} J_{a + a_1 + \ldots +
a_{l}}\right). \nonumber \end{eqnarray}
Since $\beta_{k, \lambda}^{(t)} =
\beta_{k,
\lambda}^{(\infty)}$ for $t \geq \lambda$, to all orders in $z$ we
have
\begin{eqnarray}
\lefteqn{\exp(\eps J_a) \exp(\sum_{\alpha \in \Phi_+} z_\alpha J_\alpha)
\sim}  \\ & & \exp \left(\sum_{\alpha \in \Phi_+} z_\alpha J_\alpha
- \eps \sum_{\sss k \geq 0, A_a^+(k)} \beta_{k,\lambda}^{(\infty)}
C_a(a_1, \ldots, a_k) z_{a_1} \ldots z_{a_k} J_{a + a_1 + \ldots +
a_k} \right),\nonumber
\end{eqnarray}
We will now show that $\beta_{k,\lambda}^{(\infty)} =
\beta_{k,\lambda}$.  Using the fact that $B_{2k+1} = 0$ for $k >
0$, it is not hard to determine that
\begin{eqnarray}
\beta_{k, 0}^{(\infty)} & = & - \frac{B_k}{k!}, \nonumber \\
\beta_{k, 1}^{(\infty)} & = & \delta_{k, 1}, \label{eq:alkl} \\
\beta_{k, \lambda}^{(\infty)} & = & \!\! \sum_{\sss s \geq 0, 1 < l_1
< \ldots < l_s < \lambda} \!\! (-1)^{s+k}
\frac{B_{l_1 - 1}}{(l_1 - 1) !} \frac{B_{l_2 - l_1}}{(l_2 - l_1) !}
\ldots \frac{B_{l_s - l_{s-1}}}{(l_s - l_{s-1}) !}
\frac{B_{k - l_s}}{(k - l_s) !}, \;\;{\rm for}\; \lambda > 1.
\nonumber \end{eqnarray}
When $\lambda > 1$, we can write a generating function for $\beta_{k,
\lambda}^{(\infty)}$ by
\bge
\sum_{k \geq \lambda > 1}  \beta_{k,\lambda}^{(\infty)} y^{k-\lambda}
x^k = \sum_{l > m \geq 0} y^m \frac{B_l}{l!} (-x)^{l+1} \sum_{s \geq 0}
(1-\phi(-x))^s, \label{eq:gen} \ee
where
\bge
\phi(x) = \frac{x}{{\rm e}^x - 1} = \sum_{n \geq 0} \frac{B_n}{n!}
x^n.\label{eq:bern} \ee
{}From (\ref{eq:gen}), it follows that
\bge \beta_{k, \lambda}^{(\infty)} = (-1)^k \sum_{l=k - \lambda + 1}^{k-1}
\frac{B_l}{l! \: (k-l)!} = (-1)^{k+1} \sum_{l=0}^{k-\lambda}
\frac{B_l}{l! \: (k-l)!},\ee
where we have used the fact that
\bge
\sum_{l=0}^{k-1} B_l \left( \begin{array}{cc} k \\ l \end{array}
\right) = 0,\;\;\;{\rm for}\; k > 1. \label{eq:bid} \ee
{}From (\ref{eq:alkl}) and (\ref{eq:bid}), it is also easy to verify that
$\beta_{k,0}^{(\infty)} = \beta_{k,0}$ and $\beta_{k,1}^{(\infty)}
= \beta_{k,1}$.  Thus, for all $k$ and $\lambda$, we have shown that
$\beta_{k,\lambda}^{(\infty)} = \beta_{k,\lambda}$, and
Proposition \ref{p:xcalc} is proven.$\Box$

The  result of Proposition  \ref{p:xcalc}  can be restated in terms of
the vector fields $ \xi_a$.
\begin{corollary}
The vector fields $\xi_a$ are given by
\begin{equation}
\xi_a = \sum_{\scriptscriptstyle k \geq 0, A_a^+}
\beta_{k,\lambda} C_a(a_1, \ldots, a_k) z_{a_1} \ldots z_{a_k}
\frac{\partial}{\partial_{z_{a + a_1 +\cdots +a_k}}},
\label{eq:corollary}
\end{equation}
where $\lambda$, $C_a$, $\beta$, and $A_a^+$ are defined as above.
\end{corollary}

When $a \succeq 0$, we can use (\ref{eq:alkl}) to simplify the formulae
for $\xn$
to
\bge
\xn = \!\sum_{\sss k \geq 0, a_1, \ldots, a_k \succ 0}\!
-\frac{B_k}{k!} C_a(a_1, \ldots, a_k) z_{a_1} \ldots z_{a_k} \frac{\pdv}{\pdv
z_{a+a_1+\ldots + a_k}}\;\;\; {\rm for} \; a \succ 0,
\label{eq:positivesimplification}\ee
and
\bge
\xi_h = \sum_{ \alpha\succ 0} - \ip{\alpha}{h} z_\alpha \frac{\pdv}{\pdv
z_\alpha}\; \; \; {\rm for} \; h \in \Theta.
\label{eq:cartansimplification}\ee
Note that (\ref{eq:cartansimplification}) could also have been
obtained more directly through (\ref{eq:bch3}).)

\subsubsection{Gauge fixing}

Now that we have explicit formulae for the vector fields $\xi_a$, in
order to construct the operators $\hat{J}_a$ it will suffice by the
results of the previous section to find a set of functions $f_u$ which
are linear in the parameter $ u \in \algg_\CC$ and which satisfy
conditions ({\sl i}) from Proposition~\ref{p:p1} and ({\sl ii'}) from
Proposition~\ref{p:p2}.  We will explicitly construct these functions
by making an appropriate choice of gauge; before discussing the
gauge-fixing procedure, however, it will be useful to discuss some
general features of the vector fields $\xi_a$ calculated in the
previous subsection.

The main observation to be made is that the ring $R =
\CC[\{z_\alpha\}]$ of polynomials in the variables $z_\alpha$ has a
natural grading, according to the weights $w \in \Lambda$ in the root
lattice of the group $G$.  Recall that the root lattice $\Lambda
\subset
\algt_\CC^*$ is the lattice generated by the simple roots $\alpha \in
\Delta$.  We can
associate to each variable $z_\alpha$ a degree
\begin{equation}
\deg z_\alpha = \alpha \in \Lambda.
\end{equation}
By giving the unit element $1$ a degree
\begin{equation}
\deg 1 = 0,
\end{equation}
and by  defining the degree of a product  of monomials recursively by
\begin{equation}
\deg (fg) = \deg f + \deg g,
\end{equation}
we define the degree for all monomials $f = z_{\alpha_1} \cdots
z_{\alpha_i}$ to be $\alpha_1 + \cdots + \alpha_i$.
The ring $R$ thus admits a $\Lambda$-grading
\begin{equation}
R = \bigoplus_{w \in \Lambda} R_w,
\end{equation}
where
\bge  R_w = \{f \in R : \deg(f) =  w\}, \ee
and where
\begin{equation}
R_w R_y  \subseteq R_{w + y}.
\end{equation}
We define a polynomial $f \in R$ to be {\it quasi-homogeneous} when
$f$ has a definite degree $\deg f \in \Lambda$.

Considering the action of the operators $\xi_a$  on the space $R$ of
polynomials in the $z_\alpha$, we find that
\begin{equation}
\xi_a R_w  \subseteq R_{w - a} \; \; \; {\rm when}\;a \not\approx 0,
\label{eq:action1}
\end{equation}
and
\begin{equation}
\xi_h  f = -\ip{ w}{h} f \; \; \; {\rm for}\; f \in R_w,h \in \Theta.
\label{eq:action2}
\end{equation}
Thus, the action of the vector field part of the operators $\hat{J}_a$
on $R$ has a very natural  structure with respect to the grading.

We now prove a proposition which describes further the action of the
vector fields $\xi_a$ on $R$; this proposition will be used in this
section to calculate the gauge-fixed functions $f_a$, and will also be
used in Chapter 3 where we describe in more detail the
$\algg_\CC$-module structure of ${\cal H}_b$.

\begin{prop}
The unit element 1 in $R$, which we denote by $\nul$, is the unique
function in $R$ (up to scalar multiplication) which is annihilated by
$\xi_a$ for all $ a \succ 0$; i.e., $\nul$ is the unique highest
weight state in the module $R$ under the action of the $\xn$'s.
\label{p:p4}
\end{prop}
\noindent {\it Proof.} \,\,
Assume there is another function $\phi \in R$ which is annihilated by
$\xn$ for all $a\succ 0$.  Since $R$ is graded, $\phi$ can be written as
a sum of quasi-homogeneous functions,
\bge
\phi = \sum_{w \succeq 0} \phi_w, \;\;\; \phi_w \in R_w.\ee
Take $w$ to be a minimal nonzero weight with $\phi_w \neq 0$ ($w$ is a
minimal weight satisfying this condition as long as there does not
exist a nonzero weight $y \in \Lambda$ with $w \succ y$ and $\phi_y
\neq 0$).  Now, let $\alpha$
be a maximal root such that some term in $\phi_w$ contains a
factor of $z_\alpha$.  $\phi_w$ can now be written in the form
\bge
\phi_w = \sum_{m \geq 0} z_\alpha^m g_{w-m\alpha}^{(m)}
(\{z_\beta: \beta \not\succeq \alpha\}),\ee
where for each $m \geq 0$, $g_{w-m\alpha}^{(m)}$ is a
quasi-homogeneous polynomial of degree $w -m \alpha$ in the set of variables
$z_\beta$ with $\beta \not\succeq \alpha$.    Since for $a \succ 0$, all
terms in $\xn$
except the leading term $- \pdv / \pdv z_a$ contain derivatives $\pdv
/ \pdv z_\beta$ with $\beta \succ \alpha$, we can compute
\bge
\xi_\alpha \phi_w = - \sum_{m > 0} m z_\alpha^{m-1} g_{w -
m\alpha}^{(m)} (\{\beta: \beta \not\succeq \alpha\}).
\ee
For this expression to be zero, all the functions
$g_{w-m\alpha}^{(m)}$ would have to be zero for $m > 0$.  But then
$\phi_w$ would not contain any terms with a factor of $z_\alpha$,
contradicting our assumption.  Thus, the only states in $R$
annihilated by all $\xi_a$ with $a \succ 0$ are the constant functions
in $R_0$.$\Box$

Note that the proof of this proposition also indicates that there are
not even any formal power series in $\CC[[\{z_\alpha\}]]$ other than
$\nul$ which are annihilated by all $\xi_a, a \succ 0$.  We will
mention this point again in connection with Virasoro representations.

The choice of gauge we use to construct the functions $f_a$ is a gauge
in which the structure of the action (\ref{eq:action1}),
(\ref{eq:action2}) of the operators $\xi_a$ on $R$
is preserved for the operators $\hat{J}_a$; that
is, we will find a gauge in which
\begin{equation}
\hat{J}_a R_w  \subseteq R_{w - a}
\label{eq:ansatz1}
\end{equation}
for all generators $\hat{J}_a$
($w - a = w$ when $a \in \Theta$).
A priori, it is not clear that such a gauge choice is possible.  We
will use (\ref{eq:ansatz1}) as an Ansatz for
the functions $f_a = \hat{J}_a - \xi_a$.  We now demonstrate that
there is a unique set of functions $f_a$ satisfying this Ansatz and
also conditions ({\sl i}), ({\sl ii'}), and that therefore these
functions $f_a$ correspond to a particular choice of gauge by
Proposition \ref{p:p2}.

The first consequence of the Ansatz is that the functions $f_a$ all
lie in the ring $R$, and furthermore that $f_a \in R_{- a}$.
Specifically, this means that
\begin{eqnarray}
f_a &  = &0, \; \; \; {\rm for}\; a \succ 0, \nonumber\\
f_h & \in &  \CC,\; \; \; {\rm for} \;h \in \Theta,\label{eq:consequence}\\
f_{- a} & \in &  R_a,\; \; \; {\rm for}\; a \succ 0.\nonumber
\end{eqnarray}
We have chosen the element $b$ in the coadjoint orbit $W_b$ to be such
that the stabilizer of $b$ under the coadjoint action of $\algg_\CC$
is precisely $\algt_\CC$.  In the coordinates $z_\alpha$, $b$ is the
point at which all coordinates are 0.  By condition ({\sl ii'}), it
follows that the functions $f_h$ are given at the point $b= 0$ by
\begin{equation}
f_h (0) =i \Phi_h (b) = - i \ip{ b}{h}.
\end{equation}
Since the functions $f_h$ are constant functions in $R$, it follows
that
\begin{equation}
f_h  = - i \ip{ b}{h}.
\label{eq:constantfunctions}
\end{equation}

We will now use (\ref{eq:constantfunctions}) and the consequences
(\ref{eq:consequence}) of the Ansatz  to construct functions $f_{- a}
\in R_a$ for all $a \succ 0$ which satisfy ({\sl i}).  For these
functions, the condition ({\sl i})  is given by
\begin{equation}
\xi_a f_c - \xi_c f_a = i \mixten{f}{ac}{d} f_d.
\label{eq:notparticular}
\end{equation}
For particular choices of $a$ and $c$, this equation reduces to the equations
\begin{eqnarray}
\xi_h f_{- a} & =&-\ip{ a}{h} f_{- a},
 \; \; \; \forall a \succ 0,h \in \Theta\label{eq:particular1}\\
\xi_a f_{- a} &= &\sum_{h \in \Theta} i\mixten{f}{a (- a)}{h} f_h,
 \; \; \; \forall a \succ 0 \label{eq:particular2}\\
\xi_a f_{- c} & = & 0,
 \; \; \; \forall a, c \succ 0 \; {\rm where} \; c - a \not\succeq 0
\label{eq:particular3}\\
\xi_a f_{- c} & = & i\mixten{f}{a (- c)}{a - c} f_{a - c},
 \; \; \; \forall a, c \succ 0 \; {\rm where} \; c - a \succ 0
\label{eq:particular4}\\
\xi_{- a} f_{- c} - \xi_{- c} f_{- a} & = & i\mixten{f}{(- a) (- c)}{- a -
c} f_{- a - c},
 \; \; \; \forall a,c \succ 0.\label{eq:particular5}
\end{eqnarray}
{}From (\ref{eq:action2}) and (\ref{eq:consequence}), it follows
immediately that (\ref{eq:particular1}) and (\ref{eq:particular3})
must hold.  We can use (\ref{eq:particular2}) to calculate the
functions $f_{- \alpha}$ for simple $\alpha$.  For a simple root
$\alpha$, the space $R_\alpha$ is one-dimensional and consists of all
functions of the form $f = x z_\alpha$ with $x \in \CC$.  The vector
field operator $\xi_\alpha$  has a leading term $-\partial/\partial
z_\alpha$, with all other terms containing derivatives
$\partial/\partial z_\beta$ with $\beta \succ \alpha$.  Thus, if we write
\begin{equation}
f_{- \alpha} = x z_\alpha,
\end{equation}
then
\begin{equation}
\xi_\alpha f_{- \alpha} = - x = f_{h_\alpha}= - i \ip{ b}{h_\alpha}.
\end{equation}
It follows that
\begin{equation}
f_{- \alpha} = i \ip{b}{h_\alpha} z_\alpha, \; \; \forall \alpha \in \Delta.
\end{equation}
(Note: in the case of the Virasoro algebra, this equation does not
give the function $f_{- 2}$ associated with the quasi-simple root
$\hat{L}_2$; this special case is discussed in the next section.  The
rest of the analysis in this subsection, however, is still correct for
the Virasoro algebra after one correctly fixes the function $f_{- 2}$.)

Now that we have expressions for $f_a$ where $a$ is either a simple
root or an element of the Cartan subalgebra, we can proceed to
construct the remaining functions $f_a$ inductively, using
(\ref{eq:particular5}).  Such a construction is possible because the
set of simple roots always generates the  algebra of positive roots.
We will now proceed to prove that this construction gives rise to a
unique and well-defined set of functions $f_a$ for each $a \succ 0$,
which satisfy (\ref{eq:particular1})-(\ref{eq:particular5}).
\begin{prop}
The functions $f_a \in R$ defined by the recursive formulae
\begin{eqnarray}
f_a & = & 0, \;\;\; a \succ 0, \nonumber \\
f_h & = & - i \ip{ b}{h},  \; \; \; h \in \Theta\label{eq:recursive} \\
f_{- \alpha} & = & i \ip{b}{h_\alpha} z_\alpha, \;\; \; \alpha
\in \Delta, \nonumber \\
f_{ a +  c} & = &  \frac{- i }{\mixten{f}{ac}{ a +c}}
\left[\xi_{ a} f_{ c} - \xi_{ c} f_{ a}\right], \; \; \; a,c \prec 0
\nonumber \end{eqnarray}
are well-defined, and satisfy (\ref{eq:notparticular}).
\label{p:p5}
\end{prop}
\noindent {\it Proof.} \,\,
In order to prove this proposition, it is clearly necessary to
demonstrate that for a fixed root $d\prec 0$, the last equation in
(\ref{eq:recursive}) gives a well-defined function for $f_{d} = f_{a +
(d - a)}$, independent of the choice of $a$ satisfying $0 \succ a
\succ d$.  The fact that all functions $f_d$ are defined by this
equation follows again from the fact that the simple roots generate
the algebra of all positive roots.  Because the functions $f_a$
defined through (\ref{eq:recursive}) clearly all satisfy
(\ref{eq:consequence}), we know that (\ref{eq:particular1}) and
(\ref{eq:particular3}) are satisfied by these functions.  What we must
therefore show in order to guarantee that the remaining conditions
(\ref{eq:particular2}), (\ref{eq:particular4}), and
(\ref{eq:particular5}) hold and that the functions $f_a$ are
well-defined, is that for every $a \succ 0$, there is a unique
function $\tilde{f}_{- a}\in R_a$ which satisfies the conditions
\begin{eqnarray}
\xi_a  \tilde{f}_{- a} &= &\sum_{h \in \Theta} i\mixten{f}{a (- a)}{h} f_h,
\label{eq:particularly2}\\
\xi_c  \tilde{f}_{-  a} & = & i\mixten{f}{c (- a)}{c - a} \tilde{f}_{c - a},
 \; \; \; \forall c  \; : \; a  \succ c\succ 0,
\label{eq:particularly4}\\
\tilde{f}_{- a} & = &  \frac{- i }{\mixten{f}{( c - a) ( -c)}{- a}}
\left[\xi_{ c - a} \tilde{f}_{ -c} - \xi_{ -c}
\tilde{f}_{ c - a}\right], \; \; \; \forall c : \;
a \succ c \succ 0.\label{eq:particularly5}
\end{eqnarray}

We will prove (\ref{eq:particularly2})-(\ref{eq:particularly5}) by
induction on $a$.  As a basis for the induction, we take the simple
roots $a = \alpha \in \Delta$.   We have defined the functions $f_{-
\alpha}$ for simple roots precisely such that
(\ref{eq:particularly2}) holds when $a \in \Delta$.
In this case, there are no  roots $c$ satisfying $a \succ c \succ 0$,
so (\ref{eq:particularly4}) and (\ref{eq:particularly5}) are vacuously
satisfied.  We now proceed by induction, taking as the induction
hypothesis the assertion that for all roots $a'$ satisfying $a \succ
a' \succ 0$, conditions
(\ref{eq:particularly2})-(\ref{eq:particularly5}) are satisfied.  In
order to guarantee that these conditions are also satisfied for the root
$a$, we claim that it will suffice to demonstrate that
\begin{eqnarray}
\xi_c \frac{- i }{\mixten{f}{( d - a) ( -d)}{- a}}
\left[\xi_{ d - a} f_{ -d} - \xi_{ -d}
f_{ d - a}\right] & = &\sum_{e}
i\mixten{f}{c (- a)}{e} f_{e}, \label{eq:sufficient}\\
& & \; \; \;
\forall d, c : \; a \succ d \succ 0, c \succ 0.  \nonumber
\end{eqnarray}
(where we have written $\tilde{f}_{-a'}$ as $f_{- a'}$ when $a' \prec
a$.)  That this condition implies conditions (\ref{eq:particularly2}) and
(\ref{eq:particularly4}) follows immediately when the function $f_{- a}$
is well-defined.  However, because the right hand side of
(\ref{eq:sufficient}) is independent of $d$, it also follows from this
condition that
\begin{eqnarray}
\xi_c \left[\frac{- i }{\mixten{f}{( d - a) ( -d)}{- a}}
\left(\xi_{ d - a} f_{ -d} - \xi_{ -d}
f_{ d - a}\right) -
\frac{- i }{\mixten{f}{( d' - a) ( -d')}{- a}}
\left(\xi_{ d' - a} f_{ -d'} - \xi_{ -d'}
f_{ d' - a}\right) \right] & =& 0 \nonumber\\
& &\hspace{-2.7in}
\forall d, d', c: \; a \succ d, d' \succ 0, c \succ 0.
\end{eqnarray}
But by Proposition~\ref{p:p4} this implies that the quantity in
brackets is an element of $\CC$ for fixed values of $d,d'$.  Since
this quantity is an element of $R_a$, and $a \succ 0$, the quantity
must be 0.  But this would imply that $f_{- a}$ is well-defined, and
thus that (\ref{eq:particularly5}) is satisfied for $a$.  Therefore,
it will be sufficient to prove (\ref{eq:sufficient}) in order to have
a proof of the proposition.

We shall now demonstrate that (\ref{eq:sufficient}) is satisfied,
given the induction hypothesis.  This is essentially a straightforward
algebraic computation.   Using the induction hypothesis and the Jacobi
identity, we have
\begin{eqnarray}
\lefteqn{\xi_c \frac{- i }{\mixten{f}{( d - a) ( -d)}{- a}}
\left[\xi_{ d - a} f_{ -d} - \xi_{ -d}
f_{ d - a}\right]} \\ & = &
\frac{- i }{\mixten{f}{( d - a) ( -d)}{- a}}
\left[
i \sum_{e} \mixten{f}{c (d - a)}{e} \xi_{e} f_{- d} +
\xi_{d - a} \xi_c f_{- d} -
i \sum_{e} \mixten{f}{c (- d)}{e} \xi_{e} f_{d- a} -
\xi_{- d} \xi_c f_{d- a}
\right]\nonumber \\
& = &
\frac{1}{\mixten{f}{( d - a) ( -d)}{- a}}
\sum_{e} \left[
\mixten{f}{c (d - a)}{e} \xi_{e} f_{- d} +
\mixten{f}{c (- d)}{e}  \xi_{d - a} f_{e} -
\mixten{f}{c (- d)}{e} \xi_{e} f_{d- a} -
\mixten{f}{c (d- a)}{e}  \xi_{- d} f_{e}
\right] \nonumber\\
& = &
\frac{1}{\mixten{f}{( d - a) ( -d)}{- a}}
\sum_{e} \left[
\mixten{f}{c (d - a)}{e} \left(\xi_{e} f_{- d} - \xi_{- d} f_{e}\right) +
\mixten{f}{c (- d)}{e}  \left(\xi_{d - a} f_{e} -\xi_{e} f_{d- a}\right)
\right] \nonumber\\
& = &
\frac{i}{\mixten{f}{( d - a) ( -d)}{- a}}
\sum_{e, g} \left[
\mixten{f}{c (d - a)}{e} \mixten{f}{e (- d)}{g} f_{g}+
\mixten{f}{c (- d)}{e}  \mixten{f}{(d - a) (e)}{g} f_{g}
\right] \nonumber\\
& = &
\frac{i}{\mixten{f}{( d - a) ( -d)}{- a}}
\sum_{e, g} \left[
\mixten{f}{( d - a) ( -d)}{e}
\mixten{f}{ce}{g} f_{g} \right] \nonumber\\
& = &  i\mixten{f}{c (- a)}{g} f_{g}.  \nonumber
\end{eqnarray}

We have thus shown by induction that the condition
(\ref{eq:sufficient}) is satisfied for all $a \succ 0$,
so the proposition is proven.$\Box$

We now have an explicit set of formulae which define the functions
$f_a$ in the gauge $f_a \in R_{- a}$.

\subsubsection{General formula}

We can now state the complete result of this section.
\begin{prop}
An explicit representation of the algebra $\algg$ on the space $R$ of
polynomials in the variables $\{z_\alpha: \alpha \in \Phi_+\}$ is
given by the operators
\begin{equation}
\hat{J}_a = \xi_a + f_a,
\label{eq:complete}
\end{equation}
where $\xi_a$ are the first-order differential operators given by
(\ref{eq:corollary}), and $f_a$ are the functions defined by
(\ref{eq:recursive}).   On the space of holomorphic sections ${\cal
H}_b$ of $\lb$, this representation corresponds to the
coadjoint orbit representation in the particular gauge $f_a \in R_{-
a}$.$\Box$
\label{p:complete}
\end{prop}

Because the  operators $\xi_a$ are independent of  which coadjoint
orbit $W_b$ is used, and the functions $f_a$ are 0 for $a \prec 0$, it
follows immediately from this proposition that the operators
$\hat{J}_a$  with $a \prec 0$
are independent of the chosen coadjoint orbit, and thus independent of
the highest weight of the associated representation.  This feature of
the coadjoint orbit representations will be exploited in Section
\ref{sec:CFT}, where it is used to simplify  formulae for certain
correlation functions of vertex operators in conformal field theories.

Note also that the explicit representations described in this section
can be defined for any highest weight representation of the Lie
algebra, whether or not it is integrable to a representation of the Lie
group.  Thus, the set of formulae given here are actually somewhat
more general than what we need for analysis of the coadjoint orbit
representations.

In the following sections and chapters we will give explicit examples
of these representations and study their properties further.  We give
here tables of the Bernoulli numbers (Table \ref{t:bernoulli}) and the
constants $\beta_{k,\lambda}$ which appear in (\ref{eq:corollary})
(Table \ref{t:beta}) for
convenience in explicit calculations.
\begin{table}
\begin{center}
\begin{tabular}{|c|cccccccccc|}
\hline
$\lambda\backslash k$ & 0 & 1 & 2 & 3 & 4 & 5 & 6 & 7 & 8 & 9\\ \hline
& & & & & & & & & & \\ 0 &-1 & $\frac{1}{2}$ & $-\frac{1}{12}$ & $0$ &
$\frac{1}{720}$ & $0$ & $-\frac{1}{30240}$ & $0$ & $\frac{1}{1209600}$
& $0 $ \\ & & & & & & & & & & \\ $1$ & $0$ & $1$ & $0$ & $0$ & $0$ &
$0$ & $0$ & $0$ & $0$ & $0$\\ & & & & & & & & & & \\ $2$ & $0$ & $0$ &
$-\frac{1}{2}$ & $-\frac{1}{12}$ & $0$ & $\frac{1}{720}$ & $0$ &
$-\frac{1}{30240}$& $0$ & $\frac{1}{1209600}$\\ & & & & & & & & & & \\
$3$ & $0$ & $0$ & $0$ & $\frac{1}{6}$ & $\frac{1}{24}$ &
$\frac{1}{720}$ & $-\frac{1}{1440}$ &$-\frac{1}{30240}$ &
$\frac{1}{60480}$ & $\frac{1}{1209600}$\\ & & & & & & & & & & \\ $4$ &
$0$ & $0$ & $0$ & $0$ & $-\frac{1}{24}$ & $-\frac{1}{80}$ &
$-\frac{1}{1440}$ & $\frac{1}{5040}$& $\frac{1}{60480}$ & $
-\frac{17}{3628800}$\\ & & & & & & & & & & \\ $5$ & $0$ & $0$ & $0$ &
$0$ & $0$ & $\frac{1}{120}$ & $\frac{1}{360}$ & $\frac{1}{5040}$
&$-\frac{1}{24192}$ & $ -\frac{17}{3628800}$\\ & & & & & & & & & & \\
$6$ & $0$ & $0$ & $0$ & $0$ & $0$ & $0$ & $-\frac{1}{720}$ &
$-\frac{1}{2016}$ &$-\frac{1}{24192}$ & $ \frac{1}{145152}$\\ & & & &
& & & & & & \\ $7$ & $0$ & $0$ & $0$ & $0$ & $0$ & $0$ & $0$ &
$\frac{1}{5040}$ & $\frac{1}{13440}$ &$\frac{1}{145152}$\\ & & & & & &
& & & & \\ $8$ & $0$ & $0$ & $0$ & $0$ & $0$ & $0$ & $0$ & $0$ &
$-\frac{1}{40320}$ &$-\frac{1}{103680}$\\ & & & & & & & & & & \\ $9$ &
$0$ & $0$ & $0$ & $0$ & $0$ & $0$ & $0$ & $0$ & $0$ &
$\frac{1}{362880}$\\
\hline
\end{tabular}
\caption{Values of $\beta_{k, \lambda}$}
\label{t:beta}
\end{center}
\end{table}

\begin{table}
\begin{center}
\begin{tabular}{|c|ccccccccccc|}
\hline
$l$  & 0 & 1 & 2 & 3 & 4 & 5 & 6 & 7 &
8 & 9 & 10
\raisebox{-.1in}
{\begin{picture}(1, 19)(0, - 7)
\put(0,0){\makebox(0,0){}}
\end{picture}}
\\ \hline
$B_l$ &1  &  $-  \frac{1}{2} $ & $\frac{1}{6}$   &0  &
$-\frac{1}{30}$&0  & $\frac{1}{42} $ & 0 &$ -\frac{1}{30}  $
& 0& $\frac{5}{66} $
\raisebox{-.1in}
{\begin{picture}(1, 19)(0, - 7)
\put(0,0){\makebox(0,0){}}
\end{picture}}
\\
\hline
\end{tabular}
\caption{Bernoulli numbers $B_l$}
\label{t:bernoulli}
\end{center}
\end{table}

\subsection{Examples of coadjoint orbit representations}

In this section, we calculate some explicit examples of the coadjoint
orbit representations described in the previous sections.  We give the
formulae in local coordinates for the coadjoint orbit representations
of the simple groups $SU(2)$ and $SU(3)$, the loop group $\widehat{L}
SU(2)$, and the Virasoro group.  The coadjoint orbit description of
the representations of the groups $SU(2), SU(3)$ and $\widehat{L}
SU(2)$ are already understood from a global perspective (in these
cases, the coadjoint orbit approach is equivalent to that given by the
Borel-Weil theory); however, an explicit realization for
$\widehat{L} SU(2)$ such as that given here has not previously been
described.  In the case of the Virasoro algebra, the representations
arising from the coadjoint orbit description have not been previously
analyzed in a complete fashion, since the Borel-Weil theory does not
hold for the Virasoro group.  In this case, the explicit formulae
given here provide a useful tool for understanding previously
unstudied aspects of these representations.

\label{sec:coadjointexamples}
\subsubsection{$SU(2)$}
\label{sec:coadjointexamples1}

We begin with the group $G =SU(2)$.  As described in Section
\ref{sec:compactgroups}, the Lie algebra $\algg= {\rm su}(2)$ is a
3-dimensional vector space spanned by the generators $i J_k$, $ k
\in\{1,2,3\}$.  The effect of the adjoint action of $i J_k$ is to
generate a rotation around the $x_k$-axis in $\algg$.  The dual space
$\algg^*$ is also a 3-dimensional vector space.  We choose a set of
coordinates $b_k$ in this space such that the vector
\begin{equation}
{\bf b} = (b_1, b_2, b_3) \in \alggs
\end{equation}
has coordinates
\begin{equation}
b_k= \ip{ {\bf b}}{i J_k}.
\end{equation}
In terms of these coordinates, the coadjoint action of $i J_k$
generates a rotation around the $b_k$-axis in $\algg^*$.  Given a
vector ${\bf b} = (b_1, b_2, b_3) \in \alggs$, the coadjoint orbit of
$\bb$ is given by
\bge
\wb = \{ \bb' \in \alggs: |\bb'|^2 = b^2\},\ee
which is just the 2-sphere in $\alggs$ of radius $b = |\bb|$.  We will
now explicitly calculate the 2-form $\omega$ on $\wb$.  We choose a
canonical element $\bbo = (0,0, -b) \in \wb$.  To calculate $\omega$ at
the point $\bbo$, we need only find the explicit correspondence
between elements of $\algg$ and $T_{\bbo}W_b$.  Under the Lie algebra
coadjoint action, we have
\begin{eqnarray}
\ads_{iJ_1} \bbo & = & (0,-b, 0), \nonumber \\
\ads_{iJ_2} \bbo & = & (b,0, 0),  \\
\ads_{iJ_3} \bbo & = & (0,0, 0). \nonumber \end{eqnarray}
It follows that
\bge  \omega_{12}(\bbo) = \langle \bbo,  [\frac{iJ_2}{b},
-\frac{iJ_1}{b}] \rangle =
\frac{1}{b}.\ee
Since $\omega$ is $G$-invariant, it
is easy to see that $\omega$ is defined globally on $\wb$ by
\bge \omega_{ij}(\bb) = -\frac{1}{b^2} \eps_{ijk} b_k.\ee
In order for $\omega/2\pi$ to be an integral form, we must have
$\int_{\wb} \omega / 2 \pi = 2b \in \ZZ$, so $b$ must be a
half-integer.
Thus, whenever $b \in
\ZZ/2$, we can construct a line bundle $\lb$ over $\wb$ with curvature
form $i \omega$.

Because the stabilizer of ${\bf \bo}$ under the coadjoint action is
precisely the maximal subtorus $T=\{{\rm e}^{2 i \theta J_3}:0
\leq \theta \leq 2  \pi\}$, the coadjoint orbit space $W_b$ is
homeomorphic to the space $G/T$.  Using the results of Section
(\ref{sec:compactgroups}), we have a simple complex coordinate system on
this space.  As in (\ref{eq:generalelement}), given any complex number
$z$, it is possible to find functions $\alpha(z, \bar{z})$ and
$\beta(z, \bar{z})$, with $\beta(z, \bar{z})$ real, such that
\bge
{\rm e}^{z J_+} {\rm e}^{\azz J_-} {\rm e}^{\bzz J_3} \in G.
\label{eq:spherecoordinates}\ee
The functions $\azz$ and $\bzz$ can be calculated explicitly by
working in the fundamental representation of $SU(2)$; one finds that
\begin{eqnarray}
\azz & = & \frac{-\bar{z}}{1 + |z|^2}, \label{eq:ablow} \\
\bzz & = & \ln(1 + |z|^2). \nonumber \end{eqnarray}
By the general argument given for (\ref{eq:generalelement}), $z$ gives
a natural $G$-invariant complex structure to the space $W_b$.  In
fact, $W_b$ is just the space $S^2$, and $z$ is just the usual complex
coordinate on $S^2$ given by projection from the south pole onto
$\CC$, which is naturally invariant under the rotations generated by
$SU(2)$.

We can  now rewrite the symplectic form $\omega$ in terms of the
complex coordinate $z$.  We begin by rewriting the
differentials $dz$, $d
\bar{z}$ in terms of the original coordinates $b_k$.  At $\bbo$, we have
\begin{eqnarray}
dz & = & \frac{1}{2b} (db_1 - i db_2), \\ d\bar{z} & = &
\frac{1}{2b}
(db_1 + i db_2). \nonumber \end{eqnarray}
It is now possible to express $\omega$ at $\bbo$ in terms of the
$z,\bar{z}$ coordinates; one finds that
\begin{eqnarray}
\omega_{\bar{z} z} & = & - \omega_{z \bar{z}} \; = \; 2bi,
\label{eq:omega2}\\
\omega_{z z} & = & - \omega_{\bar{z} \bar{z}} \; = \; 0. \nonumber
\end{eqnarray}
Thus, $\omega$ is indeed a (1,1)-form, and along with the
$G$-invariant complex structure given by $z$, defines a \kl structure
on $\wb$.  We can therefore restrict attention to the space $\hb$ of
holomorphic sections of $\lb$.  Since $\frac{- \omega}{2\pi}$ is the
first Chern class of $\lb$, $\lb$ is  a holomorphic line bundle over
$W_b = S^2$ of degree $2b$.
For $b \geq 0$ it is a simple result of
the Riemann-Roch theorem
that $\lb$ admits exactly $2b+1$ linearly independent holomorphic
sections (see for example Griffiths and Harris \cite{GH}).
The $2b+1$ holomorphic sections can be represented in the vicinity
of the origin $z=\bar{z}=0$ by the holomorphic monomials, $1, z, z^2,
\ldots, z^{2b}$.  $\lb$ also has a natural Hermitian metric, which we will
discuss further in the next section.
Note that  when $b < 0$, the coadjoint orbit $W_b$ is the same as the
coadjoint orbit corresponding to $- b$.  However, the complex
structure on this space defined by (\ref{eq:spherecoordinates})  has
the opposite orientation to that defined for $- b$.  The holomorphic
line bundle thus defined has negative degree, and thus no holomorphic
sections.  We  will discuss the representations associated with these
bundles again briefly in  \ref{sec:global}; for now, we simply
restrict attention to the coadjoint orbits and associated line bundles
defined for $b \geq 0$.

It is also important to observe that when $b = 0$, the coadjoint orbit
becomes singular and the above analysis breaks down.  On the level of
the coadjoint orbit, what happens is that for this exceptional value
of $b$, the stabilizing subgroup increases in size from $T$ to the
entire group $SU(2)$.  Thus, in this case the coadjoint orbit is a
single point.  In this case constructing a line bundle, polarization,
and representation over this space is trivial and leads immediately to
the trivial representation.  However, for the other groups which we
consider in this thesis, there are analogous singular coadjoint
orbits; particularly in the cases involving infinite-dimensional
groups, discovering a polarization on these coadjoint orbits is a
difficult and in some cases unsolved problem.  Thus, we will use a
slight variation on the usual coadjoint orbit method to construct
representations for such singular values of the dual space variable.

The essential observation is that when $b = 0$; that is, when the
usual coadjoint orbit is singular, we can simply take the usual space
$G/T$ as the base space upon which to construct a line bundle leading
to the appropriate representation of $G$.  We use the same holomorphic
coordinates on $W_0 = G/T = S^2$ as above, and continue to use $i
\omega$ as a curvature form where $\omega$ is defined through
(\ref{eq:omega2}).  Although $\omega$ is degenerate and thus no longer
a symplectic form on the manifold when $b = 0$, the form is still a
nonsingular $G$-invariant (1,1) form on $S^2$, and gives rise to a
K\"{a}hler structure.  The rest of the analysis here, from the
construction of a line bundle over $W_b$ with curvature $i \omega$ to
the final local formulae for a representation of $\algg$ on a
polynomial space, can be carried out directly on this ``alternative''
orbit space.  However, the resulting line bundle $\lb$ will now be
flat in certain directions (all directions in the case of $SU(2)$).
We shall see that for all the groups considered here, this general
procedure gives rise to representations for exceptional dual space
parameters similar in structure to the more generic coadjoint orbit
representations.  For the remainder of this section when we refer to a
general coadjoint orbit $W_b$, we include the alternative orbit space
$W_0 = S^2$.

We have shown that we have a K\"{a}hler polarization for ${\cal H}_b$.
We can now proceed to apply the rest of the results of the previous
section to the group $SU(2)$.  The result (\ref{eq:complete}) states
that there exist operators
\begin{equation}
\hat{J}_a = \xi_a + f_a
\end{equation}
for $a = 3, \pm$, which act on ${\cal H}_b$ to give a representation
of $\algg_\CC$, where the operators $\xi_a$ and $f_a$ are given by
(\ref{eq:corollary}) and (\ref{eq:recursive}) respectively.

We will first describe the calculation of $\xi_a$ in this case.
The operators $\xi_+$, $\xi_3$ can easily be computed from the
simplified versions of the vector field formula
(\ref{eq:positivesimplification}) and (\ref{eq:cartansimplification})
respectively.  Since $J_+$ is the only positive root for $\algg$, we
have
\begin{equation}
\xi_+ = - \partial/\partial z.
\end{equation}
{}From $\ip{J_+}{J_3}= 1$, we have
\begin{equation}
\xi_3 = - z \frac{\partial}{\partial z} .
\end{equation}
For the remaining vector field $\xi_-$, it follows from
(\ref{eq:corollary}) that
\begin{equation}
\xi_- = \beta_{2,2} C_- (+, +) z^2 \frac{\partial}{\partial z}.
\end{equation}
Computing $C_- (+, +) = - 2$, and looking up $\beta_{2,2}=- 1/2$  in
Table~\ref{t:beta}, we have
\begin{equation}
\xi_- = z^2 \frac{\partial}{\partial z}.
\end{equation}
It is easy to verify that these vector field operators satisfy the
correct commutation relations $[\xi_3, \xi_\pm]= \pm \xi_\pm$,
$[\xi_+, \xi_-] = 2 \xi_3$.

We can now calculate the functions $f_a$.
The nonzero functions are $f_3$ and $f_-$;  these functions are
immediately given by (\ref{eq:recursive}), and are
\begin{eqnarray}
f_3 & = & - i \ip{ {\bf b}}{J_3} = b,\\
f_- & = &  i \ip{{\bf b}}{2 J_3} z =- 2 b z. \nonumber
\end{eqnarray}
Combining these results together, we have
\begin{eqnarray}
\hj_3 & = & -z \frac{\pdv}{\pdv z}  +b, \nonumber \\
\hj_+ & = & - \frac{\pdv}{\pdv z}, \label{eq:representation2} \\
\hj_- & = & z^2 \frac{\pdv}{\pdv z} - 2b z. \nonumber
\end{eqnarray}
Again, it is easy to verify that these operators satisfy the correct
commutation relations $[\hat{J}_3, \hat{J}_\pm] = \pm \hat{J}_{\pm} $,
$[\hat{J}_+, \hat{J}_-] = 2 \hat{J}_3$.

The operators (\ref{eq:representation2}) give an explicit realization
of the algebra ${\rm su} (2)_\CC$ on the space $R = \CC[z]$ of
polynomials in the variable $z$.   The Hilbert space ${\cal H}_b
\subset R$ consists of all polynomials in $z$ of order $\leq 2b$, and
carries an irreducible representation of the algebra.
A basis for the Hilbert space ${\cal H}_b$ is given by the vectors
\begin{equation}
| m \rangle = z^m, \; \; 0 \leq m \leq 2b.
\end{equation}
In terms of this basis, the operators $\hat{J}_a$ act according to
\begin{eqnarray}
\hat{J}_3 | m \rangle & = &(b - m) | m \rangle \nonumber\\
\hat{J}_+ | m \rangle & = &  - m | m - 1 \rangle \\
\hat{J}_- | m \rangle & = & (m - 2b) | m + 1 \rangle. \nonumber
\end{eqnarray}

Note that the state $| 0 \rangle =\nul$ is annihilated by $\hat{J}_+$
and has weight $b$ with respect to $\hat{J}_3$, and the state $| 2b
\rangle$ is annihilated by $\hat{J}_-$; all of these conditions are
necessary in order for ${\cal H}_b$ to form an irreducible
representation of the algebra.  In Section
\ref{sec:global} we will describe in
more detail the condition of global holomorphicity on sections and the
existence of a Hermitian metric on this Hilbert space; in the
following chapters we will also discuss the general module structure
of these representations.

\subsubsection{$SU(3)$}
\label{sec:coadjointexamples1b}

We now consider the case $G = SU(3)$.  The Lie algebra $\algg = {\rm
su} (3)$ is an 8-dimensional vector space spanned by the generators $i
J_a = i \lambda_a/2, a \in\{1,2, \ldots, 8\}$ defined through
(\ref{eq:GellMann}).  As in the case of $SU(2)$, we can choose a set
of coordinates $b_k$ in the dual space $\algg^*$ such that the vector
\begin{equation}
{\bf b} = (b_1, \ldots,b_8) \in \alggs
\end{equation}
has coordinates
\begin{equation}
b_k= \ip{ {\bf b}}{i J_k}.
\end{equation}

We now choose a canonical element ${\bf b} \in \algg^*$ with
\begin{eqnarray}
b_3 & = & -p/2 \nonumber\\
b_8 & = &  -\frac{1}{\sqrt{3}}  (q + p/2) \\
b_k & = & 0, \; \; k \not\in\{3,8\}. \nonumber
\end{eqnarray}

In this subsection, we will denote the generators of $\algg$ by
\begin{eqnarray}
J_{+ \alpha} & = &  e_\alpha, \nonumber\\
J_{- \alpha} & = &  f_\alpha,\\
J_\alpha & = &  \frac{1}{2} h_\alpha, \nonumber
\end{eqnarray}
where $\alpha \in\{t,u,v\}$ (note that $J_v = J_t + J_u$).  With this
notation, the nonzero structure constants of the algebra are given by
\begin{eqnarray}
i\mixten{f}{\pm t \pm u}{\pm v} & = & -i\mixten{f}{\pm u \pm t}{\pm v} =
\pm 1 \nonumber\\
i\mixten{f}{\pm \alpha\mp\alpha}{\alpha} & = &  \pm 2 \nonumber\\
i\mixten{f}{\alpha \pm \beta}{\pm \beta} & = &
-i\mixten{f}{\pm \beta \alpha}{\pm \beta}  = \pm \ip{\beta}{J_\alpha}
\label{eq:structureconstants3}\\
i\mixten{f}{\pm u \mp v}{\mp t} &= &-i\mixten{f}{\mp v \pm u}{\mp t}  =
\pm 1 \nonumber\\
i\mixten{f}{\pm t \mp v}{\mp u} &= &-i\mixten{f}{\mp v \pm t}{\mp u}  =
\mp 1. \nonumber
\end{eqnarray}
For generic values of $p,q$ the only generators which stabilize ${\bf
b}$ under the coadjoint action are $J_u$ and $J_t$.  In this case, the
(continuous) stabilizing subgroup is the maximal subtorus $T$ of
$SU(3)$, and the coadjoint orbit space is equivalent to the
homogeneous space $SU(3)/T$.  There are exceptional values of the
parameters $p$ and $q$ (for example, $p = 0$ or $q = 0$) for which the
stabilizing subgroup of ${\bf b}$ is larger than $T$.  In this case,
the actual coadjoint orbit  space becomes a smaller quotient space of
$SU(3)$.  As in the case of an exceptional orbit for $SU(2)$ described
above, rather than dealing directly with these smaller coadjoint orbit
spaces, we use the alternative coadjoint orbit space $W_b = SU(3)/T$
for all values of $p,q$ and carry out the analysis  using the
consequent degenerate symplectic form $\omega$.

We will use the local complex coordinates
$t,u,v$ on $W_b = G/T$ defined by
\begin{equation}
\exp \left[t J_{+ t} + u J_{+ u}+v J_{+ v}\right]
\exp \left[ \sum_{\alpha \in \Phi_+} a_\alpha J_{-\alpha} \right]
\exp \left[ \sum_{\alpha \in \Delta} \beta_\alpha J_\alpha  \right]\in G,
\end{equation}
where $a_\alpha$ and $\beta_\alpha$ are functions of $t,u,v, \bar{t},
\bar{u}, \bar{v}$,  with $\beta_\alpha$ real.  In terms of these
coordinates, it is easy to calculate that $\omega$ is a (1,1) form at
the point ${\bf b}$,
with nonzero components
\begin{eqnarray}
\omega_{t \bar{t}} & = &  \ip{ {\bf b}}{[J_{+ t}, - J_{- t}]} =-ip
\nonumber\\
\omega_{u \bar{u}} & = &  \ip{ {\bf b}}{[J_{+ u}, - J_{- u}]} =-iq\\
\omega_{v \bar{v}} & = &  \ip{ {\bf b}}{[J_{+ v}, - J_{- v}]} =-i(p +
q).  \nonumber
\end{eqnarray}
Again, note that in the exceptional cases when $p = 0$ or $q = 0$,
this form is degenerate.  Because the complex structure and $\omega$
are both $G$-invariant, it follows that $\omega$ is a (1,1) form
everywhere on $W_b$, and thus that we can choose a K\"{a}hler
polarization when $\omega/2 \pi$ is an integral cohomology class.  By
using the analysis for the $SU(2)$ case in the previous subsection,
necessary conditions for the integrality condition on $\omega$ can be
derived.  Essentially, these conditions follow from taking all
subgroups $S$ of $G$ which are equivalent to $SU(2)$, and which have
the property that the stabilizer $T$ of ${\bf b}$ in $S$ is a maximal
subtorus (homeomorphic to $S^1$) of $S$.  In fact, because we are
taking the orbit space to be $SU(3)/T$ regardless of the size of the
stabilizer of ${\bf b}$, we can also take $SU(2)$ subgroups which
completely stabilize ${\bf b}$.
For each $SU(2)$ subgroup containing at least a $T$ stabilizer, the
coadjoint orbit contains a (possibly trivial) second homology class
$S/T$.  The integral of $\omega$ over this homology class must be an
integral multiple of $2 \pi$; by contracting ${\bf b}$ with the
generator of $T$ this implies an integrality condition analogous to
that derived for $SU(2)$.  For the specific group $SU(3)$ with which
we are currently concerned, these necessary conditions for integrality
can be derived by taking the two $SU(2)$ subgroups generated by the
algebra elements $\{i J_t,i J_1,i J_2\}$ and $\{i J_u,i J_6,i J_7\}$,
with stabilizing subgroups generated by $i J_t$ and $i J_u$.  The
consequent integrality condition is that $p$ and $q$ must be integral.
It is possible to show that this condition is also sufficient for
$\omega/2 \pi$ to be an integral cohomology class.  For the rest of
this subsection, we restrict to coadjoint orbits which satisfy this
integrality condition.  Just as we restricted to $b \geq 0$ for
$SU(2)$ to guarantee that the line bundle $\lb$ was of nonnegative
degree, we will also restrict here to line bundles with $p,q \geq 0$.
Negative values for these parameters again correspond to a different
choice of complex structure on the same coadjoint orbit, as was the
case for $SU(2)$;  the resulting line bundles admit no globally
holomorphic sections.

We will now proceed to construct the operators
\begin{equation}
\hat{J}_a = \xi_a + f_a.
\end{equation}
As in the previous subsection, we begin by calculating the vector
fields $\xi_a$.  From  (\ref{eq:cartansimplification}), we can
calculate
\begin{eqnarray}
\xi_t & = &  - t \derivative{t}  -
 \frac{1}{2}v  \derivative{v} +\frac{1}{2}u  \derivative{u}   \\
\xi_u & = &  - u \derivative{u}
- \frac{1}{2}v  \derivative{v} +\frac{1}{2}t  \derivative{t}.  \nonumber
\end{eqnarray}
{}From (\ref{eq:positivesimplification}), we calculate
\begin{eqnarray}
\xi_{+ t} & = & - \derivative{t} -\frac{1}{2}u \derivative{v}  \nonumber\\
\xi_{+ u} & = & - \derivative{u} + \frac{1}{2}t \derivative{v} \\
\xi_{+ v} & = &  - \derivative{v}. \nonumber
\end{eqnarray}
Finally, from the general equation (\ref{eq:corollary}), we calculate
the remaining vector fields
\begin{eqnarray}
\xi_{- t} & = & t^2 \derivative{t} +
(-v- \frac{1}{2}  tu) \derivative{u} + ( \frac{1}{2}tv -\frac{1}{4}  t^2 u)
\derivative{v}\nonumber\\
\xi_{- u} & = & (v- \frac{1}{2}  tu) \derivative{t}
+u^2 \derivative{u} + (\frac{1}{2}uv + \frac{1}{4}  t u^2)
\derivative{v}\\
\xi_{- v} & = &
(tv- \frac{1}{2} t^2 u) \derivative{t}
+(uv+ \frac{1}{2} tu^2) \derivative{u}
+(v^2+ \frac{1}{4} t^2 u^2) \derivative{ v}. \nonumber
\end{eqnarray}
It  can be verified that these
vector fields satisfy the commutation relations $[\xi_a, \xi_b] =
i\mixten{f}{ab}{c}\xi_c$, with the structure constants
(\ref{eq:structureconstants3}).

We can now proceed to calculate the functions $f_a$ from
(\ref{eq:recursive}).  The nonzero functions are $f_t,f_u$, and $f_{-
\alpha}$, $\alpha \in\{t,u,v\}$.  The functions associated with the
Cartan subalgebra are given by
\begin{eqnarray}
f_{t} & = &  - i \ip{ {\bf b}}{ J_t} = p/2\\
f_{u} & = &  - i \ip{ {\bf b}}{ J_u} = q/2. \nonumber
\end{eqnarray}
The functions associated with the simple roots are given by
\begin{eqnarray}
f_{- t} & = & i \ip{ {\bf b}}{2 J_t}t = - p t\\
f_{- u} & = & i \ip{ {\bf b}}{2 J_u}u =- q u.\nonumber
\end{eqnarray}
We can compute the final function $f_{- v}$ from the recursive
definition in (\ref{eq:recursive}), which gives
\begin{equation}
f_{- v} = [\xi_{- u}f_{- t} - \xi_{- t}f_{- u}] =-v (p + q) +
\frac{1}{2}tu (p - q).
\end{equation}
Again, it is straightforward to verify that $\xi_a f_b - \xi_b f_a= i
\mixten{f}{ab}{c} f_c$.

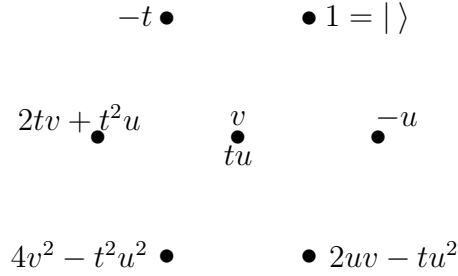
\begin{figure}
\centering
\begin{picture}(200,100)(- 100,- 50)
\put(-53,0){\circle*{5}}
\put( 0,0){\circle*{5}}
\put(53,0){\circle*{5}}
\put(27,45){\circle*{5}}
\put(-27,45){\circle*{5}}
\put(27,-45){\circle*{5}}
\put(-27,-45){\circle*{5}}
\put(60,7){\makebox(0,0){$- u$}}
\put(-60,7){\makebox(0,0){$2tv+t^2 u$}}
\put(39,45){\makebox(20,0){$1 =\nul$}}
\put(-39,45){\makebox(0,0){$- t$}}
\put(39,-45){\makebox(40,0){$2uv - tu^2$}}
\put(-90,-45){\makebox(60,0){$4v^2 -t^2 u^2$}}
\put(0,7){\makebox(0,0){$v$}}
\put(0,-7){\makebox(0,0){$tu$}}
\end{picture}
\caption[Polynomial basis for adjoint representation of $SU(3)$]{\footnotesize
Polynomial basis for adjoint representation of $SU(3)$.}
\label{f:adjoint3}
\end{figure}

We can now write the expressions for the complete operators
$\hat{J}_a$ as first-order differential operators on the space $R =
\CC[t,u,v]$ of polynomials in the variables $t,u,v$.
\begin{eqnarray}
\hat{J}_t & = &  - t \derivative{t} -
\frac{1}{2}  v \derivative{v} +\frac{1}{2} u \derivative{u} + p/2  \nonumber\\
\hat{J}_u & = &  - u \derivative{u}  -\frac{1}{2}
 v \derivative{v} +\frac{1}{2} t \derivative{t}+ q/2.  \nonumber \\
\hat{J}_{+ t} & = & - \derivative{t} - \frac{1}{2}u \derivative{v} \nonumber\\
\hat{J}_{+ u} & = & - \derivative{u} + \frac{1}{2}t \derivative{v}
\label{eq:operators3}\\
\hat{J}_{+ v} & = &  - \derivative{v}. \nonumber\\
\hat{J}_{- t} & = & t^2 \derivative{t} +
(-v- \frac{1}{2}  tu) \derivative{u} + (\frac{1}{2}tv-\frac{1}{4}  t^2 u )
\derivative{v} - pt\nonumber\\
\hat{J}_{- u} & = & (v- \frac{1}{2}  tu) \derivative{t}
+u^2 \derivative{u} + (\frac{1}{2}uv + \frac{1}{4}  t u^2)
\derivative{v}- qu\nonumber\\
\hat{J}_{- v} & = &
(tv- \frac{1}{2} t^2 u) \derivative{t}
+(uv+ \frac{1}{2} tu^2) \derivative{u}
+(v^2+ \frac{1}{4} t^2 u^2) \derivative{ v}
- v (p + q) + \frac{1}{2}  tu (p - q). \nonumber
\end{eqnarray}
We will discuss the module structure of these representations in later
sections.  For now, we consider a simple example of an $SU(3)$
coadjoint orbit representation, when $p = q = 1$.  This choice of
values for $(p,q)$ corresponds to the adjoint representation of
$SU(3)$.  The irreducible representation of ${\rm su} (3)_\CC$ on $R$
which contains the highest weight vector $\nul$ is a representation on
an 8-dimensional subspace ${\cal H}_b$ of $R$.  A basis for ${\cal
H}_b$ is graphed in Figure~\ref{f:adjoint3}, according to the weights
of the states.  Note that there are two linearly independent basis
states with weight (0,0).  It is a simple calculation to verify that
the operators (\ref{eq:operators3}) act on the basis of ${\cal H}_b$
according to the usual description of the adjoint representation.
Note that we have not normalized this basis for ${\cal H}_b$; a
Hermitian structure for this Hilbert space is discussed in Section
\ref{sec:global}.

\subsubsection{$\widehat{L} SU(2)$}
\label{sec:coadjointexamples2}

We now consider the example of the centrally extended loop group
$\widehat{L}SU(2)$.
We denote the Fourier generators of this group by $\jj{J}{a}{n}$ where
$a \in\{3, \pm\}$ and $n \in \ZZ$; the central generator will be
denoted by $J_c = C$ for uniformity of notation.
We will also abbreviate $\jj{J}{3}{n}$ to simply $\jj{J}{}{n}$
The dual space $L (\algg^*)\oplus \RR$
to the Lie algebra of a general
loop group was described in \ref{sec:loopgroups}.  In this case,  an
element $(b, - it)$ of the dual space is defined by a  function
\begin{equation}
b: S^1 \rightarrow {\rm su} (2)^*
\end{equation}
and a real number $t \in \RR$.  We will concern ourselves here only
with the elements of the dual space where $b$ is a smooth function.
The dual pairing between the algebra $\widehat{L}\algg$ and the dual
space is given by (\ref{eq:loopinnerproduct}), and the coadjoint
action of $\widehat{L}G$ on the dual space is given by
(\ref{eq:loopcoadjointaction}).  A fairly straightforward analysis of
the coadjoint orbits of a general loop group leads to the result
\cite{PS} that the coadjoint orbits are in 1-1 correspondence with
pairs $(c, - ik)$ where $c$ is a conjugacy class in $G$ and $k \in
\RR$.  In the case we are concerned with here, namely
$\widehat{L}SU(2)$, we can choose a representative of each coadjoint
orbit to be described by a pair $({\bf b}, - ik)$ with ${\bf b}$ being
a constant function taking a value in $\algg^*$ with coordinates
$(0,0, - b)$, where we use the same coordinates $b_j$ for $\algg^*$ as
in Subsection
\ref{sec:coadjointexamples1}.  We will take the dual elements of this
type to be the canonical elements of the coadjoint orbits.  We
denote a fixed coadjoint orbit of this type by $W_{b,k}$.

{}From the expression (\ref{eq:loopalgebracoadjoint}) for the coadjoint
action of the algebra, we can calculate the change in $({\bf b}, - ik)$
under the action of the generator $\jj{J}{a}{n}$.   Calculating the
components of this shift, we have
\begin{equation}
\ip{ \ad_{\jj{J}{a}{n}}({\bf b}, - ik)}{ (\jj{J}{c}{m}, 0)} =
\delta_{n, - m}\left[
i b (i \mixten{f}{ac}{3})  + i n k g_{ac} \right].
\end{equation}
{}From this equation, it is straightforward to verify that when $2b/k
\not\in \ZZ$, the
stabilizer of $({\bf b}, - ik)$ under the coadjoint action of
$\widehat{L}SU(2)$ is generated by the elements $\jj{J}{}{0}$ and $C$,
and is given by $T_{(0)}\times S^1$.  Thus, the coadjoint orbits
associated with these dual space elements are homeomorphic to
$\widehat{L}G/(T \times S^1)$.  For the exceptional values of $b,k$
where $2b = nk$ for some $n \in \ZZ$, the coadjoint orbit spaces are
smaller.  As in the previous subsections, we simply take
$\widehat{L}G/(T \times S^1)$ to be the alternative orbit space in
these exceptional cases and carry out the analysis in a uniform
fashion.

As mentioned in Section \ref{sec:loopgroups}, the expression
(\ref{eq:generalelement}) can be used to construct a holomorphic
coordinate system on $W_{b,k} = \widehat{L}G/(T \times S^1)$, with an
infinite set of coordinates $\{z_{(n, \alpha)} | (n, \alpha) \succ
0\}$.  Explicitly, we will use the coordinates $\jj{z}{a}{n}$ where $a
\in\{\pm, 3\}$, $n \in \ZZ$ and either $n > 0$ or   $(n, \alpha) = (0,
+)$.  As with the generators, we will write
$\jj{z}{}{n}=\jj{z}{3}{n}$.

In these coordinates, we can proceed to calculate the components of
the symplectic form $\omega$.  We find that $\omega$ is a (1,1)
form at the point $({\bf b}, - ik)$, with nonzero components
\begin{eqnarray}
\omega_{\jj{z}{a}{n} \jj{\bar{z}}{a}{n}} & = &
\ip{({\bf b}, - ik)}{[\jj{J}{a}{n}, - \jj{J}{-a}{-n}]} \\
& = & - ib (i \mixten{f}{a(- a)}{3}) + i k n g_{a (- a)}. \nonumber
\end{eqnarray}
In particular, we have
\begin{eqnarray}
\omega_{\jj{z}{\pm}{n} \jj{\bar{z}}{\pm}{n}} & = &
\mp 2 ib + i n k  \\
\omega_{\jj{z}{}{n} \jj{\bar{z}}{}{n}} & = &
i n k/2.\nonumber
\end{eqnarray}
This symplectic form (degenerate for exceptional values of $b,k$) is a
$G$-invariant (1,1) form on $W_{b,k}$; thus, as usual, we can choose a
K\"{a}hler polarization when $\omega/2 \pi$ is an integral cohomology
class.  As in the previous subsection, we can use $SU(2)$ subgroups of
$\widehat{L}SU(2)$ to give necessary conditions on $b$ and $k$ for the
integrality condition to be satisfied.  This analysis is parallel to
that carried out in Section
\ref{sec:loopgroups} to determine the integrality condition for a
central extension of $LG$.  Clearly, the $SU(2)$ subgroup given by the
constant loops $SU(2)_{(0)}$ is a subgroup of $\widehat{L}SU(2)$ with
a subgroup stabilizing $({\bf b}, - ik)$ given by $T_{(0)}$.  The generators
of this subgroup are the generators $I_a$ given by
(\ref{eq:subalgebra1}).  The integrality condition associated with
this subgroup is that
\begin{equation}
\ip{ ({\bf b}, - ik)}{ iI_3} = -b \in \ZZ/2.
\end{equation}
So we have the condition that $2b \in \ZZ$.  Similarly, we find that
the subalgebra  (\ref{eq:subalgebra2}) also generates an $SU(2)$
subgroup with stabilizing subgroup $T$.  The real generators of this
subgroup are given by the functions
\begin{eqnarray}
\tilde{I}_1 (\theta) & = &  \frac{1}{2} (\tilde{I}_+(\theta) +
\tilde{I}_- (\theta))
= \cos \theta J_1  +\sin \theta J_2 \nonumber\\
\tilde{I}_2 (\theta) & = & \sin \theta J_1  - \cos \theta J_2\\
\tilde{I}_3 (\theta) & = &  C/2 - J_3. \nonumber
\end{eqnarray}
The integrality condition associated with this $SU(2)$ subgroup is
that
\begin{equation}
k  - 2 b \in \ZZ.
\end{equation}
As previously, in order to have holomorphic sections in our line
bundles, we must furthermore restrict to values where $k \geq 2 b \geq
0$.  Thus, for all integers $k$ and half-integers $b$ satisfying $k
\geq 2 b \geq 0$,
we have a coadjoint orbit $W_{b,k}$ which admits a K\"{a}hler
polarization, over which we can choose a line bundle $\lb$ with
curvature form $i \omega$.

We can now proceed to construct the operators
\begin{equation}
\hat{J}_{a (n)} = \xi_{a (n)} + f_{a (n)}
\end{equation}
and
\begin{equation}
\hat{C} = \hat{J}_c = \xi_c + f_c
\end{equation}
These operators act on the space $R$ of polynomials in the infinite
set of variables $\jj{z}{a}{n}$.  This space is graded according to
$\ZZ \oplus \ZZ$,  with the  variables $\jj{z}{\pm}{n}$ having
degrees $(n, \pm 1)$ and $z_{(n)}$ having degree $(n,0)$.
{}From (\ref{eq:cartansimplification}) we can calculate the vector
fields corresponding to the generators in the stabilizer
\begin{eqnarray}
\jj{\xi}{}{0} & = &   \sum_{n \geq 0} -z_{+ (n)} \derivative{z_{+
(n)}} + \sum_{n > 0} z_{- (n)} \derivative{z_{- (n)}} \\
\xi_c & = &  0.\nonumber
\end{eqnarray}
The vector field $\xi_{(0)}$ contains an infinite number of terms, but
its action on any fixed polynomial in $R$ only involves a finite
number of these terms.  This is a characteristic property of the
generators in the polynomial coadjoint orbit representations for
infinite-dimensional algebras.  The
vector field $\xi_c$ is 0 because $C$  is central in the algebra.

We can now proceed to calculate the vector fields $\jj{\xi}{a}{n}$ for
$(n,a) \succ 0$ using (\ref{eq:positivesimplification}),  and the
remaining vector fields  using (\ref{eq:corollary}).  It is difficult
to write down these vector fields in closed-form notation; again,
there are an infinite number of terms in each vector field, only a
finite number of which contribute to the action on a specific
polynomial in $R$.  The leading terms in the vector field operators
$\jj{\xi}{a}{n}$ corresponding to positive roots are
\begin{eqnarray}
\xi_{+ (n)} = - \derivative{z_{+ (n)}} + \frac{1}{2}
\sum_{m > 0} z_{ (m)} \derivative{z_{+ (m + n)}} -
\sum_{m > 0} z_{- (m)} \derivative{z_{ (m + n)}} + \cdots  \; \; \;
(n \geq 0)\nonumber\\
\xi_{- (n)} = - \derivative{z_{- (n)}} - \frac{1}{2}
\sum_{m > 0} z_{ (m)} \derivative{z_{- (m + n)}} -
\sum_{m\geq 0} z_{+ (m)} \derivative{z_{ (m + n)}} + \cdots \; \; \; (n
> 0) \\
\xi_{ (n)} = - \derivative{z_{ (n)}} + \frac{1}{2}
\sum_{m > 0} z_{- (m)} \derivative{z_{- (m + n)}} -
\sum_{m \geq 0} z_{+ (m)} \derivative{z_{+ (m + n)}} + \cdots
 \; \; \; (n> 0). \nonumber
\end{eqnarray}

We can calculate the functions $\jj{f}{a}{n}$ using
(\ref{eq:recursive}).  The functions associated with the generators in
the stabilizer are given by
\begin{eqnarray}
f_{(0)} & = &  b, \\
f_c & = &  k. \nonumber
\end{eqnarray}
In general, the functions $\jj{f}{\pm}{n}$ are of degree $(- n, \mp 1)$,
and the function $\jj{f}{}{n}$ is of degree $(- n,0)$.  The simple
roots of $\widehat{L}SU(2)$ are associated with the generators
$\jj{J}{ +}{0}$ and $\jj{J}{-}{1}$.  The associated functions are
\begin{eqnarray}
f_{- (0)} & = & -2b z_{+ (0)}\\
f_{+ (-1)} & = &  (2b - k) z_{- (1)}. \nonumber
\end{eqnarray}

All the remaining functions $\jj{f}{a}{n}$ can be calculated using the
recursive equation (\ref{eq:recursive}).  Explicitly, we have
\begin{eqnarray}
\jj{f}{+}{-n - 1} & = &  \xi_{ (-1)}  \jj{f}{+}{-n} -
\xi_{+ (-n)}  \jj{f}{}{-1} \; \; \; (n > 0) \nonumber\\
\jj{f}{-}{-n} & = &  \xi_{- (0)}  \jj{f}{}{-n} -
\xi_{ (-n)}  \jj{f}{-}{0} \; \; \; (n > 0)\\
\jj{f}{}{-n} & = &  \frac{1}{2} \left[
\xi_{+ (-n)}  \jj{f}{-}{0} -
\xi_{- (0)}  \jj{f}{+}{ - n} \right] \; \; \; (n > 0). \nonumber
\end{eqnarray}

We can combine these functions with the vector fields described above
to get the complete set of operators $\jj{\hat{J}}{a}{n}$, $\hat{C}$
which realize the algebra $\widehat{L}{\rm su}(2)$ on the space $R$ of
polynomials.  Explicitly writing all the terms in these operators
which have nonzero matrix elements between states of degrees with $n
\leq 2$, we have
{\small
\begin{eqnarray}
\hat{C} & = & k \nonumber\\
\jj{\hat{J}}{-}{2} & = & - \derivative{z_{- (2)}}
	+ z_{+ (0)} \derivative{z_{ (2)}}
	+ \frac{1}{6} z_{+ (0)}^2 \derivative{z_{+ (2)}}
	+ \cd_3 \nonumber \\
\jj{\hat{J}}{}{2} & = &
	- \derivative{z_{ (2)}}
	- \frac{1}{2} z_{+ (0)} \derivative{z_{+ (2)}}
	+ \cd_3 \nonumber \\
\jj{\hat{J}}{+}{2} & = & - \derivative{z_{+ (2)}}
	+ \cd_3 \nonumber \\
\jj{\hat{J}}{-}{1} & = &
	- \derivative{z_{- (1)}}
	+ z_{+ (0)} \derivative{z_{ (1)}}
	+ \frac{1}{6}z_{+ (0)}^2 \derivative{z_{+ (1)}}
	- \left(\frac{1}{6} z_{+ (0)} z_{- (1)}
	+ \frac{1}{2}z_{ (1)}\right) \derivative{z_{- (2)}} \nonumber \\
& &
	+ \left(\frac{1}{6} z_{+ (0)} z_{ (1)}
	+ z_{+ (1)}\right) \derivative{z_{ (2)}}
	+ \left(\frac{1}{3} z_{+ (0)} z_{+ (1)}
	- \frac{1}{90} z_{+ (0)}^3 z_{- (1)}\right) \derivative{z_{+ (2)}}
	+ \cd_3 \nonumber \\
\jj{\hat{J}}{}{1} & = &
 	- \derivative{z_{ (1)}}
	- \frac{1}{2}z_{+ (0)} \derivative{z_{+ (1)}}
	+ \frac{1}{2}z_{- (1)} \derivative{z_{- (2)}}
	- \frac{1}{3}z_{+ (0)} z_{- (1)} \derivative{z_{ (2)}}
\nonumber \\
& &
	+ \left(\frac{1}{12} z_{+ (0)} z_{ (1)}
	- \frac{1}{2} z_{+ (1)}\right) \derivative{z_{+ (2)}}
 	+ \cd_3 \nonumber \\
\jj{\hat{J}}{+}{1} & = & - \derivative{z_{+ (1)}}
	- z_{- (1)} \derivative{z_{ (2)}}
	+ \left(\frac{1}{2} z_{ (1)}
	- \frac{1}{6} z_{+ (0)} z_{- (1)}\right) \derivative{z_{+ (2)}}
	+ \cd_3 \nonumber \\
\jj{\hat{J}}{-}{0} & = &
	-2 b z_{+ (0)}
	+ z_{+ (0)}^2 \derivative{z_{+ (0)}}
	- \left(z_{+ (0)} z_{- (1)}
	+ z_{ (1)}\right) \derivative{z_{- (1)}}\nonumber \\
& &
	+ (2 z_{+ (1)} - \frac{1}{3}2 z_{+ (0)}^2 z_{- (1)})
		\derivative{z_{ (1)}}
	+ \left(\frac{1}{6} z_{+ (0)}^2 z_{ (1)}
	+ z_{+ (0)} z_{+ (1)}\right) \derivative{z_{+ (1)}} \nonumber
\\
& &
	+ \left(\frac{1}{6} z_{+ (0)} z_{- (1)} z_{ (1)}
	- z_{+ (0)} z_{- (2)}
	- z_{ (2)}\right) \derivative{z_{- (2)}} \nonumber \\
& &
	+ \left(\frac{2}{45} z_{+ (0)}^3 z_{- (1)}^2
	- \frac{2}{3} z_{+ (0)} z_{- (1)} z_{+ (1)}
	- \frac{2}{3} z_{+ (0)}^2 z_{- (2)}
	+ 2 z_{+ (2)}\right) \derivative{z_{ (2)}} \nonumber \\
& &
	+ \left(\frac{1}{6} z_{+ (0)} z_{ (1)} z_{+ (1)}
	- \frac{1}{90} z_{+ (0)}^3 z_{- (1)} z_{ (1)}
	+ \frac{1}{6} z_{+ (0)}^2 z_{ (2)}
	+ z_{+ (0)} z_{+ (2)}\right) \derivative{z_{+ (2)}}
	+ \cd_3 \nonumber \\
\jj{\hat{J}}{}{0} & = &
	b
	- z_{+ (0)} \derivative{z_{+ (0)}}
	+ z_{- (1)} \derivative{z_{- (1)}}\nonumber \\
& &
	- z_{+ (1)} \derivative{z_{+ (1)}}
	+ z_{- (2)} \derivative{z_{- (2)}}
	- z_{+ (2)} \derivative{z_{+ (2)}}  + \cd_3 \\
\jj{\hat{J}}{+}{0} & = &
 	- \derivative{z_{+ (0)}}
	- z_{- (1)} \derivative{z_{ (1)}}
	+ \left(\frac{1}{2} z_{ (1)}
	- \frac{1}{6} z_{+ (0)} z_{- (1)}\right) \derivative{z_{+
(1)}} \nonumber \\
& &
	+ \frac{1}{6} z_{- (1)}^2 \derivative{z_{- (2)}}
	+ \left(\frac{1}{6} z_{- (1)} z_{ (1)}
	- z_{- (2)}\right) \derivative{z_{ (2)}} \nonumber \\
& &
	+ \left(\frac{1}{90} z_{+ (0)}^2 z_{- (1)}^2
	- \frac{1}{12} z_{ (1)}^2
	- \frac{1}{6}z_{- (1)} z_{+ (1)}
	- \frac{1}{6}z_{+ (0)} z_{- (2)}
	+ \frac{1}{2}z_{ (2)}\right) \derivative{z_{+ (2)}}
	+ \cd_3 \nonumber \\
\jj{\hat{J}}{-}{-1} & = &
	\frac{k -4b}{3} z_{+ (0)}^2 z_{- (1)}
	+ \frac{k -2b}{2} z_{+ (0)} z_{(1)}
	- (k +2b) z_{+ (1)}\nonumber \\
& &
	+ \left(\frac{1}{3} z_{+ (0)}^3 z_{- (1)}
	+ 2 z_{+ (0)} z_{+ (1)}\right) \derivative{z_{+ (0)}}\nonumber \\
& &
	- \left(\frac{1}{3} z_{+ (0)}^2 z_{- (1)}^2
	+ \frac{2}{3} z_{+ (0)} z_{- (1)} z_{ (1)} \right.\nonumber \\
& &\; \; \; \; \; \; \; \left.
	+ \frac{1}{2} z_{ (1)}^2
	+ z_{- (1)} z_{+ (1)}
	+ z_{+ (0)} z_{- (2)}
	+ z_{ (2)}\right) \derivative{z_{- (1)}} \nonumber \\
& &
	+ \left(2 z_{+ (2)}
	- \frac{2}{5} z_{+ (0)}^3 z_{- (1)}^2
	- \frac{1}{3} z_{+ (0)}^2 z_{- (1)} z_{ (1)}  \right.\nonumber \\
& &\; \; \; \; \; \; \; \left.
	- \frac{1}{6} z_{+ (0)} z_{ (1)}^2
	- \frac{1}{3} z_{+ (0)} z_{- (1)} z_{+ (1)}
	+ z_{ (1)} z_{+ (1)}
	+ \frac{1}{3} z_{+ (0)}^2 z_{- (2)}\right) \derivative{z_{(1)}}\nonumber \\
& &
	+ \left(\frac{1}{45} z_{+ (0)}^4 z_{- (1)}^2
	+ \frac{1}{9} z_{+ (0)}^3 z_{- (1)} z_{ (1)}
	+ \frac{1}{6} z_{+ (0)}^2 z_{- (1)} z_{+(1)}\right.\nonumber \\
& &\; \; \; \; \; \; \; \left.
	+ \frac{1}{2} z_{+ (0)} z_{ (1)} z_{+ (1)}
	+ z_{+ (1)}^2
	- \frac{1}{6} z_{+ (0)}^3 z_{- (2)}
	- \frac{1}{3} z_{+ (0)}^2 z_{ (2)}
	+ z_{+ (0)} z_{+ (2)}\right) \derivative{z_{+ (1)}}
\nonumber \\
& &
	+ \co(z_{- (3)}) \nonumber \\
\jj{\hat{J}}{}{-1} & = &
   \frac{2b - k}{2} z_{+ (0)} z_{- (1)}
	- \frac{k}{2} z_{ (1)})
	+ \left(\frac{1}{2} z_{+ (0)} z_{ (1)}
	- \frac{2}{3} z_{+ (0)}^2 z_{- (1)}
	- z_{+ (1)}\right) \derivative{z_{+ (0)}} \nonumber \\
& &
	+ \left(\frac{2}{3} z_{+ (0)} z_{- (1)}^2
	+ \frac{1}{2} z_{- (1)} z_{ (1)}
	+ z_{- (2)}\right) \derivative{z_{- (1)}} \nonumber \\
& &
	+ \left(\frac{2}{3} z_{+ (0)}^2 z_{- (1)}^2
	- z_{- (1)} z_{+ (1)}
	- z_{+ (0)} z_{- (2)}\right) \derivative{z_{ (1)}}\nonumber \\
& &
	+ \left(\frac{1}{12} z_{+ (0)} z_{ (1)}^2
	- \frac{1}{45} z_{+ (0)}^3 z_{- (1)}^2
	- \frac{1}{6} z_{+ (0)}^2 z_{- (1)} z_{ (1)} \right.\nonumber
\\& &\; \; \; \; \; \; \; \left.
	- \frac{1}{2} z_{+ (0)} z_{- (1)} z_{+ (1)}
	+ \frac{1}{6} z_{+ (0)}^2 z_{- (2)}
	+ \frac{1}{2} z_{+ (0)} z_{ (2)}
	- z_{+ (2)}\right) \derivative{z_{+ (1)}}
	+ \co(z_{- (3)}) \nonumber \\
\jj{\hat{J}}{+}{-1} & = & (2 b
	- k) z_{- (1)}
	+ \left(z_{ (1)}
	- z_{+ (0)} z_{- (1)}\right) \derivative{z_{+ (0)}}
	+ z_{- (1)}^2 \derivative{z_{- (1)}} \nonumber \\
& &
	+ \left(\frac{2}{3} z_{+ (0)} z_{- (1)}^2
	- 2 z_{- (2)}\right) \derivative{z_{ (1)}} \nonumber \\
& &
	+ \left(z_{ (2)}
	- \frac{1}{6} z_{+ (0)} z_{- (1)} z_{ (1)}
	- z_{- (1)} z_{+ (1)}\right) \derivative{z_{+ (1)}}
	+ \co(z_{- (3)}) \nonumber \\
\jj{\hat{J}}{-}{-2} & = &
	\frac{k - 4b}{15} z_{+ (0)}^3 z_{- (1)}^2
	+ \frac{k - 2b}{6} z_{+ (0)}^2 z_{- (1)} z_{(1)}
	+ \frac{k - 2b}{6} z_{+ (0)} z_{ (1)}^2\nonumber \\
& &
	+ \frac{k - 8b}{3} z_{+ (0)} z_{- (1)} z_{+ (1)}
	- b z_{(1)} z_{+ (1)}
	+ \frac{2k -4b}{3} z_{+ (0)}^2 z_{- (2)}\nonumber \\
& &
	+ (k- b) z_{+ (0)}) z_{(2)}
	- (2k + 2b) z_{+ (2)} \nonumber \\
& &
	+ \left(\frac{2}{45} z_{+ (0)}^4 z_{- (1)}^2
	+ \frac{1}{12} z_{+ (0)}^2 z_{ (1)}^2 \right.\nonumber \\
& &\; \; \; \; \; \; \; \left.
	+ z_{+ (0)}^2 z_{- (1)} z_{+ (1)}
	+ z_{+ (1)}^2
	+ \frac{1}{3} z_{+ (0)}^3 z_{- (2)}
	+ 2 z_{+ (0)} z_{+ (2)}\right) \derivative{z_{+ (0)}}
	+ \co(z_{- (3)}) \nonumber \\
\jj{\hat{J}}{}{-2} & = &
	\frac{4b - k}{6} z_{+ (0)}^2 z_{- (1)}^2
	- \frac{ k}{6} z_{+ (0)} z_{- (1)} z_{ (1)}
	 + 2b z_{- (1)} z_{+ (1)} \nonumber \\
& &
	+ (2 b - k) z_{+ (0)} z_{- (2)}
	- k z_{ (2)}\nonumber \\
& &
	+ \left(\frac{1}{6} z_{+ (0)}^2 z_{- (1)} z_{ (1)}
	- \frac{2}{15} z_{+ (0)}^3 z_{- (1)}^2
	- \frac{1}{6} z_{+ (0)} z_{ (1)}^2
	- z_{+ (2)}\right.\nonumber \\
& &\; \; \; \; \; \; \; \left.
	- \frac{4}{3} z_{+ (0)} z_{- (1)} z_{+ (1)}
	+ \frac{1}{2} z_{ (1)} z_{+ (1)}
	- \frac{2}{3} z_{+ (0)}^2 z_{- (2)}
	+ \frac{1}{2} z_{+ (0)} z_{ (2)}
\right) \derivative{z_{+ (0)}}  \nonumber \\
& &\
	+ \co(z_{- (3)}) \nonumber \\
\jj{\hat{J}}{+}{-2} & = &
	\frac{4b - k}{3} z_{+ (0)} z_{- (1)}^2
	 - b z_{- (1)} z_{(1)}
	+ (2b -2 k) z_{- (2)}\nonumber \\
& &
	+ \left(\frac{2}{3} z_{+ (0)} z_{- (1)} z_{ (1)}
	- \frac{1}{3} z_{+ (0)}^2 z_{- (1)}^2  \right.\nonumber \\
& &\; \; \; \; \; \; \; \left.
	- \frac{1}{2} z_{ (1)}^2
	- z_{- (1)} z_{+ (1)}
	- z_{+ (0)} z_{- (2)}
	+ z_{ (2)}\right) \derivative{z_{+ (0)}}
	+ \co(z_{- (3)}), \nonumber
\end{eqnarray}}
where by $\cd_3$ we denote terms containing derivatives
$\partial/\partial z_{\alpha (n)}$ with $n \geq 3$, and by $\co(z_{-
(3)})$ we denote terms containing quasi-homogeneous polynomials of
degree $(n,w) \succeq (3, -1)$.

As an example, we consider the coadjoint orbit with $(b,k) = (0,1)$.
In the representation of the algebra on $R$, the highest weight state
$\nul$ clearly must satisfy
\begin{equation}
\jj{\hat{J}}{}{0}\nul= b \nul,
\end{equation}
and
\begin{equation}
\hat{C} \nul = k \nul.
\end{equation}
{}From this, it follows that the irreducible representation of the
algebra which contains $\nul$ must be exactly the irreducible
representation with $(j,k) = (0,1)$ which was described in Section
\ref{sec:loopgroups} and whose weights are graphed in
Figure~\ref{f:affinerepresentation}.  As a check on this equivalence,
we can verify that the generator $\jj{\hat{J}}{+}{- 1}$ acting twice
on $\nul$ gives 0.  Indeed,
\begin{equation}
\jj{\hat{J}}{+}{- 1} \nul = (2b - k) z_{- (1)}= - z_{- (1)},
\end{equation}
and
\begin{equation}
\jj{\hat{J}}{+}{- 1}^2 \nul = (2b - k) z_{- (1)}^2 +(2b - k)^2 z_{-
(1)}^2= 0.
\end{equation}
We will study the structure of  the coadjoint orbit loop group
representations further in the following sections.

\subsubsection{Virasoro group}
\label{sec:coadjointexamples3}

Finally, we discuss the coadjoint orbit representations of the
Virasoro group.  As described in Subsection
\ref{sec:virasorogroup}, the generators of $\vvec_\CC$ are $L_n$ and
$C$.  The equation describing the coadjoint action of the
Virasoro algebra $\vvec$ on the dual space $\vvec^*$ is given by
(\ref{eq:virasorocoadjoint}).
By computing the stabilizer of a general dual element $(b, it)$, it
is possible to completely classify the coadjoint orbits of $\vir$
\cite{segal,lp}.  A
clear review of this analysis in the general case is given in
\cite{Witt1}.  We will only be concerned here with the simplest case,
in which the orbit contains an element $(\bo, ic)$ with $\bo(\theta) =
\bo$ a constant function.  We will refer to this orbit as $\wbc$.  In
this case, the stabilizer in $\vvec$ of the point $(\bo, ic)$ is given
by all elements $(f, -ia)$ with $f(\theta)$ satisfying
\bge
\frac{c}{24 \pi} f''' = 2 \bo f'.\ee
When $- 48 \pi \frac{\bo}{c}$ is not the square of an integer $n$, the
only solution to this equation with period $2\pi$ is $f(\theta) = 1$.
In this case, the stabilizer of $\bc$ is the subgroup generated by
$L_0$ and $C$, so the space $\wbc$ is equivalent to the space $\di$.
For the exceptional values of $\bo, c$, the generators $\L_{\pm n}$
are also stabilizers of $\bc$.  Thus, the coadjoint orbits $W_{-cn^2/
48\pi, c}$ are given by the spaces $\dis$, where $SL^{(n)}(2, \RR)$ is
generated by the elements $l_0, l_{\pm n}$ in $\vc$.  As we have done
for all the other groups, we will take the alternative orbit space to
be $\wbc =\di$ even for these exceptional values of $b,c$, and we will
proceed with the analysis using the degenerate symplectic form $\omega$
on this space.

{}From now on, we will consider a fixed orbit $\wbc$, of the $\di$ type.
We can now follow the identical procedure to that used for all the
examples in the previous subsections to construct representations of
$\vvec$ on a polynomial space.  In Subsection
(\ref{sec:virasorogroup}), we gave an argument justifying the use of
(\ref{eq:generalelement}) in constructing a holomorphic coordinate
system on the quotient space $\di$.  Explicitly, we have the formal
result that given a countable set of variables $z = \{z_1, z_2,
\ldots\}$, there exist unique functions $\mzz, \rzz, \gzz$, expressed as
formal power series in the $z_i's$, such that $\gzz$ is real and
\bge
\exp (\sum_{n>0} z_n L_n) \exp(\sum_{n>0} \mzz L_{-n}) \exp(\rzz L_0)
\exp(\gzz C) \in \dio. \label{eq:zcoord} \ee
The variables $z_n$ thus form a set of holomorphic coordinates on
$\di$ which give a $G$-invariant complex structure.

We can now compute the components of the symplectic form $\omega$ with
respect to this coordinate system.  As usual, $\omega$ is a
(1,1)-form; the nonzero components are given by
\begin{equation}
\omega_{\bar{z}_m,z_n} = - \omega_{z_n, \bar{z}_m} = i \delta_{m,n} (4\pi
m \bo + \frac{c}{12} m^3).
\end{equation}
Note again that when $\bo = - \frac{c n^2}{48 \pi}$, the 2-form
$\omega$ is degenerate.

We now make the observation that $\di$ is a contractible space.  To
see this, we use the identification of $\di$ with the group $\dio$ of
orientation-preserving diffeomorphisms of $S^1$ which fix the point 1.
Viewing an element of $\dio$ as a monotonically increasing function
$f: \RR \rightarrow \RR$ with the properties $f(0) = 0$ and $f(x + 2
\pi) = 2 \pi + f(x)$, we can explicitly give a retraction of $\dio$ to
a point by defining the one-parameter family of functions $f_t(x) =
(1-t) f(x) + tx$, for each $f \in \dio$, $t \in [0,1]$.  Since $\di$
is a contractible space, all 2-cycles are homologous to the null
2-cycle, so that $\int_a \omega = 0$ for any 2-cycle $a$.  Thus, for
any $\bo, c$, we can construct a line bundle $\lbc$ over $\wbc$ with
curvature $i \omega$.  (Note that the second cohomology of $\di$ is
nontrivial if one restricts to forms invariant under $\dif$, however
this should not affect the construction of $\lb$; it does however
imply that $\lb$ will not have a global $\dif$-invariant connection.)
We can use the K\"{a}hler structure defined by $\omega$ to restrict to
the space ${\cal H}_b$ of holomorphic sections of $\lbc$, which can be
identified with the ring $R = \CC [z_1, z_2, \ldots]$ of polynomials
in the variables $z_i$ due to the topologically trivial nature of the
orbit space (actually, for this group all power series can be
considered to be holomorphic sections; see Section
\ref{sec:holomorphicity}).  We now proceed to construct the operators
\begin{equation}
\hat{L}_{n} = \xi_{n} + f_{n}
\end{equation}
and
\begin{equation}
\hat{C} = \xi_c + f_c
\end{equation}
as differential operators on $R$.  For this group, the ring $R$ has an
integer grading with $\deg (z_n) = n$.  As usual, the  operators
$\hat{L}_n$ act on the subspace $R_m$  of degree $m$ by
\bge
\hat{L}_n R_m \subset R_{m-n}.\ee

{}From (\ref{eq:cartansimplification}) we  calculate the vector
fields corresponding to the generators   $L_0,C$;
\begin{eqnarray}
\xi_0 & = & \sum_{k > 0} k z_k \frac{\pdv}{\pdv z_k}, \\
\xi_c & = &  0. \nonumber
\end{eqnarray}
The remaining vector fields are difficult to write in closed-form; the
leading terms are given by
\bge
\xi_n = - \frac{\pdv}{\pdv z_n} + \frac{1}{2} \sum_{m > 0} (m-n) z_m
\frac{\pdv}{\pdv z_{n+m}} + \ldots, \;\;\; {\rm for} \; n > 0,\ee
and
\bge
\xi_{-n} = \sum_{m > n} (m+n) z_m
\frac{\pdv}{\pdv z_{m-n}} + \ldots, \;\;\; {\rm for} \; n > 0.\ee
Just as was the case for $\widehat{L}SU(2)$, although these operators
are all expressed as an infinite sum of terms each of which is a
first-order differential operator, the action of a fixed vector field
operator on a fixed polynomial in $R$ can be computed using only a
finite number of these terms.

To calculate the functions $f_{- n}$ and $f_c$ using
(\ref{eq:recursive}), we begin as usual by calculating the functions
for the stabilizing generators.  Writing the quantity $2 \pi\bo +
c/24$ as
\begin{equation}
h = 2 \pi\bo + c/24,
\end{equation}
we have
\begin{eqnarray}
f_c & = & - i \ip{ (b,ic)}{C} = c\\
f_0 & = & - i \ip{ (b,ic)}{L_0} =2 \pi \bo +  \frac{c}{24}  = h. \nonumber
\end{eqnarray}
For the simple root $L_1$, we have
\begin{equation}
f_{-1} = i \ip{ (b,ic)}{2L_0} z_1 = - 2hz_1.
\end{equation}
The quasi-simple root $L_2$, however, is a special case.  It cannot be
computed inductively, and does not follow from the equation in
(\ref{eq:recursive}) for simple roots.  However, we can write a pair
of equations which $f_{-2}$ must satisfy, by using the condition
(\ref{eq:notparticular}).  In this case, that condition implies that
\bge
\xi_m f_n - \xi_n f_m = (m - n) f_{m + n} + \frac{c}{12} \delta_{m,
-n} (m^3 - m). \label{eq:calpha} \ee
It follows that we must have
\begin{eqnarray}
\xi_2 f_{-2} & = & - \frac{\pdv}{\pdv z_2} f_{-2} = 4 f_0 + \frac{c}{2} = 4h
+ \frac{c}{2},  \\
\xi_1 f_{-2} & = & - \frac{\pdv}{\pdv z_1} f_{-2} = 3 f_{-1} = -6 h z_1.
\nonumber \end{eqnarray}
Since $f_{-2}$ is a linear combination of $z_1^2$ and $z_2$, these two
equations determine both coefficients, so that
\bge
f_{-2} = - (4h + \frac{c}{2}) z_2 + 3 h z_1^2. \label{eq:f2} \ee

{}From here, we  can continue the analysis according to the usual
prescription, treating the quasi-simple roots $L_1,L_2$ as simple
roots.  The remaining functions $f_{- n}$ can be calculated
inductively by
\bge
f_{-n} = \frac{1}{n-2} (\xi_{-1} f_{1-n} - \xi_{1-n} f_{-1}), \;{\rm
for}\; n > 2. \ee

We now give explicit formulae for the terms in the operators
$\hat{L}_n$ which have nonzero matrix elements between polynomials in
$R$ of degree $\leq 3$.  These terms are given by
\begin{eqnarray}
\hat{L}_3 & = & - \frac{\pdv}{\pdv z_3} + \cd_4, \nonumber \\
\hat{L}_2 & = & - \frac{\pdv}{\pdv z_2} - \frac{1}{2} z_1
\frac{\pdv}{\pdv z_3} + \cd_4, \nonumber \\
\hat{L}_1 & = & - \frac{\pdv}{\pdv z_1} + \frac{1}{2} z_2
\frac{\pdv}{\pdv z_3} + \cd_4, \nonumber \\
\hat{L}_0 & = & h + z_1 \frac{\pdv}{\pdv z_1} + 2 z_2 \frac{\pdv}{\pdv
z_2} + 3 z_3 \frac{\pdv}{\pdv z_3} + \cd_4,   \\
\hat{L}_{-1} & = & -2 h z_1 + (3z_2 - z_1^2) \frac{\pdv}{\pdv z_1} + (4
z_3 - 2 z_1 z_2) \frac{\pdv}{\pdv z_2} + \co(z_4), \nonumber \\
\hat{L}_{-2} & = & -(4h + \frac{c}{2}) z_2 + 3 h z_1^2 + (5 z_3 -
\frac{13}{2} z_1 z_2 + z_1^3) \frac{\pdv}{\pdv z_1} + \co(z_4), \nonumber \\
\hat{L}_{-3} & = & - (6h + 2c) z_3 + (13h + c) z_1 z_2 - 4h z_1^3 +
\co(z_4), \nonumber \end{eqnarray}
where $\cd_4$ denotes terms containing derivatives $\pdv / \pdv z_k$
with $k \geq 4$, and $\co(z_4)$ denotes terms containing
quasi-homogeneous polynomials
of degree at least 4.

We will  discuss in more detail the structure of these representations
in later sections.

\subsection{Global properties}
\label{sec:global}

For most of the rest of this thesis, we will concentrate on properties
of the coadjoint orbit representations which can be described in terms
of local algebraic properties of the representations.  These
properties are completely captured by the explicit formulae for
operators on spaces of polynomials described in the previous sections
of this chapter.  In this section, however, we will consider some
global properties of the coadjoint orbit representations.  We will
discuss these properties from the point of view of the local algebraic
representations, and will give results in a form which can be
represented in terms of local algebraic formulae.  In Subsection
\ref{sec:holomorphicity} we consider the question of which
polynomials (or power series) in the local holomorphic variables
$z_\alpha$ can be extended to global holomorphic sections
of the line bundle $\lb$.  In Subsection \ref{sec:unitarity} we
discuss the question of whether the holomorphic line bundle $\lb$
admits a Hermitian structure, and the related question of whether the
representation of the group $G$ on the space ${\cal H}_b$ of
holomorphic sections is a unitary representation.

\subsubsection{Global holomorphicity}
\label{sec:holomorphicity}

In the previous section, we derived formulae which described for
various groups $G$ an action of the Lie algebra $\algg$ on the ring of
polynomials in the variables $z_\alpha$ corresponding to positive
roots of $G$.  We argued that locally, this ring corresponds to the
space of holomorphic sections of the line bundle $\lb$ over the
coadjoint orbit space $W_b$.  In this description, we have ignored the
question of which of these polynomials actually corresponds to a
globally holomorphic section of $\lb$.  In this subsection, we
describe which of the polynomials in $R$ actually extends to a global
holomorphic section.  We begin with the simple case of $SU(2)$, where
the coadjoint orbit space is simply $S^2$ with the usual single
complex coordinate $z$.  We showed in Subsection
\ref{sec:coadjointexamples1} that the line bundle $\lb$ corresponding
to any half-integral value of $b$ was precisely the unique holomorphic
line bundle over $S^2$ of degree $2b$.  When $b \geq 0$, this bundle
admits precisely $2b + 1$ holomorphic sections, which
correspond to the local functions $1,z, \ldots, z^{2b}$ in $R$.  In
order to derive analogous constraints on holomorphic functions for
other groups, we will find it useful to see how this result can be
arrived at purely by local considerations.

In order to analyze which functions of $z$ are holomorphic globally,
we must begin by choosing a second coordinate chart on $S^2$ which
covers the point $z = \infty$.  The standard choice is the chart $w
\in \CC$ with the transition homomorphism $w = 1/z$.
In terms of these variables, the partial derivative operators
$\partial/\partial z, \partial/\partial w$ are related by
\begin{equation}
\frac{\partial}{\partial z}  = - w^2 \frac{\partial}{\partial w}.
\end{equation}
Given a line
bundle ${\cal L}$ over $S^2$ with a connection described by  local
gauge choices as $A_z,A_w$ in the two coordinate charts,
the connections are similarly related by
\begin{equation}
A_z = - w^2 A_w.
\end{equation}
For the gauge choice used for the local formulae
(\ref{eq:representation2}), the connection terms $A_a$, $a = 3,
\pm$ can be written in terms of $A_z$,  and are given
by
\begin{eqnarray}
A_3 & = & - z A_z \nonumber \\
A_+ & = & -  A_z \\
A_- & = & z^2 A_z. \nonumber \end{eqnarray}
The connection $A_a = - f_a + i \Phi_a$ can be explicitly calculated
by evaluating
\begin{eqnarray}
\Phi_a(z) & = & - \langle \aads_{g(z, \psi)} \bbo, J_a \rangle   \\
& = & - \langle \bbo, e^{- \azz J_-} e^{-z J_+} J_a e^{zJ_+} e^{\azz
J_-} \rangle. \nonumber
\end{eqnarray}
One finds that
\begin{eqnarray}
A_3 & = & 2b z \azz, \nonumber \\
A_+ & = & 2b  \azz, \\
A_- & = & - 2 b z^2 \azz. \nonumber \end{eqnarray}
Thus, we have
\begin{equation}
A_z = \frac{-2b \bar{z}}{1 + |z|^2},
\label{eq:zconnection}
\end{equation}
and therefore in this gauge,
\begin{equation}
A_w = \frac{2b z| z |^2}{1 + |z|^2}= \frac{2b}{w (1 + | w |^2)}.
\end{equation}
In the coordinate space $w$, we must find a choice of gauge where this
connection is nonsingular.  If we  perform the gauge transformation on
sections given by
\begin{equation}
\phi \rightarrow w^n \phi,
\label{eq:gaugetransformation}
\end{equation}
then $A_w \rightarrow A_w - \frac{n}{w} $.  Thus, when we choose $n =
2b$, we have
\begin{equation}
A_w = \frac{2b}{w} (\frac{1}{1 + | w |^2} -1 )
= \frac{-2b \bar{w}}{1 + | w |^2},
\end{equation}
which is not only nonsingular, but is of precisely the form of
(\ref{eq:zconnection}).  Under the gauge transformation
(\ref{eq:gaugetransformation}) with $n = 2b$, the monomial $z^m$
is  described in the $w$ chart by the monomial
\begin{equation}
z^m \rightarrow z^{m - 2b} = w^{2b - m}.
\end{equation}
It follows that for this choice of gauge, the functions in $R =
\CC[z]$ which extend to globally holomorphic sections of $\lb$ are
precisely the functions $z^m$ with $m \leq 2b$, just as stated
above.  For $b < 0$, this indicates that  there are no holomorphic
sections and that the associated representation of $G$ is
0-dimensional.  For $b = 0$, this result states that only the constant
function is a global holomorphic section.  This is exactly the same
result we would have gotten had we taken the correct pointlike
coadjoint orbit space in this case, rather than using the alternative
orbit space $S^2$ with a flat bundle.

The result we have derived for $SU(2)$ can be restated in the
following fashion: The only polynomials in $R$ which can extend to
globally holomorphic sections of $\lb$ are those which lie in the
orbit of the unique highest weight state $\nul$.  (For those coadjoint
orbits for which $\nul$ is not globally holomorphic, there are no
globally holomorphic sections.)  This follows for $SU(2)$ because for
a given choice of $b$, the monomial $z^{2b}$ is annihilated by the
operator $\hat{J}_-$.  We claim that in fact, for all compact simple
groups $G$ this formulation of the global holomorphicity condition is
correct not only for the group $G$ but also for $\widehat{L}G$.  An
immediate consequence of this result is that for this set of groups,
the representation on the space of holomorphic sections is an
irreducible representation.

We can prove the general condition for any $G$ by using the above
analysis for $SU(2)$ to give the necessary conditions for local
functions in $R$ to extend to holomorphic sections of the line bundle
over $S^2$ given by the restriction of $\lb$ to various $SU(2)$
subgroups of $G$.  This analysis is similar to the method used in the
previous section to ascertain the quantization condition on the
coadjoint orbit parameters using $SU(2)$ subgroups.  Basically, the
point is that for every $SU(2)$ subgroup $S$ of $G$ containing a
subgroup $T = S^1$ or $SU(2)$ which stabilizes the canonical element
$b$ of the coadjoint orbit, there is a condition on the polynomials in
$z_\alpha$ which is necessary to ensure that the polynomial can be
extended to a globally holomorphic section.  This condition is simply
that for each such subgroup $S$, a polynomial $\phi \in R$ must lie in
an irreducible representation of the ${\rm su} (2)$ algebra given by
the Lie algebra of $S$.  For finite-dimensional compact simple $G$,
all of the positive roots of $G$ are in the complexification of such
an ${\rm su} (2)$ subalgebra.  The same condition is true for
$\widehat{L}G$.  It follows then, that for all these groups the
polynomial $\phi$ can only be extended to a globally holomorphic
section if it lies in the orbit of $\nul$.

We will demonstrate this argument explicitly for $SU(3)$; the general
argument follows in an analogous fashion.  For $SU(3)$, let us choose
a general polynomial $\phi \in R = \CC[t,u,v]$.  From the  $SU(2)$
subgroup generated by $\{i J_t,i J_1,i J_2\}$, we see that $\phi$ must
be in an irreducible representation of the corresponding subalgebra,
so that applying $\hat{J}_{+ t}$ to $\phi$ some integral number of
times, we arrive at a function
\begin{equation}
\phi' = \hat{J}_{+ t}^n \phi
\end{equation}
which satisfies $\hat{J}_{+ t} \phi' = 0$.  We  now consider the $SU(2)$
subgroup generated by $\{i J_v,i J_4,i J_5\}$.  Again, we can apply
$\hat{J}_{+ v}$ some number of times to get a function $\phi''$ which
is annihilated by $\hat{J}_{+ v}$.  Because $[\hat{J}_{+ t},
\hat{J}_{+ v}]= 0$,  the function $\phi''$ is  also still annihilated by
$\hat{J}_{+ t}$.  Finally, we  take the $SU(2)$ subgroup generated
by $\{i J_u,i J_6,i J_7\}$.  Again, we apply $\hat{J}_{+ u}$ some
number of times to get a function $\psi \in R$ which is annihilated by
$\hat{J}_{+ u}$.  From the commutation relations, we again verify that
$\psi$ is annihilated also by $\hat{J}_{+ v}$ and $\hat{J}_{+ t}$.
Thus, $\psi = \nul$.  Since each of the transformations we performed
by repeated application of annihilation operators was carried out
inside an irreducible representation of the relevant $SU(2)$
subalgebra, these operations are all invertible, and it follows that
$\phi$ can be reached from $\nul$ by the application of the operators
$\hat{J}_{- \alpha}$, and thus that $\phi$ is in the orbit of $\nul$
under the action of the full algebra on $R$.

This analysis can immediately be generalized to an arbitrary group $G$
with a finite number of simple roots which lie in complexifications of
appropriate $SU(2)$ subgroup of $G$, simply by applying the simple
roots one at a time.  Thus, we have shown that for a function to be
extensible to a global section, it is necessary for the function to
lie in the orbit of $\nul$ for the groups $G$ and $\widehat{L}G$.
Furthermore, if there exists any globally holomorphic section
whatsoever, this result implies that $\nul$ is such a section, and
that the set of holomorphic sections must be precisely those sections
in the orbit of $\nul$ since the space of holomorphic sections must be
invariant under the action of $G$.  We have thus shown that for all
these groups, the representation of the algebra (and therefore of the
group) on the space of holomorphic sections of $\lb$ is irreducible.

Let us now consider the case of the Virasoro group and algebra.  In
this case, as we observed in Section \ref{sec:coadjointexamples3} the
coadjoint orbits are topologically trivial, so there is absolutely no
constraint on functions of the variables $z_n$ necessary for functions
to be holomorphic.  Thus, in this case not only are all functions in
$R = \CC[\{z_n\}]$ holomorphic, but even all power series in $
\CC[[\{z_n\}]]$ with infinite radii of convergence must be considered
as holomorphic functions on $\lbc$.  We  will not concern ourselves
with the general power series here; for the purposes of this thesis we
restrict attention to the polynomials in $R = \CC[\{z_n\}]$.  (In any
case, we must extract irreducible representations of the Virasoro
group by the resolutions which are constructed in the next chapter;
the infinite power series' do not contribute to the cohomology of
these resolutions and so vanish from consideration at that point.)
The difference between this group
and the loop groups essentially lies in the fact that although the
(complex) Virasoro algebra contains many ${\rm su} (2)_\CC$
subalgebras, these subalgebras do not correspond to compact $SU(2)$
subgroups of the Virasoro group, but rather to noncompact $SL (2,\RR)$
subgroups, as demonstrated in Section
\ref{sec:coadjointexamples3}.

Finally, we conclude this section with a brief remark on other
approaches to the question of global holomorphicity.  For
finite-dimensional compact  groups, an argument is often given
(see for instance \cite{PS}) for the
irreducibility of the representation on the space of holomorphic
sections, which  follows by demonstrating that  there is a unique
highest weight state in this representation.  In the case of compact,
finite-dimensional groups, this result follows because the space of
holomorphic sections is known to be finite-dimensional, and thus can
be written as a direct sum of irreducible representations.  For
infinite-dimensional groups, this argument breaks down because the
space of holomorphic sections is infinite-dimensional.  For example,
the space $R$ of locally holomorphic polynomials for the group
$\widehat{L}SU(2)$ carries a representation of the algebra with a
single highest weight state; this representation is, however, not
irreducible.   Although the expected conclusion holds nevertheless for
this group, in the case of the Virasoro group  not only does the
argument break down, but the conclusion is also incorrect.

\subsubsection{Hermitian structures}
\label{sec:unitarity}

We will now discuss the question of whether for a given group $G$ and
coadjoint orbit $W_b$, the line bundle $\lb$ admits a Hermitian
structure.  In general, a complex line bundle ${\cal L}$ over a
complex manifold ${\cal M}$ is defined to have a Hermitian structure
when ${\cal M}$ admits a measure ${\rm d} \mu$ and ${\cal L}$ admits
an inner product  which when integrated over  ${\cal M}$ with the
measure ${\rm d} \mu$ gives a positive-definite inner product on the
space of holomorphic sections of $\lb$.  Written in terms of a local
trivialization of $\lb$, the inner product  is described by
a real function $h:
{\cal M} \rightarrow \RR$; the inner product between two holomorphic
sections $\phi$, $\psi$ of $\lb$ is then given by
\begin{equation}
\langle \phi, \psi \rangle = \int_{\cal M} {\rm d} \mu \;{\rm e}^{h}
\phi^*\psi.
\end{equation}
In terms of such a local trivialization, there is a unique connection
$A$, the {\em metric connection}, compatible with both the Hermitian
metric and the complex structure on ${\cal M}$; written in a local set
of holomorphic coordinates $z_i$, this connection is given by
$A_{\bar{z}_i}= 0$, $A_{z_i}= \partial h/\partial z_i$ \cite{GH}.

As an example of such a Hermitian structure, let us consider again the
line bundle $\lb$ over $S^2$  of degree $2b$.  In terms of the
coordinate $z$ on $S^2$,
the rotationally invariant measure on $S^2$  is given by
\bge
d \mu = \frac{2i}{(1 + |z|^2)^2} dz d \bar{z}.\ee
Associated with this measure, there is a natural Hermitian metric on
$\lb$,  described by the function
\bge
e^{h(z, \bar{z})} = \frac{1}{(1 + |z|^2)^{2b}}.\ee
Combining these factors, the inner product $\langle, \rangle$
on $\hb$ is given by
\bge
\langle \phi, \psi \rangle = \int_{S^2} d \mu \: e^h \phi^* \psi =
\int_{S^2} \frac{2i dz d\bar{z}}{(1 + |z|^2)^{2b+2}} \phi^*(z) \psi(z),\ee
where $\phi$ and $\psi$ are arbitrary holomorphic sections of $\lb$.
Performing this integral explicitly, one finds that
\bge
\langle z^l, z^m \rangle = \delta_{l, m} {4 \pi}\left[(2b + 1)\left(
\begin{array}{c} 2b \\ l \end{array}
\right)\right]^{-1}.\label{eq:su2prod} \ee
This inner product on $\hb$ is  proportional to the usual
inner product on unitary irreducible representation spaces of
$SU(2)$.   The metric connection associated with this choice of
Hermitian structure is precisely the connection $A_z$ calculated in
the previous  subsection,
\begin{equation}
A_z = \frac{\partial h(z, \bar{z})}{\partial z}  = \frac{-2b
\bar{z}}{1 + | z |^2}.
\end{equation}

In a similar fashion, it is possible to construct explicitly a
Hermitian structure on the line bundles $\lb$ associated with
coadjoint orbits of all finite-dimensional simple compact groups $G$.
Unfortunately, such a construction does not seem to be possible when
the group under consideration is infinite-dimensional.  As an example,
we consider the Virasoro group.  If we wish to construct a Hermitian
metric $\exp(H)$ on a coadjoint orbit $\lbc$ of the Virasoro group in
the local coordinates defined in Section
(\ref{sec:coadjointexamples3}), we can begin by assuming that the
connection we have defined by our gauge-fixing procedure is the
associated metric connection, as was the case for $SU(2)$.  We can
then explicitly calculate $H$ as a formal power series in the $z$'s.
We have
\begin{eqnarray}
A_0 & = & - f_0 + i \Phi_0 \\
& = & - h -i\langle (b_0, ic), \exp(- \sum_{n>0} \mu_nL_{-n})
\exp(- \sum_{n>0} z_n L_n) L_0  \nonumber\\
& & \hspace*{1.5in} \times\exp(\sum_{n>0} z_n
L_n)\exp(\sum_{n>0} \mu_n L_{-n})  \rangle \nonumber \\
& = & \!\!\sum_{\sss k,l>0; \{n, m\}} \! \frac{(-1)^{k+l}}{k!\:l!} z_{n_1}
\ldots z_{n_k} \mu_{m_1} \ldots \mu_{m_l} \nonumber \\
& &\hspace*{.7in}
\times C_0(n_1,\ldots,n_k,-m_1,\ldots,-m_l)
(h+\frac{c}{24}(m_l^2-1)),\nonumber \end{eqnarray}
where the sum is taken over all $n_1, \ldots, n_k, m_1, \ldots, m_l >
0$ satisfying $n_1 + \ldots + n_k =  m_1+  \ldots + m_l$.
If we assume $A_{\bar{z}_n} = 0$ and $A_{z_n} = \pdv H / \pdv z_n$,
then we have
\bge
A_0 = - \sum_{n>0} n z_n A_{z_n},\ee
and
\bge
H = - \xi_0^{-1} A_0,\ee
up to a constant.  We will take this function $H$ as a candidate for the
Hermitian metric on $\lbc$.  The first few terms in a power series
expansion of $H$ are given by
\begin{eqnarray}
\lefteqn{H = -\sum_{n > 0} 2n(h + \frac{c}{24}(n^2-1)) |z_n|^2}
\label{eq:hvir}  \\
& & + \sum_{n,m>0}
\left[ (m^2 + 4mn + n^2) h + \frac{c}{24}(m^4 + 2m^3n + 2 m n^3 + n^4
- m^2 - 4 mn - n^2)\right] \nonumber \\
& & \;\;\;\;\;\;\;\;\;\;\;\;\;\times (\frac{z_m z_n \bar{z}_{m+n} + \bar{z}_m
\bar{z}_n z_{n+m}}{2}) + \co(z^4). \nonumber\end{eqnarray}
Note that $H$ is expected to be real, in order to be a Hermitian
metric.

To have a complete description of a unitary structure on $R$, it would
now be necessary to find an invariant metric on $\di$.  Unfortunately,
it is unclear whether such a metric can be found.  Attempting to
describe such a metric as a formal power series gives rise to an
expression with divergent coefficients.  The matter is complicated by
the fact that the adjoint representation of the Virasoro algebra is
not a highest weight representation.  It seems that some kind of
regularization scheme may be necessary to construct such a metric in a
sensible fashion.  We can, however, get some information about when
such a unitary structure is likely to be possible directly from
(\ref{eq:hvir}).  If we take only the first term in (\ref{eq:hvir}),
and approximate the metric with a Gaussian, we see that for $h \ll 0$
or $c \ll 0$, the metric diverges badly, and we will certainly not
find a unitary structure.  When $h, c \gg 0$, a sensible inner product
on $R$ can be found, at least in perturbation theory, by taking a
product of Gaussian integrals.  Using the Hermitian metric
(\ref{eq:hvir}) to compute anything nonperturbative, however, would be
a difficult proposition.  Further progress in this direction will be
impossible until some sort of a regularized invariant metric on $\di$
can be described explicitly.

In the case of loop groups $\widehat{L}G$, a similar problem arises.
For these groups also,  no invariant measure is known on the quotient
space $LG/T$; furthermore, it is not known whether such a measure can
exist  \cite{PS}.

The main reason  that we are  interested in Hermitian structures on
the line bundles $\lb$ is that when a line bundle admits a Hermitian
structure, the corresponding representation of the group on the space
of holomorphic sections is naturally  unitary.  Although we are unable
to construct explicitly the desired Hermitian structures for
infinite-dimensional groups, we can still investigate the question of
whether the associated representations admit unitary structures.  A
simple way of  constructing a unitary structure on a highest weight
representation is to simply define the inner product $\ip{1}{1} =
\langle \; | \; \rangle = 1$, and to assume that the creation and
annihilation operators are Hermitian conjugates under $\hat{J}_{-
\alpha}^{\dagger} = \hat{J}_\alpha$.  In fact, this is the usual
procedure used to construct unitary group representations from Verma
modules (which will be discussed in more detail in the following
chapter).
The result of this construction is that the unitary
structure is well-defined for all functions $\phi \in R$ which lie in
the orbit of the highest weight state $\nul$.  For loop groups, this
means that the complete representation on the space ${\cal H}_b$ of
holomorphic sections actually admits a unitary structure.  For the
Virasoro group, this result is slightly more subtle.  We devote the
remainder of this section to a more detailed description of this
construction in the case of the Virasoro group.

The above construction of a unitary structure is closely related to
the essential point in the proof of which irreducible Virasoro
representations admit unitary structures.  The usual approach to the
unitarity proof (see for example \cite{Gins}) is to consider all
possible polynomials in the creation operators $\hat{L}_{- n}$ acting
on a highest weight state $\nul$, and to use the above criteria to
define an inner product on this space.  (The space is a Verma module
for the Virasoro algebra, which is discussed in more detail in the
next section.)  The resulting inner product, known as the Shapovalov
form, gives a zero norm to some state in the Verma module precisely
when the determinant of the inner product matrix on some level
vanishes.  By studying the changes of sign in this ``Kac determinant''
under variations of $h$ and $c$, it was shown by Friedan, Qui, and
Shenker \cite{FQS} that for $c< 1$, a representation can only admit a
unitary structure when when conditions (\ref{eq:min1}) and
(\ref{eq:min2}) are satisfied.  The existence of a unitary
representation for all values of $h$ and $c$ satisfying these
conditions was shown in \cite{GKO}.  From the point of view of the
coadjoint orbit representations on polynomial spaces $R$, the
vanishing of the Kac determinant at level $n$ corresponds precisely to
the existence of a function in $R_n$ which does not lie in the orbit
of $\nul$ (from the point of view of the next chapter, such a function
is associated with a cosingular vector).  In general, it is impossible
to extend the unitary structure defined on the irreducible
representation associated with the orbit of $\nul$ to these extra
states.  To see an explicit example of this problem, we can compute
explicitly the lowest degree states arising from the action of the
Virasoro algebra on $\nul$.  Using a basis for $R$ of monomials in the
variables $z_i$, where $f
\in R$ is represented by the state $|f\rangle$, we have
\begin{eqnarray}
\hat{L}_{-1} \nul & = & -2 h |z_1\rangle, \nonumber \\
\hat{L}_{-1}^2 \nul & = & -6h |z_2\rangle + (4 h^2 + 2h)|z_1^2\rangle,
\label{eq:Virasorosingular}\\
\hat{L}_{-2} \nul & = & -(4h + \frac{c}{2}) |z_2\rangle + 3 h
|z_1^2\rangle. \nonumber
\end{eqnarray}
We expect that
for those values of $h,c$ with a vanishing Kac determinant at level
$2$, we   will
find a linear dependence between the states generated by combinations of
raising operators of degree $2$.  From the above
expressions for $\hat{L}_{-1}^2 \nul$ and $\hat{L}_{-2} \nul$, it is
easy to see that these states are linearly dependent when
\bge
h = \frac{5 - c \pm \sqrt{(c-1)(c-25)}}{16}.\label{eq:red} \ee
This is precisely the condition for the Kac determinant to vanish at
level 2.
For example, when $m=3, p=2, q=1$, and $h = c = 1/2$,  Equation
\ref{eq:red} is satisfied, and we have a reducible representation on
$R$, with an extra state $|z_1^2\rangle$ in $R_2$ which is not in the
orbit of $\nul$.  In this case we can compute
\bge
\langle z_1 | z_1 \rangle = \langle \: | \hl_1 \hl_{-1} | \: \rangle =
2h = 1 \ee
\[
\langle -\frac{9}{4} z_2 + \frac{3}{2} z_1^2 | -\frac{9}{4} z_2 +
\frac{3}{2} z_1^2 \rangle = \langle \: | \hl_2  \hl_{-2} | \: \rangle =
9/4. \]
If we attempt to extend this inner product to $R_2$, we get
\begin{eqnarray}
\langle 3z_2 - 2z_1^2 | z_1^2 \rangle & = & \langle z_1 | \hl_1  |
z_1^2 \rangle =
-2  \\
& = & - \frac{4}{3}\langle \: | \hl_2  | z_1^2 \rangle = 0. \nonumber
\end{eqnarray}
Thus, the inner product on the orbit space cannot be extended to one on $R$.

We have shown, then, that for compact simple $G$ and loop groups
$\widehat{L}G$, the representation on the space of holomorphic
sections of an appropriate line bundle $\lb$ is irreducible and admits
a unitary structure, in agreement with the results of the Borel-Weil
theory.  The same result holds for the coadjoint orbits associated
with unitary representations of the Virasoro group where the entire
space $R$ lies in the orbit of $\nul$.  In the case of those coadjoint
orbits of the Virasoro group corresponding to the discrete series of
unitary representations, however, there are states which correspond to
globally holomorphic sections in $R$ lying outside the orbit of
$\nul$.  The unitary structure defined in this section does not lead
to a well-defined inner product for these extra states, and thus the
complete representation of the Virasoro group does not admit a unitary
structure in these cases.  However, the irreducible representation
given by the orbit of $\nul$ does admit a unitary structure, which is
just the usual unitary structure associated with this irreducible
representation in the discrete series.



\mysection{Modules and Resolutions}

In this chapter, we define a variety of types of modules for an
arbitrary Lie algebra $\algg$, and describe how these modules can be
combined to form any irreducible representation of $\algg$ by
constructing a chain complex whose cohomology is precisely the desired
irreducible representation space.  In Section \ref{sec:modules}, we
define the concept of a $\algg$-module, and discuss some particular
types of modules, including Verma modules, dual Verma modules, and
twisted Verma modules.  We describe the singular vector structure of
these modules.  In Section \ref{sec:realizations}, we discuss Fock
space realizations of representations; these realizations are
essentially equivalent to $\algg$-modules on spaces of polynomials
where the action of the algebra is described in terms of differential
operators.  We observe that the coadjoint orbit representations
described in the previous chapter are locally equivalent to dual Verma
module representations, and we discuss the Feigin-Fuchs and Wakimoto
(free field) representations of the Virasoro and affine algebras,
which are similar to twisted Verma modules of finite-dimensional
algebras.  In Section
\ref{sec:intertwiners}, we define the notions of screening operators
and intertwiners between $\algg$-modules, and give explicit formulae
for the screening operators and intertwiners between coadjoint orbit
representations.  We find that the coadjoint orbit screening
operators, like the associated coadjoint orbit raising operators, are
independent of the highest weight of the module under consideration.
In Section
\ref{sec:resolutions}, we describe how intertwiners can be used to
form a chain complex of $\algg$-modules whose cohomology is nonzero at
a single position in the chain, and is given by an irreducible
representation of the algebra $\algg$.  The explicit formulae for
coadjoint orbit intertwiners allow us to use these resolutions to
construct irreducible representations using the coadjoint orbit
representations even in the case of the Virasoro algebra, where the
coadjoint orbit representations correspond precisely to dual Verma
modules.

For general background references on the algebraic constructions in
this section, see \cite{bmp,hs}.

\subsection{$\algg$-modules}
\label{sec:modules}

A powerful approach in the study of representation theory is given by
the algebraic formalism of modules.  Given a Lie algebra $\algg$,
every vector space $V$ which admits a representation of $\algg$ is
defined to be a $\algg$-module.  (More generally, given a ring ${\cal
R}$, an ${\cal R}$-module is defined to be an abelian group $A$ along
with a ring homomorphism from ${\cal R}$ to the ring of endomorphisms
from $A$ into itself.  In this case, the vector space $V$ has the
abelian group structure given by vector space addition, and the ring
${\cal R}$ is the universal enveloping algebra ${\cal U}(\algg)$
described below.)  We will restrict attention here to modules which
admit a grading of the natural type for the algebra $\algg$, as
described in Section \ref{sec:groups}.  There are several special
types of $\algg$-modules which are of particular interest to us here.
A useful structure in describing these special modules is the {\em
universal enveloping algebra} ${\cal U} (\algg)$ of a Lie algebra
$\algg$.  The universal enveloping algebra ${\cal U} (\algg)$ is
defined to be the ring of all polynomials (over $\RR$) in the generators
of the algebra $\algg$, modulo the relations defined by interpreting
the Lie algebra product as a commutator via the equation
\begin{equation}
x[u,v] y \sim xuvy - xvuy, \; \; \;\forall x,y \in {\cal U} (\algg),
u,v \in \algg.
\end{equation}
A $\algg$-module $V$ is said to be free over ${\cal U} (\algh)$ on a
vector $v\in V$ when each vector $w \in V$ has a unique representation
as $w = X_w \cdot v$ for some $X_w \in {\cal U} ( \algh)$.

One of the most important types of modules in the study of
representation theory is the Verma module \cite{Humphreys}.  A Verma
module for the algebra $\algg$ is a module $V$ containing a vector
$v$, such that $V$ is free over ${\cal U} (\algg_-)$ on $v$, where
$\algg_-$ is the subalgebra of $\algg_\CC$ generated by the negative
roots in $\algg_\CC$.  It is an immediate consequence of this
definition that $v$ is a highest weight state in $V$.  For a fixed
weight $\lambda\in\Lambda_w$, the Verma module $V_\lambda$ with a
highest weight vector $v_\lambda$ of weight $\lambda$ is unique.  A
natural basis for this Verma module is given by the states $X \cdot
v$, where $X$ is a member of a canonical basis for ${\cal U}
(\algg_-)$.  From the definition, it is clear that the Verma module
$V_\lambda$ has the property that the entire module is in the orbit of
the state $v_\lambda$ under the action of the algebra $\algg$.  This
is a characteristic feature of Verma modules.

We can now define a dual space to $V_\lambda$, which we denote by
\begin{equation}
V_\lambda^* = {\rm Hom} (V_\lambda, \RR).
\end{equation}
For any algebra $\algg$ which has a polarization into positive
roots $\alpha \succ 0$ and negative roots $\alpha \prec 0$, we can
define a natural dual action of $\algg$ on $V_\lambda^*$ by  the
equation
\begin{equation}
\ip{e_\alpha d}{w} = \ip{ d}{e_{- \alpha} w}, \; \; \; \forall d \in
V_\lambda^*,w \in V_\lambda.
\end{equation}
(For generators $i h$ in $\algt$, we simply take $\ip{ h d}{w} = \ip{
d}{hw}$.)   The module $V_\lambda^*$ is a dual Verma module.
Note that all the groups considered in this thesis admit
the necessary type of polarization to construct modules of this type.
In $V_\lambda^*$, there is a particular state $v_\lambda^*$ which is
defined by
\begin{equation}
\ip{ v_\lambda^*}{w} =\left\{\begin{array}{ll}
1, &  w = v_\lambda\\
0, & w \neq v_\lambda
\end{array}\right.
\end{equation}
It is clear that $v_\lambda^*$ is a highest weight state in the dual
Verma module.  Furthermore, it follows from the above definition that
there are no other highest weight states in this module besides
multiples of this state, since every other state in the dual space
must have a nonzero contraction with some state of the form $e_\alpha
w$ with $\alpha \prec 0$ and $w \in V_\lambda$.  In fact, the
existence of a single highest weight vector is a characteristic
feature of dual Verma modules.
Another way of defining dual Verma modules is through the
characteristic property of being cofree over the enveloping algebra
${\cal U} (\algg_-)$ \cite{bmp}.

In addition to the Verma and dual Verma modules defined above,
there are other types of modules of interest.  We have already
discussed irreducible highest weight representations, which have both
the features that the associated module has a single highest weight
vector, and that the entire module is in the orbit of the highest
weight state under the action of $\algg$.  There are also modules,
called ``twisted'' Verma modules, which have neither of these
properties.  These modules can be defined by the property of being
free over ${\cal U} (\algm)$ and cofree over ${\cal U} (\algm')$,
where $\algm$ and $\algm'$ are subalgebras of $\algg_-$ such that
\begin{equation}
\algg_- = \algm \oplus \algm'.
\end{equation}
Twisted Verma modules were first introduced by Feigin and Frenkel \cite{FF3}.

In order to analyze these various types of modules and their
properties, it is useful to define certain vectors in a $\algg$-module
to be {\em singular}.  Following Feigin and Fuchs, we define a vector
$v\in V$ to be singular when $\algg_+ \cdot v = 0$, and cosingular
when $v \not\in \algg_- V$.  Intuitively, when a vector $v$ is
singular it is a highest weight vector for a submodule $W \subset V$
given by acting on $v$ with ${\cal U} (\algg_-)$.  In order to construct
an irreducible representation containing the highest weight vector of
$V$, it is clearly necessary to mod out by any such submodule $W$
generated by a singular vector.  Similarly, a cosingular vector is one
which generates a submodule $X \subset V$ which lies outside the orbit
of the highest weight vector of $V$ under the action of ${\cal U}
(\algg)$.  Any irreducible representation containing the highest
weight vector of $V$ must be modded out by any such submodule
$X$ generated by a cosingular vector.  Clearly, a particular
cosingular vector is only defined modulo $\algg_- V$.  We will
frequently make use of a choice of specific cosingular vectors from
each equivalence class thereof, and will often simply refer to such
vectors as ``the'' cosingular vectors of a particular weight.  From
the above definitions, it follows that a module $V$ contains a
singular vector of weight $\lambda$ precisely when the dual module $V$
contains a cosingular vector of the same weight.  Clearly, a Verma
module contains only singular vectors and no cosingular vectors, and a
dual Verma module contains only cosingular vectors and no singular
vectors.  A useful graphical description of modules can be given by
drawing a point for each singular or cosingular vector, and then
drawing an arrow from the point associated with a vector $v$ to the
point associated with a vector $w$ whenever the submodule ${\cal U}
(\algg) v$ contains the vector $w$.  The arrow relation is assumed to
be transitive, so that when a sequence of vectors are connected by
arrows such as $v
\rightarrow w \rightarrow u$, we omit the arrow between the vectors
$v$ and $u$ in the diagram.  In this diagrammatic notation, a Verma
module has the property that all arrows lead away from the highest
weight state.  The dual of a module has the same graphical structure
as the original module, however all arrows are reversed.  In
particular, for the dual Verma module the graph has the property that
all arrows lead toward the highest weight vector.  Twisted Verma
modules are characterized by having singular or cosingular vectors
with the same weights as the singular vectors in the Verma module, but
where the graphical representation contains a different set of arrows
from the graph representing the Verma module.

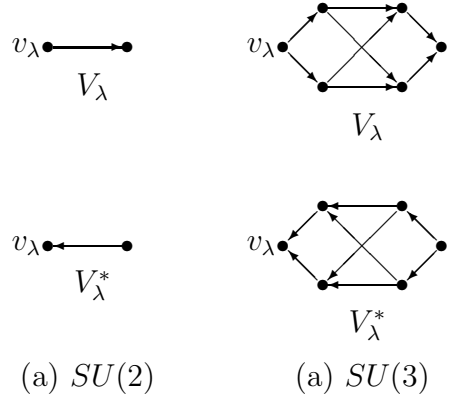
\begin{figure}
\centering
\begin{picture}(100,150)(- 50,- 80)
\put(- 15, 50){\circle*{4}}
\put(- 15, - 25){\circle*{4}}
\put(15, 50){\circle*{4}}
\put(15, - 25){\circle*{4}}
\put(- 13, 50){\vector(1,0){26}}
\put( 13, - 25){\vector( -1,0){26}}
\put(- 25, 50){\makebox(0,0){ $v_\lambda$}}
\put(- 25, - 25){\makebox(0,0){ $v_\lambda$}}
\put(0, 35){\makebox(0,0){ $V_\lambda$}}
\put(0, - 40){\makebox(0,0){ $V_\lambda^*$}}
\put(0, - 75){\makebox(0,0){(a) $ SU(2)$}}
\end{picture}
\centering
\begin{picture}(100,150)(- 50,- 80)
\put(- 30, 50){\circle*{4}}
\put( 30, 50){\circle*{4}}
\put(- 15, 65){\circle*{4}}
\put( 15, 35){\circle*{4}}
\put(15, 65){\circle*{4}}
\put(- 15, 35){\circle*{4}}
\put(- 28, 52){\vector(1,1){11}}
\put(- 28,  48){\vector(1,-1){11}}
\put(- 13, 65){\vector(1,0){26}}
\put(- 13, 35){\vector(1,0){26}}
\put(- 13, 63){\vector(1,-1){26}}
\put(- 13, 37){\vector(1,1){26}}
\put(17, 63){\vector(1, -1){11}}
\put(17, 37){\vector(1, 1){11}}
\put(- 30, -25){\circle*{4}}
\put( 30, -25){\circle*{4}}
\put(- 15, - 10){\circle*{4}}
\put( -15, - 40){\circle*{4}}
\put(15, -10){\circle*{4}}
\put( 15, - 40){\circle*{4}}
\put(28,-27){\vector( -1,-1){11}}
\put(28,-23){\vector( -1,1){11}}
\put(13,-40){\vector( -1,0){26}}
\put(13,-10){\vector( -1,0){26}}
\put(13,-38){\vector( -1,1){26}}
\put(13,-12){\vector( -1,-1){26}}
\put(-17,-38){\vector( -1,1){11}}
\put(-17,-12){\vector( -1,-1){11}}
\put(- 40, 50){\makebox(0,0){ $v_\lambda$}}
\put(- 40, - 25){\makebox(0,0){ $v_\lambda$}}
\put(0, 20){\makebox(0,0){ $V_\lambda$}}
\put(0, - 55){\makebox(0,0){ $V_\lambda^*$}}
\put(0, - 75){\makebox(0,0){(a) $ SU(3)$}}
\end{picture}
\caption[Graphs of singular vectors for $SU(2)$ and $SU(3)$
representations]{\footnotesize Graphs of singular vectors for $SU(2)$
and $SU(3)$ representations}
\label{f:graphfinite}
\end{figure}

As examples, in Figures~\ref{f:graphfinite}a and~\ref{f:graphfinite}b,
we give the graphs associated with the Verma modules and dual Verma
modules of the algebras ${\rm su} (2)$ and ${\rm su} (3)$.  The
singular vector structure of Virasoro modules was first fully studied
by Feigin and Fuchs \cite{FF}.  In general, when the Virasoro Verma
module associated with a given highest weight is reducible (contains a
singular vector), the associated representation is said to be degenerate.
For values of $h$ and $c$
corresponding to irreducible unitary representations, the Verma
modules fall into several categories.  When $c > 1$ and $h > 0$, the
resulting Verma module has no singular vectors,  and is therefore
always nondegenerate.  When $c > 1$ and $h =
0$, the resulting Verma module has a single singular vector which is
always at level 1, and the resulting graph looks like the graph for an
$SU(2)$ Verma module in Figure~\ref{f:graphfinite}a.  When $c = 1$ and
$h = m^2/4$ for an integer $m$, the Verma module contains an infinite
number of singular vectors; the resulting graph is shown in
Figure~\ref{f:graphVirasoro}a.  The remaining unitary representations
with $c = 1$ correspond to Verma modules with no singular vectors and
are nondegenerate.
The final category of Verma modules for unitary Virasoro
representations corresponds to the discrete series of representations
with $c,h$ satisfying (\ref{eq:min1}) and (\ref{eq:min2}).  The Verma
modules corresponding to these representations again have an infinite
number of singular vectors.  The graph of the singular vector
structure for these representations is shown in figure
Figure~\ref{f:graphVirasoro}b.  The graph associated with the
Feigin-Fuchs representations of the Virasoro algebra is shown in
Figure~\ref{f:graphVirasoro}c; these representations will be discussed
further in the next section.  Note that the graph corresponding to the
Feigin-Fuchs modules remains fixed under the duality transformation.
Similar graphs can be constructed describing the singular vector
structure of the affine algebras corresponding to the loop groups
$\widehat{L}G$ \cite{FF3}.

\begin{figure}
\centering
\begin{picture}(200,200)(- 100,- 75)
\multiput(- 60, 110)(30, 0){4}{\circle*{4}}
\multiput(- 57,110)( 30, 0){4}{\vector(1, 0){ 24}}
\put(65, 110){\makebox(0,0){ $\cdots$}}
\put(-60, 50){\circle*{4}}
\multiput(-45, 65)(30,0){4}{\circle*{4}}
\multiput(-45, 35)(30,0){4}{\circle*{4}}
\put(-58, 52){\vector(1,1){11}}
\put(-58,  48){\vector(1,-1){11}}
\multiput(-43, 65)(30,0){4}{\vector(1,0){26}}
\multiput(-43, 35)(30,0){4}{\vector(1,0){26}}
\multiput(-43, 63)(30,0){4}{\vector(1,-1){26}}
\multiput(-43, 37)(30,0){4}{\vector(1,1){26}}
\put( 83, 65){\makebox(0,0){$ \cdots $}}
\put( 83, 35){\makebox(0,0){$ \cdots $}}
\put(-60, -25){\circle*{4}}
\multiput(-45, - 10)(30,0){4}{\circle*{4}}
\multiput(-45, - 40)(30,0){4}{\circle*{4}}
\multiput(-43, - 40)(60,0){2}{\vector(1,0){26}}
\multiput(13, - 40)(60,0){2}{\vector(-1,0){26}}
\multiput(-43, - 38)(30,0){4}{\vector(1,1){26}}
\multiput(-17,-10)(60,0){2}{\vector( -1,0){26}}
\multiput(-13,-10)(60,0){2}{\vector(1,0){26}}
\multiput(-17,-38)(30,0){4}{\vector( -1,1){26}}
\put(-47,-38){\vector( -1,1){11}}
\put(-58, - 23){\vector(1,1){11}}
\put( 83, - 10){\makebox(0,0){ $\cdots$}}
\put(83, - 40){\makebox(0,0){ $\cdots$}}
\put(-70, 50){\makebox(0,0){ $v_\lambda$}}
\put(-70, - 25){\makebox(0,0){ $v_\lambda$}}
\put(0, 88){\makebox(0,0){(a) Verma module $V_\lambda$ for $c = 1$, $h
= m^2/4$}}
\put(0, 13){\makebox(0,0){(b) Verma module $V_\lambda$ for discrete
unitary series}}
\put(0, - 62){\makebox(0,0){(c) Feigin-Fuchs Virasoro module
$\tilde{V}_\lambda$}}
\end{picture}
\caption[Graphs of singular vectors for Verma module
and Feigin-Fuchs Virasoro representations]{\footnotesize Verma
module and Feigin-Fuchs Virasoro representations}
\label{f:graphVirasoro}
\end{figure}

\subsection{Fock space realizations}
\label{sec:realizations}

In this section we describe a particular class of $\algg$-modules
which can be described in terms of bosonic Fock spaces.  In order to
describe representations on bosonic Fock spaces, we begin by defining
a simple class of algebras, known as Heisenberg algebras.  For any set
$S$, we define the Heisenberg algebra on the set $S$ to be the Lie
algebra with generators 1 and $\{e_a,f_a| a \in S\}$ satisfying the
commutation relations
\begin{eqnarray}
\left[e_a,f_b\right] & = &  \delta_{ab} \cdot 1 \nonumber\\
\left[e_a,e_b\right] & = & \left[f_a,f_b\right] = 0
\label{eq:Heisenbergalgebra}\\
\left[1,e_a\right] & = &\left[1,f_a\right] = 0. \nonumber
\end{eqnarray}
For any set $S$, this algebra has a unique highest weight irreducible
representation.  The carrier space of this representation is the
bosonic Fock space on $S$, which has a natural basis labeled by the
occupation numbers $N_a$ of the independent  bosonic fields; these
occupation numbers are the eigenvalues of the operators
\begin{equation}
n_a = f_a e_a.
\end{equation}
There is a simple realization of the Heisenberg algebra and bosonic
Fock space on any set $S$ in terms of  differential operators acting
on a ring of polynomials.  If we identify the Fock space on $S$ with
the polynomial ring $R_S = \CC[\{z_a | a \in S\}]$, the Heisenberg
algebra on $S$ has a representation through the identifications
\begin{eqnarray}
e_a & \rightarrow &  \partial/\partial z_a\\
f_a & \rightarrow &  z_a. \nonumber
\end{eqnarray}
Thus, all representations of algebras  in terms of differential
operators acting on a space of polynomials can be naturally rewritten
in the language of bosonic Fock spaces and vice versa.

There is a natural action of the Heisenberg algebra on $S$ on the dual
space $R_S^*$.  This action is defined through the Fock space adjoint
operation $e_a^{\dagger} = f_a$, so that for any $g\in R_S^*,h \in R_S$
we have
\begin{eqnarray}
\ip{f_a g}{h} & = & \ip{g}{e_a h} \\
\ip{e_a g}{h} & = & \ip{g}{f_a h}. \nonumber
\end{eqnarray}
The space $R_S^*$ carries a highest weight irreducible representation
of the Heisenberg algebra, and therefore can be canonically identified
with $R_S$.

In terms of this bosonic Fock space language, it is clear that the
coadjoint orbit representations defined in Chapter 2 for an arbitrary
Lie group $G$ in terms of
differential operators on the space of polynomials in the variables
$z_\alpha$ with $\alpha \in \Phi_+$ are realizations on the bosonic Fock
space over $\Phi_+$.  By taking the Fock space adjoint of these
representations,  where $e_\alpha^{\dagger} \rightarrow e_{-
\alpha}$, we can construct the duals to the coadjoint
orbit representations.  For example, the dual to the coadjoint orbit
representation of $SU(3)$ described in (\ref{eq:operators3}) is
defined by the operators $\tilde{J}_a$, where
\begin{eqnarray}
\tilde{J}_t & = &  - t \derivative{t}  -
\frac{1}{2}  v \derivative{v}  +\frac{1}{2} u \derivative{u}  + p/2
\nonumber\\
\tilde{J}_u & = &  - u \derivative{u}   -\frac{1}{2}
 v \derivative{v}  +\frac{1}{2} t \derivative{t} + q/2.  \nonumber \\
\tilde{J}_{-t} & = & - t - \frac{1}{2}v \derivative{u}   \nonumber\\
\tilde{J}_{- u} & = & - u + \frac{1}{2}v \derivative{t}
\label{eq:dualoperators3}\\
\tilde{J}_{- v} & = &  - v. \nonumber\\
\tilde{J}_{+ t} & = & t \derivetwo{t}  +
u (-\derivative{v}- \frac{1}{2}  \derivative{t}\derivative{u}) +
v (\frac{1}{2}\derivative{t}\derivative{v}-\frac{1}{4}  \derivetwo{t}
\derivative{u} )   - p\derivative{t}\nonumber\\
\tilde{J}_{+ u} & = &  t (\derivative{v}- \frac{1}{2}
\derivative{t}\derivative{u})
+ u \derivetwo{u} +
v (\frac{1}{2}\derivative{u}\derivative{v} + \frac{1}{4}
\derivative{t} \derivetwo{u})- q\derivative{u}\nonumber\\
\tilde{J}_{+ v} & = &
t (\derivative{t}\derivative{v}- \frac{1}{2} \derivetwo{t}
\derivative{u}) + u (\derivative{u}\derivative{v}+ \frac{1}{2}
\derivative{t}\derivetwo{u})
+ v (\derivetwo{v}+ \frac{1}{4} \derivetwo{t} \derivetwo{u})  \nonumber\\
& & \; \; \;- (p + q) \derivative{v}  + \frac{(p - q)}{2}
\derivative{t}\derivative{u}. \nonumber
\end{eqnarray}
In the analysis of the coadjoint orbit representations in Chapter 2,
we observed that for a general group $G$, the generator
$\hat{J}_\alpha$ corresponding to a positive root $\alpha \in \Phi_+$ has
a leading term given by $-
\partial/\partial z_\alpha$, with all other terms being of the form
$\phi \; \partial/\partial z_\beta$ with $\beta \succ \alpha$ and with
$\phi$ being a polynomial in the variables $z_\alpha$ with no constant
terms.  It follows directly that in the dual representation, the
generator $\tilde{J}_{- \alpha}$ corresponding to a negative root
($\alpha \in \Phi_+$) has a leading term $- z_\alpha$ with all other
terms being differential operators of degree $\geq 1$, which have at
least as many derivatives as variables $z_\alpha$ in each term.  From
this observation, we can see that the universal enveloping algebra
${\cal U} (\algg_-)$ must act freely on the dual space to the
coadjoint orbit representation, since for any element $X$ of this
enveloping algebra there is at least one term with a maximal number
$n$ of factors (we minimize $n$ over all representations of the
element in question), and this term must contribute a characteristic
term with $n$ factors to the polynomial $X \nul$ which cannot be
canceled by any terms in $X$ with less than $n$ factors.  By a similar
argument, we see that the entire Fock space is in the orbit of $\nul$
under the action of $\algg$.  Thus, it follows that the dual
representation to the coadjoint orbit representation is a Verma module
representation, and therefore that the coadjoint orbit representation
is a dual Verma module representation for any group $G$.  The result
that coadjoint orbits give rise to dual Verma module resolutions is
already a well-known fact, at least for those groups where the
coadjoint orbit construction is equivalent to the Borel-Weil theory.
This property is perhaps somewhat simpler to understand using the
explicit representations described here, however.

It is important to note that in an explicit Fock space realization of
a particular representation of an algebra $\algg$, the expression of
the generators in terms of the Heisenberg algebra will generally
depend upon the highest weight of the particular representation being
realized.  The coadjoint orbit representations described in the
previous chapter have the unusual feature that the form of all the
raising operators is independent of the highest weight of the
representation.  This property of the coadjoint orbit representations
will simplify certain calculations in Chapter 4.

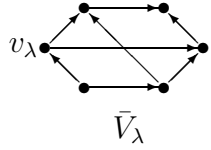
\begin{figure}
\centering
\begin{picture}(100,70)(- 50,10)
\put(- 30, 50){\circle*{4}}
\put( 30, 50){\circle*{4}}
\put(- 15, 65){\circle*{4}}
\put( 15, 35){\circle*{4}}
\put(15, 65){\circle*{4}}
\put(- 15, 35){\circle*{4}}
\put(- 28, 52){\vector(1,1){11}}
\put(- 17,  37){\vector( -1,1){11}}
\put(- 13, 65){\vector(1,0){26}}
\put(- 13, 35){\vector(1,0){26}}
\put( 13, 37){\vector( -1,1){26}}
\put(-  28, 50){\vector(1,0){56}}
\put( 28, 52){\vector(-1, 1){11}}
\put(17, 37){\vector(1, 1){11}}
\put(- 40, 50){\makebox(0,0){ $v_\lambda$}}
\put(0, 20){\makebox(0,0){ $\bar{V}_\lambda$}}
\end{picture}
\caption[Graph of twisted Verma module
representation for $SU(3)$]{\footnotesize Twisted Verma module
representation for $SU(3)$}
\label{f:singular3}
\end{figure}

We conclude this section with a brief discussion of bosonic
realizations of the Feigin-Fuchs and twisted Verma module
representations.  We have seen above that the coadjoint orbit
representation and dual coadjoint orbit representation of any group
$G$ correspond to dual Verma module and Verma module representations
respectively, and can be realized in terms of bosonic Fock space
representations.  Similarly, we can realize twisted Verma module
representations on a bosonic Fock space.  In general \cite{bmp,FF3}, a
twisted Verma module realization can be constructed by beginning with
a Verma module or dual Verma module realization on a bosonic Fock
space, and performing a Bogoliubov transformation on the Heisenberg
algebra (\ref{eq:Heisenbergalgebra}) which leaves the form of the
algebra unchanged, but requires a shift in the ground state $\nul$.
In general, such a Bogoliubov transformation corresponds to a Weyl
symmetry transformation on the Lie algebra.  As an example of this
construction, consider the Verma module representation of $SU(3)$
(\ref{eq:dualoperators3}).  Performing the Bogoliubov transformation
$t \rightarrow - \partial/\partial t$, $\partial/\partial t
\rightarrow t$, the form of this algebra is unchanged.  However, the
ground state $\nul$ is now annihilated by a different set of
operators, so to fix the eigenvalues of the diagonal operators
correctly we must replace $(p,q)$ by the vector $(-p - 2,q + p + 1)$.
This gives a new Fock space realization of the algebra of $SU(3)$ in
terms of operators $\bar{J}_a$, where
\begin{eqnarray}
\bar{J}_t & = &  t \derivative{t}  -
\frac{1}{2}  v \derivative{v}  +\frac{1}{2} u \derivative{u} -p/2
\nonumber\\
\bar{J}_u & = &  - u \derivative{u}   -\frac{1}{2}
 v \derivative{v}  -\frac{1}{2} t \derivative{t} + (q + p)/2.  \nonumber \\
\bar{J}_{-t} & = & \derivative{t} - \frac{1}{2}v \derivative{u}   \nonumber\\
\bar{J}_{- u} & = & - u + \frac{1}{2}v  t
\label{eq:twistedoperators3}\\
\bar{J}_{- v} & = &  - v. \nonumber\\
\bar{J}_{+ t} & = & -t^2\derivative{t}  -
 (\frac{1}{2}  t u+\frac{1}{4}  t^2 v)\derivative{u} +
 (\frac{1}{2}t v- u)\derivative{v}
   +p t\nonumber\\
\bar{J}_{+ u} & = &  -\derivative{t} \derivative{v}+ \frac{1}{2}
t \derivative{t}\derivative{u}
+ (u +\frac{1}{4}
tv) \derivetwo{u} +
 \frac{1}{2}v\derivative{u}\derivative{v}
- (q + p +\frac{1}{2} )\derivative{u}\nonumber\\
\bar{J}_{+ v} & = &
-t \derivative{t}\derivative{v}+ \frac{1}{2} t^2 \derivative{t}
\derivative{u} + u \derivative{u}\derivative{v}+ (\frac{1}{2}
t u+ \frac{1}{4} t^2 v) \derivetwo{u}
+ v \derivetwo{v}   \nonumber\\
& & \; \; \; - q \derivative{v}  - \frac{(2p + q + 1)}{2}
t \derivative{u}. \nonumber
\end{eqnarray}
This representation of the $SU(3)$ algebra is a twisted Verma module
representation $\bar{V}_\lambda$.  The Fock space module is free on
the state $\nul = 1$ over the subalgebra generated by $\bar{J}_{- u}$
and $\bar{J}_{- v}$, and cofree over the subalgebra generated by
$\bar{J}_{- t}$.  The graph of the singular vector structure for this
representation is shown in Figure~\ref{f:singular3}.
Note that the graphical structure of this twisted Verma module is
identical to that of the Verma and dual Verma modules; however, the
position of the highest weight vector has changed.

For any finite-dimensional Lie group, any twisted Verma module
representation of the corresponding Lie algebra on a polynomial Fock
space can similarly be constructed by performing a Bogoliubov
transformation corresponding to some Weyl transformation on the Lie
algebra.  Even the dual Verma module can be constructed in this way by
a Bogoliubov transformation on the original Verma module; note,
however, that the consequent explicit realization on the bosonic Fock
space is different from that given by the duality transformation
described above.  In the case of infinite-dimensional Lie groups,
there are additional modules similar to the twisted Verma modules
which cannot be constructed via a Bogoliubov transformation.  For
example, since the irreducible representations of the loop groups and
Virasoro group do not contain lowest weight states, the dual Verma
module representation cannot be achieved by a Bogoliubov
transformation from the Verma module representation, unlike the case
of finite-dimensional Lie groups.

Among the modules for infinite-dimensional groups which cannot be
reached by Bogoliubov transformations from the Verma module
representation, one class of representations is of particular
interest.  These are the free field realizations, which have been
studied intensively by physicists since the early days of string
theory (see \cite{bmp} and references therein).  In the case of the
Virasoro algebra, these representations were first studied by Feigin
and Fuchs \cite{FF} and were later described in terms of free fields
by Dotsenko and Fateev \cite{DF}.  In the Feigin-Fuchs
representations, there is a set of operators $\{a_n: n \in \ZZ\}$,
corresponding to the modes of a free bosonic field.  These operators
satisfy the Heisenberg-type algebra
\bge
[ a_n, a_m] = 2 n \delta_{n, -m}.\label{eq:bosonicmodes}\ee
The operators $a_n$ act on a bosonic Fock space with a vacuum
$\nul$
satisfying $a_n \nul = 0$ for $n > 0$ and $a_0
\nul = 2 \alpha
\nul$.  The Virasoro generators appear as modes
of the stress-energy
tensor, and are written in terms of the $a$'s as \cite{Feld}
\begin{eqnarray}
L_n & = & \frac{1}{4} \sum_{k = -\infty}^{\infty} a_{n-k} a_k - \alpha_0
(n+1) a_n,\;\;\;{\rm for} \; n \neq 0,   \label{eq:Feigin-Fuchs}\\
L_0 & = & \frac{1}{2} \sum_{k = 1}^{\infty} a_{-k} a_k + \frac{1}{4}
a_0^2 - \alpha_0 a_0. \nonumber \end{eqnarray}
These generators satisfy a Virasoro algebra with
\bge h = \alpha(\alpha - 2 \alpha_0),\;\;\;\; c = 1 - 24 \alpha_0^2.\ee
We can rewrite this representation in terms of the polynomial
realization of the bosonic Fock space by writing
\begin{eqnarray}
a_n & = & 2 n \frac{\pdv}{\pdv z_n}, \;\;\;{\rm for} \; n > 0,
\nonumber \\
a_0 & = & 2 \alpha, \\
a_{-n} & = & z_n, \;\;\;{\rm for} \; n > 0. \nonumber \end{eqnarray}
In this notation, the Virasoro generators are
\begin{eqnarray}
L_n & = & \sum_{k = 1}^{n-1} k (n-k) \frac{\pdv}{\pdv z_k} \frac{\pdv}{\pdv
z_{n-k}} \nonumber\\
& &\hspace{1in}+ 2 n (\alpha - \alpha_0 (n+1)) \frac{\pdv}{\pdv z_n} +
\sum_{k = n+1}^{\infty} k
z_{k-n} \frac{\pdv}{\pdv z_k}, \;\;\; {\rm for} \; n > 0 \nonumber \\
L_0 & = & \alpha(\alpha - 2 \alpha_0) +
\sum_{k = 1}^{\infty} k
z_k \frac{\pdv}{\pdv z_k},  \\
L_{-n} & = & (\alpha + \alpha_0(n-1)) z_n + \frac{1}{4} \sum_{k =
1}^{n-1} z_k z_{n-k} + \sum_{k = 1}^{\infty} k
z_{k+n} \frac{\pdv}{\pdv z_k}, \;\;\; {\rm for} \; n > 0. \nonumber
\end{eqnarray}
The module structure of this representation is precisely described by
the graph in Figure~\ref{f:graphVirasoro}c, and corresponds to a
module similar to a twisted Verma module, but which cannot be realized
by a Bogoliubov transformation from the Verma or dual Verma modules.
Note that in these Feigin-Fuchs representations, the form of all the
generators is dependent upon both the parameters $h$ and $c$ of the
representation through the parameters $\alpha$ and $\alpha_0$, unlike
the coadjoint orbit representations of the Virasoro algebra, where
only the generators $\hat{L}_{n}$ for $n \leq 0$ depend upon $h$ and $c$.

For the affine algebras corresponding to loop groups $\widehat{L}G$,
there is an analogous class of free field representations which are
similar to twisted Verma modules but again cannot be constructed from
the Verma or dual Verma module by a finite Weyl transformation.  These
representations are the Wakimoto representations, which are
constructed by taking the normal ordered product of the currents
associated with free fields having $\widehat{L}G$ symmetry.  An
equivalent mathematical description of the construction of these
representations is given by taking representations of the
finite-dimensional group $G$ and replacing all Heisenberg operators by
free fields.  The normal ordering must be enforced to give the correct
affine algebra.  This process is known as ``affinization''.  For a
review of this construction of Wakimoto representations, see
\cite{bmp,aty}.

\subsection{Intertwiners}
\label{sec:intertwiners}

In this section we derive explicit formulae for intertwining operators
between the dual Verma modules realized through the coadjoint orbit
construction.  Generally, an intertwining operator, or ``intertwiner''
between two $\algg$-modules $V$ and $V'$ is an element of the group
${\rm Hom}_{{\cal U} (\algg)}(V,V')$ of $\algg$-module homomorphisms
between $V$ and $V'$.  A $\algg$-module homomorphism is a linear map
$\phi$ between modules which intertwines with the action of $\algg$ in
the sense that
\begin{equation}
u \phi x = \phi u x, \; \; \; \forall x \in V,u \in \algg.
\end{equation}

An analysis of the construction of intertwiners in a general algebraic
context is given in \cite{bmp}.  In that work, it is proven that for
two Verma modules $V_\lambda, V_{\lambda'}$, there is a 1-1
correspondence between intertwiners in ${\rm Hom}_{{\cal U}
(\algg)}(V_\lambda,V_{\lambda'})$ and singular vectors in
$V_{\lambda'}$ of weight $\lambda$.  For each singular vector $v$ in
$V_{\lambda'}$ of weight $\lambda$, the resulting $\algg$-module
homomorphism $\phi$ satisfies $\phi \nul = v$ (here $\nul$ denotes the
highest weight state in $V_\lambda$; we will denote the highest weight
state in $V_{\lambda'}$ by $\nul'$).  This characterization of
intertwiners for Verma modules is derived by first considering
elements of the group of $\algg_-$-homomorphisms ${\rm Hom}_{{\cal U}
(\algg_-)}(V_\lambda,V_{\lambda'})$.  Since the orbit of $\nul$ under
the action of the enveloping algebra ${\cal U} (\algg_-)$ contains the
entire Verma module, such a homomorphism $\phi$ is determined uniquely
by the value of $\phi
\nul$ in $V_{\lambda'}$,  and each element $x \in V_{\lambda'}$
determines a homomorphism of this type by setting $\phi \nul = x$.
The condition that $\phi$ intertwine correctly with the Cartan
subalgebra $\algt$ requires that the weight of the state $\phi \nul$
be exactly $\lambda$.  This leads to the general definition of a set
of {\em screening operators} $s_\alpha$, which are
$\algg_-$-homomorphisms mapping
\begin{equation}
s_\alpha:V_\lambda \rightarrow V_{\lambda + \alpha}
\end{equation}
Explicitly, given an element $X \in {\cal U} (\algg_-)$, corresponding
to a state $x = X \nul$ in the Verma module $V_\lambda$, the screening
operator $s_\alpha$ acts on $x$ by $s_\alpha x = X J_{- \alpha}
\nul'$, where $\nul'$ is the highest weight state in the Verma module
$V_{\lambda + \alpha}$.  In general,  all  $(\algg_- \oplus
\algt)$-module homomorphisms between Verma modules are polynomials in
the screening operators $s_\alpha$.  The screening operators generate
an algebra isomorphic to $\algg_-$.  The condition that  an operator $\phi$
described by a polynomial in the screening operators intertwines with the
remaining $\algg$ generators simply implies that the state $\phi \nul$
is annihilated by all the generators in $\algg_+$, which is equivalent
to the assertion that $\phi \nul$ is a singular vector in
$V_{\lambda'}$,  giving the remaining part of the above result.

{}From this algebraic characterization of intertwining operators on
Verma modules, it is straightforward to consider the dual action of
the intertwiners, which gives intertwining operators on dual Verma
modules.
Clearly, for every intertwining operator $\phi$ between Verma modules
$V_\lambda$ and $V_{\lambda'}$, there is a corresponding intertwiner
$\phi^*$ from the dual Verma module $V_{\lambda'}^*$ to $V_\lambda^*$,
defined by
\begin{equation}
\ip{ \phi^* x}{w} = \ip{ x}{\phi w}, \; \; \; \forall w \in V_\lambda,
x \in V_{\lambda'}^*.
\end{equation}
This correspondence is clearly 1-1, since the same argument can be
applied on the dual space.  From the result for Verma modules, we can
derive the corresponding result for dual Verma modules, which is that
the intertwiners in ${\rm Hom}_{{\cal U}
(\algg)}(V_{\lambda'}^*,V_\lambda^*)$ are in 1-1 correspondence with
cosingular vectors in $V_{\lambda'}^*$ of weight $\lambda$, and that
the intertwiner $\phi$ corresponding to a particular cosingular vector
$v\in V_{\lambda'}^*$ satisfies $\phi v = \nul$ where $\nul$ is the
highest weight vector in $V_{\lambda}^*$.  This intertwiner must also
satisfy the condition that $\phi w = 0$ for any vector $w \in \algg_-
V_{\lambda'}^*$ of weight $\lambda$.  By taking the dual of the
screening operators $s_\alpha$, we get a set of screening operators
$\tilde{s}_\alpha = s_\alpha^{\dagger}$ for the dual Verma module.
These screening operators are maps
\begin{equation}
\tilde{s}_\alpha:V_\lambda^* \rightarrow V_{\lambda -\alpha}^*
\end{equation}
which intertwine with the algebra $\algg_+$ of generators
corresponding to positive roots acting on the dual Verma modules.  In
fact, every intertwiner on a dual Verma module can be found by
constructing an appropriate polynomial in the screening operators;
thus, in order to characterize the intertwining operators in a
particular realization, it will suffice to describe the screening
operators in that realization.

We now proceed to give explicit formulae for the screening operators
in the coadjoint orbit dual Verma module representations.
For an arbitrary group $G$, we begin by defining a set of differential
operators $D_\alpha$ for all $\alpha \in \Phi_+$ according to
\begin{equation}
D_\alpha = \sum_{\beta \in \Phi_+| \alpha + \beta \in \Phi_+} i
\mixten{f}{\alpha \beta}{\alpha + \beta} z_\beta
\frac{\partial}{\partial z_{\alpha + \beta}}.
\end{equation}
These operators act on the ring $R = \CC[\{z_\alpha\}]$ of locally
holomorphic functions as first order differential operators.
\begin{prop}
The operators $\tilde{s}_\alpha$ defined by
\begin{equation}
\tilde{s}_\alpha = \hat{J}_\alpha + D_\alpha
\end{equation}
commute with the operators $\hat{J}_\beta$ with $\beta \in \Phi_+$,
and form a set of screening operators for the dual Verma modules.
\end{prop}
\noindent {\it Proof.} \,\,
Because the only nonzero Bernoulli number $B_l$ with $l$ odd is $B_1 =
-1/2$, using (\ref{eq:positivesimplification}) we can  write these
operators explicitly as
\bge
\tilde{s}_a = \!\sum_{\sss k \geq 0, a_1, \ldots, a_k \succ 0}\!
 - \frac{B_k(-1)^k}{k!} C_a(a_1, \ldots, a_k) z_{a_1} \ldots z_{a_k}
\frac{\pdv}{\pdv z_{a+a_1+\ldots + a_k}}\;\;\; {\rm for} \; a \succ 0.
\ee
We now consider the effect of right multiplication of a general
element (\ref{eq:generalelement}) in the group by  an infinitesimal
positive generator on the right.  By taking the formal multiplication
rule
\begin{equation}
\exp (\sum_{\alpha \in \Phi_+}z_\alpha J_\alpha){\rm e}^{\epsilon J_a} =
\exp \left[\sum_{\alpha \in \Phi_+}(z_\alpha +\epsilon\tilde{v}_a^{\alpha})
J_\alpha \right] f (\{J_a | a\preceq 0\}) + {\cal O} (\epsilon^2),
\end{equation}
we define a set of vector field operators $S_a = -\tilde{v}_a^{\alpha}
\partial/\partial z_\alpha$ which must commute with the  operators $
\xi_a$.
By repeating the analysis in the proof of Proposition \ref{p:xcalc}
for these vector fields, we find that the operators $S_\alpha$ are
precisely equal to $\tilde{s}_\alpha$ for $\alpha \in \Phi_+$.  Thus,
we have proven the assertion that
\begin{equation}
[\tilde{s}_\alpha, \hat{J}_\beta] = 0, \; \; \; \forall \alpha, \beta
\in \Phi_+.
\end{equation}
Since the operator $\tilde{s}_\alpha$ lowers the degree of a
polynomial in $R$ by precisely $\alpha$, it follows that these
operators are valid screening operators for the dual Verma modules
$\Box$.

It is a consequence of this explicit formula that the screening
operators are independent of the highest weight $\lambda$ of the
module on which they act, just as are the raising operators in the
algebra, $\hat{J}_a$ with $a \succ 0$.

A similar approach to the one taken here was used by
Awata, Tsuchiya, and Yamada in \cite{aty} to construct the Wakimoto
representations of affine algebras; in their work, the result was
achieved by using a construction similar to this one for
finite-dimensional $G$ and performing an affinization to get
realizations of the
Wakimoto representations.

As an example of an intertwiner, we consider again the coadjoint orbit
dual Verma module representations of $SU(2)$ given by
(\ref{eq:representation2}).  For these representations,  there is a
single screening operator $\tilde{s}$, given by
\begin{equation}
\tilde{s} =\hj_+  =  - \frac{\pdv}{\pdv z}.
\end{equation}
This screening operator obviously commutes with $\hat{J}_+$, and
raises the weight of a state by 1, as measured by the eigenvalue of
$\hat{J}_3$ in a particular module.  Since the screening operator goes
from a dual Verma module $V_b^*$ to another dual Verma module $V_{b -
1}^*$ with a new highest weight state, we see that $\tilde{s}$
intertwines correctly with $\hat{J}_3$.  It remains to construct an
intertwiner using $\tilde{s}$.  The dual Verma module $V_b^*$ has a
single cosingular state of weight $- b - 1$, which is represented in
the coadjoint orbit realization by the polynomial $z^{2b + 1}= \alpha
\hat{J}_-^{2b + 1}
\nul$ where $\alpha$ is a proportionality constant.  Thus, we expect
that we may construct an intertwiner
\begin{equation}
\phi:V_b^*\rightarrow V_{-b - 1}^*,
\end{equation}
by taking $\phi= \tilde{s}^{2b + 1}$.  We have already verified that
$\tilde{s}$ intertwines correctly with all operators in $\algg$ except
for $\hat{J}_-$.  We can now compute directly the result of applying
$\hat{J}_-$ before and after the intertwiner $\phi$ to verify that
$\phi$ intertwines correctly with the entire algebra.  Acting on an
arbitrary monomial $z^n$, we have
\begin{equation}
\hat{J}_- \phi z^n = \phi \hat{J}_- z^n = \frac{(n + 1)!}{(n - 2b - 1)!}.
\label{eq:intertwiner2}
\end{equation}
It follows that $\phi$ is an  intertwining operator, as claimed above.
Note that the two operators $\hat{J}_-$ in (\ref{eq:intertwiner2})
have different realizations as differential operators, since they are
acting on different dual Verma modules.  In general, we will use a
parenthesized superscript to denote the highest weight of the module
on which a coadjoint orbit generator is acting when there is a
possibility of confusion.  Thus, we have
\begin{equation}
\hat{J}_-^{(b)} = z^2 \partial z - 2b z,
\end{equation}
and (\ref{eq:intertwiner2}) states that
\begin{equation}
\hat{J}_-^{(- b - 1)} \phi z^n = \phi \hat{J}_-^{(b)} z^n.
\end{equation}

As a second example of an intertwiner consider the coadjoint orbit
dual Verma module representations of $SU(3)$ given by
(\ref{eq:operators3}).  For these representations, the screening
operators are given by
\begin{eqnarray}
\tilde{s}_{+ t} & = & - \derivative{t} + \frac{1}{2}u \derivative{v}
\nonumber\\
\tilde{s}_{+ u} & = & - \derivative{u} - \frac{1}{2}t \derivative{v}
\label{eq:screening3}\\
\tilde{s}_{+ v} & = &  - \derivative{v}. \nonumber
\end{eqnarray}
A typical representation is the adjoint representation with $(p,q) =
(1,1)$.  The associated dual Verma module $V_{(1,1)}^*$ contains
cosingular vectors with weights $(3, - 3) $, $(- 3,3)$, $(1, - 5)$,
$(- 5,1)$ and $(- 3, - 3)$.  A typical intertwining operator is the
operator
\begin{equation}
\phi: V_{(1, 1)}^* \rightarrow V_{(3, - 3)}^*
\end{equation}
given by
\begin{equation}
\phi \cdot X \cdot v= \tilde{s}_{+ u}^2 X  \cdot v = X \nul,
\end{equation}
where $\nul$ is the highest weight state in the dual Verma module
$V_{(3, - 3)}^*$, $v$ is the cosingular vector of weight $(3, - 3)$ in
$V_{(1,1)}$, and $X$ is an arbitrary element of ${\cal U}
(\algg_-)$.  In the explicit coadjoint orbit realization above, $v$
is given by the polynomial $u^2/2$ (the unique vector
with this weight in the dual Verma module).  That this operator $\phi$
is an intertwining operator from $V_{(1, 1)}^*$ to $V_{(3, - 3)}^*$
follows from the fact that $\tilde{s}_{+ u}^2 (u^2/2)= 1 = \nul$.

As a final example of an intertwining operator, we consider  again
the simplest example of the discrete series of Virasoro
representations, with $m=3, p=2, q=1$, and $h = c = 1/2$.  As
was pointed out in Section \ref{sec:unitarity},  the dual Verma module
associated with this representation has a cosingular vector at level
2.  Thus, we expect to be able to construct an intertwiner $\phi$
mapping
\begin{equation}
\phi:V_{1/2}^*\rightarrow V_{5/2}^*,
\end{equation}
where $V_{h}^*$ refers to the Virasoro dual Verma module with highest
weight $h$ and central charge $c = 1/2$.  This intertwiner is a
polynomial of degree 2 in the screening operators $\tilde{s}_n =
\hat{L}_n + D_n$, and therefore can be written in the form $\phi = a
\tilde{s}_1^2+ b
\tilde{s}_2$.  Since the intertwining operator must annihilate the state
\begin{equation}
\hat{L}_{-1}^2 \nul = \frac{4}{3} \hat{L}_2 \nul = - 3z_2 + 2z_1^2,
\end{equation}
we determine that $4a - 3b = 0$.  Fixing the normalization by choosing
$z_1^2$ as the cosingular vector, we have $a = 1/2$, so the intertwining
operator is given by
\begin{equation}
\phi = \frac{1}{2}  \tilde{s}_1^2+ \frac{2}{3}  \tilde{s}_2
=\frac{1}{2}  \derivetwo{z_1}+ \frac{2}{3}   \derivative{z_2} +
\frac{1}{3}  z_1 \derivative{z_3} +
{\cal D}_4.
\end{equation}
In   Chapter 4 we will review the intertwining operators
for the Feigin-Fuchs Virasoro representations, which were first
explicitly described by Felder \cite{Feld}.  Because of the method of
construction of these intertwiners, explicit calculations of their
action on particular states are rather more involved than in the
coadjoint orbit realizations.

\subsection{Resolutions}
\label{sec:resolutions}

We now proceed to use the intertwining operators constructed in the
previous section to define algebraically a chain complex whose
cohomology is given by a single irreducible representation of an
algebra $\algg$.  The resulting resolution is essentially the dual to
the well-known BGG (Bernstein-Gelfand-Gelfand) resolution for
finite-dimensional groups \cite{BGG}.  The explicit formulae for
intertwining operators give us an explicit realization of this
resolution.  The analysis in this section is similar to that of
\cite{bmp} where the primary consideration was the understanding of
twisted type Verma modules such as the Feigin-Fuchs and Wakimoto
representations.  The primary difference between this work and the
analogous construction for free fields is the possibility of using
single-sided resolutions; in the free field approach, one is forced
into using two-sided resolutions, which are mathematically much more
difficult objects to handle.  From a mathematical point of view, the
advantage of describing irreducible representations through a
resolution of Fock space representations is that most properties of
representations are much easier to calculate for Fock space
representations than for irreducible representations.  (An example is
the character of a representation, which has a fairly straightforward
expression for the Verma module of most groups, but which is more
complicated for the irreducible representations.)  Generally, a useful
approach to computing such properties of irreducible representations
is to compute the property for Fock modules and to then combine these
results with the resolution to arrive at the result for an irreducible
representation.  From the physical point of view, the resolution is
essentially equivalent to the BRST formalism for calculating
properties of a physical Hilbert space defined through a BRST complex
of simpler Fock module-type spaces.  In the case of the free field
approach to conformal field theory, this approach was first made
explicit algebraically by Felder \cite{Feld}.

We begin by briefly reviewing the formalism of resolutions.  A chain
complex $(M_i,d_i)$ of $\algg$-modules is defined to be a sequence of
modules
\begin{equation}
\cdots\stackrel{d_{-2}}{\rightarrow} M_{-1}
\stackrel{d_{-1}}{\rightarrow} M_{0}
\stackrel{d_{0}}{\rightarrow} M_{1}
\stackrel{d_{1}}{\rightarrow}  \cdots
\end{equation}
where $d_{i} \in {\rm Hom}_{{\cal U} (\algg)}(M_i,M_{i + 1})$ is a
set of $\algg$-module homomorphisms satisfying
\begin{equation}
d_{i + 1} \cdot d_i = 0.
\label{eq:so}
\end{equation}
A chain complex can terminate on either or both ends with a trivial
module (0).
The cohomology of  the above  chain complex is defined as usual to be
\begin{equation}
H^i = \frac{{\rm Ker} \; d_i}{{\rm Im}\; d_{i - 1}}.
\end{equation}
If $L$ is an irreducible representation of $\algg$, the above complex
is said to be a resolution of $L$ when
\begin{equation}
H^i =\left\{\begin{array}{ll}
L, & i = 0\\
0, & i \neq 0
\end{array}\right.
\end{equation}
A resolution based on a complex which terminates with either $M_1 = 0$
or $M_{-1} = 0$ is said to be a one-sided resolution.  Otherwise, the
resolution is two-sided.
In physical language, $d$ is said to be a BRST operator, and the
cohomology spaces $H^i$ are the BRST cohomology spaces, which contain
all the physical states in a given representation.
The states not in the cohomology $H^0$ are called ``ghost'' states.

The classic example of a resolution is the BGG resolution for
irreducible representations of finite-dimensional semi-simple groups
$G$ \cite{BGG}.  This resolution is constructed directly from the
Verma modules of $G$.  Given a highest weight $\lambda$, we can assign
an integer $i(v)$ to each singular vector $v$ in the Verma module
$V_\lambda$ by setting $i (v_\lambda) = 0$ for the highest weight
state and $i (v) = i (v') + 1$ whenever the graph of singular vectors
contains an arrow $v' \rightarrow v$.  This ordering is equivalent to
the ``Bruhat'' ordering on the Weyl group of $G$.  We can then
construct a complex by setting
\begin{equation}
M_j = \bigoplus_{v | i (v) = -j} V_{\lambda (v)},
\end{equation}
where $\lambda (v)$ is the weight of the singular vector $v$.  The
spaces $M_j$ are thus $\algg$-modules.  The $\algg$-module
homomorphisms $d$ are constructed by combining the intertwining
operators $\phi:V_{\lambda (v)}\rightarrow V_{\lambda (v')}$ for all
singular states $v' \rightarrow v$.  It turns out that the signs on
these intertwining operators can always be chosen in such a way that
the resulting  is a chain complex, by enforcing the condition that
$d_i \cdot d_{i - 1} = 0$.

\begin{figure}
\centering
\begin{picture}(100,80)(- 50,0)
\put(- 30, 50){\circle*{4}}
\put( 30, 50){\circle*{4}}
\put(- 15, 65){\circle*{4}}
\put( 15, 35){\circle*{4}}
\put(15, 65){\circle*{4}}
\put(- 15, 35){\circle*{4}}
\put(- 28, 52){\vector(1,1){11}}
\put(- 26, 62){\makebox(0,0){$+$}}
\put(- 28,  48){\vector(1,-1){11}}
\put(- 26, 38){\makebox(0,0){$+$}}
\put(- 13, 65){\vector(1,0){26}}
\put(0, 69){\makebox(0,0){$+$}}
\put(- 13, 35){\vector(1,0){26}}
\put(0, 31){\makebox(0,0){$+$}}
\put(- 13, 63){\vector(1,-1){26}}
\put(- 9, 54){\makebox(0,0){$-$}}
\put(- 13, 37){\vector(1,1){26}}
\put(9, 46){\makebox(0,0){$-$}}
\put(17, 63){\vector(1, -1){11}}
\put(26, 62){\makebox(0,0){$+$}}
\put(17, 37){\vector(1, 1){11}}
\put(26, 38){\makebox(0,0){$+$}}
\put(- 40, 50){\makebox(0,0){$v_\lambda$}}
\put(40, 50){\makebox(0,0){$v_5$}}
\put(- 15, 75){\makebox(0,0){$v_1$}}
\put(15, 75){\makebox(0,0){$v_3$}}
\put(- 15, 25){\makebox(0,0){$v_2$}}
\put(15, 25){\makebox(0,0){$v_4$}}
\put(0, 12){\makebox(0,0){$V_\lambda$}}
\end{picture}
\caption[Singular vectors in $SU(3)$
Verma module]{\footnotesize Singular vectors in $SU(3)$
Verma module}
\label{f:graphsingular3}
\end{figure}
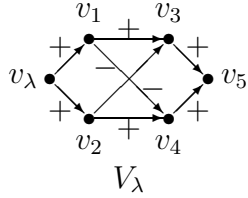

As an example of the BGG resolution, consider
again  a general Verma module for a highest weight representation of
$SU(3)$.  The graph of the singular vectors $v_1, \ldots, v_5$ is
shown in Figure~\ref{f:graphsingular3}.  The signs on the arrows
indicate the signs used in the construction of the chain complex.
The $\algg$-modules in the resulting chain complex are
\begin{eqnarray}
M_0 & = &  V_\lambda \nonumber\\
M_{-1} & = &  V_{\lambda (v_1)}\oplus V_{\lambda (v_2)} \\
M_{-2} & = &  V_{\lambda (v_3)}\oplus V_{\lambda (v_4)}  \nonumber\\
M_{-3} & = &  V_{\lambda (v_5)},\nonumber
\end{eqnarray}
and the operators are given by
\begin{eqnarray}
d_{-1} & = &  \phi_{1,0} + \phi_{2,0} \nonumber\\
d_{-2} & = &  ( \phi_{3,1} - \phi_{4,1}, - \phi_{3,2} + \phi_{4,2})\\
d_{-3} & = &  ( \phi_{5,3},  \phi_{5, 4}) \nonumber
\end{eqnarray}
where we denote by $\phi_{i,j}$ the intertwining operator from
$V_{\lambda (v_i)}$ to $V_{\lambda (v_j)}$  ($v_0 = v_\lambda$).
Because the intertwining operators satisfy the relation
\begin{equation}
\phi_{j,k} \cdot \phi_{i,j} = \phi_{j',k} \cdot \phi_{i,j'},
\end{equation}
it follows that the operator $d$ in the chain complex satisfies
(\ref{eq:so}).  That the cohomology of this complex indeed gives a
resolution of the irreducible  representation with highest weight
$\lambda$ is easy to verify by simple diagram chasing.

Just as the BGG resolution gives a chain complex whose cohomology is
an irreducible representation of a finite-dimensional group, one can
construct analogous resolutions in terms of Verma modules for the
Virasoro and loop group irreducible representations.  For the Virasoro
algebra, the relevant resolutions were essentially described by Feigin
and Fuchs \cite{FF}; for the affine algebras arising from loop groups,
the analogous result was first obtained by Garland and Lepowsky
\cite{GL}.  A proof of the existence of such a resolution in terms of
Verma modules for infinite-dimensional groups is also given in
\cite{rc}; in \cite{rc2}, these resolutions are used to calculate the
characters of irreducible representations of the Virasoro algebra.

Our interest here is in achieving an explicit realization of the dual
of the chain complex of Verma modules.  Because  the resulting
resolution is simply the dual of the resolution described in the above
works,  the existence of such a resolution in terms of dual Verma
modules follows immediately.  By combining the intertwining operators
described in the previous section, we find for any group $G$ among the
groups under consideration in this thesis, and for any highest weight
$\lambda$, a one-sided  resolution of the  irreducible representation
with highest weight $\lambda$,
\begin{equation}
0 \rightarrow V^*_\lambda \stackrel{d_0}{ \rightarrow} M_1
\stackrel{d_1}{\rightarrow} M_2
\stackrel{d_2}{\rightarrow}  \cdots.
\end{equation}
As an example of a dual Verma module resolution, consider once more
the coadjoint orbit dual Verma module realizations of $SU(2)$.  As
discussed in the previous section, when $b \geq 0$ and $b \in \ZZ/2$,
the dual Verma module $V_b^*$ contains a single cosingular state with
weight $- b - 1$.  The dual Verma module $V_{- b - 1}^*$ contains no
singular or cosingular states.  Thus, using the simple intertwiner
$\phi =
\tilde{s}^{2b + 1} = (-\partial/\partial z)^{2b + 1}$, we have a
resolution
\begin{equation}
0 \rightarrow V_b^*\stackrel{\phi}{\rightarrow}
V_{- b - 1}^*\rightarrow 0
\end{equation}
where ${\rm Im} \; \phi = V_{- b - 1}^*$, and ${\rm Ker} \; \phi =
I_b$ is the irreducible representation of $SU(2)$ with highest weight
$b$.

For the Feigin-Fuchs and Wakimoto representations of the Virasoro and
loop group algebras, it is believed that a similar resolution of
highest weight irreducible representations can be constructed in terms
of similar intertwining operators \cite{FF3}.  In the case of
Feigin-Fuchs modules, this result was essentially proven by Felder
\cite{Feld}.  For the Wakimoto representations of $\widehat{L}SU(2)$
and $\widehat{L}SU(3)$, the existence of such a resolution has also
been proven \cite{bf,ff4}.  In general, however, because the resulting
resolutions are two-sided, the usual methods of homological algebra
which are used to prove these results for one-sided resolutions must
be generalized to deal with this situation.  In \cite{bmp}, some of
the specific results regarding these two-sided resolutions are given
or conjectured, and arguments are given for the correctness of the
conjectures.  Some of the outstanding problems regarding these
two-sided resolutions were successfully dealt with in \cite{FF3}, but
the mathematical formalism necessary to resolve these difficulties is
significantly more complicated than in the case of one-sided
resolutions.  This is one of the reasons why the approach used in this
thesis, which concentrates on dual Verma modules and one-sided
resolutions, is perhaps a simpler approach to the algebraic
construction of conformal field theories.

Finally, we reiterate the point that in the case of affine algebras
arising from loop groups, in a global geometric setting the algebraic
structure of resolutions is not necessary to extract irreducible
highest weight representations from coadjoint orbit representations,
as the constraint of global holomorphicity already restricts to a
Hilbert space corresponding to an irreducible representation.  In the
case of the Virasoro algebra, however, the resolution described in
this section is necessary in order to extract irreducible
representations from the larger, reducible, dual Verma module
representations which arise from coadjoint orbits.

\mysection{Conformal Field Theories}
\label{sec:CFT}

In this chapter we describe a method for the construction of conformal
field theories using the explicit resolutions of irreducible
representations through dual Verma modules described in the previous
chapter.  Because we do not yet have a purely field-theoretic
description of the coadjoint orbit representations, we must proceed
according to the algebraic approach to conformal field theory, which
was recently developed in order to give a formal algebraic structure
to the theory and to characterize conformal field theories in terms of
free field representations.  This algebraic approach to conformal
field theory is well-described in the papers by Tsuchiya and Kanie
\cite{tk} and Bouwknegt et. al.
\cite{bmp}.

The essential element in the algebraic construction of a conformal
field theory from a set of irreducible representations of an
infinite-dimensional symmetry algebra ${\cal A}$, is the construction
of {\em vertex operators} between the irreducible representations.  In
the case of finite-dimensional groups, a vertex operator is
essentially just described by the set of Clebsch-Gordan coefficients
between three irreducible representations of the group.  In the
infinite-dimensional case, vertex operators have a similar structure;
however, they have the effect of combining two representations, one of
which is highest weight and the other of which is not highest weight,
into a third highest weight representation.

A conformal field theory is specified algebraically by giving an
infinite-dimensional symmetry algebra ${\cal A}$, a set of highest
weight irreducible representations of ${\cal A}$, and a set of vertex
operators connecting the representations.  In general, the symmetry
algebra ${\cal A}$ contains holomorphic and antiholomorphic parts, and
the vertex operators factorize into ``chiral vertex operators''
corresponding to the holomorphic and antiholomorphic parts of the
theory.  In this discussion, we will simplify by only considering the
holomorphic part of the theory; our vertex operators are therefore
what are usually referred to in the physics literature as chiral
vertex operators.  The simplest class of conformal field theories are
the {\em rational} conformal field theories, which contain a finite
set of irreducible representations of ${\cal A}$.  Each representation
is generally associated with a {\em primary field} of the theory and a
corresponding subset of the vertex operators.  The {\em fusion
algebra} of the theory is an algebra whose elements are the primary
fields of the theory, and whose product is defined by the action of
the set of nonzero vertex operators associated with one primary field
on the irreducible module associated with a second primary field.
The highest weight state of a particular representation with
exceptional symmetry is generally singled out as the physical vacuum
$| \Omega \rangle$; the physical observables of the theory (on the
sphere) are then given by acting on the vacuum with a product of
vertex operators and computing the component of the vacuum in the
resulting state.  Such an observable is referred to as a vacuum
expectation value or correlation function, and is written in the form
\begin{equation}
\langle \Omega | \Phi_1 (u_1,z_1)\ldots\Phi_k (u_k,z_k) | \Omega \rangle,
\label{eq:correlationfunction}
\end{equation}
where $\Phi_i$ are vertex operators which are parameterized by complex
coordinates $z_i$ on the sphere and enough extra parameters $u_i$ to
specify a particular vector in the representation of ${\cal A}$ (not
highest weight) associated with $\Phi_i$.

In Section \ref{sec:finitevertex}, we define vertex operators for
finite-dimensional algebras.  We show that these vertex operators can
be written in terms of the explicit screening operators for dual Verma
modules introduced in the previous chapter, and give a simple example
for the algebra of $SU(2)$.  In Section \ref{sec:generalvertex}, we
define vertex operators for the Virasoro algebra, and show that all
vertex operators between degenerate Virasoro representations satisfy a
certain set of algebraic conditions.  Furthermore, we show that any
vertex operator which has a realization as a chain map between dual
Verma module resolutions can be completely characterized by its matrix
elements.  In this section, we also give a simple class of examples of
vertex operators whose action on the vacuum can be computed in terms
of the Virasoro generator $\hat{L}_{-1}$.  In Section
\ref{sec:vertexfree}, we review the explicit construction of vertex
operators for the Feigin-Fuchs modules due to Felder, and the related
construction of intertwining operators for these modules.  In Section
\ref{sec:vertexcoadjoint} we describe the explicit construction of
vertex operators for the coadjoint orbit dual Verma modules.  We
discover a class of vertex operators whose action on any state can be
represented entirely in terms of the Virasoro generators.  The
existence of these operators is due to the independence in form of
raising operators from the highest weight of the module on which they
act in the coadjoint orbit realizations.  Finally, in Section
\ref{sec:everything} we review several approaches to calculating the
correlation functions of conformal field theories.  In principle, the
coadjoint orbit vertex operators can be used to calculate an arbitrary
correlation function for a conformal field theory on any Riemann
surface; in practice, however, a better description of these vertex
operators is necessary to make any such calculation a practical
enterprise.

For the purposes of the discussion in this chapter, we will restrict
attention to conformal field theories whose symmetry algebra ${\cal
A}$ is the Virasoro algebra.  In such a conformal field theory, a
fixed value of the central charge $c$ is chosen for all fields in the
theory.  The physical vacuum $| \Omega \rangle$ is defined to be the
highest weight state in the representation with $h = 0$.  The class of
conformal field theories corresponding to $c < 1$ contain only a
finite number of primary fields, corresponding to the discrete series
of Virasoro representations with $h$ satisfying (\ref{eq:min2}).
These rational conformal field theories are called {\em minimal
models}.  A systematic definition and study of Virasoro conformal
field theories was first performed by Belavin, Polyakov, and
Zamolodchikov (BPZ); the point of view taken in that work \cite{bpz}
and in much of the subsequent literature (see for instance
\cite{Gins}) is based on treating correlation functions
as the primary objects, and analyzing their properties using
differential equations which they can be shown to satisfy.  In the
algebraic approach to conformal field theory, the operators themselves
are taken to be the primary objects, and the correlation functions are
viewed as algebraic constructs using these operators.  We will
primarily adhere to the latter point of view; however, we will discuss
some of the parallels between structures in the algebraic formulation
and their counterparts in terms of correlation functions.  Much of the
content of this chapter, particularly Section
\ref{sec:generalvertex}, is essentially
a recasting of the original BPZ analysis of conformal field theory
into the more algebraic framework of operators, modules, and
resolutions, where the differential equations satisfied by correlation
functions are expressed as algebraic equations satisfied by vertex
operators.

Most of the results in this chapter can be generalized in a
straightforward fashion using the preceding results in this thesis to
construct conformal field theories whose symmetries are affine
algebras, such as the WZW model, in a purely algebraic fashion from
dual Verma module resolutions.  From a global geometric point of view,
however, as mentioned in the previous section, the conformal field
theories with affine algebra symmetries can be directly constructed
from the coadjoint orbit representations without the necessity of
introducing resolutions since the constraint of global holomorphicity
automatically picks out the physical states in the theory.  We discuss
briefly the possibility of extending the results in this thesis to
conformal field theories with other, extended symmetry algebras ${\cal
A}$ in the last chapter.

\subsection{Finite-dimensional vertex operators}
\label{sec:finitevertex}

Although vertex operators are primarily of interest for their
properties relative to conformal field theories and
infinite-dimensional algebras, it is possible to define analogous
objects for finite-dimensional groups \cite{bmp}.  Because these
finite-dimensional vertex operators are structurally similar to the
usual infinite-dimensional vertex operators, but are mathematically
much simpler, it is interesting to study them as toy models before
proceeding to the more complicated objects.

A vertex operator ${\cal V}_{}$ for a finite-dimensional group $G$ can
be defined for any three irreducible representations
$I_{\lambda_1}$, $I_{\lambda_2}$, and $I_{\lambda_3}$ of $G$ with
highest weights $\lambda_1, \lambda_2$, and $\lambda_3$, to be a set
of operators
\begin{equation}
{\cal V}_{u}:I_{\lambda_2} \rightarrow I_{\lambda_3} \; \; \;
\forall u \in I_{\lambda_1}
\end{equation}
which transform under $\algg$ according to
\begin{equation}
[x, {\cal V}_{u}] = {\cal V}_{xu} \; \; \; \forall x \in \algg
\label{eq:vertextransformation}
\end{equation}
Such a vertex operator is called a vertex operator of weight
$\lambda_1$.  By the Wigner-Eckart theorem, a matrix element of a
vertex operator of this form is necessarily proportional to the
corresponding Clebsch-Gordan coefficient.

Since we are concerned with irreducible representations which arise
from dual Verma module resolutions, the next problem which arises is
that of characterizing a vertex operator in terms of resolutions of
irreducible representations through chain complexes.
Generally, if the  irreducible representations $I_{\lambda_2}$ and
$I_{\lambda_2}$ are given through resolutions $(M_i,d_i)$ and
$(M'_i,d'_i)$, a bosonic realization of the vertex operator ${\cal
V}_{u}$ is a set of operators
\begin{equation}
{\cal V}_{u}^i:M_i \rightarrow M'_i
\end{equation}
which satisfy  (\ref{eq:vertextransformation}) and
\begin{equation}
d'_i \cdot {\cal V}_{u}^i = {\cal V}_{u}^{i + 1} \cdot d_i,
\label{eq:chainmapping}
\end{equation}
and where ${\cal V}_{u}^0$ agrees with ${\cal V}_{u}$ on the
cohomology of the complex.
The constraint (\ref{eq:chainmapping}) is
the statement that ${\cal V}_{u}^i$ is a map of chain complexes for
every $u \in I_{\lambda_1}$, and is
equivalent to the condition that the following diagram be commutative
\begin{center}
\begin{picture}(200,100)(- 100,- 45)
\multiput(- 90, 30)(50, 0){4}{\vector(1,0){30}}
\multiput(- 90, -30)(50, 0){4}{\vector(1,0){30}}
\multiput(- 50, 20)(50, 0){3}{\vector( 0, - 1){40}}
\put(- 50, 30){\makebox(0,0){$M_{-1}$}}
\put(0, 30){\makebox(0,0){$M_{0}$}}
\put( 50, 30){\makebox(0,0){$M_{1}$}}
\put(- 50, -30){\makebox(0,0){$M'_{-1}$}}
\put(0, -30){\makebox(0,0){$M'_{0}$}}
\put( 50, -30){\makebox(0,0){$M'_{1}$}}
\put(25, 40){\makebox(0,0){$d_0$}}
\put( -25, 40){\makebox(0,0){$d_{-1}$}}
\put(75, 40){\makebox(0,0){$d_1$}}
\put( -75, 40){\makebox(0,0){$d_{- 2}$}}
\put(25, -40){\makebox(0,0){$d'_0$}}
\put( -25, -40){\makebox(0,0){$d'_{-1}$}}
\put(75, -40){\makebox(0,0){$d'_1$}}
\put( -75, -40){\makebox(0,0){$d'_{- 2}$}}
\put(- 40, 0){\makebox(0,0){${\cal V}_{u}^{-1}$}}
\put( 10, 0){\makebox(0,0){${\cal V}_{u}^{0}$}}
\put(60, 0){\makebox(0,0){${\cal V}_{u}^{1}$}}
\put(- 110, -30){\makebox(0,0){$\cdots$}}
\put(- 110, 30){\makebox(0,0){$\cdots$}}
\put(110, -30){\makebox(0,0){$\cdots$}}
\put(110, 30){\makebox(0,0){$\cdots$}}
\end{picture}
\end{center}
A chain map ${\cal V}_{}^i$ which induces a trivial map in
cohomology is called a trivial chain map.  Some straightforward
diagram chasing suffices to verify that any chain map induces a
well-defined map in the cohomology, and that every trivial chain map
can be written in the form
\begin{equation}
{\cal V}_{}^i =d'_{i - 1} t^i + t^{i + 1} d_{i}
\end{equation}
for some set of maps
\begin{equation}
t^i:M_i \rightarrow M'_{i - 1}.
\label{eq:fcondition}
\end{equation}
It follows that every two bosonic realizations ${\cal V}_{u}^i$ and
$\bar{{\cal V}}_{u}^i$ of the same vertex operator ${\cal V}_{u}$
differ by operators of the form
\begin{equation}
{\cal V}_{u}^i- \bar{{\cal V}}_{u}^i
= d'_{i - 1} t_u^i + t_u^{i + 1} d_{i},
\end{equation}
where for each $u \in I_{\lambda_1}$ the operators $t_u^i$ satisfy
(\ref{eq:fcondition}), and furthermore we have
\begin{equation}
[x, t_{u}^i] = t_{xu}^i \; \; \; \forall x \in \algg.
\end{equation}
Since we are primarily interested in bosonic realizations of vertex
operators in terms of chain maps, we will generally not distinguish in
notation between vertex operators and their realizations.  Vertex
operators which are realized as chain maps are frequently referred to
as ``screened vertex operators''.

As mentioned above, the matrix elements of vertex operators of a
finite-dimensional group $G$ are proportional to Clebsch-Gordan
coefficients.  It is fairly straightforward to prove that a nontrivial
vertex operator  of weight $I_{\lambda_1}$ between representations
$I_{\lambda_2}$ and $I_{\lambda_3}$ exists precisely
when the tensor product rules of $G$ are satisfied, so that the
irreducible representation $I_{\lambda_3}$ occurs in the tensor
product representation
$I_{\lambda_1}\otimes I_{\lambda_2}$ \cite{bmp}.

For finite-dimensional groups, it is possible to explicitly construct
bosonic realizations of vertex operators as chain maps of dual Verma module
resolutions using the screening operators defined in Section
\ref{sec:intertwiners}.  In fact, since $I_{\lambda_1}$ is a highest
weight representation of $G$ with a highest weight state $v$, we have
\begin{equation}
[x, {\cal V}^i_{v}] = 0, \; \; \; \forall x \in \algg_+.
\end{equation}
It follows that ${\cal V}^i_{v}$, the component of the vertex operator
associated with the highest weight state, can be written as a
polynomial in the screening operators for every $i$, along with an
appropriate change in weight by $\lambda_1$.  For ${\cal V}_{v}^0$,
the degree $\mu$ of the
polynomial in the screening operators must be such that $\lambda_2 +
\lambda_1 -\mu = \lambda_3$; the degree of the polynomial for each
value of $i$ can similarly be calculated.  Thus, for example, we must
have a description of ${\cal V}_{v}^0$ as a polynomial in screening
operators $\tilde{s}_\alpha$ of degree
\begin{equation}
\mu = \lambda_1 + \lambda_2-\lambda_3.
\end{equation}
Note that we must have $\lambda_3 \prec \lambda_1 + \lambda_2$ for
such an operator to exist, in agreement with the Wigner-Eckart
theorem.
The remaining components of the vertex operator can be calculated
using the relations (\ref{eq:vertextransformation}).

As a simple example of a screened vertex operator for a
finite-dimensional group $G$, we now  give an explicit bosonic realization
of an $SU(2)$ vertex operator of weight $j$  acting on the
irreducible representation with weight $k$ and taking values in the
irreducible representation with weight $k'$.
As described above, the irreducible representation of $SU(2)$ of
weight $k$ has a resolution in terms of dual Verma modules by the
chain complex
\begin{equation}
0 \rightarrow V_k^*\stackrel{\tilde{s}^{2k + 1}}{\longrightarrow}
V_{- k - 1}^*\rightarrow 0
\end{equation}
The highest weight component ${\cal V}_{j}$ of the
screened vertex operator in question is given by two operators
${\cal V}_{j}^0:V_k^*\rightarrow V_{k'}^*$ and
${\cal V}_{j}^1:V_{- k - 1}^*\rightarrow V_{-k' - 1}^*$, as in the
following diagram
\begin{center}
\begin{picture}(200,100)(- 120,- 45)
\put(- 90, 30){\vector(1,0){30}}
\put(- 35, 30){\vector(1,0){30}}
\put( 25, 30){\vector(1,0){25}}
\put(- 90, -30){\vector(1,0){30}}
\put(- 35, -30){\vector(1,0){30}}
\put( 25, -30){\vector(1,0){25}}
\put(- 50, 20){\vector( 0, -1){40}}
\put( 10, 20){\vector( 0, -1){40}}
\put(- 50, 30){\makebox(0,0){$V_{k}^*$}}
\put(10, 30){\makebox(0,0){$V_{-k - 1}^*$}}
\put( 60, 30){\makebox(0,0){$0$}}
\put( - 100, 30){\makebox(0,0){$0$}}
\put(- 50, -30){\makebox(0,0){$V_{k'}^*$}}
\put(10, -30){\makebox(0,0){$V_{-k' - 1}^*$}}
\put( 60,- 30){\makebox(0,0){$0$}}
\put( - 100,- 30){\makebox(0,0){$0$}}
\put( -20, 40){\makebox(0,0){$\tilde{s}^{2k + 1}$}}
\put( -20, -40){\makebox(0,0){$\tilde{s}^{2k' + 1}$}}
\put(- 40, 0){\makebox(0,0){${\cal V}_{j}^{0}$}}
\put( 20, 0){\makebox(0,0){${\cal V}_{j}^{1}$}}
\end{picture}
\end{center}
{}From the above argument,   ${\cal V}_{j}^0$ and ${\cal V}_{j}^1$ must
be proportional to $\tilde{s}^{j + k - k'}$ and $\tilde{s}^{j + k' -
k}$ respectively.  In order to make the above diagram commute, we must
choose the same proportionality constant for both operators.  Note
that the form of this vertex operator depends explicitly upon the
weights $k$ and $k'$ of the initial and final modules.  This
is a general property of vertex operators defined for a particular
Fock space  realization.

\subsection{Virasoro vertex operators}
\label{sec:generalvertex}

The vertex operators associated with the Virasoro algebra are defined
in a similar fashion to the finite-dimensional vertex operators
defined in the previous section.  The main distinction is that the
representation which defines the weight of the vertex operator is not
a highest weight representation.  Within a single conformal field
theory, the central charge $c$ does not vary between irreducible
representations of the Virasoro algebra, so we shall keep $c$ fixed to
a constant value, and use the single parameter $h$ for the highest
weight of a Virasoro module throughout the rest of this chapter.
Given two irreducible highest weight representations of the Virasoro
algebra $I_{k},I_{k'}$, with highest weights $k$ and $k'$
respectively, a vertex operator of weight $h$ is defined to be an
operator-valued function of a single complex variable $z$,
\begin{equation}
\Phi(z):I_{k} \rightarrow I_{k'},
\end{equation}
which satisfies the relations
\begin{equation}
[L_n, \Phi (z)] = \left(z^{n + 1} \frac{\partial}{\partial z}
+ (n + 1) hz^n \right) \Phi (z).
\end{equation}
In the physics literature, a vertex operator is often defined to be an
operator of this type which has a nontrivial action on all irreducible
representations occurring in a given theory.  We will adhere to the
convention that a vertex operator is associated with a fixed set of
three weights $h,k,$ and $k'$, following \cite{bmp}.

To better study the structure of vertex operators, it is convenient to
expand a fixed vertex operator $\Phi (z)$ in a power series expansion
in $z$,
\begin{equation}
\Phi (z) = z^x \sum_{n \in \ZZ} \phi_{- h - x + n} z^{- n},
\label{eq:vertexexpansion}
\end{equation}
where $x = k' - h - k$.
Note that the vertex operator is not single valued unless $x \in \ZZ$.
The modes $\phi_{m}$ of the vertex operator satisfy the commutation
relations with the Virasoro algebra
\begin{equation}
[L_n, \phi_m] = (nh - n - m) \phi_{n + m}.
\label{eq:vertexcommutator}
\end{equation}
Thus, the modes of the vertex operator carry a fairly simple
representation of the Virasoro algebra.  This representation, however,
is not highest weight; the weight of the mode $\phi_{- h - x + n}$ is
precisely $h + x - n$ which is unbounded in both directions as $n \in
\ZZ$.

Just as we did for finite-dimensional vertex operators, we can define
a bosonic realization of a vertex operator as a chain map between
resolutions of the irreducible representations $I_k,I_{k'}$ in terms
of Fock modules, where the chain map satisfies the correct commutation
relations with the Virasoro generators.  A vertex operator is again
defined to be trivial when its action on the cohomology is trivial,
and two vertex operator realizations which differ by a trivial chain
map are considered to be equivalent.  Again, equivalent vertex
operator realizations $\phi_n^i, \bar{\phi}_n^i$ must differ by a
chain map of the form $d'_{i - 1} t_n^i + t_n^{i + 1} d_{i}$, where
$[L_n, t_m^i] = (nh - n - m)t_{n + m}^i$.

We are now interested in answering the question of which values of
$k,k',$ and $h$ corresponding to unitary representations of the
Virasoro group admit a vertex operator of weight $h$ from $I_k$ to
$I_{k'}$.  Moreover, we would like to know whether such a vertex
operator, when it exists as a map between irreducible representation
spaces, can be realized by a chain map between resolutions.  Because
the representation of the vertex operator itself is not highest
weight, we cannot use the methods of the previous section.  That
is, there does not exist a component of the vertex operator which
commutes with all raising operators $L_n$, $n > 0$, so we cannot
describe the vertex operators of the Virasoro algebra in terms of the
screening operators $\tilde{s}_n$.  This complication makes the
explicit description of screened vertex operators rather difficult.
Nonetheless, for those vertex operators which do exist and have
realizations as chain maps, we can give a set of equations which
uniquely characterize the action of the vertex operator on the
cohomology of the resolutions.

Unfortunately, we do not have a closed form expression for the
realizations of the existing vertex operators on the coadjoint orbit
dual Verma modules, nor do we have a proof of which vertex operators
have realizations as chain maps.  In the case where the resolutions
are trivial and the dual Verma modules $V^*_k$ and $V^*_{k'}$ contain
no cosingular vectors, we can prove the existence of a vertex operator
of arbitrary weight $h$.  Because the resolutions of these modules are
trivial, it follows immediately that these vertex operators have
realizations as chain maps; we can characterize these vertex operators
uniquely by calculating their matrix elements.  When the resolutions
are not trivial, the existence of cosingular vectors in $V^*_k$ and
singular vectors in $V_{k'}$ give a set of algebraic constraints on
the values of $h, k,k'$ for which vertex operators can exist.  In
physical language, these are constraints on the fusion algebra of the
associated CFT.  Whenever this set of constraints is satisfied, there
exists a vertex operator between the appropriate irreducible Virasoro
representations.  We believe that every such vertex operator has a
realization as a chain map between dual Verma module resolutions,
although we do not have a rigorous proof of this assertion.  We can,
however, show that if such a vertex operator exists, it is unique up
to a scaling factor and the addition of cohomologically trivial vertex
operators.  Furthermore, we can calculate the action of such a vertex
operator on the cohomology of the complex, which uniquely determines
the action of the vertex operator on the irreducible representation
and provides the essential information needed to compute (genus 0)
correlation functions in a conformal field theory.  We will now
proceed to demonstrate the existence of vertex operators between dual
Verma modules with no cosingular vectors, and to describe the
constraints on the vertex operators between nontrivial resolutions of
dual Verma modules.

In order to have a sensible algebraic description of Virasoro vertex
operators in terms of matrix elements, we find it useful to discuss
briefly several simple points about operators on bosonic Fock spaces.
Given a Heisenberg algebra on a set $S =\{\alpha \in \Phi_+
\subset\Lambda\}$ acting on a Fock space $F$ with a
$\Lambda$-grading, where $\Phi_+$ is the space of positive roots and
$\Lambda$ is the root lattice for a Lie algebra $\algg$, an operator
\begin{equation}
O:F \rightarrow F
\end{equation}
is uniquely defined by the matrix elements
\begin{equation}
\langle w | O | v \rangle, \; \; \; |v \rangle \in F,
\langle w | \in F^*.
\label{eq:matrixelements}
\end{equation}
The number of elements $v \in F$ with weight $\lambda$ is given by the
$\Lambda$-partition function $\pi_{\Lambda} (\lambda)$, which is the
number of distinct ways in which $\lambda$ can be constructed as a sum
of elements in $\Phi_+$.  In the particular case at hand, the Virasoro
group, the Fock space is the space $R =\CC[\{z_n:n > 0\}]$ with a
$\ZZ$-grading.  The dimension of the space $R_n$ of degree $n$ is just
the integer partition function  $\pi (n)$.  The canonical basis for
$R$ is the space of monomials in the variables $z_n$.  An
operator $O:R \rightarrow R$  can be described either by giving the
complete set of matrix elements (\ref{eq:matrixelements}), or by
writing the operator in the form of a differential operator
\begin{equation}
O = \sum_{n,m} \sum_{\pi_n, \pi_m}  \omega (\pi_n, \pi_m)
(\prod_{i} z_{n_i})  (\prod_{j} \derivative{z_{m_j}}),
\end{equation}
where $\pi_n$ and $\pi_m$ are summed over all partitions $\{n_i\}$ and
$\{m_j\}$ with $n_1 + \cdots + n_a = n$ and $m_1 + \cdots + m_b = m$.
The description of an operator by a set of coefficients $\omega$ is
equivalent to giving the complete set of matrix elements for  that
operator.  However, these two ways of writing the same operator are
related in a rather nontrivial fashion.  For instance, the operator
$P$ on the bosonic Fock space $\CC[z]$ which projects onto the highest
weight state $\nul= 1$ has all matrix elements equal to 0 except for
the element $\langle \: | P | \:
\rangle = 1$.  This operator has the differential operator
form
\begin{equation}
P = \; :  \exp (- z \frac{\partial}{\partial z} ):\; = \sum_{i = 0}^{\infty}
\frac{(-1)^i}{i!}   z^i \frac{\partial^i}{\partial z^i} ,
\end{equation}
where by the dots around the second expression we indicate the normal
ordering procedure, by which all annihilation operators
$\partial/\partial z$ are moved to the right of creation operators
$z$.
Up to this point in this thesis, we have primarily described operators
in terms of their description as differential operators; in the
following, we will find it more useful to describe screened vertex
operators in terms of their matrix elements.

The Virasoro vertex operators which we wish to construct are chain
maps $\Phi (z)$ between resolutions $(M_i,d_i)$ and $(M'_i,d'_i)$ of
the irreducible representation spaces $I_k$ and $I_{k'}$ of the form
\begin{center}
\begin{picture}(200,100)(- 100,- 45)
\multiput(- 90, 30)(50, 0){4}{\vector(1,0){30}}
\multiput(- 90, -30)(50, 0){4}{\vector(1,0){30}}
\multiput(- 50, 20)(50, 0){3}{\vector( 0, - 1){40}}
\put(- 100, 30){\makebox(0,0){$0$}}
\put(- 100, -30){\makebox(0,0){$0$}}
\put(- 50, 30){\makebox(0,0){$V_k^*$}}
\put(0, 30){\makebox(0,0){$M_{1}$}}
\put( 50, 30){\makebox(0,0){$M_{2}$}}
\put(- 50, -30){\makebox(0,0){$V_{k'}^*$}}
\put(0, -30){\makebox(0,0){$M'_{1}$}}
\put( 50, -30){\makebox(0,0){$M'_{2}$}}
\put(25, 40){\makebox(0,0){$d_{1}$}}
\put( -25, 40){\makebox(0,0){$d_{0}$}}
\put(75, 40){\makebox(0,0){$d_{2}$}}
\put(25, -40){\makebox(0,0){$d'_{1}$}}
\put( -25, -40){\makebox(0,0){$d'_{0}$}}
\put(75, -40){\makebox(0,0){$d'_{2}$}}
\put(- 30, 0){\makebox(0,0){$\Phi^{0}(z)$}}
\put( 20, 0){\makebox(0,0){$\Phi^{1}(z)$}}
\put(70, 0){\makebox(0,0){$\Phi^{2}(z)$}}
\put(110, -30){\makebox(0,0){$\cdots$}}
\put(110, 30){\makebox(0,0){$\cdots$}}
\end{picture}
\end{center}
We will find it useful to define a particular set of states in $V^*_k$
and $V^{**}_{k'} = V_{k'}$.  Associated with each partition $\pi_n =\{n_i |
\sum n_i = n \}$, we define the operators
\begin{eqnarray}
L_{\pi_n} & = & \prod_{i} L_{n_i}\\
L_{- \pi_m} & = & \prod_{i} L_{- m_i}, \nonumber
\end{eqnarray}
and the associated states
\begin{eqnarray}
| \pi_m \rangle & = & L_{- \pi_m}  \nul \in  V^*_k
\label{eq:states}\\
\langle \pi_n | & = &  \nuls L_{\pi_n} \in V_{k'}. \nonumber
\end{eqnarray}
Because $V_{k'}$ is always a Verma module, containing no cosingular
vectors, the set of states $\langle \pi_n |$ are always a linearly
independent basis for $V_{k'}$.

We now consider the simplest case of a Virasoro vertex operator, where
the dual Verma modules $V^*_k$ and $V^*_{k'}$ contain no cosingular
vectors and are therefore isomorphic to the irreducible
representations $I_k$ and $I_{k'}$.  In this case, the corresponding
resolutions are all trivial, with $M_1 = M'_{1} = 0$.  Because there
are no singular or cosingular vectors in the modules $M_0,M'_0$, the
states defined through (\ref{eq:states}) form complete bases for these
Fock modules.  As discussed above, to characterize the vertex operator
$\Phi (z)$ it will suffice to give all the matrix elements with
respect to these bases.  By counting weights, we see that the only
mode of the vertex operator which can contribute to a matrix element
between states $\langle \pi_n |$ and $|
\pi_m \rangle$ is $\phi_{- h - x + m - n}$, where $x = k' - h - k$, so
that
\begin{equation}
\langle \pi_n | \Phi (z) |\pi_m \rangle = z^{x + n - m}
\langle \pi_n | \phi_{- h - x + m - n} |\pi_m \rangle.
\label{eq:generalvertexelement}
\end{equation}
The only nontrivial matrix element between the highest weight states
is then
\begin{equation}
\nuls \phi_{- h - x} \nul.
\label{eq:canonicale}
\end{equation}
We will choose the normalization of the vertex operator to be such
that this matrix element is equal to 1.
Just as the Virasoro commutation relations are used to construct the
Shapovalov form on the Verma module, the Virasoro relations coupled
with (\ref{eq:vertexcommutator}) can be used to relate an arbitrary
matrix element (\ref{eq:generalvertexelement}) to the canonical
element (\ref{eq:canonicale}).  That this procedure defines each
matrix element in a well-defined fashion, and that the resulting
operator satisfies (\ref{eq:vertexcommutator}) follow from the
condition that (\ref{eq:vertexcommutator}) defines a representation of
the Virasoro algebra, which is equivalent to the Jacobi-type equation
\begin{eqnarray}
\lefteqn{[L_n,[L_m, \phi_p]]
+[L_m,[\phi_p, L_n]]+[ \phi_p,[L_n,L_m]]}\nonumber \\
& = & (m (h - 1) - p) (n (h - 1) - m - p)
- (n (h - 1) - p) (m (h - 1) - n - p)  \nonumber\\
& & \; \; \; \; \; \; \;+ (m - n) ((n + m) (h - 1) -  p)
\\
 & = &  0 \nonumber.
\end{eqnarray}
Thus, we have shown that when the dual Verma modules $V^*_k$ and
$V^*_{k'}$ contain no cosingular vectors, there exists a unique vertex
operator $\Phi (z)$ of any weight $h$.  In addition,
the matrix elements (\ref{eq:generalvertexelement}) completely
characterize $\Phi (z)$.   The resulting realization of $\Phi (z)$ as
a differential operator on a coadjoint orbit representation can be
explicitly calculated by choosing a monomial basis for $R$ and
relating this basis through a nonsingular linear transformation to the
basis $| \pi_n \rangle$.

We now consider the case where the dual Verma modules $V^*_k$ and
$V^*_{k'}$ contain cosingular vectors and the resolutions $(M_i,d_i)$
and $(M'_i,d'_i)$ are nontrivial.  As the most general case, we assume
that both weights $k$ and $k'$ are associated with representations whose
Verma modules have the singular vector structure graphed in
Figure~\ref{f:graphVirasoro}b, and we consider the resulting dual
Verma module resolutions.  For any of the unitary representations with
a simpler singular vector structure the resulting resolution is
equivalent to taking this resolution and simplifying by setting to 0
the modules which correspond to singular vectors which do not exist
in the simpler representation.

A choice of signs on the intertwining operators which allows us to
construct the dual Verma module resolutions is as follows
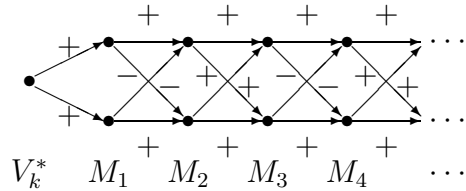
\begin{center}
\begin{picture}(200,100)(- 100,- 5)
\put(-75, 50){\circle*{4}}
\multiput(-45, 65)(30,0){4}{\circle*{4}}
\multiput(-45, 35)(30,0){4}{\circle*{4}}
\put(-73, 52){\vector(2,1){26}}
\put(-73,  48){\vector(2,-1){26}}
\multiput(-43, 65)(30,0){4}{\vector(1,0){26}}
\multiput(-43, 35)(30,0){4}{\vector(1,0){26}}
\multiput(-43, 63)(30,0){4}{\vector(1,-1){26}}
\multiput(-43, 37)(30,0){4}{\vector(1,1){26}}
\put( 83, 65){\makebox(0,0){$ \cdots $}}
\put( 83, 35){\makebox(0,0){$ \cdots $}}
\put( 83, 15){\makebox(0,0){$ \cdots $}}
\put(- 75, 15){\makebox(0,0){$V_k^*$}}
\put(- 45, 15){\makebox(0,0){$M_1$}}
\put(- 15, 15){\makebox(0,0){$M_2$}}
\put(15, 15){\makebox(0,0){$M_3$}}
\put(45, 15){\makebox(0,0){$M_4$}}
\multiput(- 30, 75)(30, 0){4}{\makebox(0,0){$+$}}
\put(- 60, 63){\makebox(0,0){$+$}}
\put(- 60, 37){\makebox(0,0){$+$}}
\multiput(- 30, 25)(30, 0){4}{\makebox(0,0){$+$}}
\multiput(- 38, 52)(60,0){2}{\makebox(0,0){$-$}}
\multiput(- 8, 52)(60,0){2}{\makebox(0,0){$+$}}
\multiput(- 22, 48)(60, 0){2}{\makebox(0,0){$-$}}
\multiput(8, 48)(60, 0){2}{\makebox(0,0){$+$}}
\end{picture}
\end{center}
In this resolution, each module $M_i,i > 0$ contains a direct sum of
two dual Verma modules $V^*_{h (v_i)},V_{h (w_i)}^*$ associated with
cosingular vectors $v_i,w_i$ in $V_k^*$ with weights $h (v_i),h
(w_i)$.  The arrows are (surjective) intertwining operators with the
given signs.

{}From the discussion in Section \ref{sec:intertwiners} on the
intertwining operators between dual Verma modules, we know that the
kernels of the operators $d_0$ and $d'_0$ are precisely the orbits
${\cal U} (\algg) \nul$ of the highest weight vectors in the modules
$V^*_k$, $V^*_{k'}$.  For every cosingular vector $| v \rangle$ in
$V^*_{k'}$ at level $n$, there exists a singular vector $\langle s |$
in $V_{k'}$, which can be written in the form
\begin{equation}
\langle s | = \sum_{\pi_n}c_{\pi_n} \langle \pi_n |,
\end{equation}
where $c_{\pi_n}$ are real coefficients.  Such a singular vector is
annihilated by all lowering operators,
\begin{equation}
\langle s | L_{- n} = 0 \; \; \; \; \forall n > 0.
\end{equation}
It follows that
\begin{equation}
\langle s | \pi_n \rangle = 0 \; \; \; \; \forall |\pi_n \rangle
\in V^*_{k'}, n > 0,
\end{equation}
and therefore that for any vector $| w \rangle \in V^*_{k'}$
satisfying $\langle s | w \rangle \neq 0$, we have
\begin{equation}
d'_0  | w \rangle \neq 0.
\end{equation}
A consequence of this condition is that the matrix element
\begin{equation}
\langle s | \Phi^0 (z) | \pi_n \rangle
\end{equation}
must vanish for any state $| \pi_n \rangle$ in the orbit of the
highest weight state in $V^*_k$,
since
\begin{equation}
d'_0 \Phi^0 (z) | \pi_n \rangle = \Phi^1 (z) d_0 | \pi_n \rangle = 0.
\end{equation}
This places a set of algebraic
conditions on the weights $h,k,k'$ for which vertex operators are
allowed to exist.  From the singular vector structure of the Verma
module, we know that there are (at most) two singular vectors $s,t$,
corresponding to the cosingular vectors $v'_1,w'_1$ in $V^*_{k'}$ with
the property that the modules generated by $s$ and $t$ contain all
other singular vectors in $V_{k'}$.  Thus, the set of all algebraic
conditions associated with singular vectors in $V_{k'}$ can be
completely described by the condition that matrix elements containing
$s$ or $t$ must vanish,
\begin{eqnarray}
\langle s | \Phi (z) | \pi_n \rangle & = & 0, \label{eq:conditions1}\\
\langle t | \Phi (z) | \pi_n \rangle & = &  0\; \; \; \forall
|\pi_n \rangle\in V^*_k,
n > 0. \nonumber
\end{eqnarray}

A further condition on the matrix elements of $\Phi (z)$ arises from
the fact that at every level $n$ at which there is a cosingular vector
in $V^*_k$, there is a linear relation between the states $| \pi_n
\rangle$.  Each such linear relation gives a constraint to the matrix
elements of $\Phi$
of the form
\begin{equation}
\sum_{\pi_n}c_{\pi_n} \langle \pi_m | \Phi (z) | \pi_n \rangle = 0.
\label{eq:conditions2}
\end{equation}
Again, because of the cosingular vector structure, this infinite set
of constraints can be completely described by the two families of
constraints arising from the cosingular vectors $v_1,w_1$.

Thus, we have derived a set of algebraic constraints on the weights
$h,k,k'$ which must be satisfied for a vertex operator to exist as a
map of dual Verma module resolutions.  Although we have derived these
constraints from the point of view of chain complexes, the constraints
(\ref{eq:conditions1}) and (\ref{eq:conditions2})
also hold for the simpler construction of a vertex operator of weight
$h$ between irreducible representations with highest weights $k,k'$.
In the case of irreducible representations, the dual elements $\langle
s |$ corresponding to singular vectors must vanish.  Similarly, the
linear conditions on vectors $| \pi_n \rangle$ associated with
cosingular vectors in the dual Verma module still hold in the
irreducible representation.

Just as the set of all matrix elements defines a vertex operator
uniquely in the case where the resolutions are trivial, the
determination of all matrix elements of the form
\begin{equation}
\langle \pi_n | \Phi (z) | \pi_m \rangle
\end{equation}
determines the action of the vertex operator on the cohomology
uniquely, and thus defines a vertex operator on the associated
irreducible representation (the uniqueness of this representation
essentially follows from the fact that the Shapovalov form is
positive-definite on an irreducible unitary representation).  Since
the constraints (\ref{eq:conditions1}) and (\ref{eq:conditions2}) are
the only extra conditions placed on a vertex operator between
irreducible representations, we have essentially proven the existence
of such vertex operators whenever these conditions are satisfied.  As
mentioned above, we do not have a proof that in all cases these vertex
operators can be realized as chain maps, but we believe this to be the
case.

We will now discuss some special cases of the constraints
(\ref{eq:conditions1}) and (\ref{eq:conditions2}), and describe the
resulting conditions on the fusion algebras of the related conformal
field theories.  For the purposes of this discussion, we shall find it
useful to have explicit calculations of various matrix elements of a
vertex operator $\Phi (z) = z^x \sum \phi_{- h - x + n}
z^{- n}$ taking a representation with weight $k$ to one with weight
$k' = k + h + x$.
\begin{eqnarray}
\nuls \phi_{- h - x}  \nul & = & 1\label{eq:explicit1}\\
\nuls \phi_{- h - x + 1} L_{-1} \nul & = & - x\label{eq:explicit2}\\
\nuls \phi_{- h - x + 2} L_{-2} \nul & = & h -
x\label{eq:explicit3}\\
\nuls \phi_{- h - x + 2} L_{-1}^2 \nul & = & x (x -
1)\label{eq:explicit4}\\
\nuls \phi_{- h - x + 3} L_{-3} \nul & = & - 2h -
x\label{eq:explicit5}\\
\nuls \phi_{- h - x + 3} L_{-2} L_{-1} \nul & = & x (h + x -
1)\label{eq:explicit6}\\
\nuls \phi_{- h - x + 3} L_{-1}^3 \nul & = &  - x (x - 1) (x - 2)
\label{eq:explicit7}\\
\nuls L_1\phi_{- h - x - 1} \nul & = & 2h +x\label{eq:explicit9}\\
\nuls L_2\phi_{- h - x - 2} \nul & = & 3h + x\label{eq:explicit10}\\
\nuls L_1^2\phi_{- h - x - 2} \nul & = & (2h + x)(2h + x+ 1)
\label{eq:explicit11}\\
\nuls L_3\phi_{- h - x - 3} \nul & = & 4h + x\label{eq:explicit12}\\
\nuls L_1 L_2\phi_{- h - x - 3} \nul & = & (3h + x + 1) (2h +
x)\label{eq:explicit13}\\
\nuls L_1^3 \phi_{- h - x - 3} \nul & = &  (2h + x) (2h + x + 1) (2h +
x + \label{eq:explicit14}2)
\end{eqnarray}

First, we consider the case where $c > 1$.  For any values of $h,k$
and $k'$, where $k,k' \neq 0$, all of the representations have trivial
resolutions, and a vertex operator can be constructed through the
above reasoning.  It is interesting to note that although vertex
operators between modules of this form are well-known in terms of
multiple bosonic fields (we will discuss this further in the following
section), we can use the above results to construct any vertex
operator of this type in terms of a single bosonic field with modes
$a_n,n \in \ZZ$.  The representation of such a vertex operator is
rather complicated algebraically, however, and does not appear to have
as simple an expression as when multiple bosonic fields are used.

Now, let us consider the case where either $k = 0$ or $k' = 0$,
indicating a vertex operator which either acts on or maps to the
vacuum module.  The dual Verma module with highest weight $h = 0$ has
precisely one cosingular vector, at level 1.  Thus, if $k' = 0$, the
constraint that (\ref{eq:explicit9}) is 0 indicates that $x =- 2h$, so
$h = k$.  Similarly, if $k = 0$, setting (\ref{eq:explicit2}) to 0
gives $k' = h$.  Thus, we have the result that a vertex operator of
weight $h$ acting on the vacuum in either direction gives the module
of weight $h$.  This is precisely what we would expect from standard
physical theories.

Finally, we consider a slightly more complicated example,  that of a
rational conformal field theory with $c = 1/2$.  From the equations
(\ref{eq:min1}), (\ref{eq:min1}) describing the discrete series of
Virasoro representations, one finds that the only unitary
representations corresponding to this choice of central charge  are
those with highest weights $h =0,h =1/16,$ and $h = 1/2$.  We know
that there is a cosingular state in the module $V^*_{1/2}$ at level 2.
The linear combination of states in this module which vanishes as a
result  can be calculated from (\ref{eq:Virasorosingular}) to be
\begin{equation}
(L_{- 2} - \frac{3}{4} L_{-1}^2) \nul = 0.
\end{equation}
If we take $k = 1/2$, this equation leads to the constraint on $h$ and
$k'$,
\begin{equation}
3 x^2+ x - 4h = 0.
\end{equation}
The solutions for this where $k'$ and $h$ both correspond to unitary
representations are $(h,k') =(0,1/2)$, $(h,k') =(1/2,0)$, and $(h,k')
=(1/16,1/16)$.    We can perform a similar analysis using the initial
module with $k = 1/16$.  The consequent vanishing condition in the
dual Verma module is
\begin{equation}
(L_{- 2} - \frac{4}{3} L_{-1}^2) \nul = 0.
\end{equation}
The pairs $h,k'$ which satisfy this vanishing condition are $(h,k')
=(0,1/16)$, $(h,k')=(1/16,0)$, $(h,k') =(1/16,1/2)$, and $(h,k')
=(1/2,1/16)$.  We could
go on and consider the constraints imposed by the singular vectors in
the Verma module $V_{k'}$ at level 2.  In order to ensure that all
constraints are satisfied, it is also necessary to include the
constraints arising from the second cosingular vector in the dual
Verma module, which in this case appears at level 3, and also the
constraints from the corresponding singular vectors at level 3 in the
Verma module.  However, none of these other constraints places any
further restriction upon the vertex operators which are allowed in a
$c = 1/2$ conformal field theory.  We have thus calculated the
restrictions on the fusion algebra of this CFT.  Traditionally, a CFT
fusion algebra is written in the form
\begin{equation}
[h][k] = \sum_{k'}[k'],
\end{equation}
where $k'$ is summed over all weights for which there exists a
nontrivial vertex operator of weight $h$ connecting the irreducible
representations with weights $k,k'$.  In this notation, our results
for the $c = 1/2$ fusion rules are
\begin{eqnarray}
\left[1/16\right]\left[1/16\right] & = & \left[0\right]
+\left[1/2\right] \nonumber\\
\left[1/16\right]\left[1/2\right]  =
\left[1/2\right]\left[1/16\right] & = &\left[1/16\right]\\
\left[1/16\right]\left[1/16\right] & = & \left[0\right].\nonumber
\end{eqnarray}
(We do not write the fusion rules for the vacuum module [0]
explicitly, since these fusion rules are standard for all CFT's.)
This is precisely the fusion algebra for the set of operators in the
$c = 1/2$ CFT which corresponds to the Ising model \cite{bpz}.

We will now say a few words about the relationship between the
analysis in this section and the differential equations approach to
CFT.  In their initial work on conformal field theory
\cite{bpz}, BPZ defined vertex operators primarily as objects which appear
inside correlation functions, and not in terms of an explicit action
on a representation space.  From this point of view, they analyzed the
operator product expansion (OPE) of a product of two vertex operators,
represented as a sum over other vertex operators and so-called
``descendant fields'' which are defined by applying Virasoro operators
to the primary fields (vertex operators).  They showed that the
resulting coefficient for every descendant field in a given conformal
family (descending from a particular primary field) could be written
in terms of the coefficient of the primary field, multiplied by a
constant depending only upon the conformal weights of the three
primary fields involved.  This set of equations is precisely analogous
to the determination of the matrix elements
(\ref{eq:generalvertexelement}).  This equivalence can be made precise
by observing that these matrix elements are essentially components of
a correlation function of 3 vertex operators of the form
(\ref{eq:correlationfunction}).  In their work, BPZ also derive a set
of constraint equations on the fusion algebra analogous to
(\ref{eq:conditions1}).  They write these equations in terms of a set
of differential operators which must be satisfied by certain
correlation functions.  Although these constraints are effectively
equivalent to those imposed here, from the field theory formalism it
is somewhat less clear that these constraints can all be consistently
imposed, and what their effect is on the structure of the relevant
Virasoro representations.

Finally, we conclude this section with a simple example of a vertex
operator whose action can be computed purely in terms of the Virasoro
generators.
A standard result in conformal field theory is that any vertex
operator $\Phi (z)$ satisfies the equation
\begin{equation}
\Phi (z) ={\rm e}^{z L_{-1}} \Phi (0) {\rm e}^{- z L_{-1}}.
\label{eq:simplevertex}
\end{equation}
This follows immediately from the differential equation
\begin{equation}
[L_{-1}, \Phi (z)] = \frac{\partial}{\partial z}  \Phi (z).
\end{equation}
A result of (\ref{eq:simplevertex}) is that the action of this vertex
operator on the physical vacuum is described completely in terms of
the Virasoro generator $L_{-1}$ by
\begin{equation}
\Phi (z) | \Omega \rangle = {\rm e}^{z L_{-1}} \nul = \sum_{i = 0}^{
\infty}  \frac{z^i L_{-1}^i}{i!} \nul,
\label{eq:simpleaction}
\end{equation}
where $\nul = \phi_{- h} | \Omega \rangle$ is the highest weight state
in the module associated with the primary field $\Phi$.  This result
follows because the physical vacuum $| \Omega \rangle$ is annihilated
by $L_{-1}$.  In general, acting on a highest weight state of an
arbitrary module with the vertex operator (\ref{eq:simplevertex}) does
not give a result which has such a simple description in terms of the
Virasoro generators.  We shall demonstrate in Section
\ref{sec:vertexcoadjoint} that the explicit realization of the
Virasoro generators $\hat{L}_n$ on coadjoint orbit dual Verma modules
leads to a more general version of this result.

\subsection{Free field vertex operators}
\label{sec:vertexfree}

In this section we briefly review the results of Felder \cite{Feld}
and others on the construction of screened vertex operators in terms
of resolutions over Feigin-Fuchs modules.
The fundamental object used in this construction is the standard
bosonic vertex operator
\begin{equation}
V_{\beta} (z) =\;: {\rm e}^{\beta\phi (z)}: \;
= z^{\beta a_0} \exp \left(\beta \sum_{n = 1}^{ \infty}  \frac{a_{-
n}}{n}z^n \right)  \left( -\beta \sum_{n = 1}^{ \infty}  \frac{a_{
n}}{n}z^{- n} \right),
\label{eq:freefieldvertex}
\end{equation}
where the bosonic field
\begin{equation}
\phi (z) = a_0 \ln z + \sum_{n = 1}^{\infty} \frac{a_{
n}}{n}z^{- n} + \sum_{n = 1}^{\infty} \frac{a_{-n}}{n}z^{n}
\end{equation}
is expressed in terms of the bosonic modes (\ref{eq:bosonicmodes}).
The operator $V_{\beta} (z)$ is a vertex operator of weight $\beta
(\beta - 2 \alpha_0)$ with respect to the  realization of
the Virasoro algebra (\ref{eq:Feigin-Fuchs}) on  Feigin-Fuchs type
Virasoro modules.
This operator  changes the Fock space of the Feigin-Fuchs module by
increasing the eigenvalue ($2 \alpha $) of $a_0$ by $2 \beta$.

Setting the background charge $\alpha_0$ to 0, we can combine a set of
$D$ bosonic fields to give a realization of the Virasoro algebra on a
Fock space with central charge $c = D$, with vertex operators carrying
a momentum vector $\stackrel{\rightharpoonup}{\beta}$ and having a
conformal weight of $\stackrel{\rightharpoonup}{\beta}
\cdot\stackrel{\rightharpoonup}{\beta}$.  These are the original
vertex operators from string theory \cite{gsw}.  From these vertex
operators, for $D > 1$ it is possible to construct a vertex operator
with arbitrary weight $h$, acting on a module with highest weight $k$
and resulting in a module with highest weight $k'$ for any $k,k'> 0$.
This can be accomplished by simply choosing an initial state of
momentum $\stackrel{\rightharpoonup}{\beta}$ where $k = |\beta |^2$
and a vertex operator with momentum
$\stackrel{\rightharpoonup}{\gamma}$ where $h = | \gamma |^2$ and $k'
- h - k =2 \stackrel{\rightharpoonup}{\beta}\cdot
\stackrel{\rightharpoonup}{\gamma}$.
Note, however, that not only does this construction only work for
integral $c$, but also that a larger Fock space, containing at least
two sets of bosonic field modes, is necessary.  The construction
described in the previous section realizes the same vertex operators
in turns of a single bosonic field.  Whether this observation can be
used to a useful end in conformal field theory is  an open question;
however, it is a useful example of the general applicability of the
methods in this thesis.

Using the above description of a free field vertex operator, Dotsenko
and Fateev were able to explicitly construct vertex operators on the
discrete series of Virasoro modules in terms of Feigin-Fuchs
representations \cite{DF}.  This work was subsequently explained in
the language of BRST cohomology and resolutions by Felder, who
introduced explicitly the associated intertwining (BRST) operators
\cite{Feld}.  In this approach, a value of $c=1 - 24 \alpha_0^2$
corresponding to a particular rational CFT is fixed, and a set of
intertwining operators $Q_m$ are defined in terms of the vertex
operators (\ref{eq:freefieldvertex}) by
\begin{equation}
Q_m = \frac{1}{m}  \int V_{\alpha_+} (z_0) \cdots  V_{\alpha_+} (z_{m
- 1}) \prod_{i = 0}^{m - 1}  {\rm d} z_i
\left[ \frac{{\rm e}^{2 \pi i \alpha^2_+ m} - 1}{{\rm e}^{2 \pi i
\alpha^2_+} - 1}\right].
\end{equation}
where
\begin{equation}
\alpha_\pm = \alpha_0 \pm\sqrt{1 + \alpha_0^2},
\end{equation}
and the contours of the integration variables $z_0, \ldots, z_{m - 1}$
are concentric circles about $z = 0$.
Similarly, the screened vertex operators are defined by
\begin{equation}
V^{r' r}_{n' n} =   \int V_{1 - n' \alpha_-/2 - n \alpha_+/2} (z)
V_{\alpha_-} (z_0) \cdots  V_{\alpha_-} (z_{r'})
V_{\alpha_+} (w_0) \cdots  V_{\alpha_+} (w_{r})
\prod_{i = 1}^{r'}  {\rm d} z_i
\prod_{i = 1}^{r}  {\rm d} w_i,
\label{eq:screenedvertex}
\end{equation}
with similar integration contours.  The integer indices $n,n', r,r'$
in the screened vertex operators, and $m$ in the intertwining
operators, specify which representations in the discrete series these
operators map to which other representations.  We will not study these
operators in any significant detail, but we will give a brief synopsis
of the results achieved using this approach, and compare to the
methods used here.

The vertex operators and intertwiners above do not exactly commute;
heuristically,  these operators obey a relation of the form
\begin{equation}
Q V (z) = {\rm e}^{2 \pi i \theta} V (z) Q,
\end{equation}
with a generally nonzero phase $\theta$.  This relation is highly
reminiscent of the $q$-commutation relations used to define quantum
groups.  In fact, in the related context of free field theories using
Wakimoto realizations of affine algebras, a similar nontrivial
commutation relation gives precisely the form of a particular class of
quantum group relations, and leads into a rather subtle relationship
between the representation theory of the affine algebras and quantum
groups \cite{bmp,bmp2}.  In the case of the Virasoro algebra, these
nontrivial phases can simply be removed by a redefinition of the
phases of the vertex operators.  It can then be shown \cite{Feld} that
the operators $Q_m$ act on the Feigin-Fuchs modules associated with
the discrete series Virasoro representations, such that $Q^2 = 0$, and
in such a way that the resulting chain complex has a cohomology
resolving the corresponding irreducible representation.  The operators
$V^{r' r}_{n' n}$ are then vertex operators describing a chain map
between two resolutions of irreducible representations.  These
screened vertex operators can then be used to construct CFT
correlation functions in the fashion discussed in Section
\ref{sec:everything}

The elementary objects of the form
\begin{equation}
Q_1 = \int  V_{\alpha_+} (z)  {\rm d} z
\end{equation}
play a role in this theory analogous to the screening operators
defined in Section \ref{sec:finitevertex} (in fact, this operator has
the effect of screening a single unit of charge, and was the initial
motivation for the term ``screening operator'').  This object clearly
commutes with the Virasoro generators, since the conformal weight of
the bosonic vertex operator in the integral is 1.  However, not only
do the modes of this operator not obey the expected commutation
relations among themselves, but the operator itself is not generally
well-defined since in most cases the initial power of $z$ (which is
equal to $2 \alpha_+ \alpha$), is not an integer.  Only for the proper
combinations of $m$ and $\alpha$ does the operator $Q_m$ have a
sensible definition as an operator on the bosonic Fock space.   The
connection between the field-theoretic point of view and the algebraic
perspective in this case remains rather unclear.

To conclude this review of the free field approach to constructing
vertex operators for Feigin-Fuchs representations, we now give an
example of a specific computation of the action of an intertwining
operator $Q_m$.  This explicit calculation hopefully demonstrates
clearly the connection between the language of fields and the more
algebraic language of modules and resolutions, and illustrates the
computational complexity of moving between these descriptions.

We consider again our favorite example, the discrete series Virasoro
representation with $c = h = 1/2$.  In terms of Feigin-Fuchs modules,
this irreducible Virasoro representation has a two-sided resolution;
denoting by $F_{h}$ the Feigin-Fuchs representation with highest
weight $h$, the central part of this resolution looks like
\begin{equation}
\cdots \rightarrow
F_{7\frac{1}{2}} \stackrel{Q_1}{\rightarrow}
F_{2\frac{1}{2}} \stackrel{Q_2}{\rightarrow}
F_{\frac{1}{2}} \stackrel{Q_1}{\rightarrow}
F_{3\frac{1}{2}} \stackrel{Q_2}{\rightarrow}
F_{17\frac{1}{2}}  \rightarrow
\end{equation}
We will analyze the particular intertwining operator
\begin{equation}
Q_2:F_{5/2} \rightarrow F_{1/2}
\end{equation}
as it appears in this module (in the place of the usual operator
$d_{-1}$).  From the discussion earlier in this chapter, we expect
that the image under $Q_2$ of the module $F_{5/2}$ will be a submodule
of $F_{1/2}$ with the highest weight vector $\nul'$ from $F_{5/2}$
being mapped to a singular vector at level 2 in $F_{1/2}$.  As we
know, this singular vector must be the vector
\begin{equation}
(L_{- 2} - \frac{3}{4} L_{-1}^2)\nul
\end{equation}
where $\nul$ is the highest weight vector in $F_{1/2}$.

The appropriate values of $\alpha_0$, $\alpha$, and $\alpha_+$ for the
Feigin-Fuchs module $F_{5/2}$ are
\begin{eqnarray}
\alpha_0 & =&  \frac{1}{4 \sqrt{3}}   \nonumber\\
\alpha & = &   - 10 \alpha_0\\
\alpha_+ & = &  8 \alpha_0\nonumber
\end{eqnarray}
Writing the action of the operator $Q_2$ on the highest weight vector
of this module explicitly, we have
\begin{eqnarray}
Q_2 \nul'  & = &
\frac{1}{2}  \int {\rm d}u \; {\rm d}v \; v^{- 10/3} u^{- 2/3}
	\exp (\alpha_+ \sum_n \frac{a_{- n}}{n} u^n) \\
& & \hspace{.5in}	\exp (-\alpha_+ \sum_n \frac{a_n}{n} u^{- n})
	\exp (\alpha_+ \sum_n \frac{a_{- n}}{n} v^n) \nul. \nonumber
\end{eqnarray}It is a standard exercise to commute the annihilation
operators from
$V_{\alpha_+} (u)$ with the creation operators from $V_{\alpha_+}
(u)$.  This exchange gives an extra factor of $(1 - v/u)^{2
\alpha_+^2}$, so we have
\begin{equation}
Q_2 \nul' = \frac{1}{2}  \int {\rm d}u \; {\rm d}v \; v^{- 10/3} u^{- 2/3}
(1 - v/u)^{8/3}
	\exp (\alpha_+ \sum_n \frac{a_{- n}}{n} (u^n + v^n)) \nul.
\end{equation}
Changing variables to $u = z,v = wz$ we have
\begin{equation}
Q_2 \nul' = \frac{1}{2}   \int {\rm d}z \; {\rm d}w \; z^{- 3} w^{- 10/3} (1
- w)^{8/3}
	\exp \left(\alpha_+ \sum_n \frac{a_{- n}}{n} (z^n (1 +
w^n))\right) \nul.
\end{equation}
We can now explicitly integrate over $z$; the remaining integral over
$w$ is around a contour surrounding the branch points $w =0$ and $w
=1$,
\begin{equation}
Q_2 \nul'  = \pi i  \int {\rm d}w \; w^{- 10/3} (1 - w)^{8/3}
\left[\alpha_+ \frac{1 + w^2}{2}  a_{- 2} \nul
+ \alpha_+^2 \frac{(1 + w)^2}{2} a_{-1}^2\nul\right].
\end{equation}
Deforming the contours to run directly from $w = 0$ to $w = 1$ (these
are the original contours used by Dotsenko and Fateev in \cite{DF}),
the integrals become standard $\beta$-function integrals.  Writing the
states $a_{-2} \nul$ and $a_{-1}^2 \nul$ in the polynomial forms $|
z_2 \rangle$, $|z_1^2 \rangle$ respectively, we have a result
proportional to
\begin{eqnarray}
\frac{\Gamma (- 7/3) \Gamma (11/3)}{\Gamma (4/3)}
\left( | z_2 \rangle + 8 \alpha_0 | z_1^2 \rangle \right)  &+ &
\frac{\Gamma (- 4/3) \Gamma (11/3)}{\Gamma (7/3)}
\left( 16 \alpha_0 | z_1^2 \rangle \right) \\ & + &
\frac{\Gamma (- 1/3) \Gamma (11/3)}{\Gamma (10/3)}
\left( | z_2 \rangle + 8 \alpha_0 | z_1^2 \rangle \right) \nonumber
\end{eqnarray}
Using the usual property $x \Gamma (x) = \Gamma (x + 1)$, we see that
this result is proportional to the vector
\begin{equation}
| z_2 \rangle  -  6 \alpha_0 | z_1^2 \rangle.
\end{equation}
{}From the explicit Feigin-Fuchs realization of the Virasoro algebra
(\ref{eq:Feigin-Fuchs})  with the value $\alpha = 6 \alpha_0$, we can
compute the singular vector
\begin{eqnarray}
(L_{- 2} - \frac{3}{4} L_{-1}^2)\nul & = &
(\alpha + \alpha_0) | z_2 \rangle + \frac{1}{4}  | z_1^2 \rangle)
- \frac{3}{4}  (\alpha^2 | z_1^2 \rangle + \alpha | z_2 \rangle)
\nonumber\\
 & = &  \frac{5}{2} \alpha_0 | z_2 \rangle - \frac{15}{48}  | z_1^2 \rangle \\
& = & \frac{5}{2}  \alpha_0 (| z_2 \rangle  -  6 \alpha_0 | z_1^2 \rangle).
\nonumber
\end{eqnarray}
Thus, the vector $Q_2 \nul'$ is indeed proportional to the singular
vector in $F_{1/2}$.

This example demonstrates that although the vertex operators and
intertwining operators for the free field theory realizations of
minimal models are expressed in a very different language from the
algebraic concepts of modules and resolutions developed in the
previous sections, the effective content of the two formalisms is
equivalent.  The field-theoretic approach used by authors such as
Felder is more useful for computing certain properties of correlation
functions and partition functions which are expressable in terms of
that language.  On the other hand, as this example calculation
demonstrates, certain underlying mathematical properties of the theory
are much more easily accessible using the algebraic formulation.
Hopefully, in the future these disparate formalisms will merge
somewhat, clarifying aspects both of conformal field theory and the
underlying group theory.  Recent work which will be discussed in the
next chapter indicates that this convergence of perspectives is indeed
occurring.

\subsection{Coadjoint orbit vertex operators}
\label{sec:vertexcoadjoint}

We will now discuss the specific form of the vertex operators of
weight $h$
\begin{equation}
\Phi (z): V^*_k \rightarrow V^*_{k'}
\label{eq:vertexdv}
\end{equation}
which are defined on the coadjoint orbit realizations of the Virasoro
dual Verma modules.  As we showed in Section
\ref{sec:generalvertex}, these vertex operators exist for all values
of the highest weights $k,k'$ which correspond to Virasoro dual Verma
modules without cosingular vectors.  In addition, whenever such vertex
operators can be defined on chain complexes of dual Verma modules with
cosingular vectors, their action on all physical states (states in the
orbit of the highest weight vector of $V^*_k$) is completely
determined by the matrix elements (\ref{eq:generalvertexelement}).
Our first goal in this section is to translate these matrix elements
into the coefficients of the modes of the vertex operator, when these
modes are expressed as differential operators on the polynomial space
$R$.  We then show that when $x = 0$ the action of these operators on
any highest weight state are closely related to the operator
$\hat{L}_{-1}$ associated with the Virasoro generator $L_{-1}$, and
that the action of these operators on any state can be defined
completely in terms of Virasoro generators $\hat{L}_n$.

Explicitly, we can write the modes of a vertex operator
(\ref{eq:vertexdv}) as individual differential operators
\begin{equation}
\phi_{- h - x - n} = \sum_{m,l > 0:m -l = n}^{\infty}
\sum_{\pi_ m, \pi_l}p (\pi_m,\pi_l)
(\prod_{i} z_{m_i} \prod_{j}\derivative{z_{l_j}})
\end{equation}
which act directly on the space $R$.  To  calculate the coefficients
$p(\pi_m, \pi_l)$ in these differential operators in terms of the
matrix elements  (\ref{eq:generalvertexelement}), we must relate the
vectors $| \pi_n \rangle$ to the canonical basis for $R$ of monomials.
Similarly, we  must relate the basis $\langle \pi_m |$ of the Verma
module $V_{k'}$ to the set of differential operators
\begin{equation}
\partial_{\pi_n} =
\left( \prod_{i} \derivative{z_{n_i}}\right) |_{z_j = 0}
\end{equation}
where the evaluation at $z_j = 0$ for all $j  > 0$ has the effect of
extracting the constant term in a polynomial after the action of the
product of partial derivatives.

The linear relationship between the vectors $| \pi_n \rangle$ and the
monomials in $R$ of degree $n$ up to level 2 was given by
(\ref{eq:Virasorosingular}); we expand this calculation explicitly to
level 3,
\begin{eqnarray}
\nul & = & 1 \nonumber \\
\hat{L}_{-1} \nul & = & -2 h |z_1\rangle, \nonumber \\
\hat{L}_{-1}^2 \nul & = & -6h |z_2\rangle + (4 h^2 + 2h)|z_1^2\rangle
\nonumber\\
\hat{L}_{-2} \nul & = & -(4h + \frac{c}{2}) |z_2\rangle + 3 h
|z_1^2\rangle. \label{eq:linearrelations1} \\
\hat{L}_{-1}^3 \nul & = & -24 h |z_3\rangle
+ (24 h + 36 h^2) | z_1 z_2 \rangle
 + (- 8h^3 - 12h^2 - 4h)|z_1^3\rangle  \nonumber\\
\hat{L}_{-1} \hat{L}_{-2} \nul & = & -(16 h + 2c) |z_3\rangle
+ (26 h + c + 8h^2 + ch) | z_1 z_2 \rangle
 + (- 6h^2 + 6h) |z_1^3\rangle  \nonumber\\
\hat{L}_{-3} \nul & = & -(6 h + 2c) |z_3\rangle
+ (13 h +  c) | z_1 z_2 \rangle
 + 4h |z_1^3\rangle  \nonumber.
\end{eqnarray}
The relationship between the states $\langle \pi_n |$ and the
derivative operators $\partial_{\pi_m}$ is easy to calculate; up to
level 3, we have
\begin{eqnarray}
\nuls & = & 1  |_{z_j = 0} \nonumber \\
\nuls \hat{L}_{1}  & = & -\derivative{z_1}  |_{z_j = 0} \nonumber \\
\nuls \hat{L}_{1}^2  & = & \derivetwo{z_1}
 |_{z_j = 0} \nonumber \\
\nuls \hat{L}_{2}  & = &  - \derivative{z_2}  |_{z_j = 0}
\label{eq:linearrelations2}  \\
\nuls \hat{L}_{1}^3  & = &  - \frac{\partial^3}{\partial z_1^3}
 |_{z_j = 0} \nonumber\\
\nuls \hat{L}_{1} \hat{L}_{-2}  & = &  \frac{1}{2}  \derivative{z_3} +
\derivative{z_1} \derivative{z_2} |_{z_j = 0} \nonumber \\
\nuls \hat{L}_{3}  & = &
- \derivative{z_3} |_{z_j = 0} \nonumber
\end{eqnarray}
Clearly, at each level $n$, inverting these relationships involves
calculating the inverse of a $\pi (n) \times \pi (n)$ matrix.  The
matrix relating the basis $\langle \pi_n |$ to the operators
$\partial_{\pi_m}$ is always invertible; the matrix relating $| \pi_n
\rangle$ to monomials in $R$ is only invertible when $V^*_k$ contains
no cosingular vectors.
We do
not have any kind of general prescription for calculating this
inverse; for the relations (\ref{eq:linearrelations1}) at level 2 the
inverse relations are given by
\begin{eqnarray}
| z_2 \rangle & = &  \frac{1}{\gamma}
3h  \hat{L}_{-1}^2\nul - 2h (2h + 1) \hat{L}_{-2} \nul  \\
| z_1^2 \rangle & = &  \frac{1}{\gamma}
(4h + \frac{c}{2} ) \hat{L}_{-1}^2\nul -  6h \hat{L}_{-2} \nul,    \nonumber
\end{eqnarray}
where
\begin{equation}
\gamma = 16 h^3 + (2c -10) h^2 + ch
\end{equation}
When $h = 0$ or the condition (\ref{eq:red}) is satisfied the
determinant $\gamma$ is 0 and the inverse transformation is undefined.

In Section \ref{sec:generalvertex}, we calculated in
(\ref{eq:explicit1}-\ref{eq:explicit13}) some of the  matrix elements
of $\Phi (z)$ between states of low level.  The remaining matrix
elements between states of level $\leq 2$ are given by
\begin{eqnarray}
\nuls L_1 \phi_{- h - x} L_{-1}  \nul & = &
x + 2k - 2hx - x^2
\label{eq:matrix1}\\
\nuls L_1 \phi_{- h - x + 1} L_{-1}^2 \nul & = &
(2h + x - 2)x (x - 1) - x (4k + 2)
\label{eq:matrix2}\\
\nuls L_1 \phi_{- h - x + 1} L_{-2} \nul & = &
(h - x - 1) (2h + x)
\label{eq:matrix3}\\
\nuls L_1^2 \phi_{- h - x - 1} L_{-1} \nul & = &
8 h k + 4 k x - 4 x h^2 - 4 h x^2 - x^3 + 2 h x + x^2
\label{eq:matrix4}\\
\nuls L_2 \phi_{- h - x - 1} L_{-1} \nul & = &
x - 3 h x - x^2
\label{eq:matrix5}\\
\nuls L_1^2 \phi_{- h - x} L_{-1}^2 \nul & = &
8 k^2 + 4 k + 8 k x - 16 h k x - 8 k x^2 + 2 x - 4 x h^2 - 2 h x + x^2
\nonumber\\
&&\;\;\;      + 4 h^2 x^2 + 4 h x^3 + x^4 - 10 h x^2 - 4 x^3
\label{eq:matrix6}\\
\nuls L_1^2 \phi_{- h - x} L_{-2} \nul & = &
6 k + 4 h^3 - 3 h x^2 - x^3 - 6 h^2 - 9 h x - 3 x^2 + 2 h + 4x
\label{eq:matrix7}\\
\nuls L_2 \phi_{- h - x} L_{-1}^2 \nul & = &
6 k + 3 h x^2 + x^3 + 2 x - 3 h x - 3 x^2
\label{eq:matrix8}\\
\nuls L_2 \phi_{- h - x} L_{-2} \nul & = &
c/2 - 2h + 4k + 2x + 3h^2 - 2h x - x^2
\label{eq:matrix9}
\end{eqnarray}

{}From (\ref{eq:explicit1}-\ref{eq:explicit14}),
(\ref{eq:linearrelations1}-\ref{eq:linearrelations2}) and
(\ref{eq:matrix1}-\ref{eq:matrix9}), we can write the explicit
formulae for the low-order terms in $\phi_{- h - x - n}$ for small $|
n |$.  Using again the notation that ${\cal D}_n$ indicates terms
containing derivatives $\derivative{z_m}$ with $m \geq n$, we have
\begin{eqnarray}
\phi_{- h - x+ 2} & = &
\frac{1}{\gamma} \left[2h (2h + 1) (x - h) + 6h x (x - 1)\right]
 \derivative{z_2} \nonumber \\
& &\hspace{.5in} +
\frac{1}{\gamma} \left[6h (x - h)+ (c/2 + 4h)x (x - 1)\right]
 \derivetwo{z_1} + {\cal D}_3
 \nonumber \\
\phi_{- h - x + 1} & = & \frac{x}{2h} \derivative{z_1} + {\cal D}_2
 \nonumber \\
\phi_{- h - x} & = & 1 + \frac{x + 2k - 2hx - x^2}{2h}
z_1 \derivative{z_1} + {\cal D}_2
\label{eq:vertexterms} \\
\phi_{- h - x - 1} & = & - (2h + x) z_1 + {\cal D}_1
 \nonumber \\
\phi_{- h - x - 2} & = & - (3h + x) z_2 +
\frac{1}{2} (2h + x)(2h + x+ 1) z_1^2
 + {\cal D}_1
 \nonumber \\
\phi_{- h - x -  3} & = &
- (4h + x) z^3
+ (4h + 3 x/2 + 6h^2 + 5hx + x^2) z_1 z_2
 \nonumber \\
& & \;- \frac{1}{6} (2h + x) (2h + x + 1) (2h + x + 2) z_1^3 + {\cal D}_1
\nonumber
\end{eqnarray}

Although this explicit computation of the initial terms in the vertex
operators is in general rather unenlightening, we can nevertheless
extract some interesting information from this calculation.

The first observation we can make is that the general form of the
vertex operator has a nontrivial dependence on all three parameters
$k,h,$ and $x$ (generally, this dependence appears for all terms after
the leading order; here, we have only bothered to compute leading
order terms except for $\phi_{- h - x}$).  Although one might
naturally expect this dependence based on the method of construction,
a cursory examination of the free field vertex operator
(\ref{eq:freefieldvertex}) seems to indicate that $V_{\beta} (z)$
depends only upon the parameter $\beta$, and thus upon the conformal
weight $h=\beta (\beta - 2 \alpha_0)$ of the vertex operator.  Upon
closer inspection, this problem becomes slightly less acute; the
exponent $\beta a_0$ is equal to the quantity we denote by $x$, and
the vertex operator $V_{\beta} (z)$ always acts from a space of weight
$k = \alpha (\alpha - 2 \alpha_0)$ to a space of weight $k' = (\alpha
+ \beta) (\alpha + \beta - 2 \alpha_0)$, reducing the dimensionality
of the space of allowed vertex operators to 2.  Nonetheless, the
action of this vertex operator on the Fock space certainly depends on
one parameter fewer than the vertex operators we have constructed on
the coadjoint orbit dual Verma modules.  This may be an indication
that finding a closed form expression for these coadjoint orbit vertex
operators will be a particularly difficult task.  Certainly, the
construction of a field theory from such operators will be complicated
by this fact; we would like to have a construction in which  the
vertex operators associated with a single primary field depend only
upon the weight of that primary field.

It may be, however, that the extra dependencies which this operator
exhibits are a result of two features of the explicit coadjoint orbit
representations.  The first feature is that the Virasoro operator
realizations on these representations are somewhat {\em less}
dependent upon the weight of the representation, in the sense that all
the raising operators $L_n, n > 0$ take an identical form
independently of the weight of the associated representation.  The
second feature is that the coadjoint orbit vertex operators
can be defined for arbitrary triplets of weights $h,k,k'$ when $c >
1$.  All known free field representations on a single bosonic Fock
space place strong restrictions on the weights and central charges for
which vertex operators in this regime can be constructed.  Thus, it
may be that by increasing the generality of this type of vertex
operator, we have lost some structure in the dependence of the vertex
operator itself.  In any case, further research is necessary to
determine whether this formalism will actually lead to results which
go beyond those accessible through the free field formalism.

The second result which we can see from the explicit construction of
the initial vertex operator terms (\ref{eq:vertexterms}) is that these
operators simplify considerably when we take $x = 0$ or $x = - 2h$.
When we set $x =  0$, for instance, the low-order terms in the vertex
modes look like
\begin{eqnarray}
\phi_{- h - x+ 2} & = &
-\frac{1}{\gamma} 2h^2 (2h + 1)
 \derivative{z_2}  +
-\frac{1}{\gamma} 6h^2 \derivetwo{z_1} + {\cal D}_3
 \nonumber \\
\phi_{- h - x + 1} & = & {\cal D}_2
 \nonumber \\
\phi_{- h - x} & = & 1 + \frac{k}{h} z_1 \derivative{z_1}+{\cal D}_2
\label{eq:vertextermsx} \\
\phi_{- h - x - 1} & = & - 2h z_1 + 2k z_1^2 \derivative{z_1} +
{\cal D}_2
 \nonumber \\
\phi_{- h - x - 2} & = & - 3h z_2 + h(2h+ 1) z_1^2
 + {\cal D}_1
 \nonumber \\
\phi_{- h - x -  3} & = &
- 4h z^3
+ (4h  + 6h^2) z_1 z_2
- \frac{1}{3} h (2h + 1) (2h + 2) z_1^3 + {\cal D}_1 \nonumber
\end{eqnarray}
Comparing to (\ref{eq:linearrelations1}), we see that for $n < 4$ at
least, the polynomial terms in $\phi_{- h - x - n}$ are precisely the
result of acting on the highest weight vector of weight $h$ with the
operator $L_{-1}^n/n!$.  In fact, this correspondence is true to all
orders; when $k' = k + h$, the action of a vertex operator of weight
$h$ on the highest weight state $\nul$ of a module of weight $k$ can
be described by
\begin{equation}
\Phi (z) \nul = {\rm e}^{z \hat{L}_{-1}^{(h)}} \nul' = \sum_{i = 0}^{
\infty}  \frac{z^i \hat{L}_{-1}^{(h)i}}{i!} \nul',
\label{eq:generalsimpleaction}
\end{equation}
where $\nul'$ is the highest weight state in $V^*_{k'}$ and we denote
by $\hat{L}_{-1}^{(h)}$ the realization of the operator $\hat{L}_{-1}$
on the module of highest weight $h$.  From the simplification
(\ref{eq:simpleaction}) of the action of the vertex operator on the
vacuum, we expect that this relation should hold for a vertex operator
acting on the vacuum.  However, when a vertex operator acts on a
highest weight state with weight $k \neq 0$, there is no similar
reason to expect a similar result.  Using the vertex operator
construction (\ref{eq:freefieldvertex}) on a single bosonic Fock
space, no analogous situation can arise due to the relation $x = 2
\alpha \beta$.  We would not  expect this type of result  to
be possible for the free field realizations in any case, however,
since for the realizations (\ref{eq:Feigin-Fuchs}),
\begin{equation}
[L_n^{(h + k)}, L_{-1}^{(h)}] \neq [L_n^{(h)}, L_{-1}^{(h)}]
\end{equation}
when $k \neq 0$.  Because the coadjoint orbit realizations
have raising operators independent of the conformal weight, we do
have  the relation
\begin{equation}
[\hat{L}_n^{(h + k)}, \hat{L}_{-1}^{(h)}] =
[\hat{L}_n^{(h)}, \hat{L}_{-1}^{(h)}] \; \; \; \forall n > 0.
\label{eq:raisingequivalence}
\end{equation}
for these representations.

For the coadjoint orbit representations,
(\ref{eq:generalsimpleaction}) follows from
(\ref{eq:raisingequivalence}) and the result in the special case $k =
0$.  We can also demonstrate this property explicitly.  Generally, we
wish to show that
\begin{equation}
\hat{L}_n \frac{\hat{L}_{-1}^{(h)m}}{m!}\nul = (n (h - 1) + h + m)
\frac{\hat{L}_{-1}^{(h)m - n}}{(m - n)!}\nul,
\label{eq:strange}
\end{equation}
where $\nul$ is a highest weight state of any weight $k$.  From $[\hat{L}_1,
\hat{L}_{-1}^{(h)}] = 2h + 2\xi_0$, it is easy to see that this equation
holds for $n = 1$, since
\begin{equation}
\hat{L}_1 \frac{\hat{L}_{-1}^{(h)m}}{m!}\nul =
\left[2h + (2h + 2) + \cdots + (2h +2m - 2) \right]
\frac{\hat{L}_{-1}^{(h)m - 1}}{(m - 1)!}\nul,
\end{equation}
We can now prove (\ref{eq:strange}) by induction on $n$.  Performing
the commutator of $\hat{L}_n$ with each factor of $\hat{L}_{-1}$, and
assuming the result for all values $n' < n$, we
have
\begin{eqnarray}
\hat{L}_n \frac{\hat{L}_{-1}^{(h)m}}{m!}\nul & = &
\sum_{i = n - 1}^{m - 1} (n + 1) \hat{L}_{-1}^{(h)m -i - 1} \hat{L}_{n - 1}
\frac{\hat{L}_{-1}^{(h)i}}{m!}\nul,  \\
 & = &  \sum_{i = n - 1}^{m - 1}  (n + 1)
\left[ nh - n  + i + 1\right]\frac{i!(m - n)!}{(i -n + 1)!m!}
\frac{\hat{L}_{-1}^{(h)m - n}}{(m - n)!}\nul,\nonumber
\end{eqnarray}
For the desired result to be proven, it suffices to demonstrate that
\begin{equation}
\sum_{i = n - 1}^{m - 1}  (n + 1)
\left[ nh - n  + i + 1\right] \frac{i!}{(i -n + 1)!}
= \frac{m!}{(m - n)!}  [h (n + 1) + m - n].
\end{equation}
But we can again use induction to prove this relation.  The relation
is trivial for $m = n$.  Fixing $n$ and assuming the relation for $m'
< m$, we   have
\begin{eqnarray}
\lefteqn{\sum_{i = n - 1}^{m - 1}  (n + 1)
\left[ nh - n  + i + 1\right] \frac{i!}{(i -n + 1)!} } \\
 & = & \frac{(m - 1)!}{(m - n - 1)!}  [h (n + 1) + m - n - 1]
+ (n + 1) \left[ nh - n + m \right] \frac{(m - 1)!}{(m - n)!}  \nonumber
\\
 & = &  \frac{(m - 1)![mh (n + 1) + m^2 - mn]}{(m - n)!}. \nonumber
\end{eqnarray}
Thus, we have proven (\ref{eq:strange}).  This gives us a formula for
the action of a  vertex operator on any highest weight state when $x =
0$ in terms of the Virasoro generator $\hat{L}_{-1}^{(h)}$.  We can
extend this result to give an expression for an arbitrary mode of the
vertex operator acting on an arbitrary state in $V^*_k$ of the form $|
\pi_n \rangle$ by using the relations (\ref{eq:vertexcommutator}) to
move the creation operators to the left, and then expressing the
result in terms of modes of the vertex operator acting on the highest
weight state $\nul$ of $V^*_k$.  For example, we have (still assuming
$k' = k + h$),
\begin{eqnarray}
\phi_{- h - n} \hat{L}_{- m} \nul & = & \hat{L}_{- m} \phi_{- h - n}
\nul + (h (m - 1)- n - m) \phi_{- h - n - m} \nul
\\
 & = & \hat{L}_{- m}^{(h + k)}  \frac{\hat{L}_{-1}^{(h)n}}{n!}
\nul + (h (m - 1)- n - m)
\frac{\hat{L}_{-1}^{(h)n+ m}}{m!}  \nul. \nonumber
\end{eqnarray}
This gives us an alternative approach to calculating the action of the
vertex operators in this special case, and indicates an interesting
relationship between the vertex operators and the Virasoro generators.
However, it is still rather difficult to see how this relationship
might be used to construct any sort of closed form expression for the
vertex operators.  It is also unclear whether this relationship has
any natural geometric interpretation  when we describe the vertex
operators as global differential operators acting on the holomorphic
sections of various line bundles.

\subsection{Correlation functions}
\label{sec:everything}

We end this chapter with a short discussion of correlation functions
in conformal field theories.  For any conformal field theory, the
correlation functions (\ref{eq:correlationfunction}) are the primary
quantities necessary to describe the physics of the theory.  The free
field approach and the use of differential equations give a fairly
powerful set of tools for the computation of correlation functions in
a wide variety of physical theories.  The approach presented in this
thesis gives a simple mathematical structure to the definition of such
correlation functions, but so far has not given rise to any
particularly useful approach to the explicit calculation of these
objects.  In this section, we review briefly some of the main methods
and simplest results about correlation functions, check that in the
simplest cases our formalism is in agreement with these results, and
finally present some speculation about possible approaches to
computing correlation functions using this approach.

The simplest correlation functions in any Virasoro conformal field
theory are the 2-point and 3-point correlation functions.  It is easy
to see, using the scaling properties of vertex operators under the
transformation $z \rightarrow cz$, that the 2-point functions
are of the form
\begin{equation}
\langle \Omega | \Phi_1 (z_1) \Phi_2 (z_2) | \Omega \rangle
\sim  (z_1 - z_2)^{- 2h},
\label{eq:correlation2}
\end{equation}
up to an undetermined proportionality constant, where the conformal
weights $h_1,h_2$ of the vertex operators $\Phi_1, \Phi_2$ satisfy $h
= h_1 = h_2$.
Similarly, the 3-point functions are generally of the form
\begin{equation}
\langle \Omega | \Phi_1 (z_1) \Phi_2 (z_2)\Phi_3 (z_3) | \Omega \rangle
\sim  (z_1 - z_2)^{h_3 - h_2 - h_1}  (z_1 - z_3)^{h_2 - h_3 - h_1}
(z_2 - z_3)^{h_1 - h_2 - h_3}.
\label{eq:correlation3}
\end{equation}
For 4 or more vertex operators, the form of the correlation functions
can become more complex.  For example, the 4-point function can depend
upon an arbitrary function of the anharmonic quotient
\begin{equation}
x = \frac{(z_1 - z_2) (z_3 - z_4)}{(z_1 - z_3) (z_2 - z_4)}.
\end{equation}
There are several main approaches currently used to calculate these
more complicated correlation functions.  One approach, which was first
developed in the original work of BPZ \cite{bpz} on conformal field
theories, is to use the differential equations analogous to the
constraint equations (\ref{eq:conditions1}) as conditions on the
correlation functions.  Applying this method to the 4-point functions
of the minimal models, BPZ showed that these correlation functions
could be expressed in terms of hypergeometric functions.  This method
has subsequently been generalized and applied to a wide variety of
conformal field theories.  Perhaps the most famous example are the
Knizhnik-Zamolodchikov differential equations which apply to the
correlation functions for  rational conformal field theories with
affine algebra symmetries \cite{kz}.

Another approach to the explicit calculation of conformal field theory
correlation functions comes from the free field approach.  The
correlation function of an arbitrary product of (unscreened) vertex
operators (\ref{eq:freefieldvertex}) is known to have the form
\begin{equation}
\langle \Omega | V_{\beta_1} (z_1)\cdots V_{\beta_n} (z_n) | \Omega
\rangle = \prod_{i < j} (z_i - z_j)^{2 \beta_i \beta_j},
\end{equation}
when $0 = \sum \beta_i$, and to vanish when this condition is not
satisfied.  Using this result and the definition of the screened
vertex operators (\ref{eq:screenedvertex}), the correlation functions
of minimal models can be calculated in terms of a set of contour
integrals \cite{DF,FF}.

The correlation functions we have discussed so far are restricted to
the case of conformal field theories on the sphere.  In particular,
the free field calculation described above does not need to explicitly
deal with resolutions of the irreducible representations in terms of
a chain complex of Feigin-Fuchs modules because the vacuum $| \Omega
\rangle$ is a BRST invariant state, and therefore any matrix element
of the form of (\ref{eq:correlationfunction}) will contain no
contribution from the ``ghost'' states in the module which are
cohomologically trivial.  One may also consider a general conformal
field theory on a higher genus Riemann surface.  For example, a
correlation function on a torus could be calculated by taking a trace
over the irreducible representation space of a product of vertex
operators acting on that irreducible representation space.  It was
pointed out by Felder that the free field approach gives a simple way
of computing a correlation function of this type; a theorem due to
Lefschetz states that when a module $I$ has a resolution of the form
$(M_i,d_i)$, and an operator ${\cal O}:I \rightarrow I$ has a
realization as a chain map
\begin{equation}
{\cal O}^i:M_i \rightarrow M_i,
\end{equation}
then the trace of ${\cal O}$ is given by
\begin{equation}
{\rm Tr}\; {\cal O} = \sum_{i = - \infty}^{\infty}  (-1)^{i} {\rm
Tr}\; {\cal O}^i.
\label{eq:Lefschetz}
\end{equation}
A simple example of such a calculation is the formula for the
character of an irreducible Virasoro representation $I$,
\begin{equation}
\chi_{I} (q) = {\rm Tr}_{I}\; q^{L_{0}}.
\end{equation}
Although the character formula for a member of the discrete series of
Virasoro representations is somewhat nontrivial, the character of a
Verma, dual Verma, or Feigin-Fuchs module $M$ of highest weight $h$ is
easily computed to be
\begin{equation}
{\rm Tr}_{M}\;q^{L_0} = q^h \prod_{i = 1}^{\infty}
\left( \frac{1}{1 - q^i}  \right) = q^{h + 1/24} \eta (q)^{-1}.
\label{eq:bosonicpartition}
\end{equation}
Using this formula, along with (\ref{eq:Lefschetz}) and a resolution
of the discrete series irreducible representations in terms of Verma
modules, the characters of the discrete series were first computed by
Rocha-Caridi \cite{rc2}.

In the last few years, there has been a fairly large body of work in
which the free field method for calculating correlation functions has
been generalized to higher-genus Riemann surfaces \cite{bg,fs,flms}.  The
approach taken in these papers is essentially to sew together
correlation functions on the sphere by using 3-point functions to
define a ``vertex'' connecting spheres with punctures.  Some care must
the taken to ensure that the  ghost states combine correctly with this
sewing procedure such that their contribution to physical correlation
functions vanishes.

In principle, correlation functions for any conformal field theory on
an arbitrary Riemann surface can be constructed using the coadjoint
orbit representations in much the same fashion that the free field
representations were used in the work described above.  Because we do
not yet have a general closed form expression for a vertex operator
between coadjoint orbit realizations on dual Verma modules, we cannot
use this abstract construction to calculate any particularly
sophisticated correlation functions using the methods of this thesis
at this time.  We can however, do some elementary computations to
verify that for the simple correlation functions such as the 2- and
3-point functions, our formalism gives the expected results
(\ref{eq:correlation2}) and (\ref{eq:correlation3}).  In the following
calculations, we denote by $\phi^{(h,k,k')}_{- h - x + n}$ the $n$th
mode of the Virasoro vertex operator of weight $h$ mapping $V^*_k
\rightarrow V^*_{k'}$ (\ref{eq:vertexexpansion}).  In the case of the
2-point function, the expression for a correlation function of two
vertex operators of weight $h$ in terms of modes is
\begin{equation}
\langle \Omega | \Phi^{(h)} (z) \Phi^{(h)} (w) | \Omega \rangle
= \sum_{n = 0}^{\infty}
w^{n} z^{-2h - n} \phi^{(h,0,h)}_{- h - n}
\phi^{(h,h,0)}_{h - n}
\end{equation}
Expanding around $w = 0$ and using the result (\ref{eq:vertexterms}),
we have the expression
\begin{eqnarray}
\langle \Omega | \Phi^{(h)} (z) \Phi^{(h)} (w) | \Omega \rangle
 & = &  z^{-2h} \left(1 +  \frac{w}{z} \phi^{(h,0,h)}_{- h -1}
\phi^{(h,h,0)}_{h  + 1}+\frac{w^2}{z^2} \phi^{(h,0,h)}_{- h - 2}
\phi^{(h,h,0)}_{h  + 2} + {\cal O} (\frac{w^3}{z^3} ) \right)  \nonumber\\
& = & z^{-2h} \left(1 + 2h \frac{w}{z} + h (2h + 1)\frac{w^2}{z^2}
+{\cal O} (\frac{w^3}{z^3} ) \right),
\end{eqnarray}
in agreement with (\ref{eq:correlation2}).
A similar argument shows that the 3-point function computed in an
expansion around $z_2 = z_3 = 0$ agrees with (\ref{eq:correlation3}).
In fact, from the general result describing the action of a vertex
operator on the vacuum  in terms of $\hat{L}_{-1}$ and $\hat{L}_1$
when acting on the left and right respectively, we can calculate
(\ref{eq:correlation2}) and (\ref{eq:correlation3}) to all orders for
the coadjoint orbit vertex operators.  It is reassuring, however, to
see that the explicit calculations in (\ref{eq:vertexterms}) are in
agreement with the  results of other methods.

The calculation of a correlation function of 4 or more  vertex
operators using the approach of this thesis would be a rather arduous
task.  In fact, without a more general or abstract description of the
coadjoint orbit vertex operators, there is absolutely no reason to
pursue such a computation.  It is hoped, however, that eventually a
better description of these operators will be possible, with which the
calculation of such correlation functions will be simplified; since
this approach gives a description of conformal field theories which is
in some ways much simpler than the free field picture, it is possible
that this viewpoint may even give insights into the theory which are
inaccessible from the free field point of view.

For now, however, we can only speculate about what form a more concise
description of the vertex operators described in this thesis might
take.  One possible approach is to attain a better geometrical picture
of the nature of the vertex operator as a set of differential
operators on line bundles over the coadjoint orbits.  From this point
of view, there is a very rich geometrical and group-theoretical
structure to these operators, and it seems likely that a simple
intrinsic definition of the vertex operators might be given in a
geometrical language so that index theorems and other powerful
geometrical tools might be of assistance in the explicit calculation
of correlation functions.  It is also possible that by purely
algebraic manipulations, one might be able to construct a closed form
expression for the vertex operators similar to the expressions derived
in this thesis for the differential operators associated with the
action of the algebra generators and the screening operators and
intertwiners.  If such a closed form expression were found, it might
be possible to use this result to study more deeply certain
geometrical, algebraic, and physical properties of conformal field
theories.

\mysection{Conclusions}

\subsection{Results}

In this thesis, we have presented a basic conceptual framework in
which conformal field theory has a completely geometric
interpretation in terms of mathematical objects with well-understood
geometric, algebraic, and group-theoretic structures.

We began by considering a general Lie group $G$ which could be either
a finite-dimensional group, the Virasoro group, or a centrally
extended loop group.  We outlined the coadjoint orbit construction,
which associates with each irreducible unitary representation $I_b$ of
$G$ of highest weight $b$ a complex line bundle $\lb$ over a complex
homogeneous space of the form $G/T$ where $T$ is a maximal subtorus of
$G$.  The group $G$ has a natural action on the space ${\cal H}_b$ of
holomorphic sections of $\lb$.  We described the associated action of
the algebra $\algg$ explicitly in terms of a set of (first-order)
differential operators acting on ${\cal H}_b$ in a particular choice
of local gauge.  For finite-dimensional groups and centrally extended
loop groups, the space ${\cal H}_b$ carries precisely the irreducible
representation $I_b$ under the natural action of $G$.  For the
Virasoro group, we find that the resulting representation on the space
of holomorphic sections is that of the dual Verma module associated
with the highest weight $b$.  In those cases where this representation
is reducible, we can use an infinite sequence of line bundles
associated with different highest weights $b_i,b'_i$ to construct a
chain complex of line and vector bundles
\begin{equation}
0 \rightarrow {\cal L}_b
\stackrel{d_{0}}{\rightarrow}  {\cal L}_{b_1}\oplus {\cal L}_{b'_1}
\stackrel{d_{1}}{\rightarrow}  {\cal L}_{b_2}\oplus {\cal L}_{b'_2}
\stackrel{d_{2}}{\rightarrow}  \cdots
\label{eq:finalcomplex}
\end{equation}
where the operators $d_i$ are combinations
of intertwining operators which map holomorphic sections of one line
bundle to holomorphic sections of another line bundle in such a way
that the intertwining operator commutes with the $\algg$-action on the
spaces of sections.  The cohomology of this complex is nontrivial only
in the first bundle, and gives precisely the irreducible
representation $I_b$ of the Virasoro algebra.

Thus, we have a geometric construction of every irreducible unitary
representation of the groups of interest as either a space of all
holomorphic sections of a line bundle, or as the space of sections of
a line bundle which are in the kernel of a particular differential
operator.  We then proceeded to define vertex operators, which map
between irreducible representation spaces by acting on the spaces of
holomorphic sections of the associated bundles.  Given three weights
$b_1,b_2,b_3$, a vertex operator of weight $b_1$ is defined by a set
of maps
\begin{equation}
\Phi_u: {\cal L}_{b_2} \rightarrow {\cal L}_{b_3},
\label{eq:finalvertex}
\end{equation}
where $u$ takes a value in a representation of the algebra $\algg$
associated with the weight $b_1$.  In the case of finite-dimensional
groups, this space is merely the irreducible representation space with
highest weight $b_1$; in the case of the Virasoro and affine algebras,
the   space is again an irreducible representation space of $\algg$,
but is no longer highest weight.  The operators $\Phi_u$ satisfy the
vertex operator condition
\begin{equation}
[x, \Phi_u] = \Phi_{xu} \; \; \; \forall x \in \algg.
\end{equation}
For the Virasoro algebra, the vertex operator construction is slightly
more complex when $b_2$ or $b_3$ are associated with degenerate
representations.  In this case, the vertex operator is actually a
chain map between complexes of the form (\ref{eq:finalcomplex}).

In order to make the connection to conformal field theory, a vacuum is
chosen which corresponds to a particular line bundle ${\cal L}_0$ with
extra symmetries under the group action.  We denote the holomorphic
section of ${\cal L}_0$ which is highest weight under the $\algg$
action by $| \Omega \rangle$.  Generally, a conformal field theory
correlation function is then written in the form
\begin{equation}
\langle \Omega | \Phi^{(1)}_{u_1} \Phi^{(2)}_{u_2} \cdots
\Phi^{(k)}_{u_k}| \Omega \rangle
\end{equation}

Although through most of this thesis, we dealt most directly with
local expressions for the differential operators associated with
vertex operators, intertwiners, and the $\algg$ action on ${\cal
H}_b$, the fact that these operators take globally holomorphic
sections of one line bundle to globally holomorphic sections of
another indicates that these are actually globally defined
complex-analytic differential operators on the bundles in question.

Thus, the picture of conformal field theory which emerges is that of a
fairly simple geometrical structure which is closely related to the
geometry of the Lie symmetry group of a particular conformal field
theory.  Although we have completed the basic skeleton of this
geometrical picture of conformal field theory in this thesis, clearly
much more work is necessary to arrive at a real understanding of the
geometric significance of the vertex operators (\ref{eq:finalvertex})
and the associated correlation functions.  In particular, the local
formulae we have derived in this thesis are useful for analyzing and
understanding the basic structures involved, but are not as powerful
for specific computations as the related algebraic construction of
free field theories.  It is hoped that a better understanding of the
underlying geometry of this construction will lead to more powerful
tools for the analysis of conformal field theories in general.  In the
next section we discuss briefly some recent work which is related to
the material, outlook, and/or methods of this thesis.  In the final
section of this chapter, we discuss some open questions and possible
directions for future research.

\subsection{Related work}

In this section we discuss several issues involving a variety of
recent work which is related to the subject matter of this thesis, but
which was not mentioned or significantly referenced in the main
discussion.

\subsubsection{Virasoro coadjoint orbits}

One of the main developments which motivated the  part of this
thesis in which coadjoint orbit representations of infinite
dimensional groups were explicitly constructed was the paper by Witten
\cite{Witt1}, in which progress was  made in
understanding the structure of the coadjoint orbit representations of
the Virasoro group.  There has been a great deal of previous work in
the mathematics literature on the structure of the Virasoro group, its
representations, and the homogeneous space $\di$, which we do not
attempt to survey here; as starting points, the interested reader is
referred to the works of Kirillov \cite{k1}, Goodman and Wallach
\cite{GW}, Bowick and Rajeev \cite{BR}, and Freed \cite{fr}.
The coadjoint orbits of $\vir$ were first
classified by Segal \cite{segal} and Lazutkin and Pankrotova
\cite{lp}.
These coadjoint orbits were further studied by Kirillov \cite{k1}.  In
\cite{Witt1}, Witten presented some partial results relating the
Virasoro coadjoint orbits to the irreducible unitary representations
of $\vir$.  By using perturbative techniques and the fixed point
version of the Atiyah-Singer index theorem, Witten was able to
calculate the characters of the representations associated with the
$\di$ orbits, which he found to be the standard bosonic partition
function (\ref{eq:bosonicpartition}) associated with a nondegenerate
Virasoro representation.  The perturbative methods used by Witten,
however, were only valid in the semi-classical $c \gg 1$ domain.  In
particular, the structure of the $c \leq 1$ discrete series of unitary
representations could not be understood in terms of coadjoint orbits
using these techniques.  Witten conjectured that these representations
would be found in the $\dis$ orbits, but since these spaces do not
admit \kl structures, it has not yet been possible to perform
geometric quantization in these cases, and the representations
associated with these orbits are still not understood.

More recently, related investigations provided further clues to the
structure of the Virasoro coadjoint orbit representations.  By using a
technique involving quantization on a group manifold, Aldaya and
Navarro-Salas were able to construct representations of the Virasoro
group on spaces of polarized functions on the group manifold $\vir$
itself \cite{ANS1}.  For those values of $c$ and $h$ where the Kac
determinant vanishes, they made the observation that the
representation constructed through their method is reducible, yet
contains only a single highest weight vector.  By taking the orbit of
the highest weight vector under the Virasoro action, they found a
subspace of the original representation space which corresponds
exactly to the appropriate irreducible unitary representation in the
$c \leq 1$ discrete series.  Furthermore, they introduced an auxiliary
set of operators whose kernel gave the irreducible representation.  By
analogizing their techniques to the coadjoint orbit method, Aldaya and
Navarro-Salas conjectured that a similar situation would arise in the
$\di$ coadjoint orbit representations with $c \leq 1$.

The results in this thesis are in agreement with Witten's results for
all $c$, since the dual Verma module indeed has the partition function
(\ref{eq:bosonicpartition}).  Instead of finding the discrete series
of representations in $\dis$ orbits as Witten suggested, however, we
find that they can be constructed through resolutions of the dual
Verma module representations arising from the $\di$ orbits.  In addition,
the structure of the dual Verma module resolutions described in
Chapter 3 corresponds precisely to the results of Aldaya and
Navarro-Salas.  The dual Verma modules indeed contain a single highest
weight vector whose orbit is the irreducible representation for the
discrete series of representations.  Furthermore, the operators
defined by Aldaya and Navarro-Salas are closely related to the
screening operators and intertwiners introduced to construct the dual
Verma resolution.

\subsubsection{Coadjoint orbits and actions}

Recently, the coadjoint orbits of infinite-dimensional groups have
been studied in connection with conformal field theories for reasons
which ostensibly are rather distinct from those underlying the
discussion in this thesis.  Specifically, coadjoint orbits have been
used to construct the actions for Lagrangian descriptions of a wide
variety of conformal field theories.  The basic idea behind this
construction was first developed in the context of
infinite-dimensional groups by Alekseev and Shatashvili
\cite{al-sh1,al-sh2}, and essentially follows from the observation
that the symplectic form $\omega$ on a coadjoint orbit space can be
treated as the natural symplectic form associated with a classical
mechanical system.  Under this interpretation, the orbit space is
interpreted as a classical phase space, and the process of geometric
quantization is simply the usual quantization procedure on the
associated classical system.  By writing the symplectic form $\omega$
as the exterior derivative of a 1-form, $\omega = {\rm d} \alpha$, a
candidate action $S = \int
\alpha$ (corresponding to a vanishing Hamiltonian) can be given for
any trajectory of a particle in the classical phase space.  Because
$\omega$ is closed, but not generally an exact form, this procedure
can only be applied locally.  However, in the quantum theory the fact
that $2 \pi \omega$ is constrained to be an integral homology class
means that the action is always well-defined up to a phase of ${\rm
e}^{2 \pi i}$.  Using this approach, it was shown by Alekseev and
Shatashvili that the resulting action when this procedure is applied
to a loop group coadjoint orbit is precisely the action of the WZW
model.  Similarly, the result of applying this procedure to the
Virasoro $\di$ orbits is precisely the action originally derived by
Polyakov in \cite{Poly} for two-dimensional quantum gravity.  From
this point of view, the hidden $SL(2,\RR)$ symmetry found by Polyakov
has a natural geometrical significance.  Similar results have been
achieved by Aldaya, Navarro-Salas, and Navarro using their approach of
quantization on the group manifold \cite{ann}.

More recently, the coadjoint orbit method for constructing field
theory actions has been applied to a wide variety of
infinite-dimensional algebras, including the super Virasoro and $N =
2$ super Virasoro algebras \cite{a,anps,dnr}, $W_n$ and $W_{\infty}$
algebras \cite{aj,np,npv}, and the algebras associated with groups of
maps from 2-dimensional and 4-dimensional manifolds into a target
space
\cite{ef,m}.  For a review of this work, see \cite{dnr}.

Another interesting and related development originated with the work
of Bershadsky and Ooguri \cite{ber-og}.  In their paper, they
considered the quantum version of Hamiltonian reduction on a theory of
gauge fields coupled to the WZW model.  They found that the resulting
constrained $SL(2,\RR)$ WZW model is closely related to the quantum
field theory on Virasoro coadjoint orbits which is described by the
Polyakov gravitational WZW model action.  Similar constructions for
$SL(n,\RR)$ current algebras give rise to $W_n$-algebra actions.  In
their construction, Bershadsky and Ooguri explicitly used the bosonic
free field realizations of the affine and Virasoro algebras, along
with the relevant BRST complexes.  Since the original paper, there has
been a large amount of literature extending this work; see for example
\cite{sev} and references therein.

The Hilbert spaces associated with the quantum field theories defined
by the natural coadjoint orbit actions should in fact be precisely the
spaces of polarized wave-functions on the orbits, which are the states
in the irreducible representations studied in this thesis.  It is
natural to speculate that there may be a deeper relationship between
the infinite-dimensional symmetry groups of conformal field theories
and the theories themselves than is so far understood using either the
methods of this thesis or the methods used to construct field theory
actions.  Such a relationship was hinted at in \cite{anp}, where the
coadjoint orbit action construction was extended to give Ward
identities for correlation functions.  Clearly, what is needed is some
synthesis of these viewpoints in which the action and the Hilbert
space associated with a coadjoint orbit enter into a unified geometric
picture of a conformal field theory.  Possibly, this could be
accomplished by having a better understanding of how a Hamiltonian
picture of conformal field theories relates to the action which
naturally arises from coadjoint orbits.  In this case, it is possible
that the formal structure defined in this thesis which gives an
association between coadjoint orbits and conformal field theory
correlation functions and vertex operators may be more than just a
convenient way of expressing the physical theory in a consistent
mathematical fashion.  In fact, it is possible that the vertex
operators constructed here may be in some way naturally connected with
the intrinsic action associated with coadjoint orbits.  This could be
a very promising direction for future research.

\subsection{Remaining questions}

In this section, we discuss some of the open questions which remain
regarding the work presented in this thesis.
\vspace{.15in}

\noindent $\bullet$ {\bf Field theoretic formulation}

Although we have given explicit realizations of the affine and
Virasoro algebras in terms of a bosonic Fock space, the associated
description in terms of bosonic free fields is not completely clear.
Unlike the Feigin-Fuchs and Wakimoto realizations, where the
expressions for the generators of the algebra are quadratic in the
field modes, the coadjoint orbit realizations have generators which
contain an arbitrarily large number of modes in a single term.  In the
field theory picture, this implies that the current algebras of the
corresponding physical theory must have an analogous representation,
which seems to indicate that any associated field theory would not
only have interactions, but might contain interaction terms of
arbitrarily high order which are related in a
rather nontrivial fashion.  Although a simpler formulation  of such
a theory might be possible (such as the action formulation described
in the previous section), the essential distinction between these
representations and the free field representations remains.  Because
the structure of the Fock space is that of a dual Verma module for the
coadjoint orbits, it is probably impossible to write the theory purely
as a free field theory.  A better understanding of the connection with
field theory will probably be an essential component of any
application of the methods in this thesis.

\vspace{.15in}

\noindent $\bullet$ {\bf Vertex operators}

There are several questions remaining regarding the vertex operators
constructed in this thesis.  First, we have not proven that any
resolution of an irreducible representation through dual Verma modules
admits a set of vertex operators of the desired form, even when such a
vertex operator exists for the irreducible representations.  It would
be nice to have a formal proof of this result.  Second, we have given
here a rather inconvenient characterization of the vertex operators
by defining them through their matrix elements on the bosonic Fock
space.  For a better understanding of the structure of these
operators, it is desirable to have a more abstract geometric or
group-theoretic description.  Such a description might well involve
relating these vertex operators more directly to the associated field
theory and the action arising from the coadjoint orbit.
\vspace{.15in}

\noindent $\bullet$ {\bf Coset models}

Another interesting question is whether the class of coset conformal
field theories has a natural description in the language of coadjoint
orbit representations.  Some work has been done using the free field
representations of affine algebras to construct the representations
associated with a coset theory (see for instance \cite{bmp}).  Due to
the complications involved with the associated two-sided resolutions,
however, there are some difficulties with this construction.
Hopefully, using the one-sided resolutions of dual Verma modules could
simplify this construction.
\vspace{.15in}

\noindent $\bullet$ {\bf Other groups}

A natural extension of this work would be to consider a larger class
of groups and algebras, such as the super algebras and $W$-algebras.
Just as free field $W$-algebra representations can be constructed
through the process of Hamiltonian reduction from an affine algebra
using free field realizations \cite{ber-og}, and $W$-algebra actions
can be constructed from coadjoint orbits, one might expect to find
similar constructions using the coadjoint orbit dual Verma modules.

\subsection{Conclusion}

In conclusion, we have constructed an alternative approach to
conformal field theory based on fundamental principles of geometry and
group symmetry.  This construction unifies the coadjoint orbit
structure associated with a group and the description of vertex
operators through their action on algebraic resolutions.  The main
problems remaining with this construction are the connection with a
field theory and the challenge of finding a more convenient geometric
description of the vertex operators.  This construction may provide a
link between the construction of field theory actions using coadjoint
orbits and the physics of the associated field theory which  is
described in terms of correlation functions.


\end{document}